\documentclass{aa}  
\usepackage{graphicx}
\usepackage{xcolor}
\usepackage{lscape}
\newcommand{\lacc}{$\rm \langle L_{acc} \rangle$}
\newcommand{\macc}{$\rm \dot{M}_{acc}$}

\usepackage{txfonts}
\begin{document}

   \title{GIARPS High-resolution Observations of T Tauri stars (GHOsT)\thanks{Based on observations made with the Italian {\it Telescopio Nazionale
Galileo} (TNG) operated by the {\it Fundaci\'on Galileo Galilei} (FGG) of the {\it Istituto Nazionale di Astrofisica} (INAF) at the {\it  Observatorio del Roque de los Muchachos} (La Palma, Canary Islands, Spain), under programs A36TAC\_22, A38TAC\_8, A40TAC\_6, A42TAC\_12 (PI: S. Antoniucci).}}

   \subtitle{IV. Accretion properties of the Taurus-Auriga young association}

   \author{M. Gangi            \inst{1}
          \and  S. Antoniucci  \inst{1}
          \and  K. Biazzo      \inst{1}
          \and  A. Frasca      \inst{2}
          \and  B. Nisini      \inst{1}
          \and  J. M. Alcal\'a \inst{3}
          \and  T. Giannini    \inst{1}
          \and  C. F. Manara   \inst{4}
          \and  A. Giunta      \inst{5}
          \and  A. Harutyunyan \inst{6}
          \and  U. Munari      \inst{7}
          \and  F. Vitali      \inst{1}
          }

   \institute{INAF - Osservatorio Astronomico di Roma, Via Frascati 33, I-00078 Monte Porzio Catone, Italy \\
              \email{manuele.gangi@inaf.it}
         \and
             INAF - Osservatorio Astrofisico di Catania - Via S. Sofia 78, 95123 Catania, Italy
        \and
             INAF - Osservatorio Astronomico di Capodimonte - Salita Moiariello 16, 80131 Napoli, Italy
        \and
            European Southern Observatory, Karl-Schwarzschild-Strasse 2, 85748 Garching bei M\"unchen, Germany
        \and
            ASI - Agenzia Spaziale Italiana, Via del Politecnico, I-00133 Roma, Italy
        \and
             Fundaci\'on Galileo Galilei - INAF - Telescopio Nazionale Galileo, 38700 Bre$\rm \tilde n$a Baja, Santa Cruz de Tenerife, Spain
        \and
            INAF–Osservatorio Astronomico di Padova, via dell’Osservatorio 8, 36012 Asiago (VI), Italy
             }

   \date{Received 18 May, 2022; accepted 31 August, 2022}

 
  \abstract
   {}
   {In the framework of the GIARPS High-resolution Observations of T Tauri stars (GHOsT) project, we study the accretion properties of 37 classical T Tauri stars of the Taurus-Auriga star-forming region (SFR) with the aim of characterizing their relation with the properties of the central star, with jets and disk winds, and with the global disk structure, in synergy with complementary ALMA millimeter observations.}
   {We derive the stellar parameters, optical veiling, the accretion luminosity ($\rm L_{acc}$), and the mass accretion rate ($\rm \dot M_{acc}$) in a homogeneous and self-consistent way using high-resolution spectra acquired at the Telescopio Nazionale Galileo with the HARPS-N and GIANO spectrographs that are flux-calibrated based on contemporaneous low-resolution spectroscopic and photometric ancillary observations.}
   {The $\rm L_{acc}$-$\rm L_{\star}$, $\rm \dot{M}_{acc}$-$\rm M_{\star}$ and $\rm \dot{M}_{acc}$-$\rm M_{disk}$ relations of the Taurus sample are provided and compared with those of the coeval SFRs of Lupus and Chamaeleon I. We analyzed possible causes for the observed large spreads in the relations. We find that (i) a proper modeling in deriving the stellar properties in highly spotted stars can reduce the spread of the  $\rm \dot{M}_{acc}$-$\rm M_{\star}$ relation, (ii) transitional disks tend to have lower $\rm \dot{M}_{acc}$ at a given $\rm M_{\star}$, (iii) stars in multiple systems have higher $\rm \dot{M}_{acc}$ at the same $\rm M_{disk}$, (iv) the $\rm \dot{M}_{acc}$ versus disk surface density has a smaller spread than the $\rm \dot{M}_{acc}$-$\rm M_{disk}$, indicating that opacity effects might be important in the derivation of $\rm M_{disk}$. Finally, the luminosities of the [\ion{O}{i}] 630 nm narrow low-velocity component and high-velocity component (HVC) and the deprojected HVC peak velocity were found to correlate with the accretion luminosity. We discuss these correlations in the framework of the currently accepted models of jets and winds.}
   {Our results demonstrate the potential of contemporaneous optical and near-infrared high-resolution spectroscopy to simultaneously provide precise measurements of the stellar wind and accretion wind properties of young stars.}

   \maketitle

\section{Introduction}

Pre-main-sequence star evolution and planet formation are connected to the interplay of mass accretion onto the star, ejection of outflows, and photoevaporative disk winds \citep[][]{Hartmann2016,Ercolano2017}. The mass loss from outflows and slow disk winds leads to the extraction of angular momentum, which in turn enables the accretion of matter onto the star and alters the disk density, which affects the formation and migration of protoplanets.

In this context, the mass accretion rate $\rm \dot{M}_{acc}$ \citep[e.g.,][]{Hartmann1998} is a key parameter for the study of young stellar objects (YSOs) during their first several million years of evolution toward the main sequence. Reliable measures of $\rm \dot{M}_{acc}$ for YSOs located in different star-forming regions (SFR) and at different evolutionary phases are necessary to set important constraints on disk evolutionary models and disk clearing mechanisms \citep[e.g.,][]{Manara2022, Pascucci2022}.

In the current magnetospheric accretion (MA) paradigm \citep[e.g.,][]{Hartmann2016}, strong large-scale stellar surface magnetic fields of a few hundred to a few kilogauss truncate the inner disk at a few stellar radii. Driven by the stellar magnetic field lines, gas flows from this truncation radius onto the star and produces localized shocks. The characteristic emission line spectrum of classical T Tauri Stars (CTTs), including the Balmer and Paschen series, is then partly formed in the aforementioned accretion funnel flows \citep{Hartmann1994}. Accretion shocks are also responsible for continuum excess emission mainly from the UV to the optical wavelength range. This continuum emission is overimposed on the photospheric spectrum and thus weakens the stellar photospheric lines. This effect is usually referred to as veiling \citep{Calvet1998}, and it is particularly remarkable in the UV Balmer continuum region at $\rm \lambda \leq 400$ nm. A proper modeling of the continuum excess emission is necessary to derive the energy that is released in the accretion shocks, the so-called accretion luminosity $\rm L_{acc}$, which in turn allows measuring the mass accretion rate, $\rm \dot{M}_{acc}$, given the stellar mass and radius \citep{Gullbring1998}. 

On the other hand, $\rm L_{acc}$ and hence $\rm \dot{M}_{acc}$ can also be measured from the well-known correlations between $\rm L_{acc}$ and the luminosity of specific lines in emission ($\rm L_{line}$) of \ion{H}{i}, \ion{He}{i,} and \ion{Ca}{ii}, which have been derived for a wide spectral range from the UV to the near-infrared (NIR) \citep[e.g.,][]{Calvet2004,Herczeg2008,Rigliaco2012,Alcala2014,Alcala2017}. The simultaneous use of many diagnostic lines in a wide spectral region turned out to be of enormous importance in reducing the uncertainties in the $\rm L_{acc}$ and $\rm \dot{M}_{acc}$ measurements at typical values of 0.2-0.3 dex \citep[e.g.,][]{Rigliaco2012,Alcala2014}.

Accurate determination of $\rm L_{acc}$ and $\rm \dot{M}_{acc}$ is essential to derive the dependence of accretion on the properties of the central star and its disk. $\rm \dot{M}_{acc}$ and $\rm L_{acc}$ are usually compared to the stellar and disk properties, and empirical $\rm \dot{M}_{acc}$-$\rm M_{\star}$, $\rm \dot{M}_{acc}$-$\rm M_{disk}$, and $\rm L_{acc}$-$\rm L_{\star}$ scaling relations are used as benchmark tests for theoretical predictions of disk evolution \citep[e.g.,][]{Manara2016, Rosotti2017, Lodato2017, Mulders2017, Tabone2022}. Previous surveys have shown that these scaling relations present large scatters (more than 2 dex in log $\rm \dot{M}_{acc}$ or log $\rm L_{acc}$ at a given YSO mass and luminosity) that cannot be explained alone in terms of the high variability characterizing the accretion processes \citep[e.g.,][]{Costigan2012, Costigan2014, Biazzo2012}. One of the current challenges is then to understand the nature of this dispersion. Even though purely viscous evolution disk models predict the above correlations \citep[e.g.,][]{Lynden1974}, the observed high dispersion suggests a more complex scenario for disk evolution, in which magnetohydrodynamics (MHD) winds, external photoevaporation, and local dust processing may play a crucial role \citep[e.g.,][and references therein]{Manara2022}. In addition, recent works have pointed out that stellar multiplicity can be another ingredient at the basis of the high accretion observed in some sources \citep[][]{Zagaria2022}.

From an observational point of view, this type of investigation requires simultaneously and in a self-consistent fashion deriving all the relevant parameters of as many as possible statistically complete samples of YSO in SFRs, avoiding methodological systematics and biases due to temporal variability. This was attempted in several SFRs such as Lupus \citep{Alcala2014,Alcala2017}, Chamaeleon I \citep{Manara2016, Manara2017}, $\rm \eta-$Chamaeleon \citep{Rugel2018}, TW Hydra \citep{Venuti2019}, and Upper Scorpius \citep{Manara2020} using the capabilities of the X-Shooter instrument, which is the medium-resolution spectrograph \citep[R=10000-20000;][]{Vernet2011} covering the wide spectral range between 0.3 and 2.5 $\rm \mu m$ at the ESO Very Large Telescope (VLT). However, a similar survey is lacking for Taurus-Auriga, although it is a paradigmatic SFR.


To cover this gap, we started the GIARPS High-resolution Observations of T Tauri stars (GHOsT) project, a flux-limited survey of T Tauri Stars in the Taurus-Auriga SFR based on data obtained with the GIARPS instrument. This high-resolution spectrograph simultaneously covers the optical and NIR wavelength ranges and is located at the Italian Telescopio Nazionale Galileo (TNG).

The GHOsT project aims at providing reliable measurements of stellar and accretion parameters of Taurus-Auriga members and to place them in relation with the disk properties, in synergy with complementary ALMA observations that are available for the majority of the targets \citep{Andrews2018,Long2019}. In addition, the high sensitivity and spectral resolution of GIARPS also allows us to investigate the sub-au disk and jets environments by studying specific diagnostic lines. A first study of the GHOsT program was focused on the jet line emission for a subsample of bright sources \citep[][henceforth Paper I]{Giannini2019}, and then was followed by an investigation of the link between atomic and molecular disk winds \citep[][henceforth Paper II]{Gangi2020}. The definition and assessment of the methods for determining stellar and accretion properties was then presented in \citet[][henceforth Paper III]{Alcala2021}.

In this fourth work of the GHOsT series, we present the results of the accretion measurements for the complete sample.
The paper is organized as follows. In Sect. \ref{sec:sample} we present the source selection and the sample properties, while observations and data processing are reported in Sects. \ref{sec:obs}
and \ref{sec:data_reduct}, respectively. In Sect. \ref{sec:methods} we describe the method we used to derive the stellar and accretion properties, while results are shown in Sect. \ref{sec:results}. Discussion and conclusions are finally presented in Sects. \ref{sec:discussion} and \ref{sec:conclusions}, respectively.

\section{Sample}\label{sec:sample}
The GHOsT original sample was selected based on the most recent Taurus-Auriga young population census at the time of the beginning of the GHOsT project, that is, the one presented by \citet{Esplin2014}. 
The initial selection was driven by the GIARPS sensitivity. We considered sources with $\rm J< 11$ mag and $\rm R < 13$ mag, which reduced the total sample to about 70 objects with spectral type (SpT) earlier than M5. We then started the observational campaign giving priority to the sources with known jets and with outer disks that were already characterized by ALMA observations. 

It was possible to observe 46 of the 70 objects with the telescope time available to GHOsT. Their masses are in the range between $\sim 0.2$ and $\sim 2.2$ $\rm M_{\odot}$, the spectral types are between G1 and M3, and the luminosities are between $\sim 0.05$ and $\sim 8.9$ $\rm L_{\odot}$. Although the observed sample is not complete, it is representative of the distribution in the complete sample with SpT earlier than M5. We observed 9 class III sources in the sample as classified in the new census by \citet{Esplin2019}, with the purpose of refining the empirical relations between the equivalent widths (EWs) and fluxes of the H$\rm \alpha$ and Pa$\rm \beta$ lines and the effective temperature ($\rm T_{eff}$) for a preliminary selection between class II and class III sources to be observed in the NIR.

Information such as disk inclination, dust mass, effective disk radius, and disk structure is available for the majority of the observed sources. In Table \ref{tab:disk_prop} we report the respective data collected from the most recent ALMA literature. Disk classification was made preferentially on the basis of ALMA images. In addition to the full (i.e., no substructures in the dust density profile detected at ALMA resolution) and transitional (TD; i.e., with an inner cavity around the star) disk categories, we indicate as substructured the class of disks that are characterized by the presence of gaps and rings. 
Finally, $\sim 35 \%$ of the sample consists of multiple systems. The type of multiplicity, separation, and the relevant references are reported in columns 4 and 5 of Table \ref{tab:sources_param}.

\begin{table*}
\small
\center
\caption{\label{tab:disk_prop} Disk properties of the selected accreting CTTs for this study, based on ALMA observations.}
\begin{tabular}{lcccc|lcccc}
\hline
\hline
\noalign{\smallskip}
Source    & $\rm i_{disk}^{a}$ & $\rm M_{disk, dust}$ & $\rm R_{eff,68\%}$ & Disk type$^{\rm (b)}$  & Source    & $\rm i_{disk}^{a}$ & $\rm M_{disk, dust}$ & $\rm R_{eff,68\%}$ & Disk type$^{\rm (b)}$  \\
          &  [deg]             & $\rm [M_{\oplus}]$               &  [au] &                  &         &          [deg]           &                $\rm [M_{\oplus}]$  & [au] &                    \\
\hline  
\noalign{\smallskip}
 BP Tau  & $38.2^{(1)}$  & $7.8^{(1)}$    & $29^{(1)}$  & Full$^{(\rm 1)}$ & GI Tau  & $43.8^{(1)}$  & $3.1^{(1)}$  & $18^{(1)}$  & Full$^{(\rm 1)}$ \\
 CI Tau  & $50^{(1)}$    & $37.0^{(1)}$   & $111^{(1)}$ & Sub$^{(1)}$ 	    & GK Tau  & $40.2^{(1)}$  & $0.89^{(1)}$ & $8^{(1)}$   & Full$^{(\rm 1)}$ \\ 
 CQ Tau  & $29^{(3)}$    & $41.0^{(3)}$   & $63^{(3)}$  & TD$^{(3)}$       & GM Aur  & $53^{(3)}$    & $43^{(3)}$   & $76^{(3)}$ & TD$^{(3)}$    	 \\   
 CW Tau  & $59^{(4)}$    & $11^{(4)}$    & $46^{(4)}$  & Full$^{(4)}$      & HN Tau  & $69.8^{(1)}$  & $2.7^{(1)}$  & $15^{(1)}$  &	Full$^{(\rm 1)}$ \\     
 CY Tau  & -     						  & $14^{(5)}$    & $48^{(5)}$  & Full$^{(\rm c)}$ & HQ Tau   & $53.8^{(1)}$  & $1.1^{(1)}$  & $20^{(1)}$  & Full$^{(\rm 1)}$ \\     
 DE Tau  & -     						  & $5.2^{(7)}$    & $22^{(7)}$  & Full$^{(\rm c)}$ & IP Tau  & $45.2^{(1)}$  & $2.6^{(1)}$  & $30^{(1)}$  & TD$^{(1)}$       \\     
 DF Tau  & $52^{(6)}$    & $0.65^{(6)}$   &  -          & Full$^{(\rm c)}$ & IQ Tau  & $62.1^{(1)}$  & $11^{(1)}$  & $55^{(1)}$  & Sub$^{(1)}$	     \\     
 DG Tau  & $37^{(4)}$    &  -             &   -         & Full$^{(4)}$     & LkCa 15 &	-   						   & $33^{(3)}$  & $124^{(3)}$ & TD$^{(3)}$   	 \\     
 DH Tau  & $16.9^{(1)}$  & $5.1^{(1)}$    & $14^{(1)}$  & TD$^{(9)}$       & MWC480  & $36.5^{(1)}$  & $73^{(1)}$  & $55^{(1)}$  & Sub$^{(1)}$		 \\     
 DK Tau  & $12.8^{(1)}$  & $5.2^{(1)}$    & $11^{(1)}$  & Full$^{(\rm 1)}$ & RW Aur A & $55.1^{(1)}$ & $7.2^{(1)}$ & $14^{(1)}$  &	Full$^{(\rm 1)}$ \\     
 DL Tau  & $45.0^{(1)}$  & $45^{(1)}$    & $111^{(1)}$ & Sub$^{(1)}$      & RY Tau  & $65^{(1)}$     & $39^{(1)}$ & $50^{(1)}$  & TD$^{(1)}$	     \\     
 DN Tau  & $35.2^{(1)}$  & $15^{(1)}$    & $39^{(1)}$  & Sub$^{(1)}$ 	   & SU Aur  & -                               & $7.1^{(8)}$  & $50^{(8)}$  &	Full$^{(\rm c)}$ \\     
 DO Tau  & $27.6^{(1)}$  & $25^{(1)}$    & $25^{(1)}$  & Full$^{(\rm 1)}$ & UX Tau  & $39^{(5)}$ 	   & -            & -            & TD$^{(9)}$       \\     
 DQ Tau  & $16.1^{(1)}$  & $28^{(1)}$    & $24^{(1)}$  & Full$^{(\rm 1)}$ & UY Aur  & $23.5^{(1)}$ & $5^{(1)}$   & $5^{(1)}$   & Sub$^{(10)}$ 	 \\     
 DR Tau  & $5.4^{(1)}$   & $50^{(1)}$    & $36^{(1)}$  & Full$^{(\rm 1)}$ & UZ Tau E & $56.1^{(1)}$ & $24^{(1)}$  & $59^{(1)}$  & Sub$^{(1)}$      \\     
 DS Tau  & $65.2^{(1)}$  & $5.8^{(1)}$    & $59^{(1)}$  & Sub$^{(1)}$      & V409 Tau & $69.3^{(1)}$ & $3.6^{(1)}$ & $31^{(1)}$  & Full$^{(\rm 1)}$ \\     
 FT Tau	 & $35.5^{(1)}$  & $15^{(1)}$    & $33^{(1)}$  & Sub$^{(1)}$      & V807 Tau & -    & $1.6^{(8)}$ & $12^{(8)}$   & TD$^{(9)}$	     \\       
 GG Tau A & $57^{(3)}$   &  -             &  -          & Full$^{(\rm c)}$ & V836 Tau & $43.1^{(1)}$ & $7.8^{(1)}$ & $21^{(1)}$   & Full$\rm ^{(1)}$   \\         
 GH Tau  &  -    						& $0.69-0.51^{(8)}$ & $12^{(8)}$ & Full$\rm ^{(9)}$ &         &      &             &              &                     \\   
 \hline                            
\end{tabular}                      
\begin{quotation}                  
\textbf{Notes.} 

$^{(a)}$ Disk inclinations are measured from ALMA observations and refer to the outer disk. An exception is made for GG Tau, where the inclination of the resolved inner disk  is reported.

$^{\rm(b)}$  \textbf{Full}: disk without substructures, \textbf{Sub}: disk with substructures (e.g. inner hole, gaps and rings), \textbf{TD}: transitional disk.

$^{\rm(c)}$ Classification adopted from \citet{Esplin2019}, based on the mid-IR SED.

\textbf{References.}$\rm ^{(1)}$\citet{Long2019}, $\rm ^{(2)}$\citet{Antonellini2020}, $\rm ^{(3)}$\citet{Francis2020}, $\rm ^{(4)}$\citet{Bacciotti2018}, $\rm ^{(5)}$\citet{Tripathi2017}, $\rm ^{(6)}$\citet{Andrews2013},  $\rm ^{(7)}$\citet{Simon2019}, $\rm ^{(8)}$\citet{Akeson2019}, $\rm ^{(9)}$\citet{Currie2011}, $\rm ^{(10)}$\citet{Tang2014}.
\end{quotation}
\end{table*}   

\section{Observations}\label{sec:obs}
Observations were carried out from October 2017 to January 2020. They consist of high-resolution optical and IR spectra obtained with the GIARPS instrument that were flux-calibrated through ancillary low-resolution spectroscopy and photometric data. The logbook of observations is given in Table \ref{tab:sources_param}.

\subsection{GIARPS observations}
The sample was observed in four distinct runs, hereafter run I in October - November 2017, run II in December 2018, run III in November 2019 - January 2020, and run IV in October - December 2020. The GIARPS observing mode consists in the simultaneous use of the HARPS-N \citep[][resolving power $\rm R=115000$]{Pepe2002, Cosentino2012} and GIANO-B \citep[][$\rm R=50000$]{Oliva2012, Origlia2014} spectrographs. HARPS-N is a fiber-fed echelle with a $\rm FoV=1''$ and covers the spectral range between 390 and 690 nm while GIANO-B is a near-infrared cross-dispersed echelle with a slit on-sky with dimensions of $\rm 6''\times 0.5''$ and a spectral range between 940 and 2400 nm.

\subsection{Ancillary observations}
To accurately flux-calibrate the GIARPS spectra we carried out contemporaneous ancillary observations consisting of absolute flux-calibrated low-resolution spectroscopy and photometry. In particular we acquired optical low-resolution spectra (R=2400, 330-790 nm) with the 1.22 m telescope of the University of Padova at the Asiago observatory (Italy). Spectra were reduced and flux-calibrated against spectrophotometric standards observed on the same night. We also checked and refined the flux zeropoint with $\rm BVR_cI_c$ photometry acquired with the ANS Collaboration telescopes \citep{Munari2012} in runs I-III, while in run IV, $griz$ photometry was taken with the ROS2 instrument of the REM telescope \citep{Molinari2014}.
In the NIR we obtained $\rm JHK$ photometry that was acquired in run I, II, III-2019 with the REMIR instrument at the REM telescope \citep{Vitali2003} and in run III-2020 and IV with the NICS camera \citep{Baffa2001} at the TNG telescope. In addition, we also acquired low-resolution (R$\rm \sim 50$) NIR spectra in run IV using the NICS Amici prism.

Details of the observation date and instruments we used are listed in Table \ref{tab:Logbook}. The photometry of runs I-III is reported in \citet{Gangi2020}, while that of run IV is reported in Table \ref{tab:phot_last_run} for completeness.

\begin{table*}
\small
\center
\caption{\label{tab:sources_param} Logbook of observations, Gaia EDR3 distances, multiplicity as reported in the literature and stellar properties derived in this work.}
\begin{tabular}{lcccclcccccc}
\hline
\hline
\noalign{\smallskip}
Source    & Obs Date   & d & M$\rm ^{(f)}$  & Sep. & $\rm T_{eff}$ $\rm (\pm err)$ & r600 & SpT$\rm ^{(g)}$ & $\rm A_v (\pm err) $ & $\rm L_\star^\ddagger$ & $\rm R_\star^\ddagger$ & $\rm M_\star^\ddagger$ \\
          &        & [pc] & & [''] & [K] & &  & [mag] & [$\rm L{\odot}$] & [$\rm R_{\odot}$] &  [$\rm M_{\odot}$]        \\
\hline
\multicolumn{12}{c}{Accreting sources} \\
\hline
\noalign{\smallskip}
 BP Tau$^\dagger$ & 26 Gen 2020 & 127 & S & - & 3717 (150)	 & 0.5    & K5  & 0.20 (0.10) & 0.37 & 1.47 & 0.48  \\
 CI Tau           & 09 Dec 2018 & 160 & S & - &  4559 (114)    & 0.8    & K3  & 2.60 (0.20) & 1.04 & 1.63 & 1.29  \\      
 CQ Tau           & 13 Nov 2017$\rm ^{(e)}$  & 149 & S & - &  6823 (136) & 0.0 & F4 & 0.50 & < 15.13 & < 2.75 & < 1.99    \\
 CW Tau			  & 08 Dec 2018 & 131 & S & - &  4854 (95)	 & 2.0    & K0  & 3.60 (0.20) & 0.40 & 0.89 & 0.90 \\
 CY Tau           & 13 Dec 2020 & 126 & S & - &  3514 (34)     & 0.0    & M2  & 0.00 (0.10) & 0.22 & 1.27 & 0.36 \\
 DE Tau           & 24 Oct 2020 & 128 & S & - &  3499 (57)     & 0.1    & M2  & 0.00 (0.10) & 0.49 & 1.90 & 0.26 \\  
 DF Tau           & 08 Dec 2018 & 176 & B & 0.1$\rm ^{(1)}$ &  3565 (92)	 & 0.7    & M2  & 0.00 (0.20) & 0.35 & 1.55 & 0.39 \\
 DG Tau           & 29 Oct 2017$\rm ^{(a)}$   & 125 & S & - &  4004 (153) & 1.5 & K7 & 1.50 & 0.44 & 1.38 & 0.70  \\
 DH Tau           & 02 Nov 2019 & 133 & B & 1.5$\rm ^{(2)}$  &  3631 (87)	 & 1.0    & M1  & 0.60        & 0.20 & 1.13 & 0.41 \\
 DK Tau$^\dagger$ & 08 Dec 2018 & 132 & B & 2.4$\rm ^{(3)}$ & 3923 (150)	 & 0.1    & K5  & 1.20 (0.20) & 0.78 & 1.91 & 0.62 \\
 DL Tau           & 29 Oct 2017$\rm ^{(a)}$   & 160 & S & - &  4188 (100) & 1.2 & K5 & 1.50 & 0.40 & 1.20 & 0.90 \\
 DN Tau$^\dagger$ & 01 Nov 2019 & 129 & S & - &  3823 (150)	 & 0.0    & K7  & 0.50 (0.20) & 0.49 & 1.60 & 0.55  \\
 DO Tau           & 13 Nov 2017$\rm ^{(a)}$ & 138 & S & - &  3694 (104) & 1.0 & M1 & 1.40 & 0.42 & 1.58 & 0.50  \\
                  & 26 Gen 2020 & 138 & S & - &  3630 (132)	 & 1.2    & M1  & 1.40 (0.20) & 0.37 & 1.54 & 0.42  \\          
 DQ Tau$^\dagger$ & 02 Nov 2019 & 195 & T & 7.6/2.3 $\rm ^{(4)}$	 &  3795 (150)	 & 0.3    & K7  & 1.50 (0.20) & 0.98 & 2.29 & 0.52  \\
 DR Tau           & 25 Gen 2020 & 193 & S & - &  4443 (190)	 & 3.5    & K4  & 1.20 (0.20) & 0.29 & 0.91 & 0.83  \\
 DS Tau$^\dagger$ & 01 Nov 2019 & 158 & S & - &  3876 (150)	 & 1.0    & K5  & 0.40 (0.10) & 0.29 & 1.19 & 0.61  \\
 FT Tau           & 25 Gen 2020 & 130 & S & - &  3407 (73)	 & 1.5    & M3  & 0.60 (0.20) & 0.05 & 0.64 & 0.25  \\
 GG Tau A$^\dagger$ & 09 Dec 2018 & 116 & B & 1.5$\rm ^{(1)}$ &  4034 (150)	 & 0.2    & K6  & 1.40 (0.20) & 0.80 & 1.83 & 0.73  \\
 GH Tau           & 25 Gen 2020 & 130 & B & 0.3$\rm ^{(1)}$ &  3632 (67)	 & 0.2    & M1  & 0.80 (0.20) & 0.79 & 2.24 & 0.42  \\
 GI Tau$^\dagger$ & 26 Gen 2020 & 129 & B & 13.1$\rm ^{(3)}$ &  3815 (150)	 & 0.7    & K7  & 1.90 (0.20) & 0.27 & 1.19 & 0.55  \\
 GK Tau           & 02 Nov 2019 & 129 & B & 13.1$\rm ^{(3)}$ &  4439 (89)	 & 0.2    & K4  & 1.90 (0.10) & 0.87 & 1.58 & 1.20  \\
 GM Aur           & 09 Dec 2018 & 158 & S & - &  4564 (76)	 & 0.0    & K3  & 1.30 (0.20) & 1.16 & 1.72 & 1.19  \\
 HN Tau           & 29 Oct 2017$\rm ^{(e)}$ & 134 & B & 3.1$\rm ^{(3)}$ &  4617 (97) & 0.8 & K4 & 1.25 & 2.63 & 2.57 & 1.58   \\
 HQ Tau           & 01 Nov 2019 & 161 & S & - &  5005 (86)	 & 0.0    & G8  & 3.70 (0.10)     & 5.83 & 3.21 & 2.00    \\
 IP Tau$^\dagger$ & 09 Dec 2018 & 129 & S & - &  3770 (150)	 & 0.1    & K7  & 0.60 (0.20)     & 0.31 & 1.30 & 0.51  \\
 IQ Tau           & 02 Nov 2019 & 131 & S & - &  3811 (65)$\rm ^{(b)}$  & 1.5$\rm ^{(c)}$ & M0     & 1.60 (0.10) & 0.19 & 1.00 & 0.55 \\
 LkCa 15          & 14 Dec 2020 & 157 & S & - &  4588 (73)     & 0.1    & K4  & 0.90 (0.20) & 0.60  & 1.23 & 1.09   \\
 MWC480           & 01 Nov 2019 & 156 & S & - &  8500     	 & 0.0    & A5  & 0.00 (0.10) & 10.79 & 1.51 & 1.71  \\
 RW Aur A         & 13 Nov 2017$\rm ^{(a)}$  & 183 & B & 1.4$\rm ^{(3)}$ &  4870 & 1.2 & K0 & 1.00 & 1.64 & 1.80 & 1.50  \\
 RY Tau           & 13 Nov 2017$\rm ^{(a)}$   & 138 & S & - &  5856 (151) & 0.0 & G1 & 2.25 & 8.87 & 2.89 & 1.80 \\
 SU Aur           & 24 Oct 2020 & 157 & S & - &  5414 (171)  & 0.0    & G5    & 0.90 (0.10) & 8.85 & 3.38 & 2.22  \\
 UX Tau A         & 26 Gen 2020 & 142 & Q & 5.6/2.6$\rm ^{(5)}$ &  5191 (79)	 & 0.0    & G8  & 1.20 (0.10) & 1.76 & 1.64 & 1.36 \\
 UY Aur$^\dagger$ & 08 Dec 2018 & 152 & B & 8.8$\rm ^{(3)}$ &  3773 (150)	 & 0.4    & K7  & 0.50 (0.20) & 0.46 & 1.59 & 0.51 \\
 UZ Tau E         & 09 Dec 2018 & 130 & Q & 3.6/0.34$\rm ^{(5)}$ &  3609 (112)	 & 0.7    & M2  & 0.70 (0.20) & 0.45 & 1.72 & 0.41 \\
 V409 Tau         & 26 Gen 2020 & 130 & S & - &  3649 (106)	 & 0.5    & M1  & 0.30 (0.20) & 0.24 & 1.23 & 0.43 \\
 V807 Tau A$\rm ^{(d)}$ & 24 Oct 2020 & 184 & T & 1.9/0.04$\rm ^{(7)}$ &  4550 (150) & 0.2    & K4 & 1.30 (0.20) & 1.69 & 2.08 & 1.41     \\
 V836 Tau$^\dagger$ & 02 Nov 2019 & 167 & S & - &  3333 (150)	 & 0.1    & K5  & 0.90 (0.20)  & 0.54     & 2.20   & 0.22  \\
                   & 26 Gen 2020 & 167 & S & - &  3313 (150)    & 0.1    & K5  & 1.00 (0.20)  & 0.63     & 2.41   & 0.28  \\                            
\hline
\multicolumn{12}{c}{Non-accreting sources} \\
\hline
\noalign{\smallskip}
DI Tau$^\dagger$	   & 13 Dec 2020 & 138 & B & 1.5$\rm ^{(4)}$ &  3695 (150)  & 0.0    & K7    & 0.60 (0.10) & 0.61 & 1.91  & 0.46  \\
IW Tau$^\dagger$	   & 13 Dec 2020 & 142 & B &   &  3678 (150)  & 0.0    & K7    & 0.40 (0.20) & 0.70 & 2.06  & 0.45  \\
LkCa 4$^\dagger$	   & 13 Dec 2020 & 130 & S & - &  3379 (150) & 0.0    & K7    & 0.20 (0.10) & 0.50 & 2.06  & 0.31  \\ 
LkCa 21		    	   & 13 Dec 2020 & 117 & B & &  3638 (92)  & 0.0    & M1.5  & 0.60 (0.20) & 0.37 & 1.53  & 0.43  \\
V819 Tau$^\dagger$ 	   & 14 Dec 2020 & 129 & S & - &  4293 (150)  & 0.0    & K4    & 1.90 (0.20) & 0.72 & 1.53  & 1.05  \\
V827 Tau$^\dagger$     & 13 Dec 2020 & 164 & B & &  3648 (150) & 0.1    & K5    & 0.50 (0.20) & 0.88 & 2.35  & 0.43  \\
V1070 Tau$^\dagger$    & 13 Dec 2020 & 125 & S & - &  3897 (150)  & 0.0    & K5    & 0.60 (0.20) & 0.71 & 1.85  & 0.60  \\
V1098 Tau       	   & 24 Oct 2020 & 124 & B & 0.5$\rm ^{(6)}$ &  3837 (138) & 0.1    & M0    & 0.70 (0.20) & 1.11 & 2.38  & 0.55  \\
V1115 Tau       	   & 14 Dec 2020 & 128 & S & - &  4546 (52)  & 0.0    & K4    & 0.40 (0.10) & 0.65 & 1.32  & 1.03  \\
\hline\end{tabular}	
\begin{quotation}
\textbf{Notes.} 

$\rm ^{\dagger}$ Heavily spotted source. $\rm T_{eff}$, $\rm L_{\star}$, $\rm R_{\star}$ and $\rm M_{\star}$ values are estimated following the methodology described in Section \ref{sec:params_spotted}.

$\rm ^{\ddagger}$ Average uncertainties are 0.2 in $\rm log\ L_{\star}$, 0.3 in $\rm log\ R_{\star}$ and 0.15 in $\rm log\ M_{\star}$.

$\rm ^{(a)}$ Parameters derived in \citet{Alcala2021}. 

$\rm ^{(b)}$ Value taken from APOGEE-2 DR16 \citep{Jonsson2020}. 

$\rm ^{(c)}$ Estimated using GES and PALLA models. 

$\rm ^{(d)}$ Triple system: $\rm L_{\star}$, $\rm R_{\star}$ and $\rm M_{\star}$ taken from \citet{Schaefer2012} and scaled with the new Gaia EDR3 distance.
 
$\rm ^{(e)}$ Subluminous YSO: the reported values are corrected for obscuration effects (See Section 4.3.1 of \citet{Alcala2021}).

$\rm ^{(f)}$ Multiplicity: single (S), binary (B), tertiary (T), quadruple (Q).

$\rm ^{(g)}$ Derived from the SpT-$\rm T_{eff}$ conversion of \citet{Herczeg2014}.

\textbf{References.} $\rm ^{(1)}$ \citet{Kraus2012}, $\rm ^{(2)}$ \citet{Kraus2011},  $\rm ^{(3)}$ \citet{Akeson2019}, $\rm ^{(4)}$ \citet{Daemgen2015}, $\rm ^{(5)}$ \citet{Zapata2020},  $\rm ^{(5)}$ \citet{Martin2005},  $\rm ^{(6)}$ \citet{Cieza2009},  $\rm ^{(7)}$ \citet{Schaefer2012}.

\end{quotation}
\end{table*}

\section{Data reduction}\label{sec:data_reduct}
The data reduction processes are described in detail in Papers I-III, but are briefly summarized here for completeness.
The HARPS-N spectra were reduced according to the standard procedures with the HARPS-N data reduction software pipeline \citep{Pepe2002}. Spectra were corrected for heliocentric and radial velocity ($\rm RV$) using the \ion{Li}{I} photospheric profile, assuming the weighted $\rm \lambda_{air}=670.7876$ nm. We estimate a final accuracy on the wavelength calibration of about 2 $\rm km$ $\rm s^{-1}$. We then extracted and normalized to the continuum the 16 spectral portions, 50-100 \AA\ wide, that contained the diagnostics that we used for the accretion measurement, namely the Balmer recombination lines H3, H4, H5, H6, H7, H8, the helium lines \ion{He}{II}469, \ion{He}{I}403, \ion{He}{I}447, \ion{He}{I}471, \ion{He}{I}492, \ion{He}{I}502, \ion{He}{I}588, \ion{He}{I}668, and the \ion{Ca}{II} H and K doublet.

The GIANO-B spectra were reduced following the prescriptions given in \citet{Carleo2018}. We then considered the five spectral segments containing the accretion diagnostics, namely the hydrogen recombination lines of the Paschen series (i.e. Pa5, Pa6, Pa7 and Pa8) and the \ion{He}{I}1083 nm line. They were continuum-normalized and corrected for heliocentric and $\rm RV$. For this latter task, we used the three NIR \ion{Al}{I} lines at $\rm \lambda_{vac}=2109.884, 2116.958, 2121.396$ $\rm nm$ as reference after verifying that they agreed well with the velocity measured from the optical \ion{Li}{I}  line. We obtained a typical final accuracy on the wavelength calibration of about 1-2 $\rm km$ $\rm s^{-1}$. Finally, the five spectral segments were corrected for telluric features. For this purpose, we used the \textsc{molecfit} tool \citep{Smette2015} to compute the synthetic telluric spectrum and the IRAF task \textsc{telluric} to properly correct the observed spectral segments for telluric contribution.

We flux-calibrated the GIARPS spectra on the basis of the ancillary data. They were acquired close in time with the GIARPS observations, with a maximum temporal distance of ten days. On these timescales, the typical flux variability of CTTs can reach values of 10\% \citep[e.g.,][]{Venuti2021}. We then conservatively considered this latter as the typical uncertainty to the flux calibration due to the source variability.

For each source, we computed a polynomial curve representative of the continuum flux within the GIARPS spectral range on the basis of the available ancillary data. In particular, for runs I-III, we built this curve through a polynomial fit of the continuum of the Asiago spectrum and an interpolation of the $IJHK$ photometric points. In run IV, we fit the total spectrum resulting from the sum of the Asiago and Amici spectra after refining the continuum level with the photometric points.

We then multiplied the 21 continuum-normalized spectral segments for the computed curve. Considering both the source variability and the errors on the photometric points, we estimate an accuracy of 20\% on the final flux calibration.

\section{Methods for the derivation of stellar and accretion properties}\label{sec:methods}

In this section we present the methods we adopted for measuring the stellar and accretion parameters. The general method has been extensively discussed in Paper III, to which we refer. Here, we briefly present it and complement it with a detailed discussion of the case of heavily spotted stars.

\subsection{Stellar properties}
\subsubsection{General method}
Photospheric and kinematical parameters (i.e. effective temperature $\rm T_{eff}$, surface gravity $\rm log g $, $\rm RV$, and projected rotational velocity $\rm v \sin i$) and veiling as a function of wavelength $\rm r_{\lambda}$ were determined with the ROTFIT code \citep{Frasca2003}. In short, the code performs a $\chi^2$ minimization between observed and template spectra in specific optical spectral segments and has been successfully applied both on medium- and high-resolution data \citep[e.g.,][]{Frasca2015,Frasca2017}. We used the HARPS-N spectra and a grid of templates retrieved from the ELODIE archive \citep{Moultaka2004} of real spectra of slowly rotating and low-activity stars with well-known atmospheric parameters. 

$\rm T_{eff}$ and veiling at 600 nm, $\rm r_{600}$, which are the only two parameters used for our analysis, are reported in Table \ref{tab:sources_param}, while the complete set of photospheric parameters derived with ROTFIT will be presented and discussed in a future dedicated work. Spectral types were determined using the SpT-$\rm T_{eff}$ conversion of \citet{Herczeg2014}, and they are also reported in Table \ref{tab:sources_param}.

As discussed in Paper III, the visual extinction $\rm A_v$ was estimated by computing the ratios of the flux-calibrated low-resolution Asiago spectra and artificially reddened templates of the same spectral type. We adopted the grid of nonaccreting YSO templates with negligible extinction of \citet{Manara2013,Manara2017} reddened by $\rm A_v$ in the range 0.0-5.0 mag in steps of 0.10 mag using the extinction law by \citet{Weingartner2001} with $\rm R_v=5.5$. The $\rm A_v$ was then estimated as that of the reddened template that minimized the slope of the corresponding ratio. This analysis was limited to the portion of the spectra between 550 and 800 nm, which is least affected by both the optical and infrared veiling \citep[e.g.,][]{Fischer2011}. Values derived in this way are reported in column 7 of Table \ref{tab:sources_param}.

To compute the stellar luminosity $\rm L_{\star}$ , we first considered the BTSettl model \citep{Allard2012} for each source that best matched the photospheric parameters derived with ROTFIT and normalized it to the extinction-corrected Asiago spectrum at $\rm \lambda = 600\ nm$. We then integrated the BTSettl spectrum over all wavelengths and multiplied it by the factor 1/(1+$\rm r_{600}$) to take the veiling contribution into account, thus obtaining the bolometric flux of the object photosphere. The stellar luminosity was then computed as $\rm L_{\star} = 4 \pi d^2 F $, where $\rm F$ is the bolometric flux and $\rm d$ is the Gaia EDR3 distance reported in Table \ref{tab:sources_param}. As reported in Paper III, we estimated an average uncertainty of 0.2 in $\rm log\ L_{\star}$ on the basis of the typical errors in flux calibration of the Asiago spectra and in veiling correction. The stellar radius, $\rm R_{\star}$, was computed from $\rm L_{\star}$ and $\rm T_{eff}$ , and an average uncertainty of 0.3 in $\rm log\ R_{\star}$ is estimated.

Finally, the mass, $\rm M_{\star}$, was computed from the pre-main-sequence evolutionary tracks by \citet{Siess2000}. From the typical uncertainties of $\rm L_\star$ and $\rm T_{eff}$ , we estimated an average uncertainty of about 0.15 in $\rm log\ M_{\star}$. Values of $\rm L_\star$, $\rm M_{\star}$, and $\rm R_{\star}$ are also listed in Table \ref{tab:sources_param}.

\subsubsection{Stellar parameters in spotted sources}\label{sec:params_spotted}
Magnetic activity in young stars can generate starspots over a large area of the stellar surface that can significantly contribute to the total stellar flux, with a spectrum that is distinct from the ambient photosphere. Observationally, this can lead to a strong spread in the effective temperature and luminosity derived by different methods and wavelength ranges. It has been often found that effective temperatures measured in young stars with NIR spectra are systematically lower than those derived at optical wavelengths \citep[e.g., ][]{Cottaar2014, Flores2021}. This offset is particularly remarkable in the temperature range between $\sim 3800$ and $\sim 4500$ K where it can reach average values of 500 K. This discrepancy can also introduce a significant uncertainty in the mass and age values estimated from the position of the star in the HR diagram \citep[e.g.,][]{Gully2017}. \citet{Flores2021} compared the masses estimated from optical and NIR temperatures with the dynamical masses measured from ALMA for a sample of YSOs, finding that neither the optical nor the NIR temperatures reproduce the stellar dynamical masses. 

A detailed analysis of the influence of starspots on the measurement of photospheric properties and a discussion of whether the NIR-optical $\rm T_{eff}$ offset can be fully explained in terms of starspots are beyond the scope of this paper. Here we limit our analysis to heavily spotted stars in our sample and evaluate the best strategy to properly estimate their luminosity and mass because these latter are the parameters that can directly affect the $\rm L_{acc}$-$\rm L_{\star}$ and $\rm \dot{M}_{acc}$-$\rm M_{\star}$ distributions. 

\begin{figure}
\begin{center}
\includegraphics[trim=0 100 80 30,width=1\columnwidth]{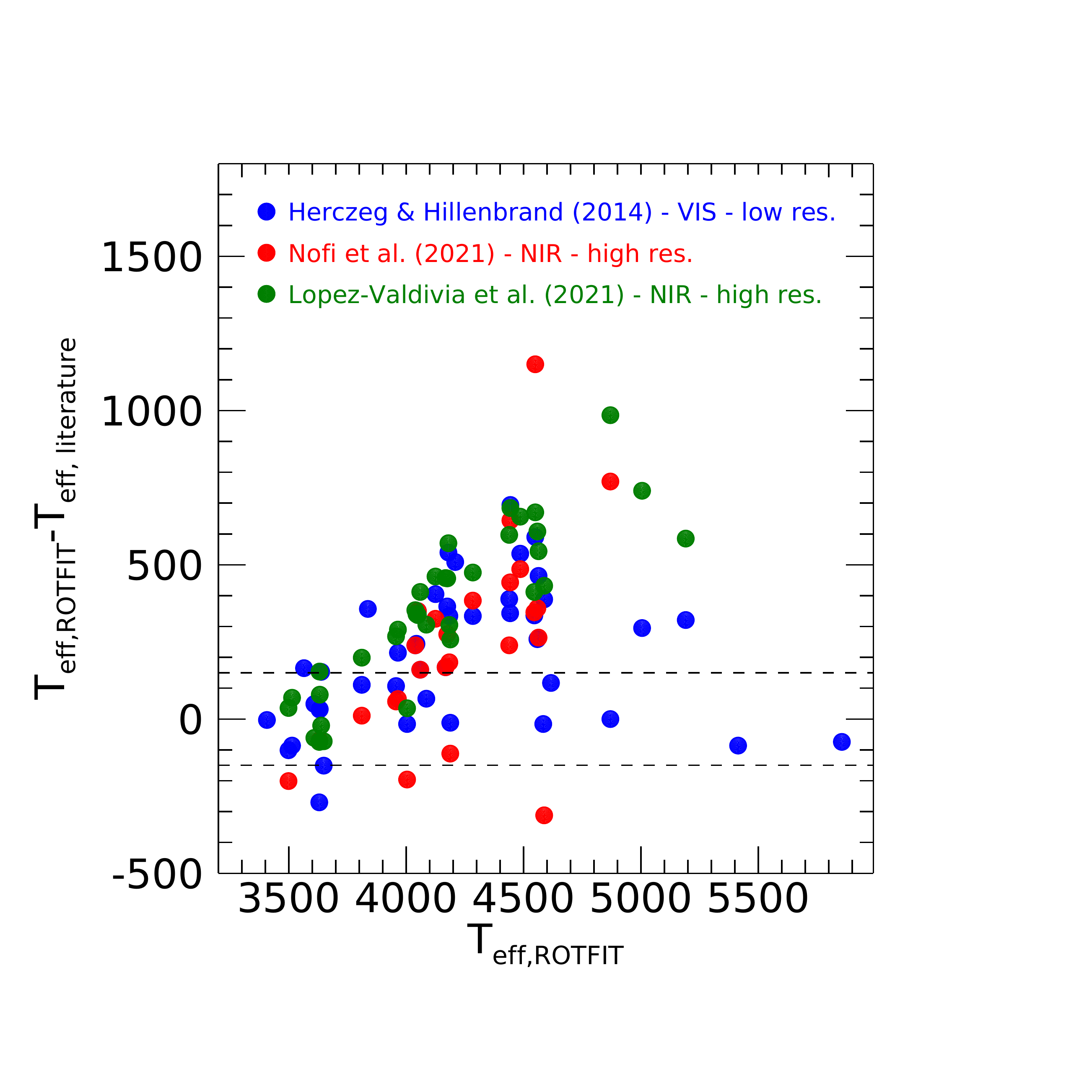}
\end{center}
\caption{\label{fig:teff_vs_literature} Comparison of the measured effective temperature with literature values for the same sources. Symbols with different colors have been used for the three literature sources of $\rm T_{eff}$, as indicated in the legend. Dashed lines represent typical errors on the $\rm T_{eff}$ measurements.}
\end{figure}

Fig. \ref{fig:teff_vs_literature} shows the offset between the $\rm T_{eff}$ we measure in the HARPS-N range with ROTFIT and the $\rm T_{eff}$ values reported in the literature measured from an analysis mostly based on TiO bands \citep{Herczeg2014} and from NIR spectral features \citep{Nofi2021, Lopez2021}. Similarly to what was found in other YSOs samples, we confirm that the main $\rm T_{eff}$ discrepancies lie in the temperature range between $\sim 4000$ and $\sim 4500$ K. In particular, the detection of optical TiO bands (which are typically sensitive to $\rm T_{eff} \leq 3800 $ K) in this temperature range strongly suggests the presence of regions colder than 4000 K.

\begin{figure*}
\begin{center}
\includegraphics[trim=20 40 100 10,width=1\columnwidth]{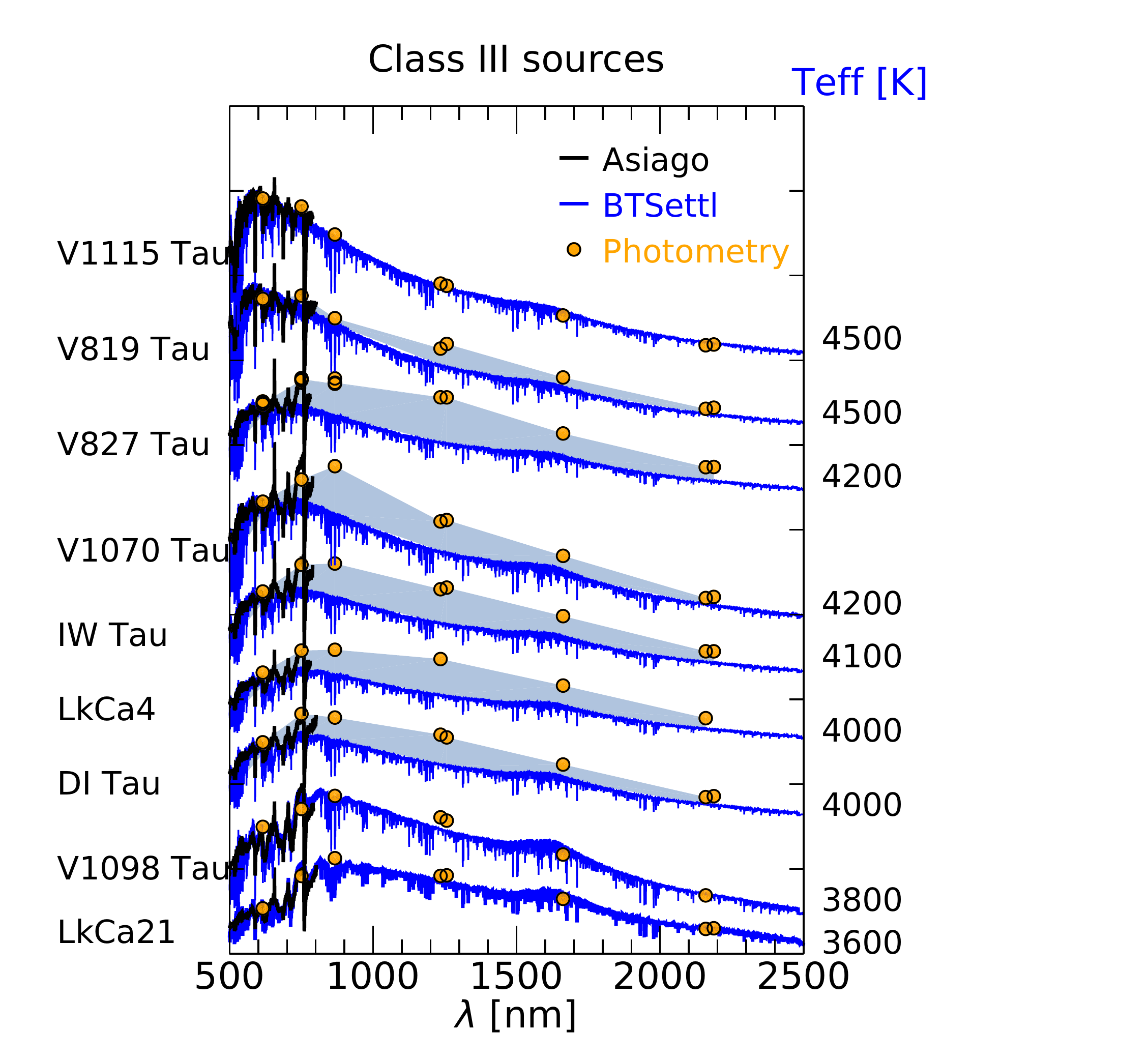}
\includegraphics[trim=20 40 100 10,width=1\columnwidth]{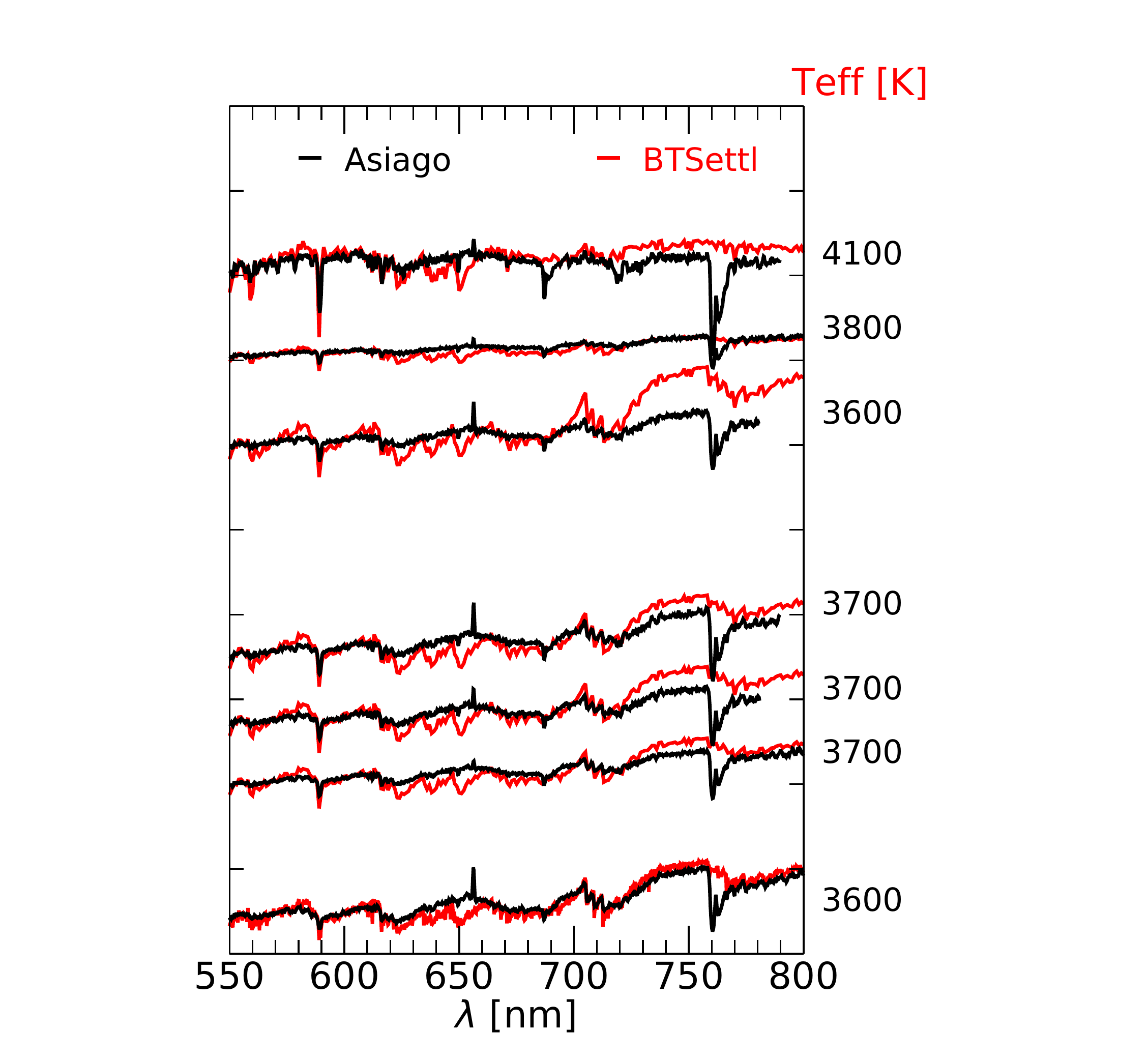}
\end{center}
\caption{\label{fig:SpT_comparison} Comparison between observed spectra of the class III sample and their respective BTSettl models.} Left: Extinction-corrected Asiago spectra (black) with contemporaneous photometry (orange points). BTSettl models corresponding to the optical $\rm T_{eff}$ are shown in blue. The filled light blue regions highlight the NIR flux that is underestimated by the models. Right: As in the left panel but assuming Av=0 and BTSettl models corresponding to the $\rm T_{eff}$ measured from NIR high-resolution spectra \citep{Lopez2021}. Source names and $\rm T_{eff}$ are labeled for each spectrum.
\end{figure*}

\begin{figure*}
\begin{center}
\includegraphics[trim=0 0 0 0,width=1.8\columnwidth]{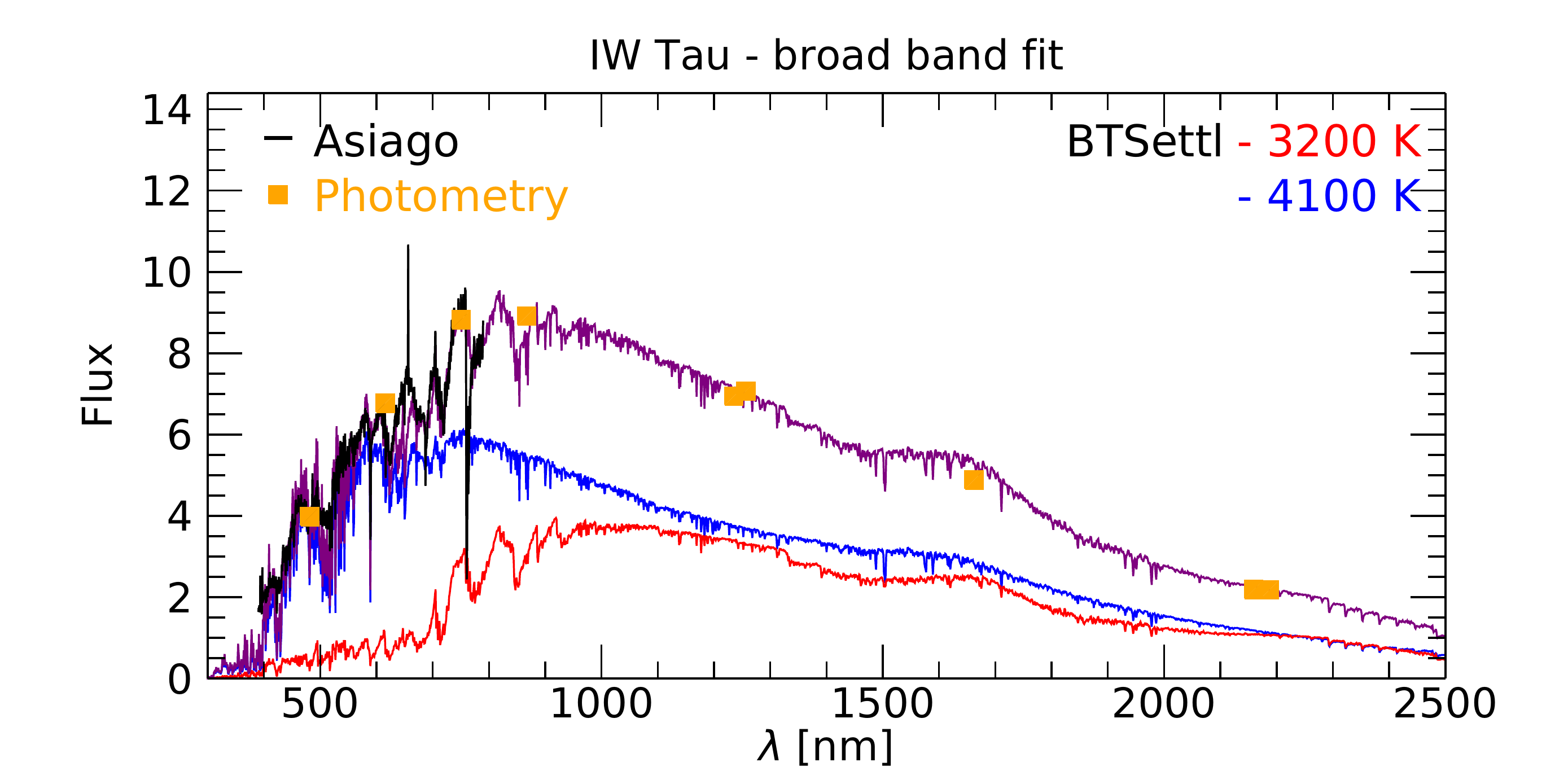}
\includegraphics[trim=0 0 0 0,width=1.8\columnwidth]{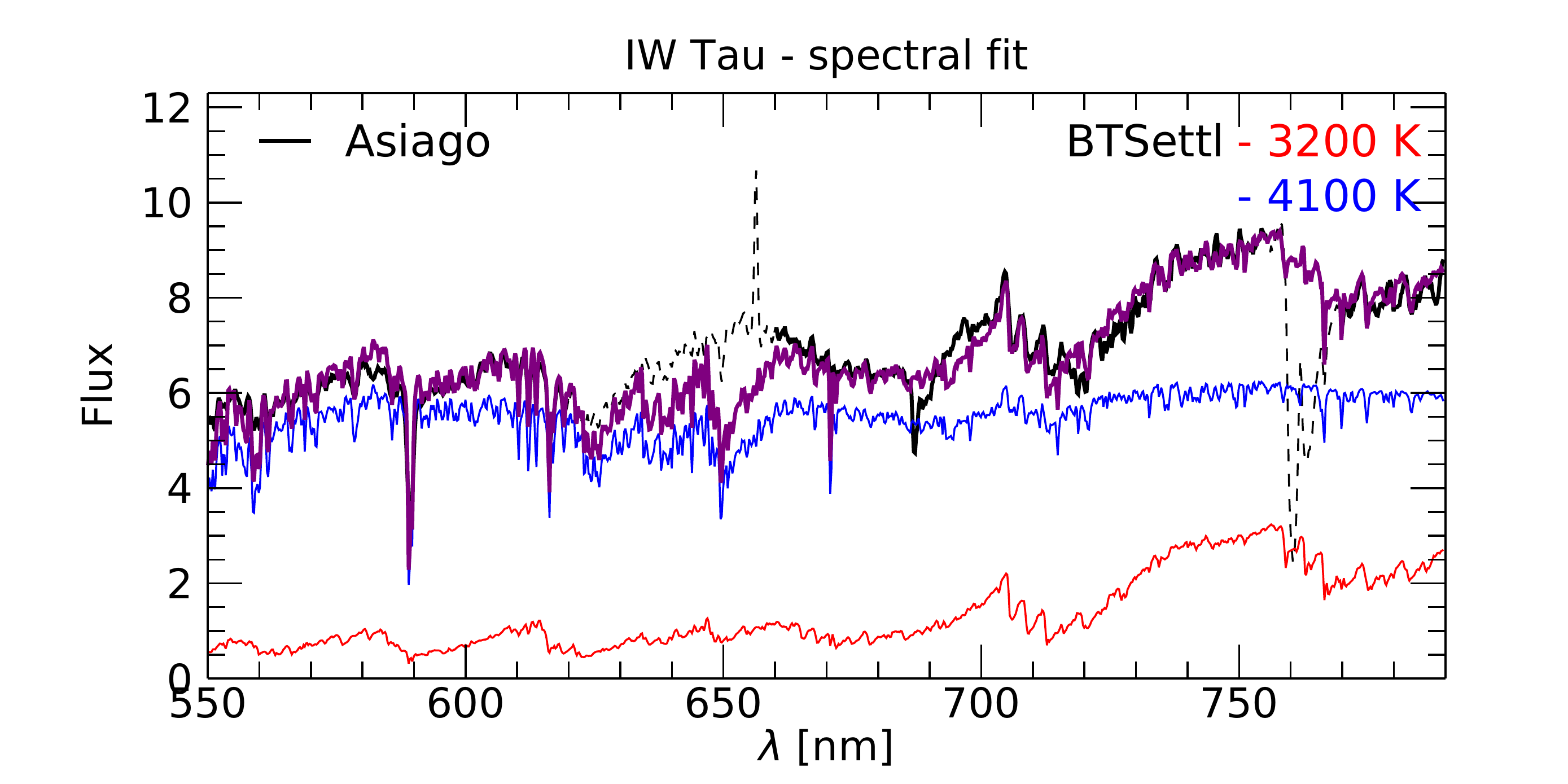}
\end{center}
\caption{\label{fig:IW_Tau_example} Example of the 2 $\rm T_{eff}$ fitting procedures for the class III source IW Tau. The extinction-corrected Asiago spectrum is shown in black, and contemporaneous photometry is shown as filled orange squares. The best-fit spectrum is shown in purple, and the corresponding $\rm T_{hot}$=4100 K and the $\rm T_{cool}$=3200 K BTSettl models are reported in blue and red, respectively. Top: Broadband fitting. Bottom: Spectral fitting. Flux units are $\rm 10^{-13}\ ergs^{-1}\ cm^{-2}\ nm^{-1}$. Fits to the complete subsample of heavily spotted sources are reported in Appendix \ref{fig:Spectral_fitting} and \ref{fig:Broad_band_fitting}.} 
\end{figure*}

\begin{figure*}
\begin{center}
\includegraphics[trim=0 250 0 0,width=1.0\columnwidth, angle=0]{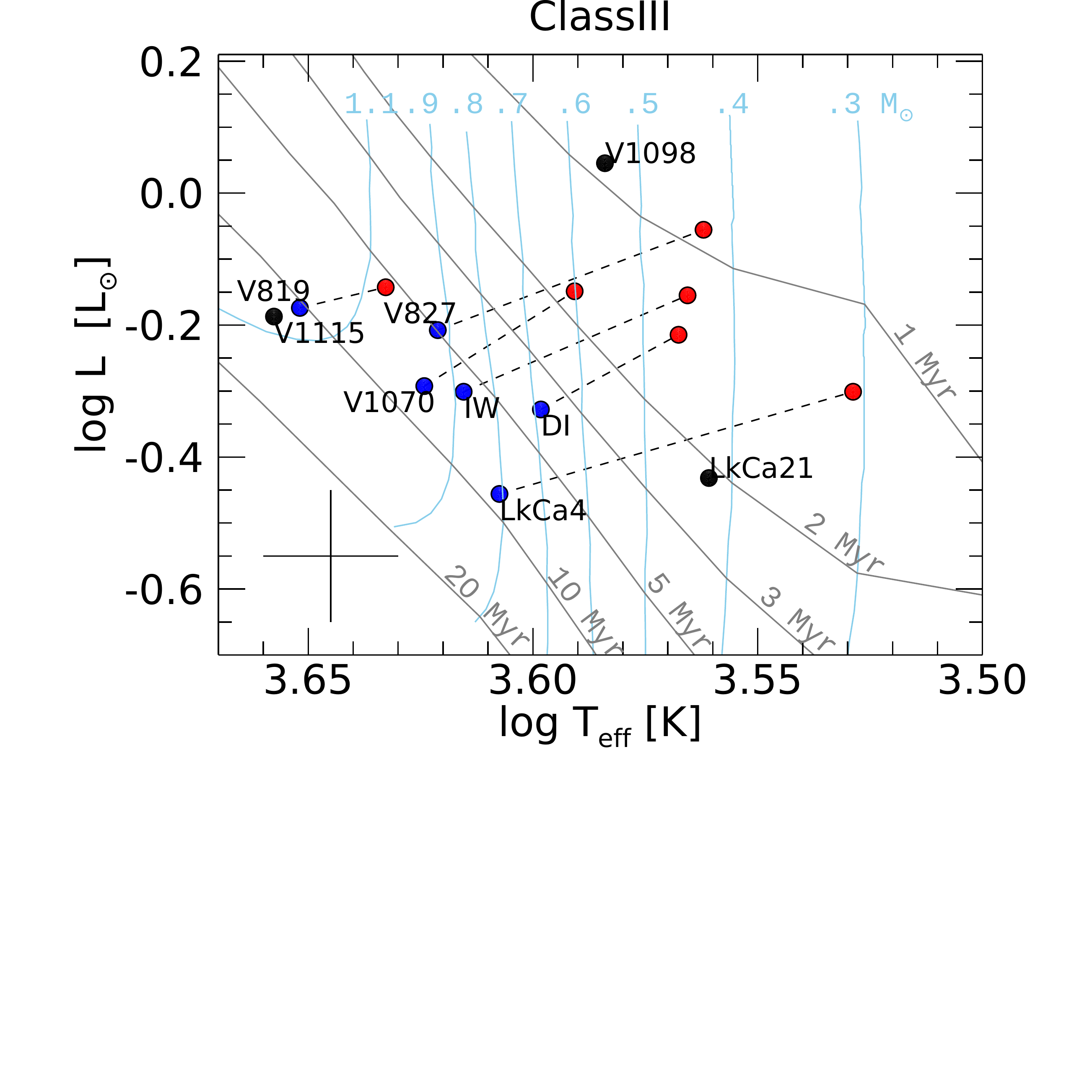}
\includegraphics[trim=0 250 0 0,width=1.0\columnwidth, angle=0]{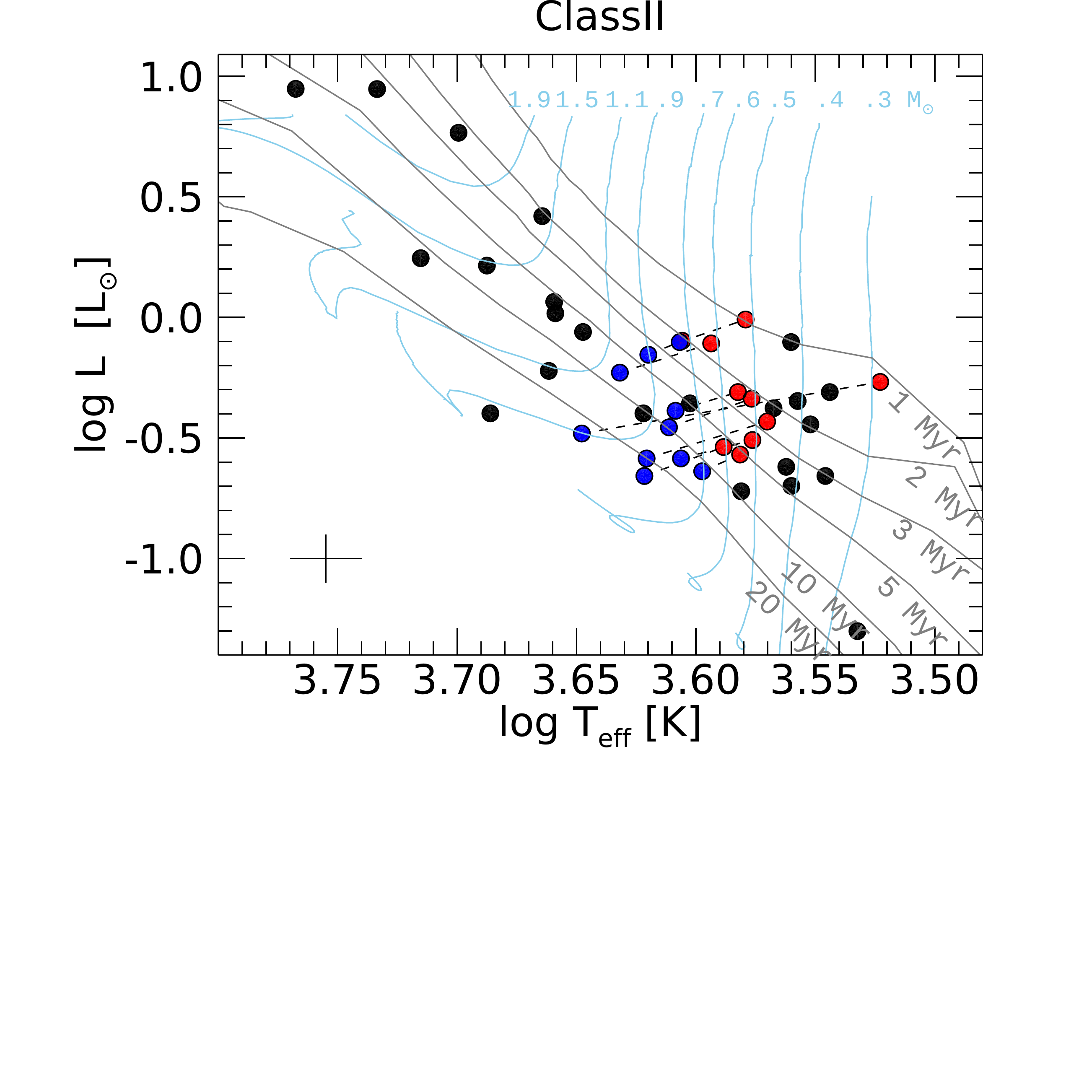}
\end{center}
\caption{\label{fig:HR_classIII} Hertzprung-Russel diagram of class III sources (left) and class II sources (right), with evolutionary tracks by \citet{Siess2000}. The isochrones are for ages of 1, 2, 3, 5, 10, and 20 Myr. Positions of heavily spotted sources with $\rm T_{eff}$ and $\rm L_{\star}$ based on the 2 $\rm T_{eff}$ fit are shown as red points, and those obtained assuming a single temperature based on many optical spectral features are shown as blue points. Positions of sources without evidence of spots are shown as black points. The average errors are drawn in the bottom left of each panel.}
\end{figure*}

\subsubsection{Class III sources}
To evaluate the impact of multiple temperature components on the broadband emission of the source, we first focused on the subsample of class III sources, whose analysis is greatly simplified because they are not affected by significant veiling. Fig. \ref{fig:SpT_comparison} (left panel) shows the extinction-corrected low-resolution Asiago spectra and the contemporaneous broadband photometry compared with the BTSettl model corresponding to the $\rm T_{eff}$ measured from the analysis of many optical features and scaled to the Asiago flux, as explained in the previous section. For the six sources lying in the temperature range between $\sim 4000$ and $\sim 4500$ K (namely DI Tau, LkCa 4, IW Tau, V1070 Tau, V827 Tau, and V819 Tau), the synthetic models significantly underestimate the NIR fluxes.

This discrepancy is not attributable to the variability because class III sources are known to be quite stable with small changes of amplitude of about a few tenths of magnitude \citep[e.g.,][]{Grankin2008,Lanza2016} while the mismatch we found reaches values of more than 1 mag in the NIR bands. An additional source of bias could be the assumption of an incorrect extinction value. In principle, we can still reproduce the broadband spectrum with a single model corresponding to the optical $\rm T_{eff}$ by significantly increasing the $\rm A_v$ value with respect to that determined with the method described in the previous subsection. However, as reported in Sec. 5.2, for class II sources, we can adopt a independent way, free from the choice of the SpT, to compute the extinction based on the minimization of the spread around the $\rm L_{acc}$-$\rm L_{line}$ relations. The results of this latter approach are always consistent within the errors with that obtained with the former method, confirming the robustness of our measures. On the other hand, the synthetic models corresponding to the $\rm T_{eff}$ measured from the NIR are not able to reproduce the optical region of the spectra, showing deeper TiO molecular bands and an incorrect extinction (Fig. \ref{fig:SpT_comparison}, right panel).  In conclusion, our evidence indicates that a single $\rm T_{eff}$ model is not sufficient to reproduce the total flux of a subsample of class III sources and suggests the need of multiple-temperature modeling.

In these sources, large spotted regions have been revealed on V819 Tau, V827 Tau, and V1070 Tau from photometric long-term time-series analysis \citep{Grankin2008}. The models show that the observed light-curve amplitudes are compatible with spotted regions covering from 17\% to 73\% of the visible stellar surface with mean temperatures 500-1400 K lower than the ambient photosphere. In addition, spectral features of LkCa4 have been interpreted in terms of 2 $\rm T_{eff}$ components by \citet{Gully2017}, who found a hot photosphere of $\sim 4100$ K and a cool contribution of $\sim 2700-3000$ K due to starspots covering a surface area of  $\sim$ 80\%. 

For the six class III sources with evidence of starspots we have followed the method described in \citet{Gully2017} which is similar to that proposed by \citet{Frasca2005}. In short, we assumed that the stellar photosphere can be represented by two components, one with a hot temperature $\rm T_{hot}$ and one with a cool temperature $\rm T_{cool}$. Each of these components contributes to the total flux with appropriate weighting factors, $\rm w_{hot}$ and $\rm w_{cool}$. The effective temperature best representing the total stellar flux was computed as

\begin{eqnarray}\label{eqn:teff}
\rm T_{eff} = (T_{hot}^4 w_{hot} + T_{cool}^4 w_{cool})^{0.25},
\end{eqnarray}

while the luminosity was computed from the integral of the total flux in the whole spectral range. We used BTSettl models by fixing $\rm T_{hot}$ to the closest value found by ROTFIT, and we varied the $\rm T_{cool}$ models from 2500 to 4000 K in steps of 100 K. The weighting factors were left as free parameters of the fit. 
The adopted method (hereafter broadband fit) allowed us to reproduce the broadband spectrum of the six class III sources in the $\rm T_{eff}$ range between 4000 and 4500 K, and hence to properly estimate the corresponding luminosities. Fig. \ref{fig:IW_Tau_example} (top panel) shows an example of the fitting procedure for IW Tau, and the whole subsample of class III sources and the fitting results (i.e., temperatures and weighting factors) are reported in Fig. \ref{fig:Broad_band_fitting} and Table \ref{tab:sources_param_2Teff}, respectively.

We found cool components with temperatures between $\sim 3100$ and $\sim 3500$ K and weighting factors between $\sim 27\%$ and $\sim 80$\%, leading to a decrease in the effective temperature between $\sim 5\%$ and $\sim 18$\%. LkCa4 has the largest contribution ($\sim 80\%$) from a cool component ($\rm T_{cool} \sim 3100$ K), which perfectly agrees with the results of \citet{Gully2017}.

However, we stress that the physical meaning of the $\rm T_{cool}$ and $\rm T_{hot}$ components and their relative weighting factors cannot be directly linked to starspots, ambient photosphere, and filling factor. Localized hotspots such as plages can additionally contribute to the total spectrum, which is indistinguishable from the ambient photosphere with the current technique. More advanced methods such as the Zeeman Doppler Imaging \citep{Semel1989, Brown1991} or line-bisector analysis \citep[e.g.,][]{Prato2008} are necessary for this purpose. 

Fig. \ref{fig:HR_classIII} (left panel) shows the position of these sources in the HR diagram with an effective temperature and luminosity based on the 2 $\rm T_{eff}$ broadband fit compared with that obtained assuming a single temperature. The  2 $\rm T_{eff}$ analysis leads to a systematic shift toward younger ages and lower masses. 

\subsubsection{Class II sources}
The broadband fitting procedure cannot be directly applied to class II sources because of the accretion excess, which can contribute significantly to the observed flux.
Our strategy for these sources was to perform a 2 $\rm T_{eff}$ spectral fitting of the low-resolution Asiago spectra in the spectral range between 550 and 800 nm. This latter is in fact less affected by the veiling and contains the optical TiO and VO molecular bands which are particularly sensitive to the presence of a cool $\rm T_{eff}$ component. 

The adopted procedure is the same as for the broadband fit with the difference that we now consider specific spectral features. For this reason the spectral resolution of the models was degraded to that of the Asiago spectra and the portion of the spectrum containing the $\rm H\alpha$ line was excluded with a specific mask.

We first applied the spectral fitting procedure to the subsample of class III sources (an example is shown in the bottom panel of Fig. \ref{fig:IW_Tau_example}) and compared the results with those obtained with the broadband fit. The results shown in Figure \ref{fig:Compare_L_Teff} show that differences between the two methods are less than $\sim 10\%$ for both luminosities and temperatures, thus indicating that the spectral fitting procedure unambiguously allows a proper estimation of the stellar luminosity and can also be applied in the case of accreting sources.

We then applied this method to the whole class II sample by again fixing $\rm T_{hot}$ to the closest value found by ROTFIT. The second $\rm T_{cool}$ component was considered non-negligible only when it improved the $\rm \chi^2$ of the spectral fit by more than 20\% of its minimum value. Following this criterion, we found that a second $\rm T_{cool}$ component was necessary for ten class II sources with optical $\rm T_{eff} \geq 4000$ K, which we therefore identified as heavily spotted sources. Independent evidence of large spotted regions in these sources has already been found for DQ Tau \citep{Kospal2018} and GI Tau \citep{Guo2018} from the analysis of multiband light curves, for DN Tau from RV variations \citep{Prato2008}, and for V836 Tau from both multiband light curves \citep{Grankin2008} and RV variations \citep{Prato2008}. The results of our modeling are listed in Table \ref{tab:sources_param_2Teff} and shown in Fig.  \ref{fig:Spectral_fitting}. They show that cool regions with mean temperatures 900-1300 K lower than the optical $\rm T_{eff}$ contribute as much as 24\% to 90\% to the total flux.

Fig. \ref{fig:HR_classIII} (right panel) shows the HR diagram for these sources and the comparison with the positions obtained from a single temperature. Sources are now placed in the region of the HR diagram corresponding to the average expected age of the cloud \citep[$\sim $ 2.5 Myr; e.g., ][and references therein]{Lopez2021}.

\subsection{Accretion properties}
\label{sec:accretionpropeties}
The accretion luminosity of the 37 class II sources was estimated as described in Paper III, that is, by using the empirical relations between $\rm L_{acc}$ and the luminosity of emission lines ($\rm L_{line}$) given in \citet{Alcala2017} and determined from X-shooter observations of a sample of class II sources in Lupus. The
$\rm L_{line}$ was computed as $\rm L_{line}=4 \pi d^2 F_{line}$, where $\rm d$ is the Gaia EDR3 distance reported in Table \ref{tab:sources_param} and $\rm F_{line}$ is the extinction-corrected line flux. This latter was in turn determined through the integration of the line profile with the IRAF task \textsc{splot}. For each line we performed three independent measurements at the lowest, highest and middle position of the local continuum, to take the uncertainties introduced by the local noise into account. $\rm F_{line}$ and its error were then computed as the average and standard deviation of these three measurements, respectively. The correction for the extinction was performed using the $\rm A_v$ value, which was determined as described in the previous section.

For sources with spectral types earlier than K0 the $\rm F_{line}$ measurement is strongly influenced by photospheric absorptions. For these sources, we therefore performed the photospheric subtraction using appropriate spectral templates as described in Paper III. In addition to CQ Tau and RY Tau, which were analyzed in Paper III, we successfully performed this subtraction in another four sources, namely HQ Tau, MWC480, UX Tau and SU Aur. Details are reported in Appendix \ref{sec:ew_and_sub}. We also note that regardless of the level of accretion, the strong photospheric continuum expected for these sources can limit the number of detected accretion diagnostics.

In all sources the \ion{Ca}{II} H and H7 lines were found to be blended. We attempted to separate the two contributions following the Gaussian decomposition procedure of \citet{Gangi2020}. We obtained reliable results for nine sources (i.e., CI Tau, CY Tau, DE Tau, DN Tau, DQ Tau, GH Tau, IP Tau, UY Aur, and V836 Tau) for which the limited broadening of the profiles allowed us to unambiguously disentangle the two lines at the HARPS-N resolution.

When no diagnostic line was detected, we estimated a 3$\rm \sigma$ upper limit as $\rm 3 \times$ \textsc{rms} $\times \Delta \lambda$, with \textsc{rms} the local flux noise and $\rm \Delta \lambda$ the expected line width. The latter was estimated from the other detected lines and assumes typical values between 0.1 and 0.2 nm.

\begin{figure}
\begin{center}
\includegraphics[trim=0 0 0 0,width=0.48\columnwidth, angle=90]{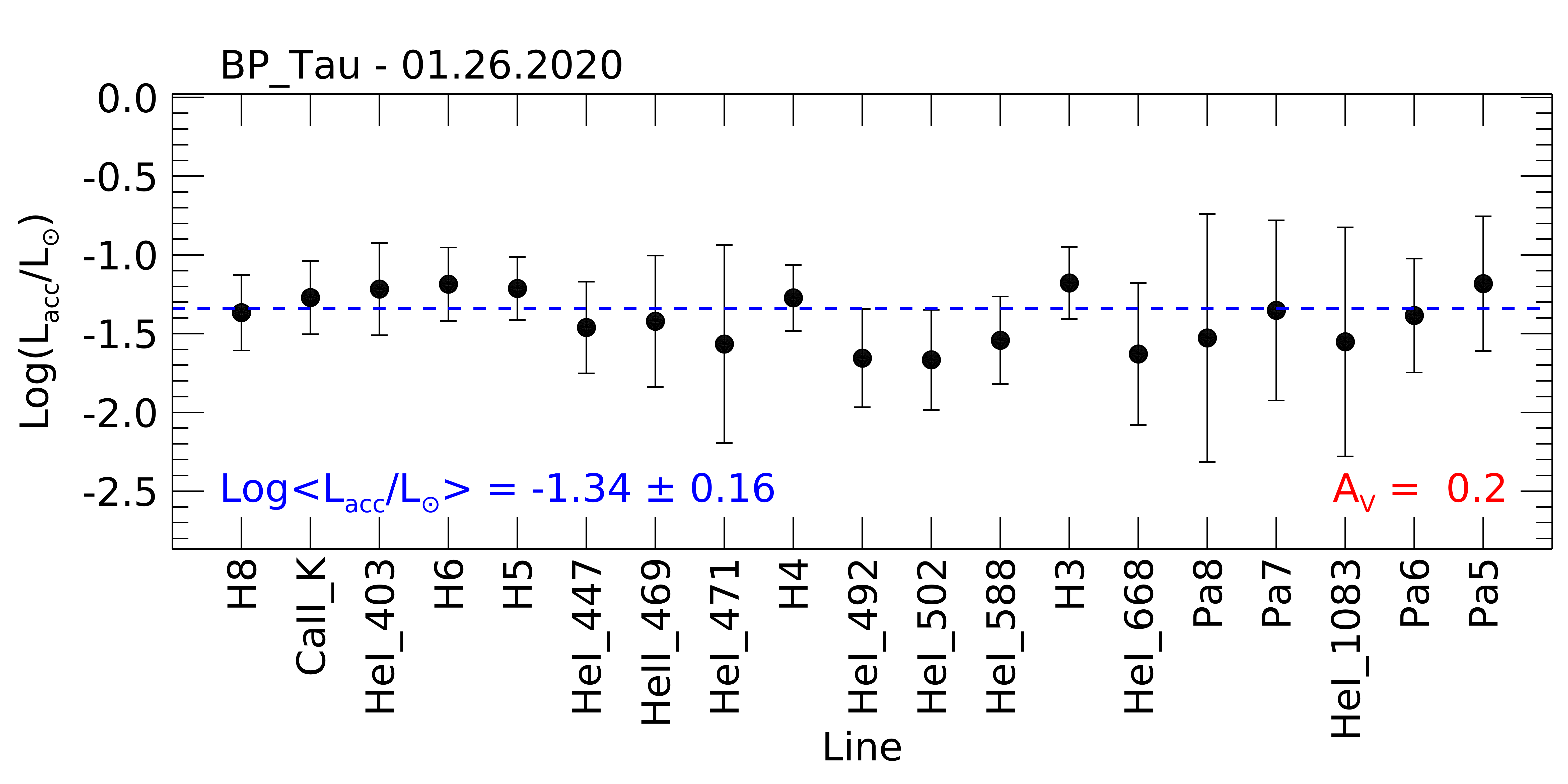}
\end{center}
\caption{\label{fig:Lacc_vs_lines_example} Example of $\rm L_{acc}$  as a function of the different accretion diagnostics. The dashed horizontal blue line represents the median \lacc. The target name, the observation date (MM.DD.YYYY), and the computed \lacc\ and $\rm A_v$ are also indicated. The plots for the complete sample are reported in Fig. \ref{fig:Lacc_vs_lines}.}
\end{figure}

Fig. \ref{fig:Lacc_vs_lines_example} shows an example of the derived $\rm L_{acc}$ values plotted as a function of the line diagnostics, while the plots for the complete sample are reported in Fig. \ref{fig:Lacc_vs_lines}. We note that the $\rm A_v$ calculated in the previous section is always consistent with that obtained by minimizing the spread of $\rm L_{acc}$, confirming the consistency of the adopted methods. For each source, we assumed the median value \lacc\ and its standard deviation as the accretion luminosity and error, respectively.

The mass accretion rate \macc\ was then computed as
\begin{eqnarray}
\rm \dot{M}_{acc} = \Biggr{(}1- \frac{R_{\star}}{R_{in}}\Biggl{)}^{-1} \frac{\langle L_{acc} \rangle R_{\star}}{G M_{\star}} \approx 1.25 \frac{\langle L_{acc} \rangle R_{\star}}{GM_{\star}}
,\end{eqnarray}

with $\rm R_{star}$ and $\rm R_{in}$ the stellar and the inner disk radius, respectively. We assumed $\rm R_{star}/R_{in} \approx 1/5$ as in \citet{Gullbring1998} and \citet{Hartmann1998}, and used the stellar mass $\rm M_{\star}$ and radius $\rm R_{\star}$ derived in the previous section and reported in Table \ref{tab:sources_param}.
As discussed in Paper III, we estimate an average uncertainty of about 0.4 in $\rm log$ \macc , which derive from the uncertainties on \lacc, $\rm M_{\star}$, $\rm R_{\star}$, and Gaia EDR3 distances and from the differences in the adopted evolutionary tracks.
The obtained values of \lacc\ and \macc\ are reported in Table \ref{tab:accretion_prop}, together with the number of lines used for the determination of \lacc.

Finally, we analyzed the EWs and fluxes of $\rm  H\alpha$ and $\rm  Pa\beta$ lines as a function of effective temperature for the complete sample and provided a new empirical criterion to distinguish class III from class II sources based on the $\rm EW_{Pa\beta}$-$\rm T_{eff}$ and $\rm  F_{Pa\beta}$-$\rm T_{eff}$ planes. Details of the adopted method and results are reported in Appendix \ref{sec:ew_and_sub}.

\begin{table*}
\small
\center
\caption{\label{tab:accretion_prop} Accretion properties of the selected accreting CTTs derived in this work and number of diagnostics used to compute the \lacc.}
\begin{tabular}{lccc|lccc}
\hline
\hline
\noalign{\smallskip}
Source    & $\rm log $ \lacc\ $\rm (\pm err)$ & $\rm log $ \macc$^{\ddagger}$ & No. lines  & Source    &  $\rm log $ \lacc\ $\rm (\pm err)$                    & $\rm log $ \macc$^{\ddagger}$ & No. lines \\
          &  [$\rm L_{\odot}$]              & $\rm [M_{\odot}$ $\rm yr^{-1}]$  &   &  & [$\rm L_{\odot}$]              & $\rm [M_{\odot}$ $\rm yr^{-1}]$                     \\
\hline  
 BP Tau$^\dagger$     &   -1.34 (0.16)  & -8.25    & 19 & GI Tau$^\dagger$   & -1.37 (0.20) & -8.43   & 19 \\
 CI Tau               &    0.02 (0.29)  & -7.28    & 19 & GK Tau   & -1.39 (0.17) & -8.67   & 8  \\ 
 CQ Tau$\rm ^{(a,b)}$ &    < -0.42      & < -7.68  & 9  & GM Aur   & -1.05 (0.26) & -8.29   & 9  \\   
 CW Tau              &   -0.07 (0.20)  & -7.47    & 15 & HN Tau$\rm ^{(a,b)}$   & -0.12 (0.23) & -7.31   & 15 \\     
 CY Tau              &   -2.00 (0.17)  & -8.85    & 17 & HQ Tau   & -1.24 (0.17) & -8.43   & 6  \\     
 DE Tau              &   -1.45 (0.17)  & -7.99    & 13 & IP Tau$^\dagger$   & -2.16 (0.28) & -9.14   & 9  \\     
 DF Tau              &   -0.75 (0.17)  & -7.55    & 19 & IQ Tau   & -1.40 (0.26) & -8.54   & 15 \\     
 DG Tau$\rm ^{(a)}$  &   -0.25 (0.18)  & -7.35    & 14 & LkCa 15  & -1.89 (0.24) & -9.24   & 14 \\     
 DH Tau              &   -1.53 (0.15)  & -8.49    & 18 & MWC480   & -0.19 (0.27) & -7.64   & 5  \\     
 DK Tau$^\dagger$    &   -1.42 (0.19)  & -8.33    & 15 & RW Aur A$\rm ^{(a)}$ & +0.39 (0.30) & -6.93   & 15 \\     
 DL Tau$\rm ^{(a)}$  &   -0.35 (0.18)  & -7.62    & 17 & RY Tau$\rm ^{(a)}$   & -0.38 (0.15) & -7.57   & 12 \\     
 DN Tau$^\dagger$    &   -2.10 (0.42)  & -9.04    & 11 & SU Aur   & -0.54 (0.23) & -7.76   & 8  \\     
 DO Tau$\rm ^{(a,c)}$ &  -0.93 (0.25)  & -7.80    & 17 & UX Tau   & -1.29 (0.34) & -8.58   & 2  \\     
 DQ Tau$^\dagger$     &  -0.92 (0.18)  & -7.68    & 16 & UY Aur$^\dagger$   & -1.27 (0.13) & -8.18   & 21 \\     
 DR Tau               &  -0.57 (0.31)  & -7.93    & 15 & UZ Tau E & -1.27 (0.25) & -8.05   & 14 \\     
 DS Tau$^\dagger$     &  -1.46 (0.18)  & -8.57    & 17 & V409 Tau & -2.64 (0.19) & -9.58   & 7  \\     
 FT Tau	              &  -1.93 (0.21)  & -8.92    & 16 & V807 Tau & -0.91 (0.07) & -8.14   & 9  \\       
 GG Tau A$^\dagger$   &  -0.76 (0.18)  & -7.76    & 12 & V836 Tau$^\dagger$ $\rm ^{(c)}$  & -1.93 (0.20) & -8.37   & 14 \\        
 GH Tau               &  -2.31 (0.15)  & -8.98    & 12 &    & 			 &         & \\ 
 \hline                            
\end{tabular}     
\begin{quotation}
\textbf{Notes.}

$^\dagger$ Heavily spotted sources. 

$\rm ^{\ddagger}$ Average uncertainties are 0.4 in $\rm log$ \macc.

$\rm ^{(a)}$ Parameters derived in \citet{Alcala2021}. 

$\rm ^{(b)}$ Subluminous YSO: the reported values are corrected for obscuration effects (see Section 4.3.1 of \citet{Alcala2021}). 

$\rm ^{(c)}$ Source observed in different epochs; the reported measurements are the averaged values.
\end{quotation}                 
\end{table*}

\begin{figure*}[!]
\begin{center}
\includegraphics[trim=40 40 20 0,width=1\columnwidth]{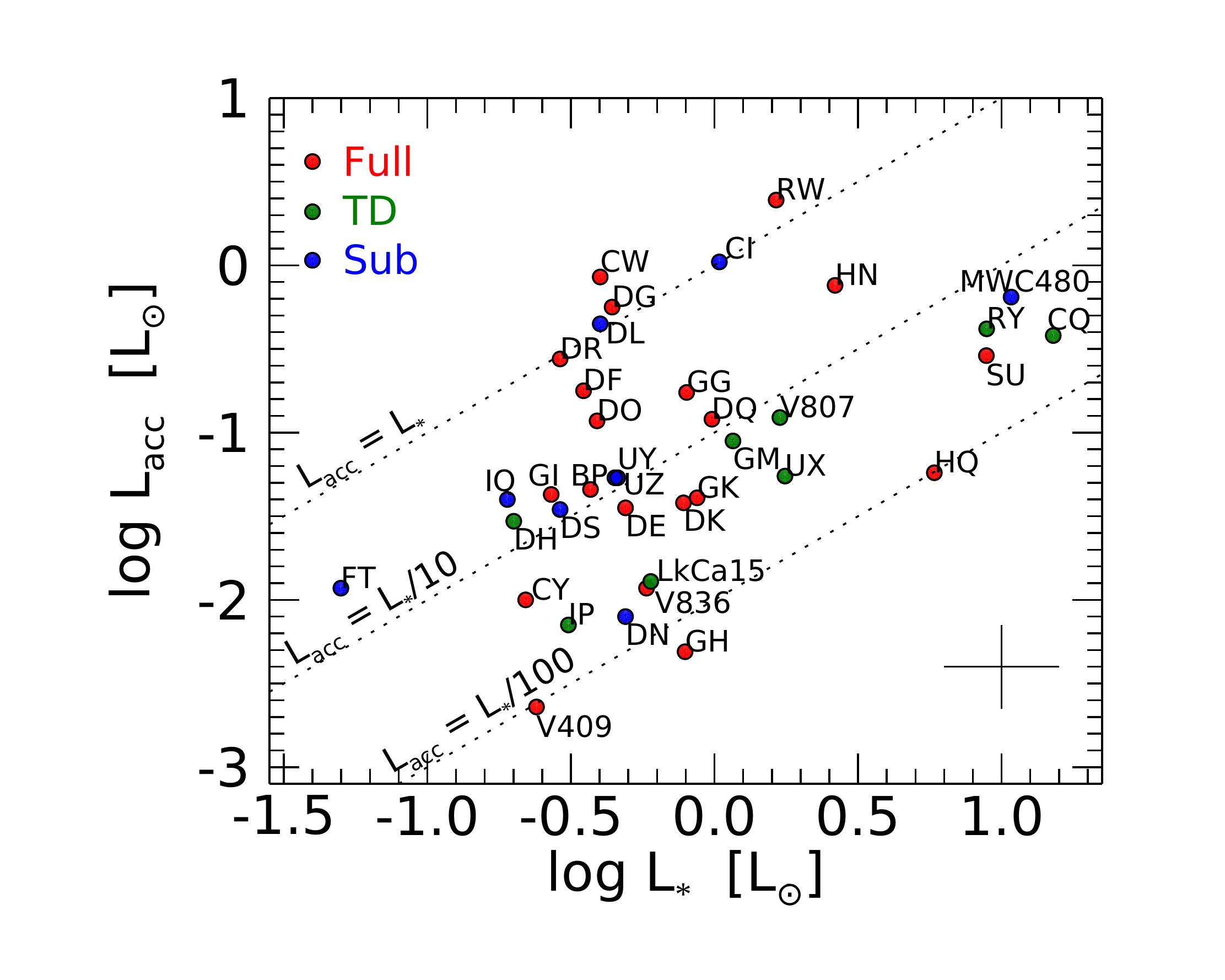}
\includegraphics[trim=40 40 20 0,width=1\columnwidth]{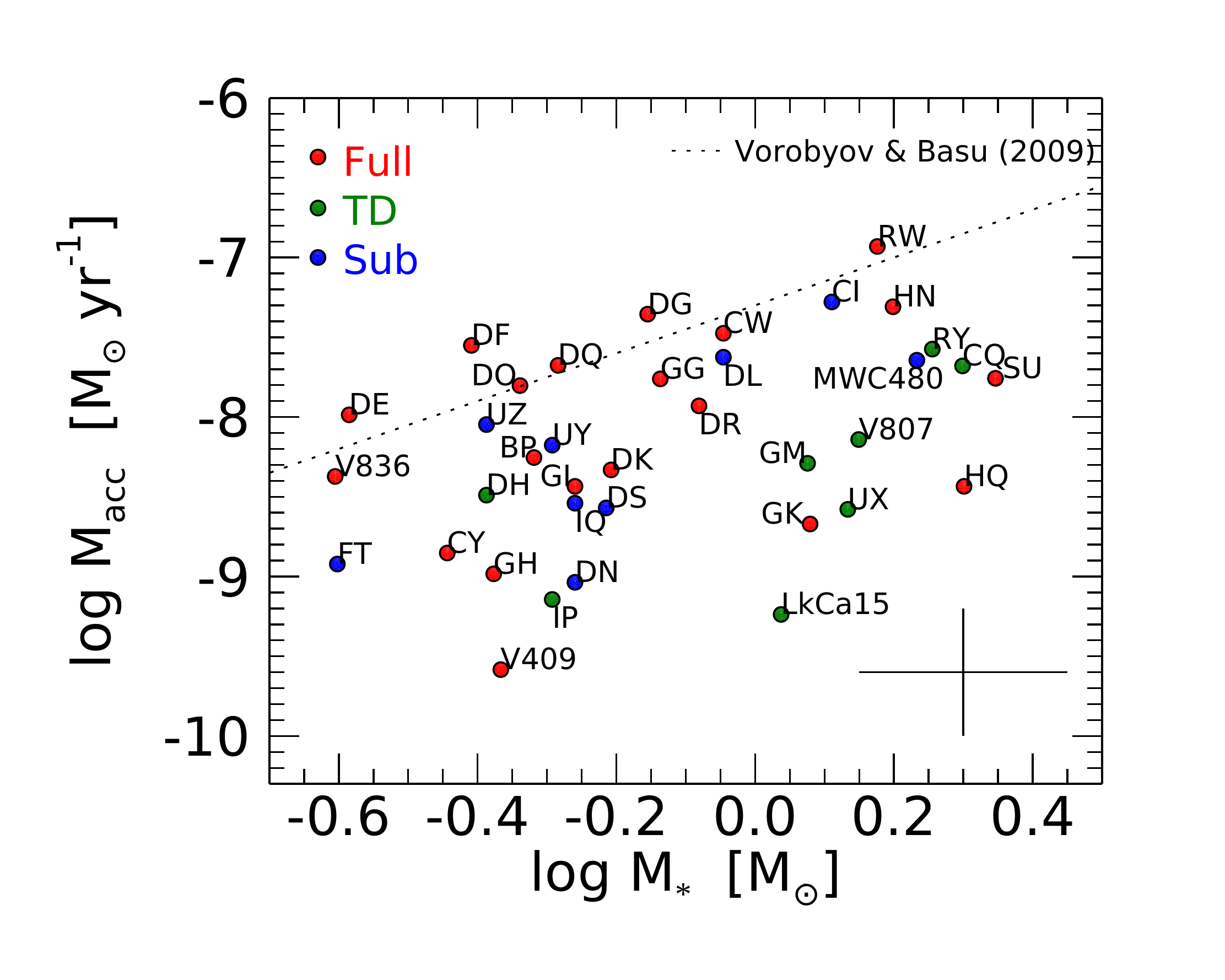}
\includegraphics[trim=40 40 20 0,width=1\columnwidth]{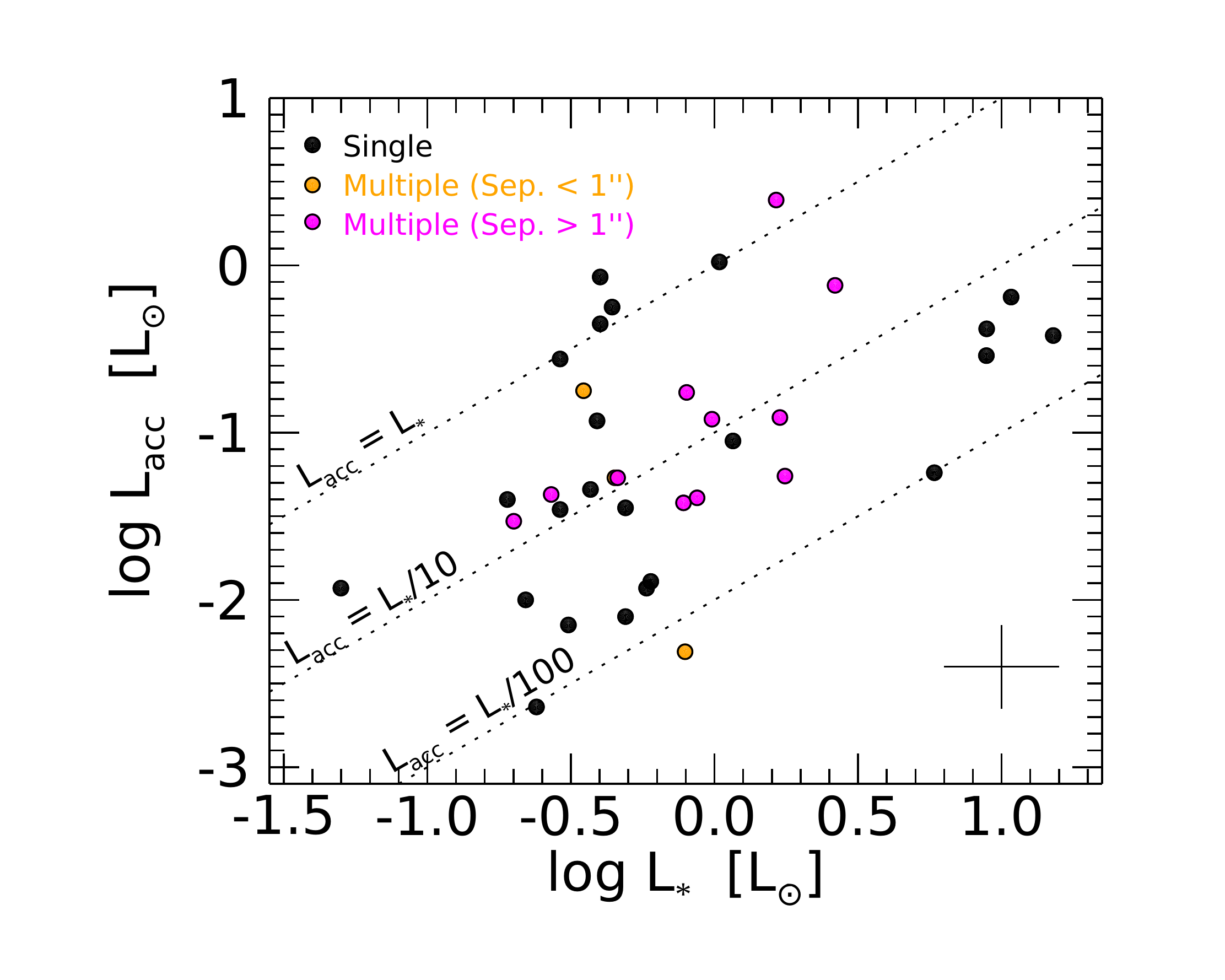}
\includegraphics[trim=40 40 20 0,width=1\columnwidth]{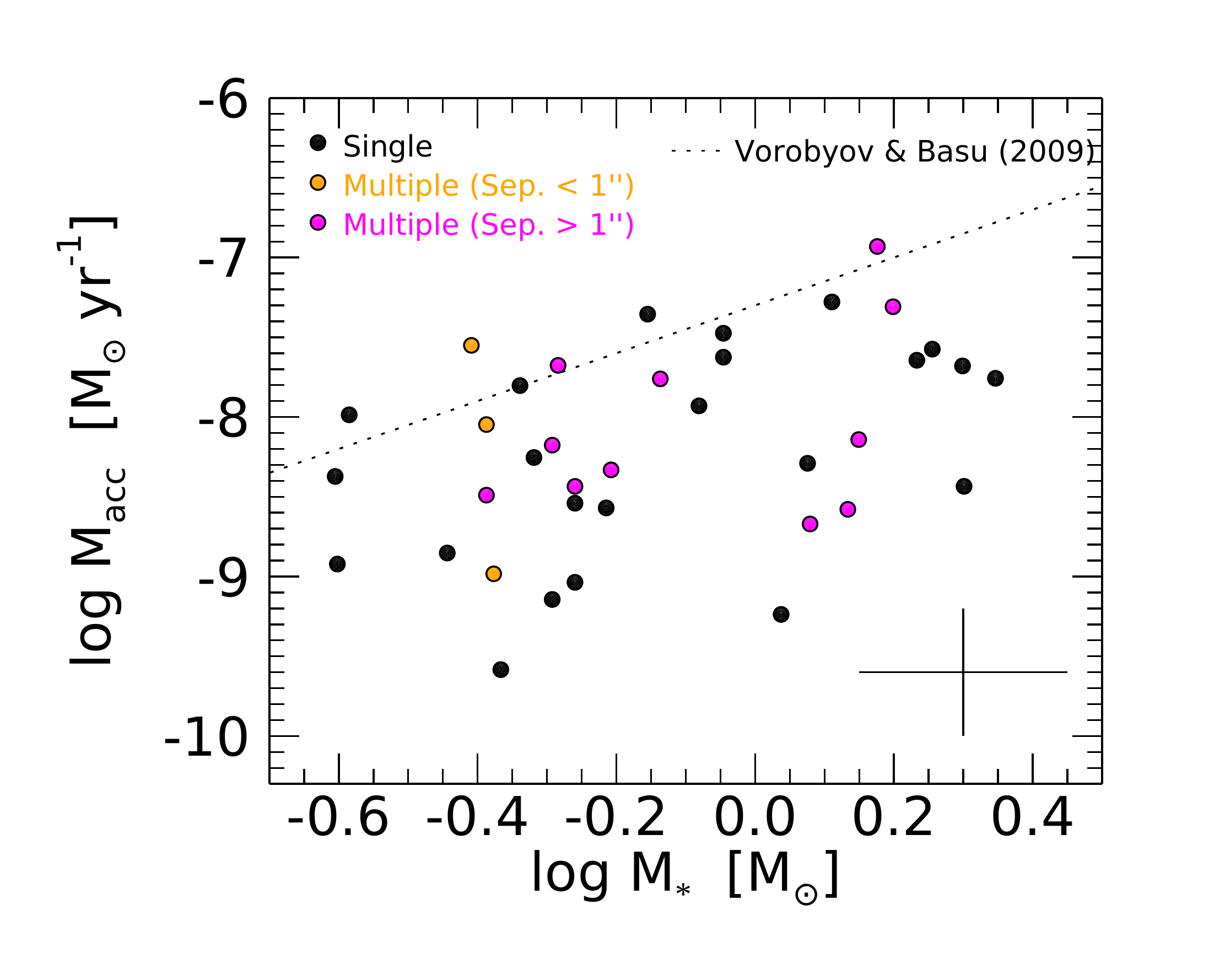}
\includegraphics[trim=40 40 20 0,width=1\columnwidth]{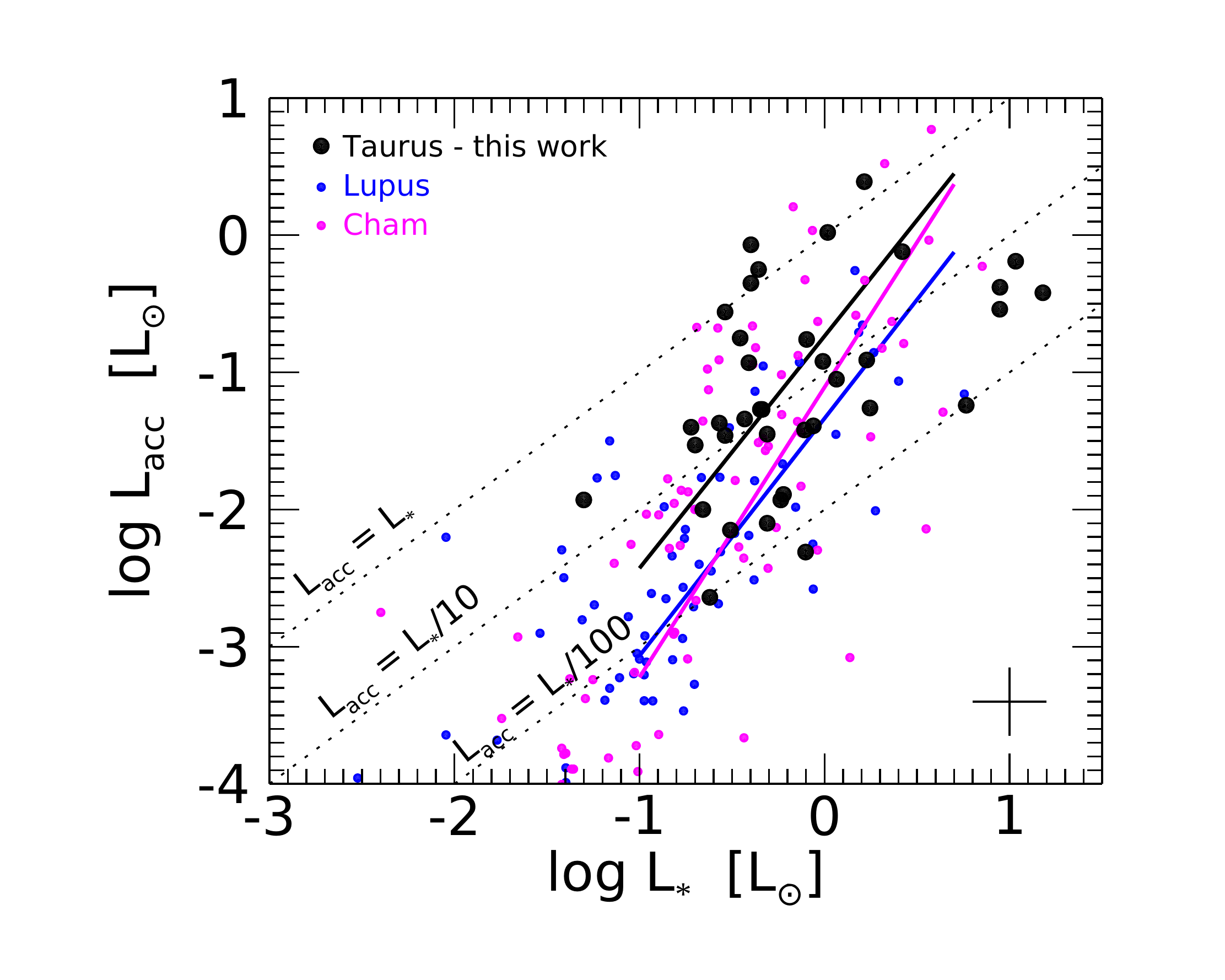}
\includegraphics[trim=40 40 20 0,width=1\columnwidth]{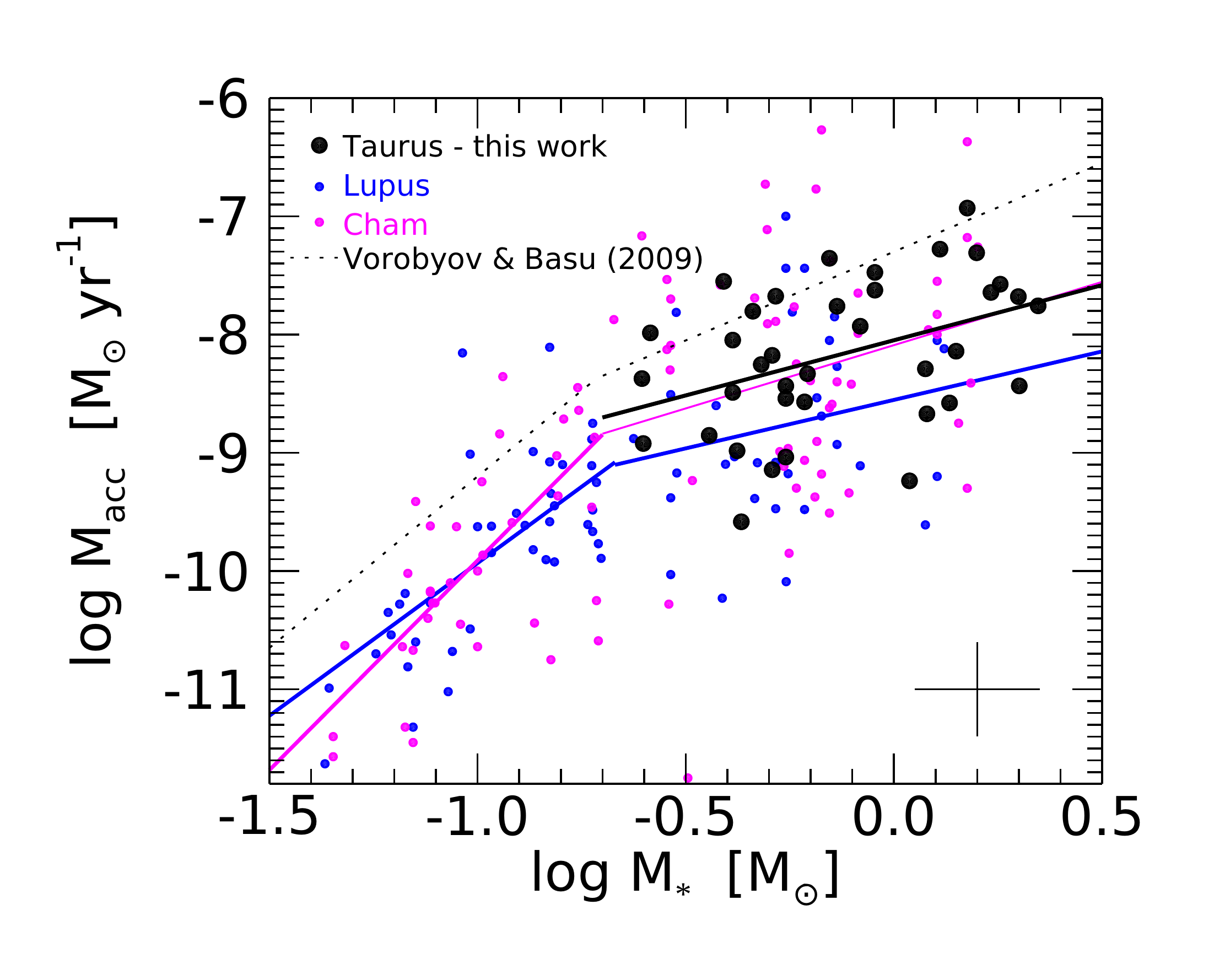}
\end{center}
\caption{\label{fig:accretion_results} Accretion luminosity as a function of stellar luminosity (left column) and mass accretion rate as a function of stellar mass (right column) for the Taurus sample (large circles). In the top panels the symbols are distinguished according to the disk structure and in the lower panels the symbols are allotted according to the multiplicity. The bottom panels provide a comparison with Lupus (blue) and Chamaeleon I (magenta). Solid lines are the best fit to data of the same color. The average errors for the Taurus sample are drawn in the lower corner.}
\end{figure*}

\begin{figure*}
\begin{center}
\includegraphics[trim=40 20 20 0,width=1\columnwidth]{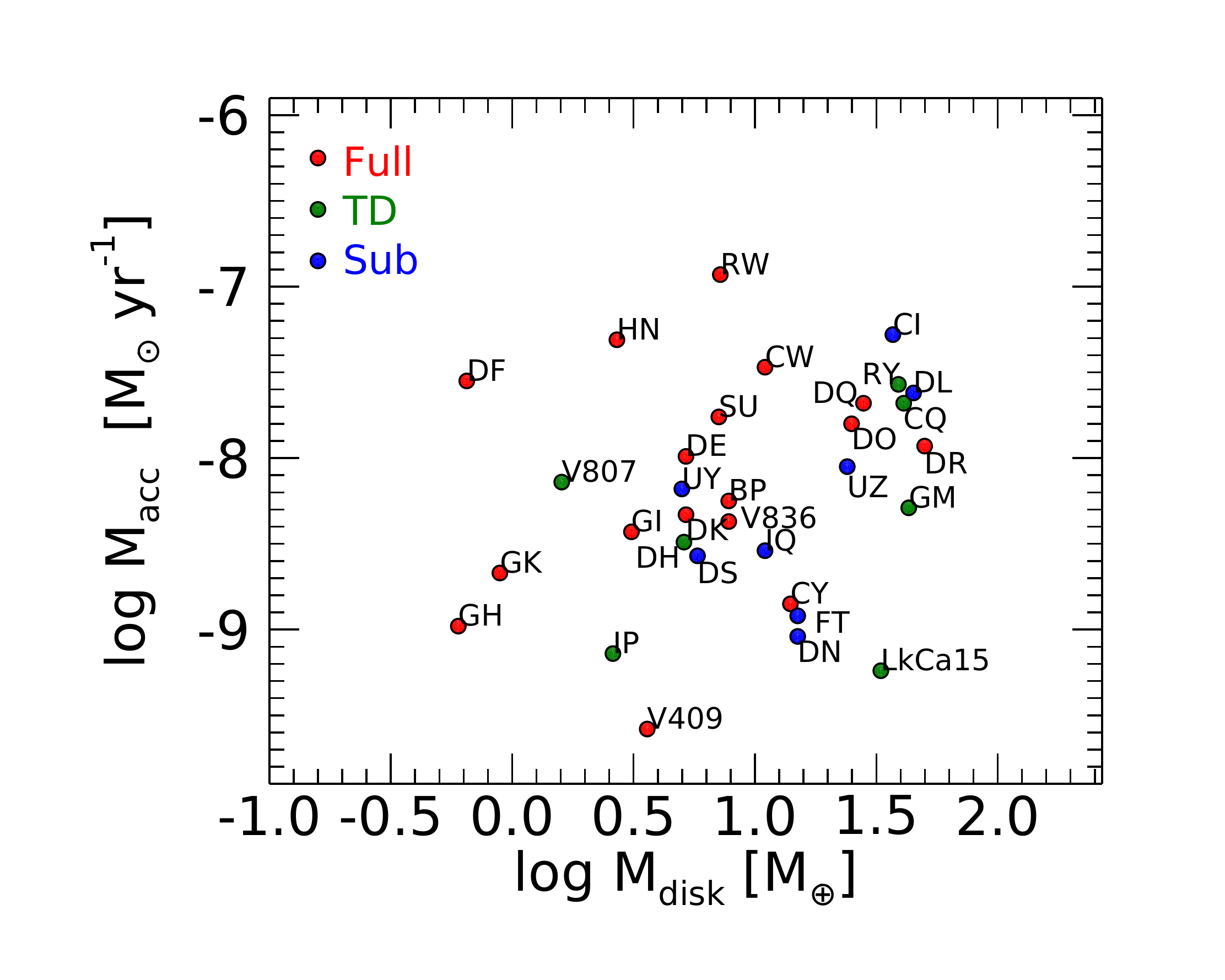}
\includegraphics[trim=40 20 20 0,width=1\columnwidth]{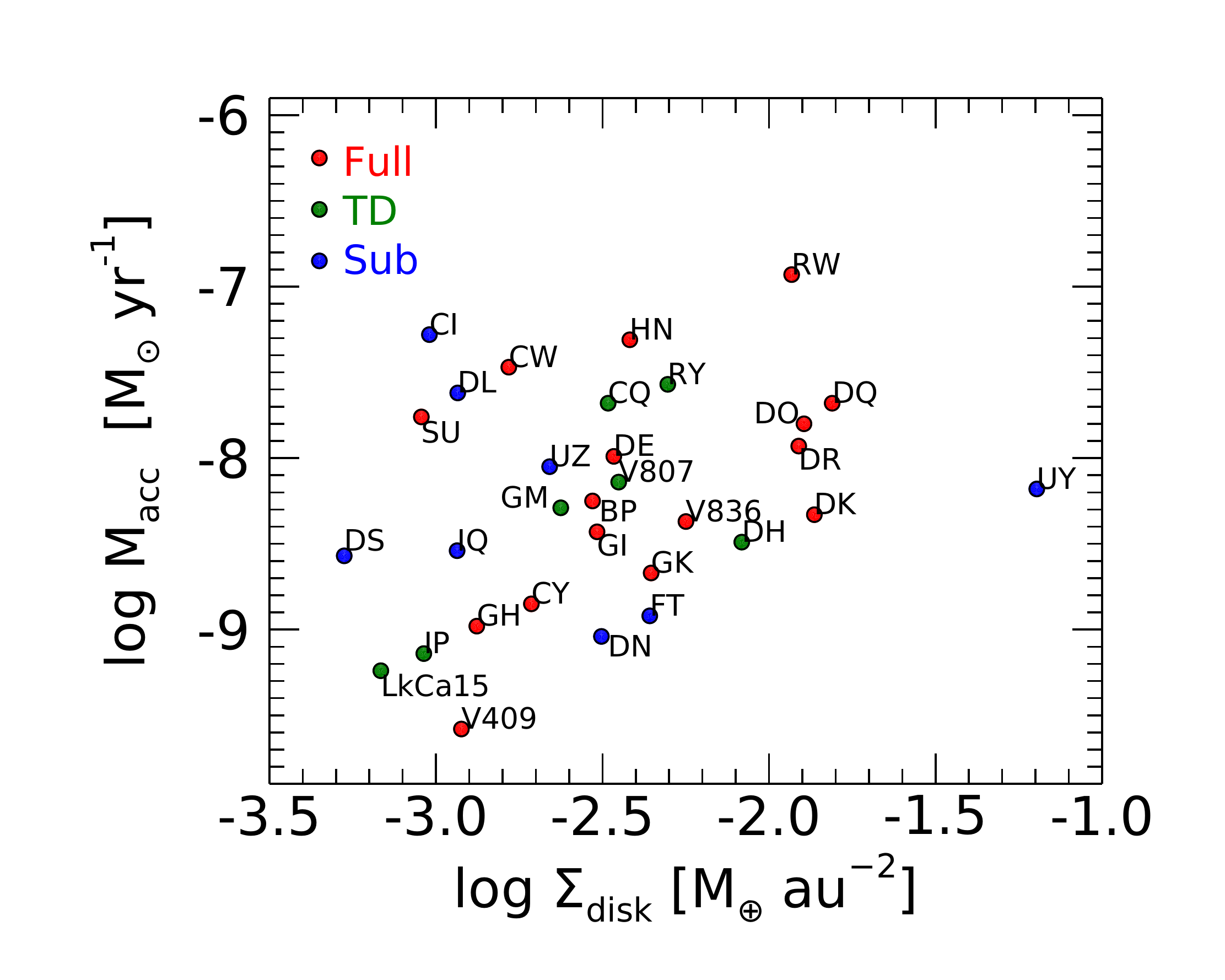}
\includegraphics[trim=40 20 20 0,width=1\columnwidth]{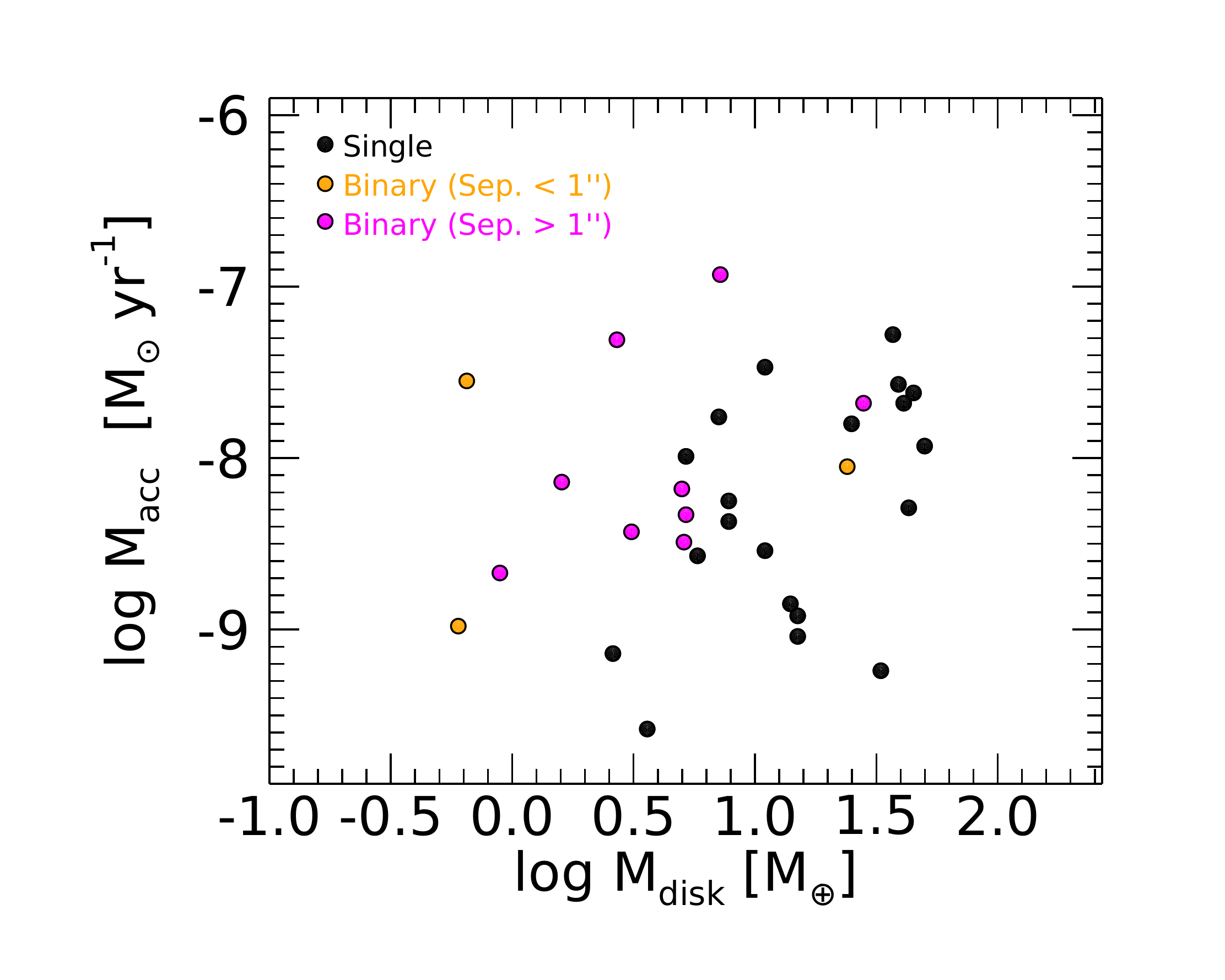}
\includegraphics[trim=40 20 20 0,width=1\columnwidth]{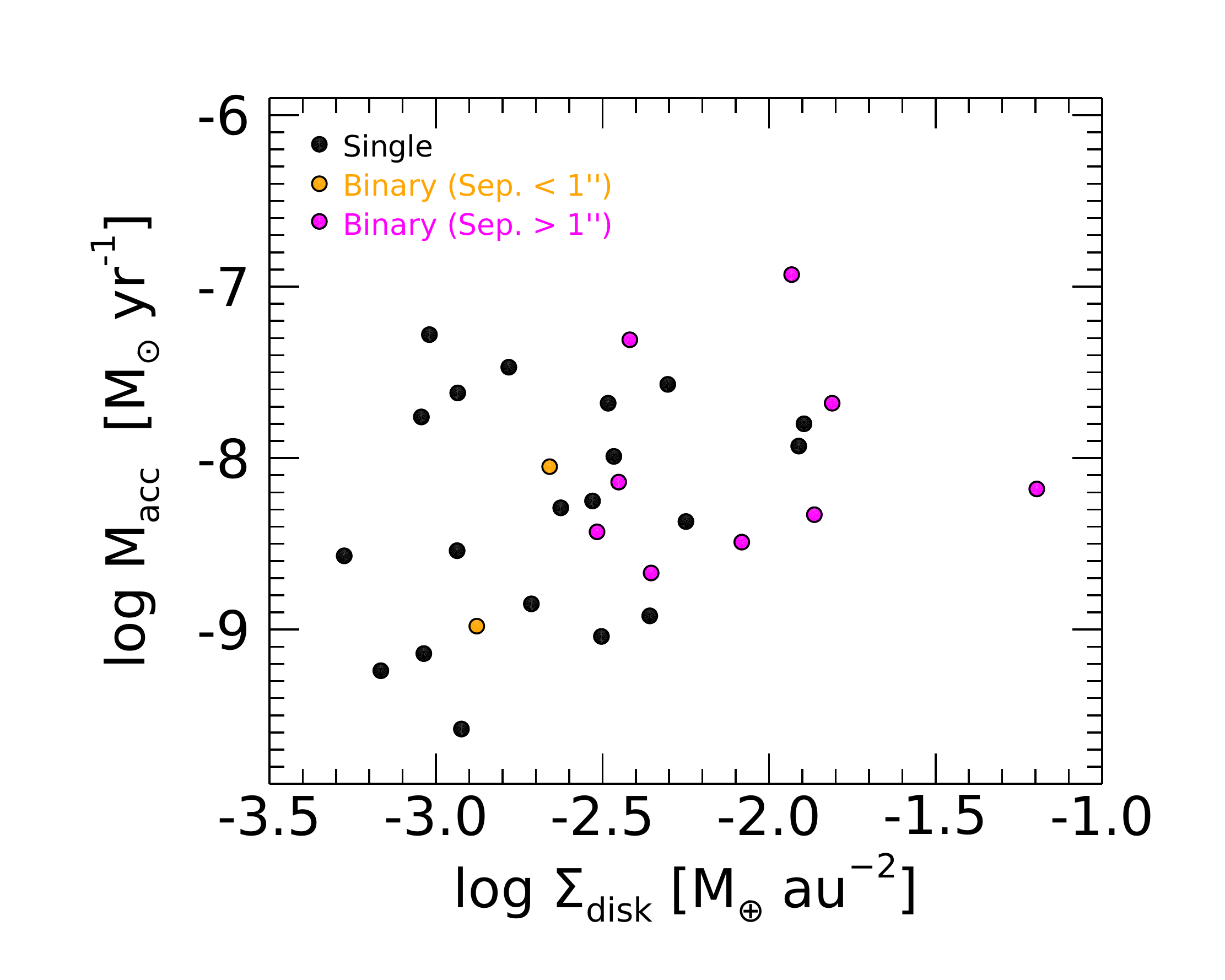}
\includegraphics[trim=40 20 20 0,width=1\columnwidth]{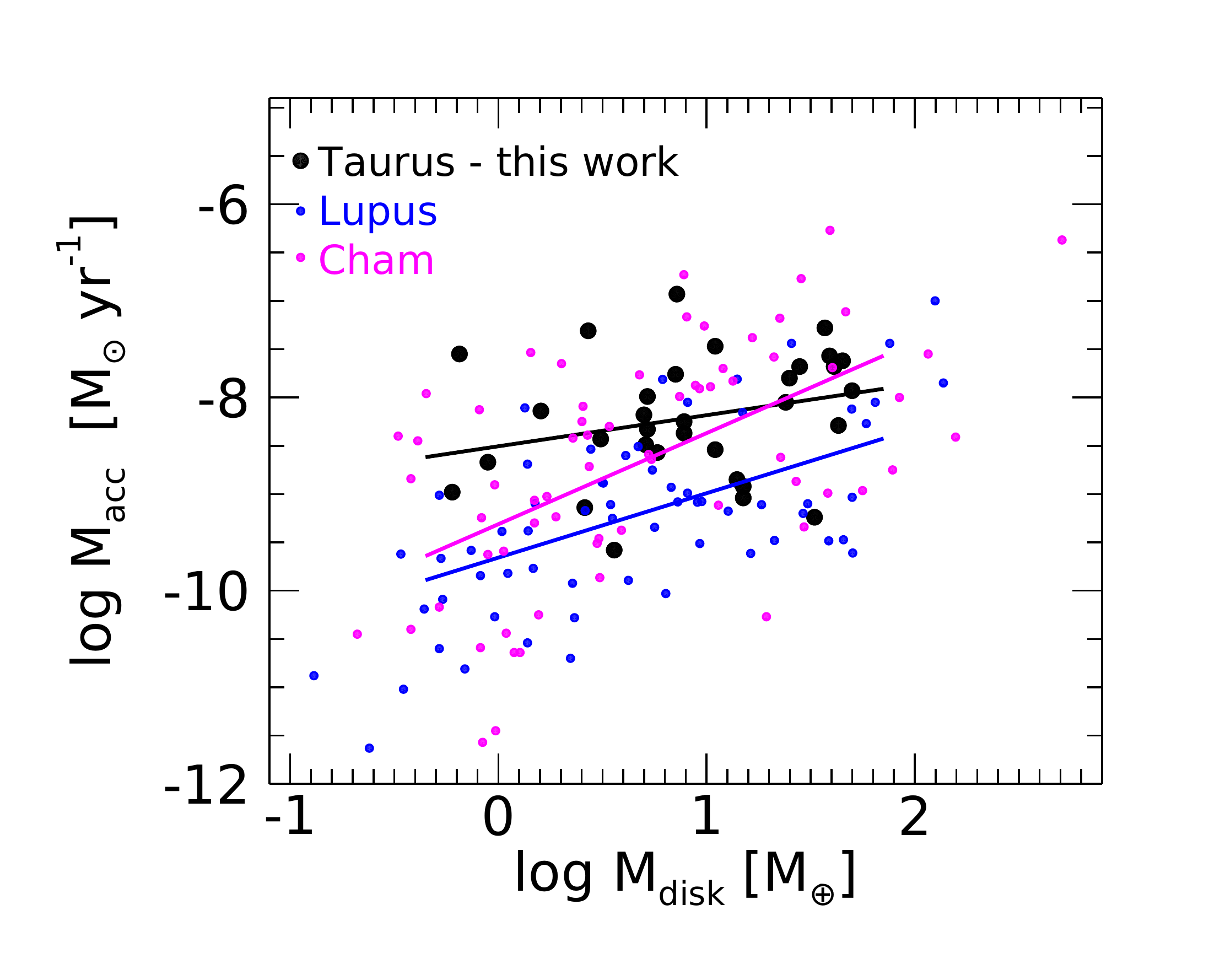}
\includegraphics[trim=40 20 20 0,width=1\columnwidth]{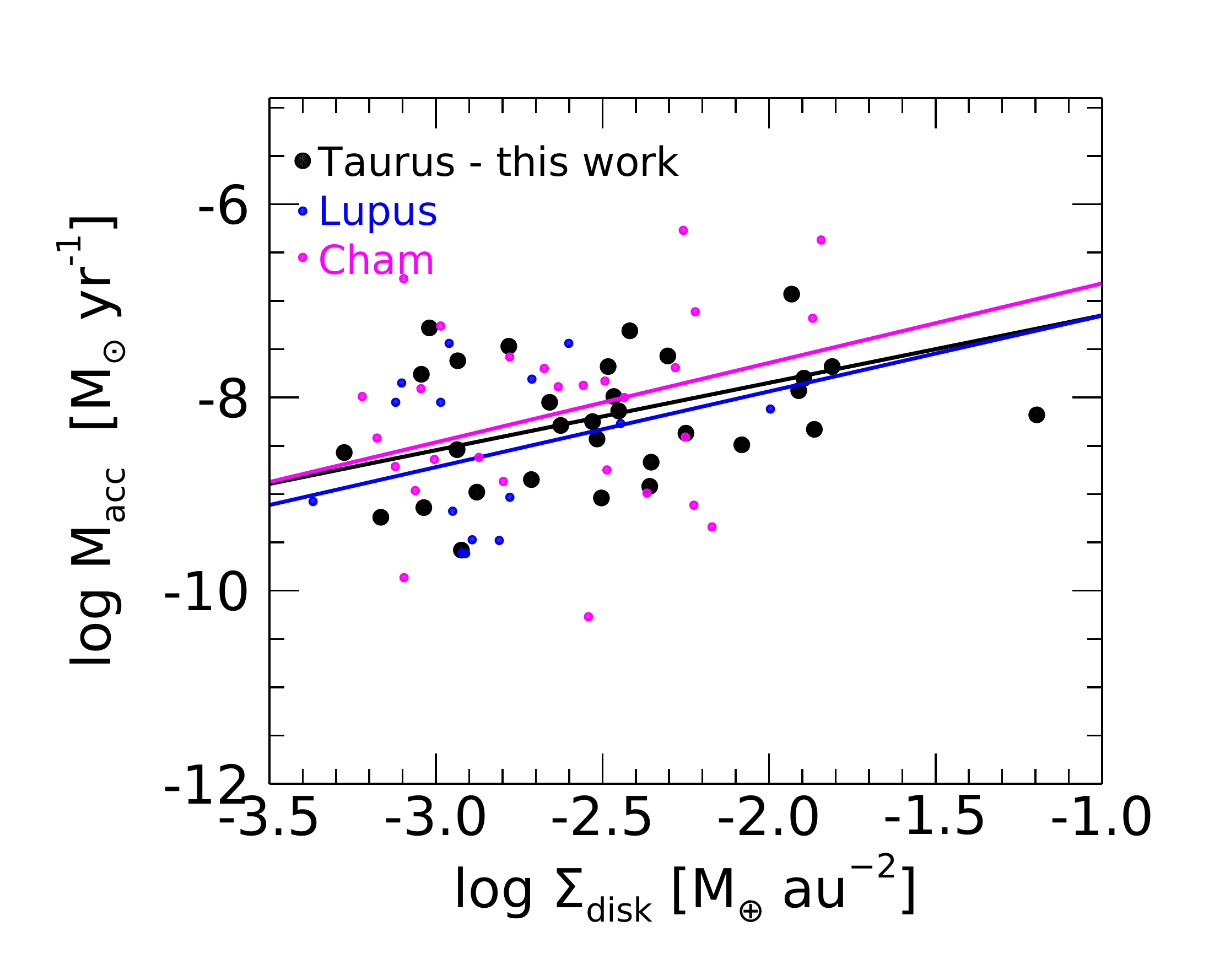}
\end{center}
\caption{\label{fig:Macc_Mdisk} Mass accretion rate as a function of disk dust mass (left column) and disk dust density (right column) for the Taurus sample (large circles). In the top panels the symbols are distinguished according to the disk structure and in the lower panels the symbols are allotted according to the multiplicity. The bottom panels provide a comparison with Lupus (blue) and Chamaeleon I (magenta). Solid lines are the best fit to data of the same color.}
\end{figure*}

\section{Correlations}\label{sec:results}
\subsection{Accretion luminosity versus stellar luminosity}
Fig. \ref{fig:accretion_results} (left panels) shows the distribution of the GHOsT sources in the $\rm L_{acc}-L_{\star}$ plane, color-coded according to the disk structure classification and multiplicity. For the latter, we distinguish between sources with a separation $> 1''$ and $< 1''$ based on the on-sky dimension of the HARPS-N fiber. We also compare our results with the distribution of the similar age star-forming regions of Lupus and Chamaeleon I (bottom right) taken from previous surveys by \citet{Alcala2019} and \citet{Manara2017} and rescaled to the new Gaia EDR3 distances \citep{Manara2022}.

The dotted lines in the $\rm L_{acc}-L_{\star}$ plane show the $\rm L_{acc}/L_{\star}=1, 0.1, 0.01$ relations. Only a few sources (i.e., DR Tau, DL Tau, CW Tau, DG Tau, CI Tau, and RW Aur) are located close to the $\rm L_{acc}=L_{\star}$ line, while the other sources fall below this boundary. None of the sources show $\rm L_{acc}$ values significantly below the $\rm 0.01L_{\star}$ locus.

Inspecting the disk classification, we note that sources with full and substructured disks show the largest spread both in $\rm L_{acc}$ and $\rm L_{\star}$. In contrast, sources with transitional disks fill the area between the $\rm L_{acc}=0.1 L_{\star}$ and $\rm L_{acc}=0.01 L_{\star}$ boundaries. No clear trend between different source multiplicities is found with the current statistics.

Although the completeness level of our sample is lower than that of Chamaeleon I and Lupus, there is an overall similarity in the distribution of the three SFRs in the slope and in the spread (Fig. \ref{fig:Accretion_hist}). We performed a linear fit of the data with the \textsc{idl} \textsc{linmix\_err} tool, which performs a linear regression using the Bayesian method of \citet{Kelly2007} and allows us to take the uncertainties on both axes into account. We adopted the median values and standard deviations of the chains as best-fit coefficients and errors, respectively. The fitting procedure, applied to the same $\rm L_{\star}$ range as was probed with our data, confirms this agreement:

\begin{eqnarray}
\rm Taurus:\ log\ L_{acc} = 1.69 (\pm 0.93)\ log\ L_{\star} - 0.74 (\pm 0.29), \\
\rm Lupus:\ log\ L_{acc} = 1.73 (\pm 0.27)\ log\ L_{\star} - 1.33 (\pm 0.16), \\
\rm Cham:\ log\ L_{acc} = 2.11 (\pm 0.23)\ log\ L_{\star} - 1.10 (\pm 0.22).
\end{eqnarray}

The fitting coefficients found here for the Lupus and Chamaeleon I distributions are compatible within the errors with those obtained using Gaia DR2 distances by \citet{Alcala2017,Alcala2019} and \citet{Manara2017}. 

\subsection{Mass accretion rate versus stellar mass}
The dependence of $\rm \dot{M}_{acc}$ on $\rm M_{\star}$ in the Taurus sample is shown in the right panels of Fig. \ref{fig:accretion_results}. Again, there is a general similarity between the Taurus, Lupus, and Chamaeleon I distributions. For the latter two, $\rm M_{\star}$, and thus $\rm \dot{M}_{acc}$, was determined basing on the nonmagnetic models of \citet{Baraffe2015} for targets with $\rm T_{eff} \leq 3900$ K and of \citet{Feiden2016} for hotter stars \citep[see][for details]{Manara2022}. 

The GHOsT points do not fill the region at masses lower than $\sim 0.2$ $\rm M_{\sun}$ , and hence it is not possible to investigate whether a double power trend, as theoretically predicted by \citet{Vorobyov2009} (dashed black line of Fig. \ref{fig:accretion_results}) and already suggested for the Lupus and Chamaeleon I datasets \citep[][]{Alcala2017,Manara2017}, is also present in Taurus.

A best-fit linear regression to the complete GHOsT sample yields
\begin{eqnarray}
\rm Taurus:\ log\ \dot{M}_{acc} = 0.93 (\pm 0.67)\ log\ M_{\star} -8.05 (\pm 1.20).
\end{eqnarray}

We therefore find a lower slope than was proposed in the past \citep[$\sim$ 1.6-2.2; e.g.,][]{Antoniucci2014, Alcala2014, Manara2016, Venuti2019}, which is probably due to the small $\rm M_{\star}$ range we probed. Our result is comparable within the errors with the values obtained for the distributions of Lupus and Chamaeleon I when the analysis is restricted to the high-mass (i.e., $\rm M_{\star} \geq 0.2\ M_{\odot}$) regime,
\begin{eqnarray}
\rm Lupus:\ log\ \dot{M}_{acc} = 0.81 (\pm 0.79)\ log\ M_{\star} -8.55 (\pm 1.21), \\
\rm Cham:\ log\ \dot{M}_{acc} = 0.67 (\pm 0.58)\ log\ M_{\star} -8.11 (\pm 1.21).
\end{eqnarray}

Again, the fitting results for the Lupus and Chamaeleon I distributions agree well with those reported in \citet{Alcala2017,Alcala2019} and \citet{Manara2017}. 

We also note that except for a few cases, the Taurus distribution falls below the theoretical prediction of \citet{Vorobyov2009}. This is in line with the Lupus distribution, while the Chamaeleon I dataset presents a substantial fraction of points placed above this locus. 
Finally, transitional disks tend to be placed on the lower side of the distribution, while no particular trend is seen between the full and substructured disks, nor for the source multiplicity.

\subsection{Mass accretion rate versus disk mass}
The dependence of $\rm \dot{M}_{acc}$ on the disk dust mass ($\rm M_{disk,dust}$) is shown in the left column of Fig. \ref{fig:Macc_Mdisk}, where again the disk structure and the source multiplicity is highlighted and the Taurus distribution is compared with the distributions of the Lupus and Chamaeleon I populations. 
$\rm M_{disk,dust}$ were retrieved from the most recent literature based on ALMA 3.1 mm flux measurements (see Table \ref{tab:disk_prop}). When only the flux measurement has been reported, we consistently computed the $\rm M_{disk,dust}$ as described by \citet{Hildebrand1983},
\begin{eqnarray}
\rm M_{disk,dust} = \frac{F_{\nu} d^2}{k_{\nu} B_{\nu}(T_{dust})},
\end{eqnarray}

with $\rm F_{\nu}$ the integrated 3.1 millimeter flux, $\rm d$ the Gaia EDR3 distance, $\rm k_{\nu}$ the dust opacity, and $\rm B_{\nu}(T_{dust})$ the blackbody emission at temperature $\rm T_{dust} = 20\ K$. We adopted a power-law opacity $\rm k_{\nu} = 2.3 (\nu / 230\ GHz)^{0.4}$ $\rm cm^2$ $\rm g^{-1}$ as in \citet{Long2019}.

We find a moderate correlation between $\rm M_{disk,dust}$ and $\rm \dot{M}_{acc}$, which is similar to what is observed in the Lupus and Chamaeleon I samples. As pointed out by \citet{Manara2016}, for example, a large range of $\rm M_{disk,dust}$ needs to be probed to find a stronger correlation.
Considering the disk classification, we note that transitional and substructured disks tend to have higher values of $\rm M_{disk,dust}$, while sources with full disks homogeneously fill the $\rm \dot{M}_{acc}$-$\rm M_{disk,dust}$ plane.

In the right column of Fig. \ref{fig:Macc_Mdisk}, we consider the surface mass density defined as $\rm \Sigma_{disk,dust} = M_{disk,dust}/4\pi R_{eff}^2$ instead of the disk mass, with $\rm R_{eff}$ the disk radius enclosing $\rm 68\%$ of the 3.1 mm flux. In this case, the distribution of transitional and substructured disks becomes similar to that of full disks. We discuss this different behavior in Section \ref{sec:discussion}.

The distribution of sources with multiplicity is well distinct from that of single sources in the $\rm \dot{M}_{acc}$-$\rm M_{disk,dust}$ plane. This trend can be interpreted both in terms of lower $\rm M_{disk,dust}$ at a given $\rm \dot M_{acc}$ or higher $\rm \dot M_{acc}$ at a given $\rm M_{disk,dust}$ in multiple stars compared with single stars. In any case, stellar multiplicity provides a substantial contribution to the observed spread in the $\rm \dot{M}_{acc}$-$\rm M_{disk,dust}$ relation.

A best-fit linear regression to our sample yields a lower slope than that of the Lupus and Chamaeleon distributions, again considering the same $\rm M_{disk}$ range as was probed with our data: 

\begin{eqnarray}
\rm Taurus:\ log\ \dot{M}_{acc} = 0.32 (\pm 0.25)\ log\ M_{dust} - 8.50 (\pm 0.26), \\
\rm Lupus:\ log\ \dot{M}_{acc} = 0.67 (\pm 0.23)\ log\ M_{dust} - 9.66 (\pm 0.19), \\
\rm Cham:\ log\ \dot{M}_{acc} = 0.94 (\pm 0.18)\ log\ M_{dust} - 9.31 (\pm 0.18).
\end{eqnarray}

However, this difference is strongly subject  to the lower number of points in our distribution, and hence we consider it not statistically significant. In contrast, when we considered the $\rm \dot{M}_{acc}$-$\rm \Sigma_{disk,dust}$ distribution, which has a comparable number of points, we find similar dependences:

\begin{eqnarray}
\rm Taurus:\ log\ \dot{M}_{acc} = 0.70 (\pm 0.40)\ log\ \Sigma_{dust} -6.45 (\pm 1.03), \\
\rm Lupus:\ log\ \dot{M}_{acc} = 0.78 (\pm 1.53)\ log\ \Sigma_{dust} -6.37 (\pm 4.37), \\
\rm Cham:\ log\ \dot{M}_{acc} = 0.82 (\pm 0.67)\ log\ \Sigma_{dust} -6.00 (\pm 1.77).
\end{eqnarray}

\begin{figure}
\begin{center}
\includegraphics[trim=100 60 120 10,width=0.7\columnwidth, angle=0]{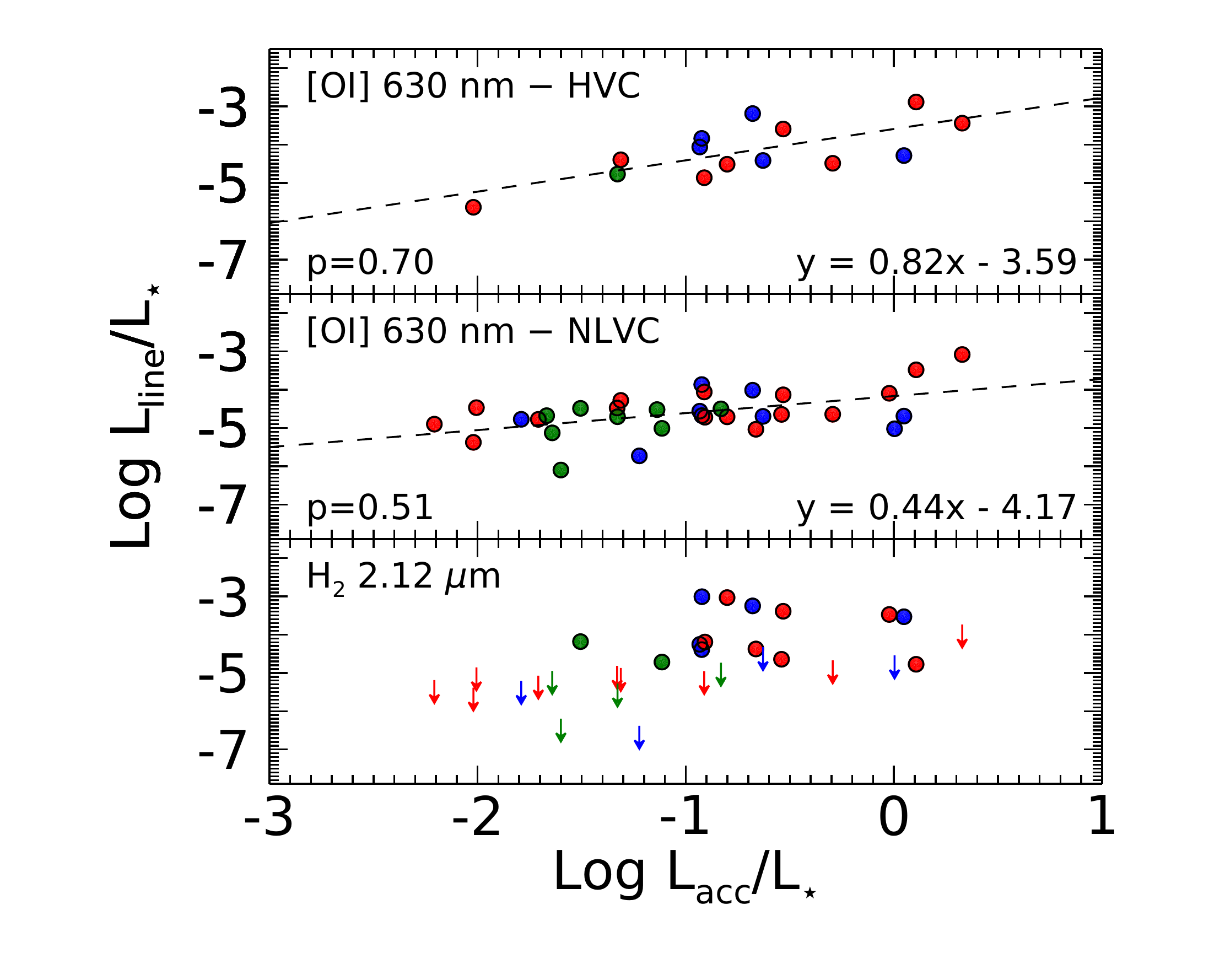}
\end{center}
\caption{\label{fig:compare_lum_lacc} Line luminosity as a function of accretion luminosity. Both quantities are normalized to the stellar luminosity. From top to bottom: \ion{[O}{i]} 630 nm high-velocity component, \ion{[O}{i]} 630 nm narrow low-velocity component, and $\rm H_2$ 2.12 $\rm \mu m$. Linear fits to the data are marked by the dashed lines. Analytical solutions and Pearson coefficients are also labeled. Upper limits are drawn with red arrows.}
\end{figure}

\begin{figure}[!h]
\begin{center}
\includegraphics[trim=100 60 120 10,width=0.7\columnwidth, angle=0]{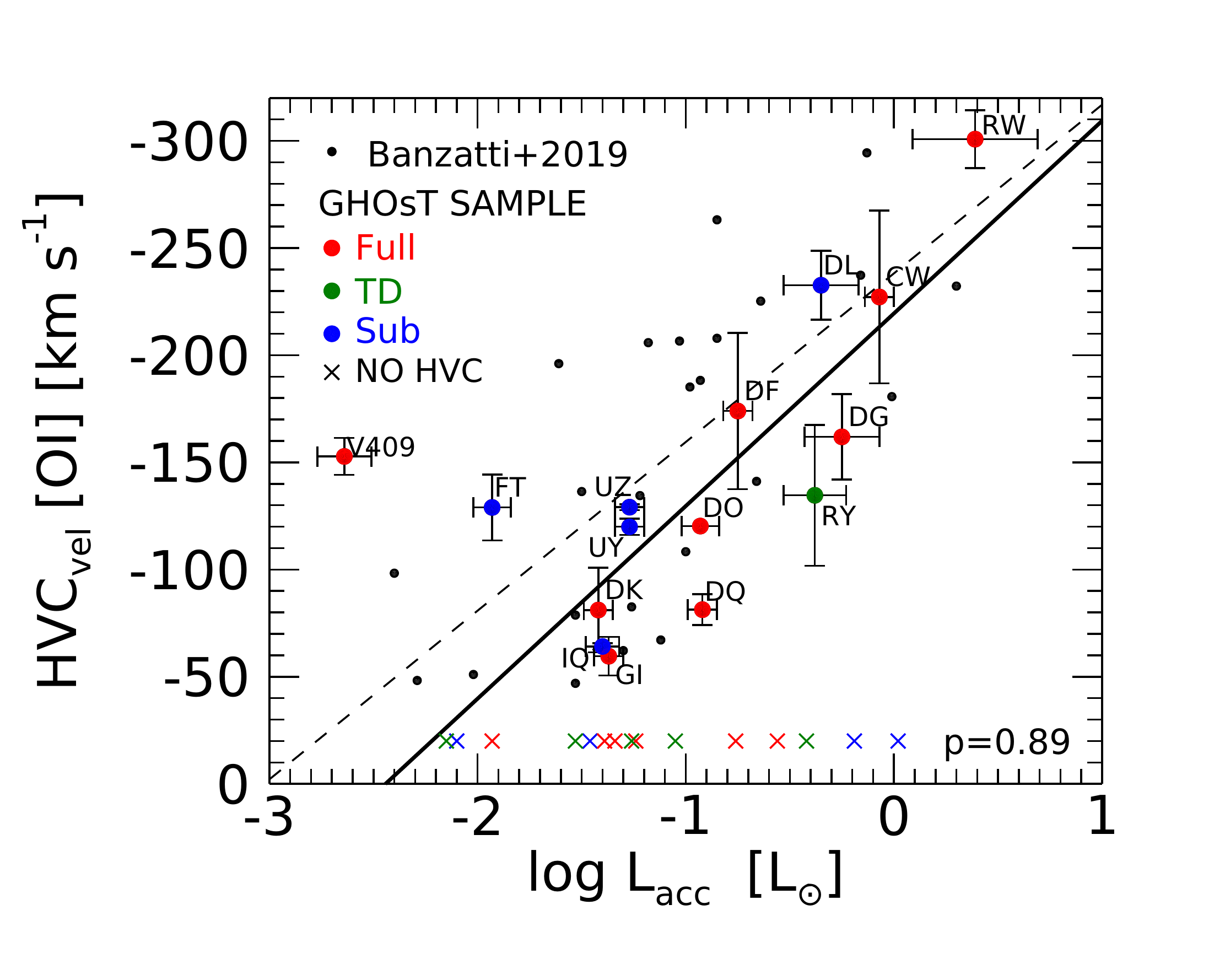}
\end{center}
\caption{\label{fig:Lacc_vs_HVC} Deprojected \ion{[O}{i]}630nm HVC peak velocities as a function of $\rm L_{acc}$ for the GHOsT subsample driving jets (colored points) compared with a CTTs sample reported in \citet{Banzatti2019} (black points). A fit to both data samples is shown as the dashed black line (\citet{Banzatti2019} sample) and as the continuous black line (GHOsT sample). The latter was computed by excluding the outliers (V409 Tau and FT Tau). The Pearson coefficient for the GHOsT distribution is labeled in the figure.}
\end{figure}

\begin{figure*}[!]
\begin{center}
\includegraphics[trim=40 40 20 0,width=1\columnwidth]{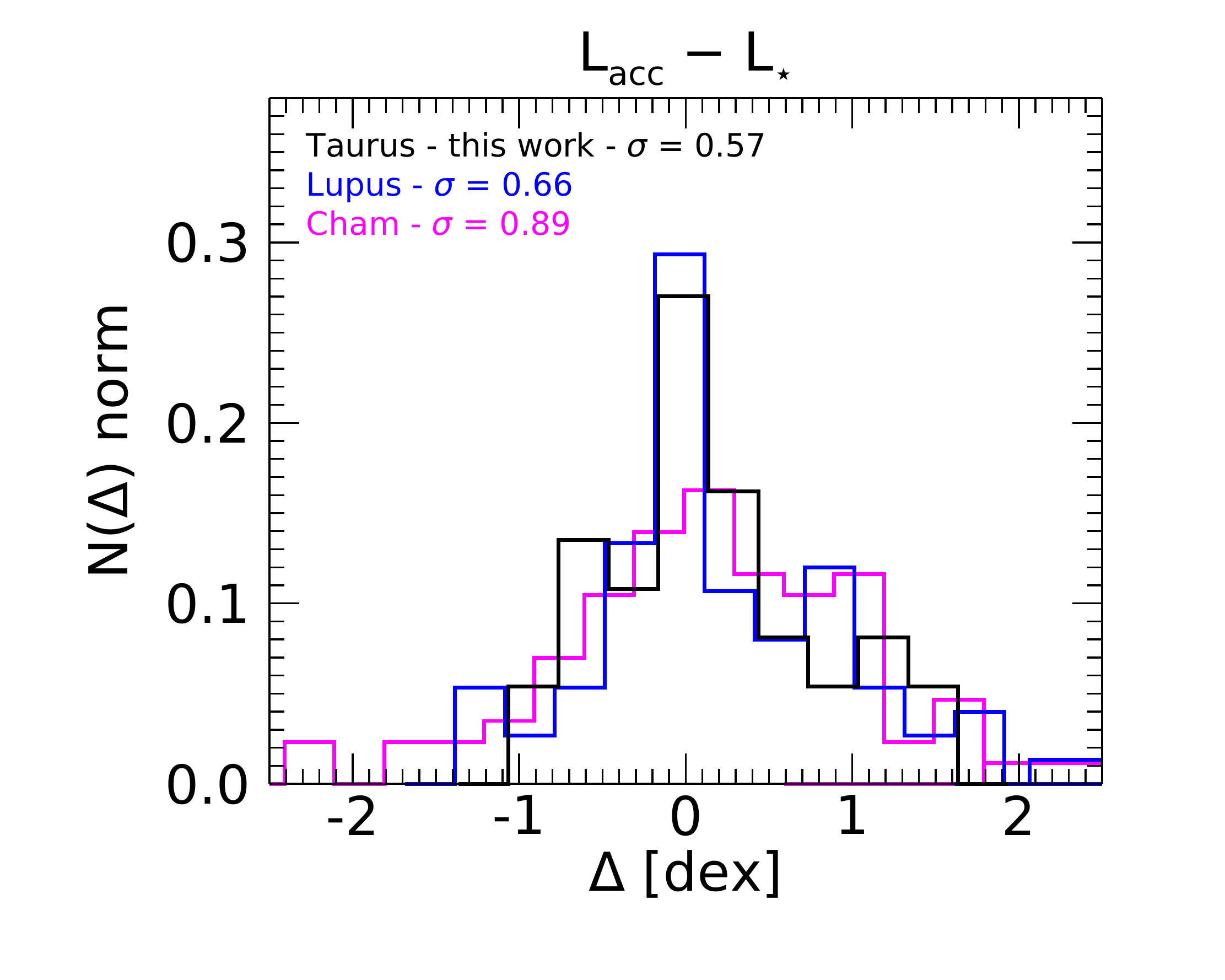}
\includegraphics[trim=40 40 20 0,width=1\columnwidth]{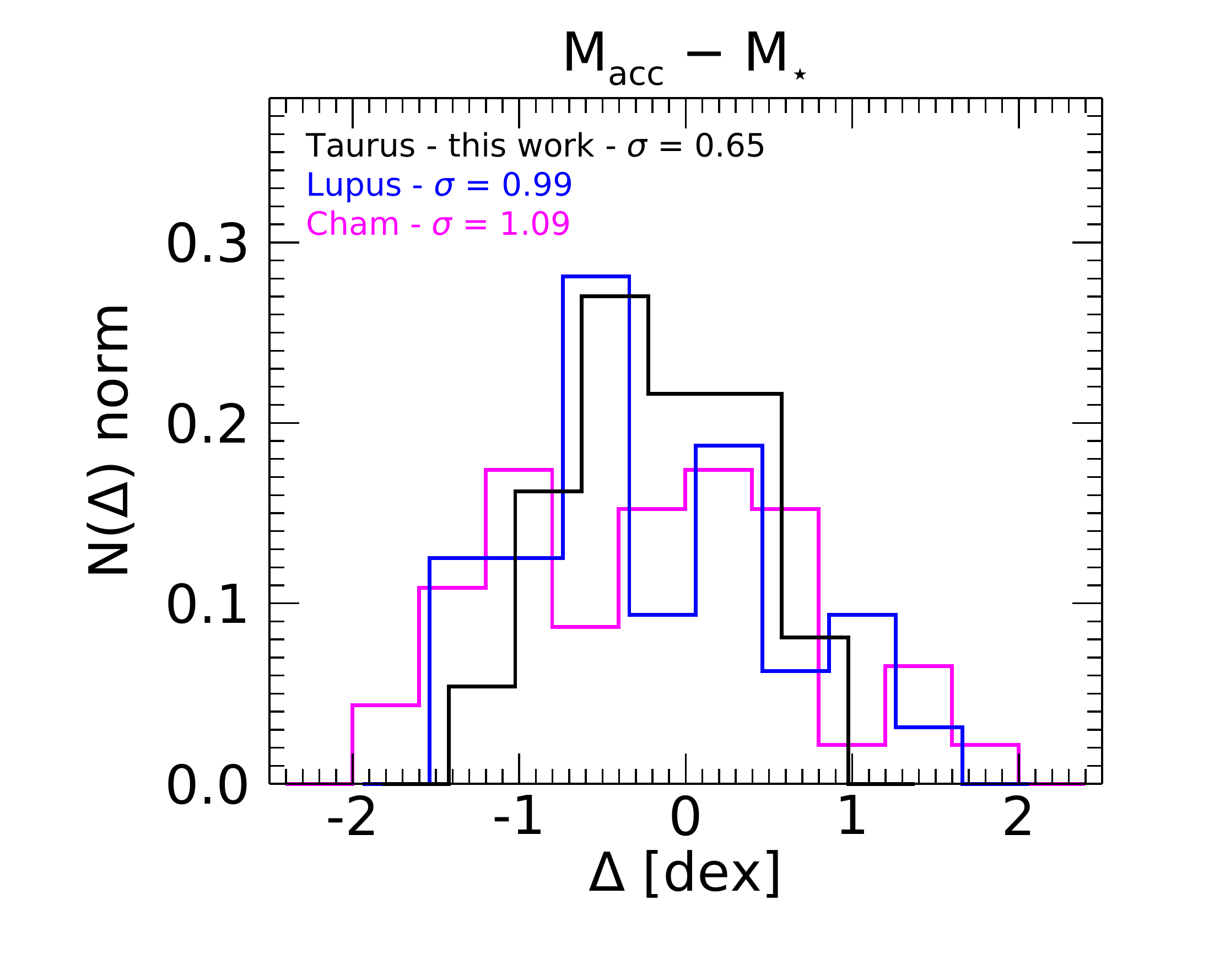}
\end{center}
\caption{\label{fig:Accretion_hist} Normalized distribution of the residuals from the best-fitting $\rm L_{acc} - L_{\star}$ (lef panel) and $\rm \dot{M}_{acc} - M_{\star}$ (right panel) relation,  computed within the same $\rm L_{\star}$ and $\rm M_{\star}$ ranges as were probed with the GHOsT data. In black we show the distribution for the Taurus sample, and the distributions of Lupus and Chamaeleon I are reported in blue and magenta, respectively. The spread of the distributions is labeled in the panels. They are computed as the 1$\rm \sigma$ of the Gaussian fits to the distributions.}
\end{figure*}

\subsection{Accretion luminosity versus wind and jet properties}
The physical properties of protoplanetary-disk (PPD) winds and jets can be investigated through the emission of atomic or weakly ionized forbidden lines in the optical and infrared spectral ranges \citep[e.g.,][]{Giannini2019}. These lines usually display a composite profile, with a high-velocity component (HVC, $\rm |v| >$  $\rm 40$ $\rm km$ $\rm s^{-1}$), associated with extended collimated jets, and a low-velocity component (LVC, $\rm |v| < $ $\rm 40$ $\rm km$ $\rm s^{-1}$), attributed to 0.5-10 AU PPD winds \citep{Hartigan1995}. In some cases, the LVC shows a composite profile where a broad low-velocity component (BLVC; FWHM $\rm \gtrsim 40\ km\ s^{-1} $) and a narrow low-velocity component (NLVC; FWHM $\rm \lesssim 40\ km\ s^{-1} $) can be distinguished \citep{Rigliaco2013, Simon2016, Banzatti2019, Gangi2020}. Of the main tracers of atomic winds, the [\ion{O}{I}] $\rm 630$ $\rm nm$ line is the brightest and is most frequently studied to infer the properties of PPD winds. Molecular winds have also been observed, and from the ground, they are usually traced through high-resolution spectroscopy from the UV to the IR by studying specific bands such as those of $\rm H_2$, CO, $\rm H_2O$, and OH \citep[e.g.,][]{Najita2007,France2012,Banzatti2022}. The band that falls within the GIARPS spectral range is the ro-vibrational 1-0 S(1) transition of molecular hydrogen at 2.12 $\rm \mu m$.

The properties of the PPD atomic and molecular winds and jets have previously been analyzed for some subsamples of the GHOsT program in Paper I and II, while those of the complete sample will be presented in a forthcoming work (Nisini et al., in preparation). Here we investigate the connection with the accretion properties. The [\ion{O}{I}] line luminosity ($\rm L_{line}$) was found to correlate with $\rm L_{acc}$ and was used to measure the mass accretion rate \citep[e.g.,][]{Herczeg2008}. This correlation is still valid when the LVC and HVC luminosities are considered separately, as found by \citet{Rigliaco2013} and \citet{Simon2016} in a sample of Taurus sources with an inhomogeneous set of data, and by \citet{Nisini2018} for a large sample of CTTs from the Lupus, Chamaeleon, and $\sigma$-Orionis SFRs. This supports the scenario in which the physical origin of the two line components is related to the accretion mechanism. In Fig. \ref{fig:compare_lum_lacc} we report the line luminosities of the [\ion{O}{I}] NLVC (which is the most frequently detected LVC, see Paper II) and HVC and $\rm H_2$ as a function of the accretion luminosity. Line luminosities were computed from the deconvolved fluxes of Paper II, corrected for the values of $A_v$ reported in Table \ref{tab:sources_param} and rescaled to the new Gaia EDR3 distances. In addition, as suggested by \citet{Mendigutia2015}, both quantities were normalized at the stellar luminosity to take off the well-known $\rm L_{acc} - L_{\star}$ and $\rm L_{line} - L_{\star}$ correlations. Despite the low statistics and the small $\rm L_{acc}$ range that was probed, we can confirm the correlation between the luminosity of the [\ion{O}{I}] NLVC and HVC  and $\rm L_{acc}$. The best linear regression fits in log scale give
\begin{eqnarray}
\rm log{\frac{L_{\ion{[O}{i]},NLVC}}{L_{\star}}} = 0.44 (\pm 0.13)\ log{ \frac{L_{acc}}{L_{\star}}} -4.17 (\pm 0.16), \\
\rm log{\frac{L_{\ion{[O}{i]},HVC}}{L_{\star}}} = 0.82 (\pm 0.24)\ log{ \frac{L_{acc}}{L_{\star}}} -3.59 (\pm 0.22).
\end{eqnarray}

For both components we then find a slope that agrees with the results of \citet{Nisini2018}, namely 0.4 for the LVC and 0.7 for the HVC, respectively, confirming that the HVC line luminosity decreases more steeply for low accretion luminosities. This behavior was also pointed out in \citet{Rigliaco2013}. Regarding the hydrogen molecule component, we find no correlation between the $\rm H_2$ line luminosity and $\rm L_{acc}$. As we pointed out in paper II, the detection of $\rm H_2$ 2.12 $\rm \mu m$ depends on the degree of $\rm H_2$ dissociation in the wind, a process that might depend on several causes (i.e., stellar luminosity or radial wind extension) that are not necessarily connected with accretion.

In addition to the $\rm L_{line} - L_{acc}$ correlation, \citet[][henceforth B19]{Banzatti2019} found a correlation between the deprojected HVC peak velocity ($\rm HVC_{vel}$) and $\rm L_{acc}$. In Fig. \ref{fig:Lacc_vs_HVC} we plot $\rm HVC_{vel}$ as a function of $\rm L_{acc}$ for our sample and for the B19 sample. The latter is based on the Taurus, Lupus, Ophiucus, Corona Australis, and TW Hya star-forming regions, and their deprojected $\rm HVC_{vel}$ values were updated here considering the latest measurements of disk inclinations available in the literature. Our data confirm the trend found in B19. 
The best-fit linear regression gives
\begin{eqnarray}
\rm GHOsT: dHVC_{vel} = -90 (\pm 26)\ log\ L_{acc} -219 (\pm 29), \\
\rm B19: dHVC_{vel} = -124 (\pm 46)\ log\ L_{acc} -295 (\pm 55).
\end{eqnarray}

Moreover, we find a smaller scatter ($\sim 40$ $\rm km$ $\rm s^{-1}$) than B19, which could be explained by the homogeneity and simultaneous measurements of $\rm L_{acc}$ and $\rm dHVC_{vel}$ in our sample. 

Finally, in Fig. \ref{fig:Lacc_vs_HVC} we also indicate with a cross the $\rm L_{acc}$ for the sources for which the \ion{[O}{i]} HVC was not detected. The distribution of these points within the wide range of $\rm L_{acc}$ supports the idea,  already suggested in \citet{Nisini2018}, that the presence of a jet only marginally depends on the accretion level of the source.

\section{Discussion}\label{sec:discussion}
Our survey of the accretion properties of YSOs in the Taurus-Auriga population confirms the results found in other star-forming regions with similar age. Although the incompleteness of our sample prevents us from deriving global results for the whole Taurus population, we can analyze some general properties of the accretion process at this stage of stellar evolution, taking advantage of (i) the self-consistency of the method we adopted to derive both stellar and accretion properties, (ii) the detailed knowledge of the disk structures from the complementary ALMA observations that are available for the majority of our sample, and (iii) the knowledge of the jet and disk wind properties derived in Paper I and II using the same set of data.

\subsection{Dispersion on the $\rm L_{acc}-L_{\star}$ and $\rm \dot{M}_{acc} - \rm M_{\star}$ planes}
\begin{figure}
\begin{center}
\includegraphics[trim=15 40 20 0,width=1\columnwidth]{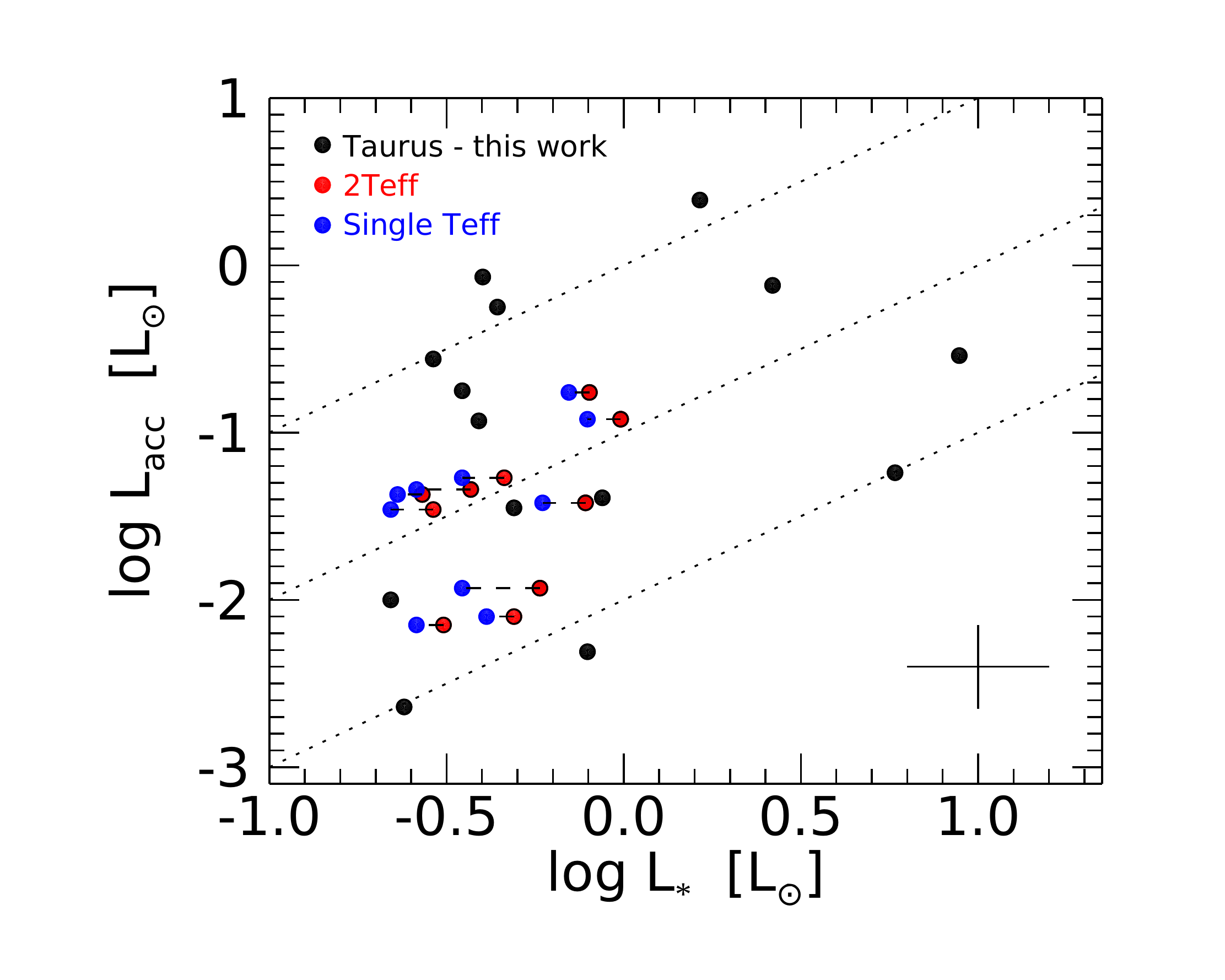}
\end{center}
\caption{\label{fig:Lacc_vs_Lstar_disp2Teff} Accretion luminosity as a function of stellar luminosity for the Taurus sample studied in this work. The position of heavily spotted sources, with $\rm T_{eff}$ and $\rm L_{\star}$ based on the 2 $\rm T_{eff}$ fit, is shown as red points, and the positions obtained assuming a single $\rm T_{eff}$ are shown as blue points. The position of sources without evidence of spots is shown as black points. Differences between the two methods are within the averaged errors in $\rm L_{\star}$ and do not affect the $\rm L_{acc}-L_{\star}$ distribution.}
\end{figure}

\begin{figure*}
\begin{center}
\includegraphics[trim=30 40 20 0,width=1\columnwidth]{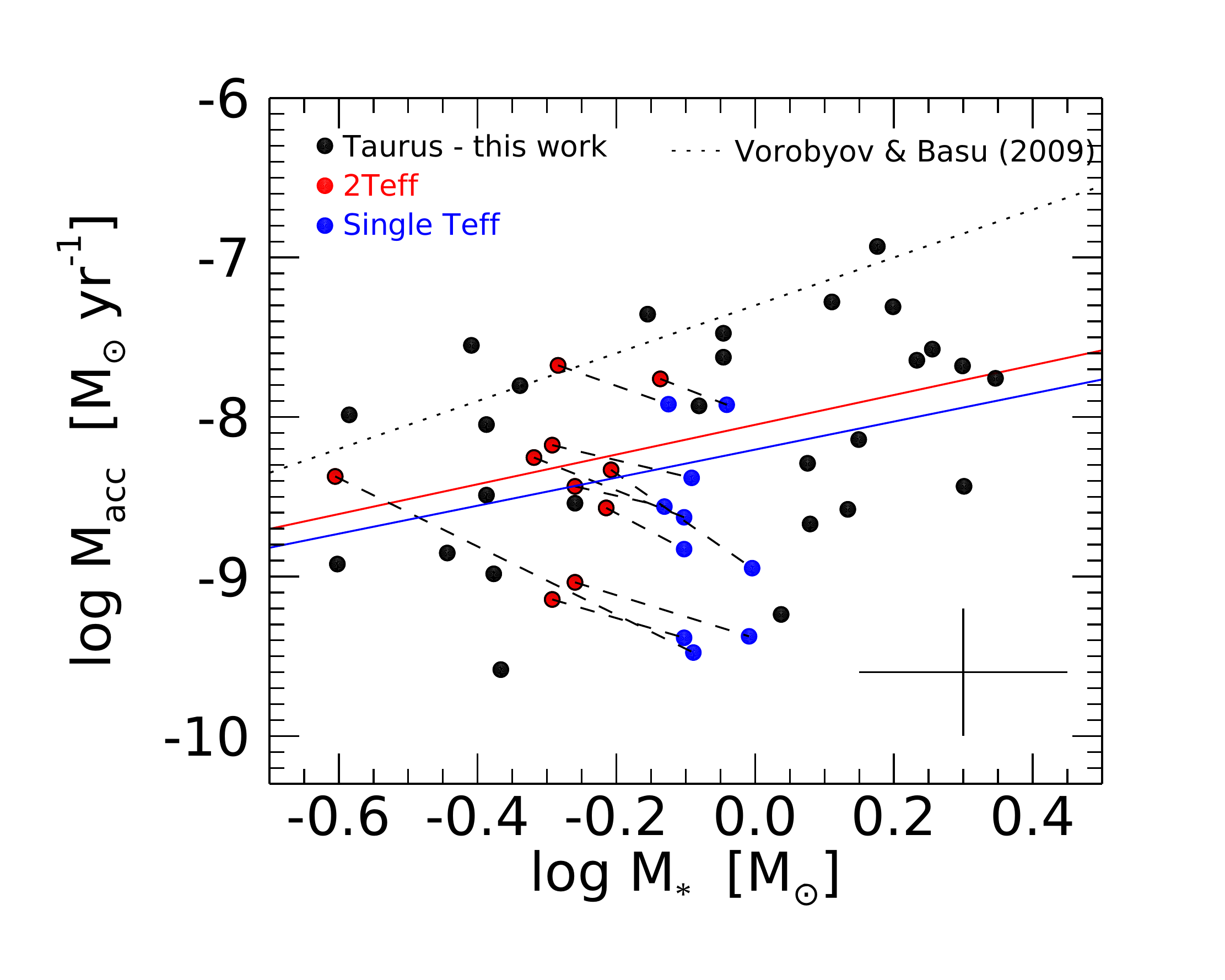}
\includegraphics[trim=30 40 20 0,width=1\columnwidth]{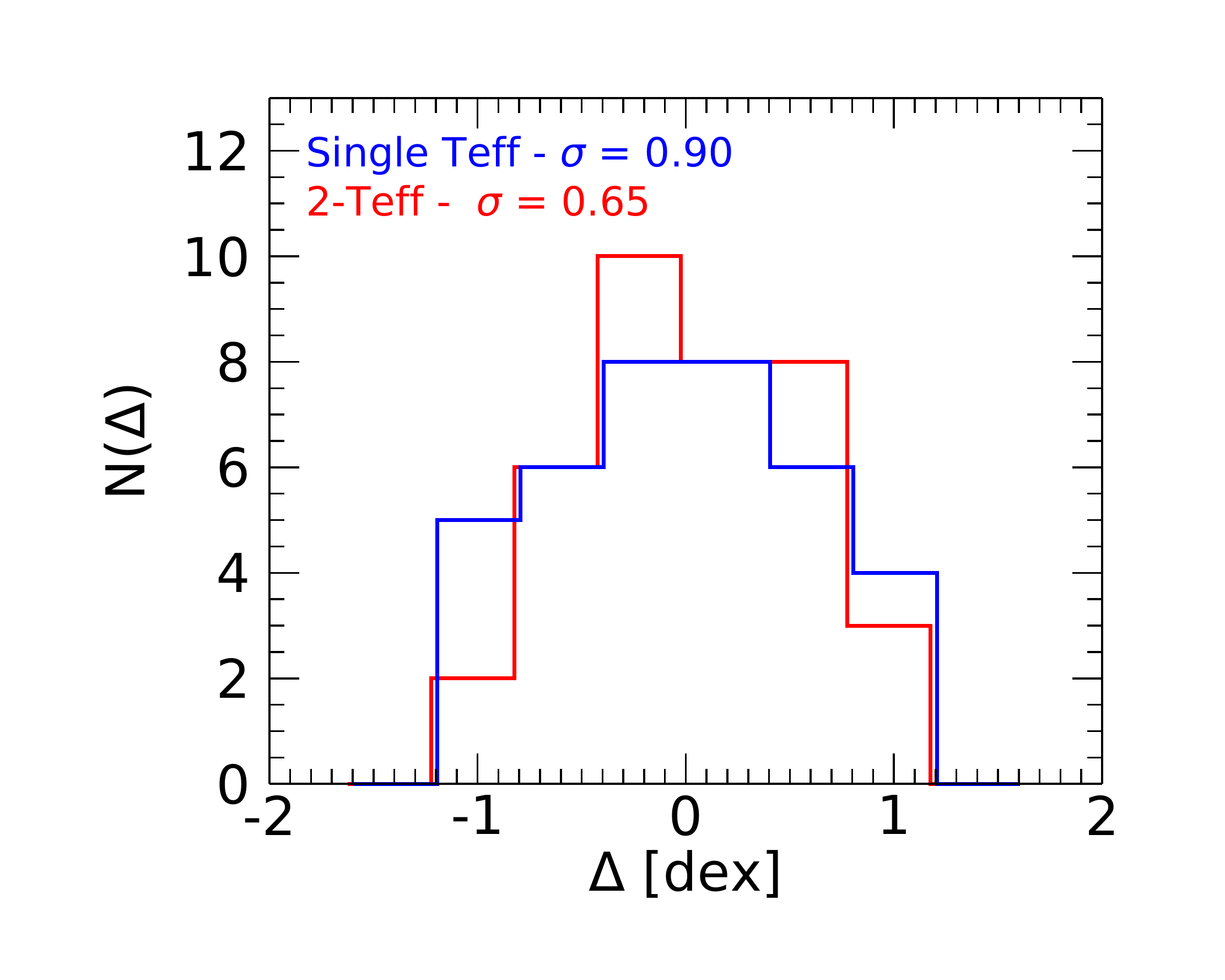}
\end{center}
\caption{\label{fig:Macc_vs_Mstar_disp2Teff} Left: As in Figure \ref{fig:Lacc_vs_Lstar_disp2Teff}, but for mass accretion rate and stellar mass. Differences between the two methods affect the $\rm \dot M_{acc}$ and $\rm M_{\star}$ directions.  Fits to the black and red distributions and black and blue distributions are shown as solid red and blue lines, respectively.  Right: Distribution of the dispersion from the individual best-fitting $\rm M_{acc}-M_{\star}$ relation considering the total sample with the 2 $\rm T_{eff}$ (red) and the single $\rm T_{eff}$ (blue) modeling for the spotted source. The dispersion of the distributions is labeled in the panel.}
\end{figure*}

We find a dispersion of $\rm \sim 0.6$ dex for the $\rm L_{acc}-L_{\star}$ and $\rm \dot{M}_{acc} - \rm M_{\star}$ relations. The dispersion is computed as the 1$\rm \sigma$ of the Gaussian fit to the distribution of the residuals, which are defined as the distances of the individual $\rm L_{acc}$ and $\rm \dot{M}_{acc}$ values from the respective best-fit relations. Fig. \ref{fig:Accretion_hist} shows the normalized distributions of the residuals obtained for the Taurus sample and compared with those of the Lupus and Chamaeleon I. For these latter we limited the computation of the residuals to the same $\rm L_{\star}$ and $\rm M_{\star}$ ranges as were probed with the GHOsT data to avoid possible biases due to the adoption of different abscissa ranges. While the dispersion on the $\rm L_{acc}-L_{\star}$ relation is similar for the three samples, the dispersion around the $\rm \dot{M}_{acc} - \rm M_{\star}$ relation in our sample appears to be lower than those measured in the Lupus and Chamaeleon I datasets. 
We emphasize that the stellar and accretion parameters were measured in a self-consistent way in all three samples, so that this discrepancy cannot be an effect of the inhomogeneity of the data. On the other hand, for Taurus, we adopted a proper modeling to derive the stellar properties in case of heavily spotted sources by performing a 2 $\rm T_{eff}$ component fit for the photospheric emission. In Fig. \ref{fig:Lacc_vs_Lstar_disp2Teff} and \ref{fig:Macc_vs_Mstar_disp2Teff} we compare the $\rm L_{acc}-L_{\star}$ and $\rm \dot{M}_{acc} - \rm M_{\star}$ distributions obtained assuming both a single and a 2 $\rm T_{eff}$ modeling for the same subsample of sources with a strong evidence of spots. The $\rm L_{acc}-L_{\star}$ relation is not significantly influenced by the choice of the method because this latter only affects the $\rm L_{\star}$ measurements. As already noted in Sect. \ref{sec:params_spotted} the stellar luminosities derived with a 2 $\rm T_{eff}$ modeling are systematically higher than those measured from a single $\rm T_{eff}$, but the differences are always within the typical errors. In contrast, the dispersion in the $\rm \dot{M}_{acc} - \rm M_{\star}$ plane seems to be affected in a significant way: the adoption of a single $\rm T_{eff}$ modeling for the determination of stellar parameters in spotted sources tends to increase the dispersion in the $\rm \dot{M}_{acc} - \rm M_{\star}$ plane by $\sim 0.25 $ dex (Fig. \ref{fig:Macc_vs_Mstar_disp2Teff}) in such a way that we obtain a dispersion value that is more in line with those measured in the Lupus and Chamaeleon I samples. This suggests that the lower dispersion found in the $\rm \dot{M}_{acc} - \rm M_{\star}$ relation for the Taurus sample might be the result of our accurate determination of the stellar parameters in sources with a strong presence of spots.

Even considering possible biases in the $\rm \dot{M}_{acc}$ measurements, it is well established that large dispersions are a common characteristics of different SFR populations, with typical values of $\rm \sim 0.75$ dex and a total range in log $\rm \dot{M}_{acc}$ and log $\rm L_{acc}$ of about two orders of magnitude at a given stellar mass and luminosity \citep[e.g.,][]{Hartmann2016, Testi2022}. All of the observational efforts currently conclude that much of this dispersion is real, that is, not due to observational and methodological biases, and it cannot be entirely attributed to the scatter due to the variability of the accretion processes \citep[e.g.,][]{Biazzo2012,Venuti2014}. 

Several processes are usually invoked to explain the observed spreads. In the framework of a pure viscous evolution of the disk, it has been suggested that the mass accretion rate should decline with age and would be prevented by the stellar magnetosphere at ages $\rm \gtrsim 10$ Myr \citep[][]{Hartmann1998bis}. A tentative decrease in mass accretion rates with age has been found in samples from SFR with different ages \citep[e.g.,][]{Sicilia2010, Antoniucci2014, Testi2022}, but the uncertainties in measuring reliable individual ages make it very difficult to test any dependence of the mass accretion rates on stellar age in the individual SFR. 

A different way to indirectly investigate the effect of disk evolution on the accretion process is to search for any trend with the disk dust structure. In the current idealized picture of the protoplanetary disk evolution, the disk dispersal is predominantly guided by the interplay between accretion and mass loss through disk winds. While the disk evolution is initially driven by accretion, mass-loss through low-velocity winds starts to significantly alter the disk density and dominates the accretion on a timescale of a few million years. As a consequence the disk gas is rapidly cleared inside-out leading to the formation of transitional and substructured disks \citep[e.g.,][]{Ercolano2017}. In this framework, at a given stellar mass, it is expected that transitional disks and/or disks with substructures have relatively lower accretion rates than full disks. In our sample, with the limited statistics we have on hand, we note that TD sources tend to have lower values of accretion of $\rm L_{acc}$ (they fill the region of the plane where $\rm L_{acc}=0.1 L_{\star}$ and $\rm L_{acc}=0.01 L_{\star}$) and $\rm \dot{M}_{acc}$ on average, in line with this scenario and with other previous observations \citep[e.g.,][and references therein]{Espaillat2014}. However, this does not effectively contribute to an increase in spread in the relations because sources with full disks present the largest spread and homogeneously fill the $\rm L_{acc}-L_{\star}$ and $\rm \dot{M}_{acc} - M_{\star}$ planes. This indicates that the complexity in disk structure is not the main tracer of the accretion level.

\subsection{Accretion and disk mass}
Viscous evolution models predict a correlation between disk mass and mass accretion rate \citep[e.g.,][]{Lynden1974,Hartmann1998bis,Rosotti2017} that can be expressed as
\begin{eqnarray}
\rm M_{disk} = t_{disk} \dot M_{acc} \propto ( t + t_{v}) \dot M_{acc}
,\end{eqnarray}

with $\rm t_{disk}$ the disk lifetime and $\rm t_{v}$ the viscous time. The $\rm M_{disk}/\dot M_{acc}$ ratio is then driven by the disk age if $\rm t \gg t_v $ or by the viscous time if $\rm t \ll t_v$ \citep[e.g.,][]{Lodato2017}. 

From an observational point of view, the $\rm \dot{M}_{acc} - \rm M_{disk}$ correlation was confirmed for the 1-3 Myr old SFRs of Lupus \citep{Manara2016} and Chamaeleon I \citep{Mulders2017}. However, these studies found large spreads of $\rm \sim 1$ dex around the $\rm \dot{M}_{acc} - \rm M_{disk}$ relation, which could be compatible with a purely viscous model only assuming large spreads on the viscous timescales and typical values of about the stellar population age. As a consequence, the spread around the $\rm \dot{M}_{acc} - \rm M_{disk}$ relation in the old populations of SFRs is expected to be reduced to the typical uncertainties of the $\rm \dot{M}_{acc}$ measurements. However, this was not the case for the > 5Myr old SFR of Upper Scorpius, where a large spread > 0.9 dex and median high values of  $\rm \dot{M}_{acc}$ were found by \citet{Manara2020}. This evidence has driven the development of different models for disk evolution, alternative or in parallel to the simple viscous-driven model. For example, recent models predict that disk winds might have a large influence in driving accretion by removing the disk angular momentum excess \citep[e.g.,][]{Bai2016}. These wind-driven models moreover appear to be able to reproduce the large spreads observed in the $\rm \dot{M}_{acc} - \rm M_{disk}$ correlation. In addition, both internal and external photoevaporation are expected to play an important role in the evolution of the $\rm M_{disk}$/$\rm \dot{M}_{acc}$ ratio with the stellar age and might contribute to increase the scatter of the correlation \citep[]{Jones2012,Rosotti2017,Somigliana2020}. Our finding that TD shows lower $\rm \dot{M}_{acc}$ values than sources with the same $\rm M_{disk}$ supports the hypothesis that photoevaporation can play an important role. 

In addition, \citet{Zagaria2022} recently found that binaries or high-order multiple star systems present higher values of $\rm \dot{M}_{acc}$ than single stars at a given $\rm M_{disk}$ for the SFR populations of Lupus, Chamaeleon I and Upper Scorpius. This can be explained by the fact that disks in multiple stars are subject to tidal truncation, which increases the grain radial drift and reduces the disk lifetime. In our sample we have found two distinct distributions in the $\rm \dot{M}_{acc} - \rm M_{dust,disk}$ plane between single and multiple sources, which is evidence that can be compatible with the scenario described in \citet{Zagaria2022}.

Finally, an additional cause for the high dispersion observed in the $\rm \dot{M}_{acc} - \rm M_{disk}$ plane can be found in the lower spread in the $\rm \dot{M}_{acc} - \Sigma_{disk,dust}$ correlation with respect to the $\rm \dot{M}_{acc} - \rm M_{disk}$ (Fig. \ref{fig:Macc_Mdisk}). Despite the lower statistics of our sample and the small $\rm M_{disk}$ range we probed, it seems that plotting the mass accretion rate as a function of the disk surface density significantly reduces the spread in the correlation. We speculate that the reason might be that millimeter emission is not totally optically thin, which can cause $\rm M_{dust,disk}$ to be not properly estimated. A number of works have already pointed out that $\rm M_{dust,disk}$ might be underestimated when measured from millimeter flux density alone \citep[e.g.,][]{Ballering2019, Zhu2019, Ribas2020}. Millimeter emissions partially or totally dominated by optically thick structures have been invoked to explain the size-luminosity relation found in 50 nearby PPDs \citep{Tripathi2017} as well as in multiband ALMA observations for a sample of 26 bright PPDs in the Lupus SFR \citep{Tazzari2021}. \citet{Ueda2022} found that in the case of CW Tau, the disk mass can be underestimated due to opacity effects if the measurement is carried out from ALMA bands 6-8 (1.3-0.75 mm), highlighting again the importance of multiband observations and a proper modeling of the dust opacity.

In conclusion, the large spread observed in the $\rm \dot{M}_{acc} - \rm M_{dust,disk}$ plane can be partially due to source multiplicity, as found in \citet{Zagaria2022}. However, it might also be due to systematics in the computation of $\rm M_{dust,disk}$ stemming from the adoption of the same dust opacity for all three disk categories. In this case, considering the surface density instead of the disk mass would mitigate this systematics.

\subsection{Accretion, wind and jets}
We have shown the clear correlation between the luminosity of the \ion{[O}{i]} LVC and HVC and $\rm L_{acc}$ for the Taurus sample with the use of self-consistent and homogeneous data. We have found a relation similar to that found by \citet{Nisini2018} for the sources in Lupus, Chamaeleon, and $\rm \sigma$-Ori in terms of slope and spread, suggesting a common behavior among different SFRs. As pointed out by \citet{Nisini2018}, this correlation, which persists even when both luminosities are normalized to the stellar luminosity, suggests a common mechanism connected to the accretion process for these two components. However, we note that the slight difference in the slope between the $\rm L_{\ion{[O}{i]},LVC}$-$\rm L_{acc}$ and $\rm L_{\ion{[O}{i]},HVC}$-$\rm L_{acc}$ relations seems to be real and might support a scenario in which the two components originate from distinct mechanisms, but both are related to the accretion. In particular, while MHD jet-formation models predict a direct correlation between $\rm L_{\ion{[O}{i]},HVC}$ and $\rm L_{acc}$ driven by the $\rm \dot M_{jet}/\dot M_{acc}$ relation, the LVC could be dominated by photoevaporation, as suggested by \citet{Weber2020}. Models that simultaneously address the physics of PPD winds and jets are necessary to confirm this scenario.

Regarding the kinematics of the jets, the correlation between the \ion{[O}{i]}630 nm HVC velocity and accretion luminosity confirms the finding of \citet{Banzatti2019}, that is, faster jets are driven by stronger accretion. However, we note that the opposite is not always valid because we found strong accretors with a lack of HVC in the \ion{[O}{i]} profile. Two possible interpretation can be at the basis of the observed correlation.

In MHD jets, the mass ejection rate, $\rm \dot M_{jet}$, is directly related to the mass accretion rate and thus to the accretion luminosity. $\rm \dot M_{jet}$ can be expressed as $\rm \dot M_{jet} = M_{gas} v_{jet}/l_{jet}$, where $\rm M_{gas}$ is the mass of the gas, and $\rm v_{jet}$ and $\rm l_{jet}$ are the velocity and length of the jet, respectively. Thus, a direct correlation between $\rm v_{jet}$ and $\rm L_{acc}$ can be found provided that the mass of the gas in the considered jet section is constant. This latter condition is hardly met, however. Studies of the physical conditions in some of the considered sources (e.g., Paper I) show that they can vary significantly from one source to the next, with the total density ranging between $10^5$ $\rm cm^{-3}$ and more than $10^7$ $\rm cm^{-3}$, making a direct proportionality between $\rm L_{acc}$ and $\rm v_{jet}$ difficult.

\begin{figure}[!h]
\begin{center}
\includegraphics[trim=100 60 120 10,width=0.7\columnwidth, angle=0]{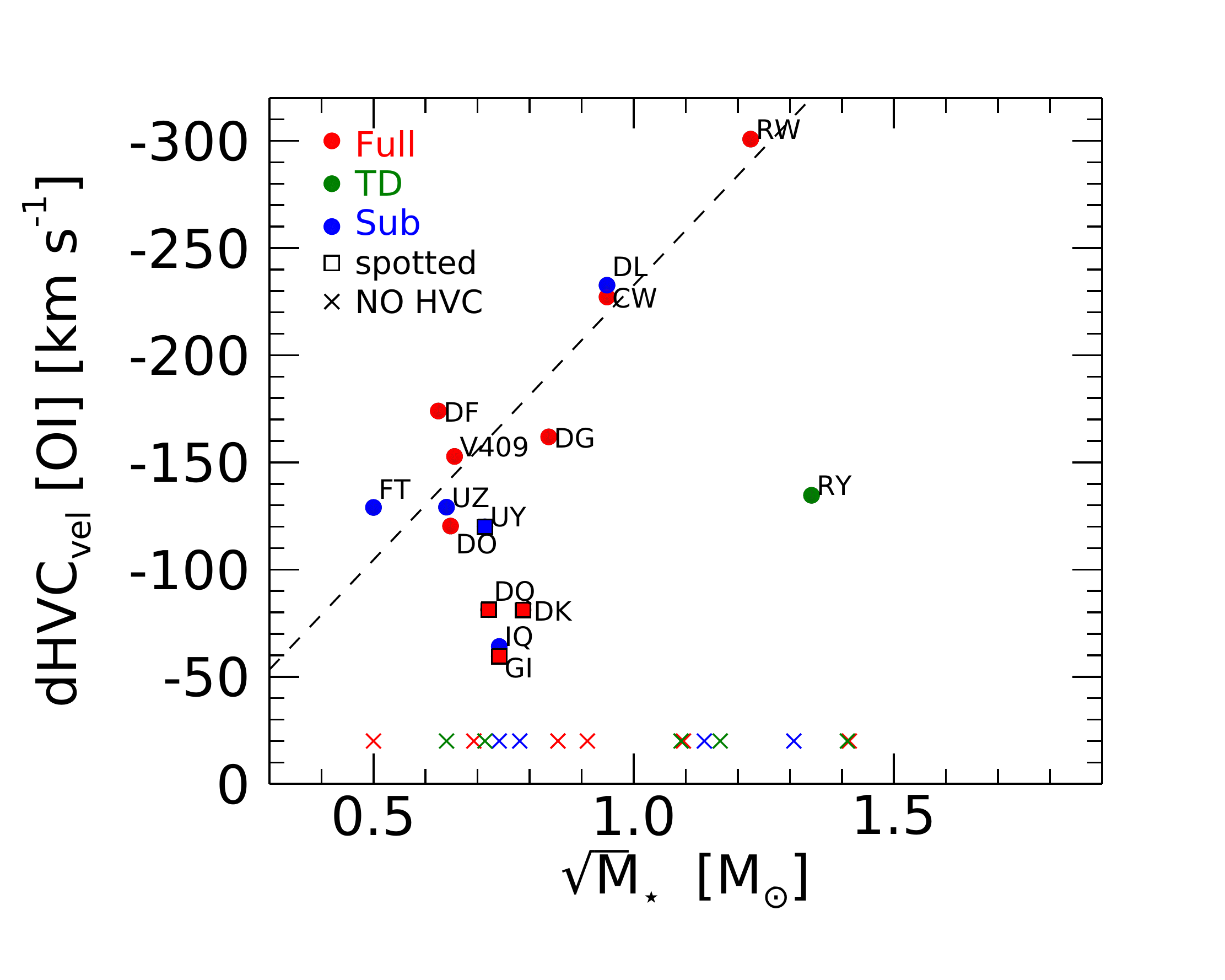}
\end{center}
\caption{\label{fig:Mstar_vs_HVC} Deprojected \ion{[O}{i]}630nm HVC peak velocities as a function of $\rm \sqrt{\rm M_{\star}}$ for the GHOsT subsample showing jets. The relevant properties of the sources are indicated in the legend.}
\end{figure}

On the other hand, in the framework of the standard MHD model, the poloidal jet velocity $\rm v_{jet}$ is proportional to $\rm (r_A/r_0)v_k$, with $\rm r_A$ the Alv\'en radius, $\rm r_0$ the footprint radius (i.e., the radius of the disk where the jet originates), and $\rm v_k$ the disk Keplerian velocity at distance $\rm r_0$. The latter can be expressed as $\rm v_K = \sqrt{\rm GM_{\star}/r_0}$, thus making $\rm v_{jet}$ proportional to $\rm \sqrt{\rm M_{\star}}$. In this case, the observed HVC velocity and $\rm L_{acc}$ dependence would be a consequence of the fact that both the $\rm v_{jet}$ and $\rm \dot M_{acc}$ correlate with $\rm M_{\star}$. In support of this, Fig. \ref{fig:Mstar_vs_HVC} shows the deprojected HVC peak velocity as a function of $\rm \sqrt{\rm M_{\star}}$. Except for a few exceptions, we find a moderate correlation between the two quantities, which might be compatible with this scenario. We note that RY Tau, which presents the largest discrepancy with respect to the supposed correlation, is the only source with a transitional disk showing an HVC in the considered sample. In this case, the lower deprojected HVC velocity could be the result of a larger footprint radius, which is expected if the very inner disk region is devoid of both dust and gas. Other cases that deviate from the suggested correlation include four sources (i.e. DQ Tau, DK Tau, GI Tau and IQ Tau) for which the deprojected HVC peak velocities are particularly low (lower then $\sim$ -80 km/s), thus making the interpretation of these line components as tracers of jets doubtful. On the other hand, three of them were identified in this work as heavily spotted sources. In this case, we might also speculate that the MHD jet acceleration could be significantly reduced due to the higher-order complexity of the stellar magnetic field expected in these sources. A higher statistics and a detailed knowledge of the magnetic field topology are necessary to draw any firm conclusion on this latter point.

\section{Conclusions}\label{sec:conclusions}
In the framework of the GHOsT project we have presented a study of the accretion properties of 37 CTTs of the Taurus-Auriga population with masses $\rm > 0.2$ $\rm M_{\sun}$. We used high-resolution spectroscopic observations taken with the GIARPS instrument in a wide range from the optical to the NIR to measure the stellar and accretion parameters in a homogeneous and self-consistent fashion. The complementary ALMA data also enabled us to compared these parameters with the physical properties of the disks. The main results of our study are summarized below.
\begin{itemize}
    \item The standard approach of inferring effective temperatures solely from high-resolution optical or infrared spectra could be inappropriate in case of heavily spotted sources because it can lead to effective temperatures that are systematically too hot or too cold when the measurement is performed from the visible and IR spectral range, respectively. As a consequence, the determination of the stellar luminosity, mass, and age will be strongly affected by this systematics. We showed that the adoption of a 2-$\rm T_{eff}$ modeling of the photospheric emission can mitigate this discrepancy and provides reliable estimates of stellar parameters for stars in which large areas are affected by spots.
    \item The $\rm L_{acc}-L_{\star}$ and $\rm \dot{M}_{acc} - \rm M_{\star}$ relations derived for the Taurus sample are consistent with those measured in SFRs of similar age such as Lupus and Chamaeleon I but with a lower spread in the $\rm \dot{M}_{acc} - \rm M_{\star}$ relation. This latter could be a consequence of properly addressing the stellar properties of heavily spotted sources.
    \item Transitional disk sources are placed on the lower side of the distributions of accretion luminosity and mass accretion rate when compared to the bulk of the CTTSs with full disks. This is in line with the currently accepted scenario for the protoplanetary disk evolution. However, the complexity of the disk structure is not the main tracer of the accretion level because the largest spread around the $\rm L_{acc}-L_{\star}$ and $\rm \dot{M}_{acc}-M_{\star}$ relations is dominated by sources with a full disk. 
    \item Multiple star systems display higher values of $\rm \dot{M}_{acc}$ than single stars at a given $\rm M_{dust,disk}$, which partially explains the observed $\rm \dot{M}_{acc} - \rm M_{dust,disk}$ dispersion. The high dispersion can also be partially due to a bias in computing the $\rm M_{dust,disk}$ with the same disk opacity for all sources. This latter can be mitigated inspecting the $\rm \dot{M}_{acc} - \rm \Sigma_{dust,disk}$ dependence, with $\rm \Sigma_{dust,disk}$ the surface dust mass density of the disk, which is less sensitive to the assumptions on the optical thickness of the emitting dust.
    \item The luminosities of the \ion{[O}{i]} 630 nm NLVC and HVC were found to correlate with the accretion luminosity, suggesting that the excitation of the two components is related to the accretion mechanisms. In contrast, we did not find any clear correlation between the  $\rm H_2$ 2.12 $\rm \mu m$ line luminosity and $\rm L_{acc}$, supporting the conclusion of Paper II that the detection and luminosity of $\rm H_2$ depend on local conditions for $\rm H_2$ survival in the upper disk surface under the action of FUV stellar photons.
    \item The deprojected peak velocity of the \ion{[O}{i]} 630 nm HVC correlates with the accretion luminosity, indicating that faster jets are driven by stronger accretion. We discussed different explanations for this correlation and concluded that the most plausible explanation is related to the fact that both $\rm v_{jet}$ and $\rm L_{acc}$ correlate with the square root of the stellar mass.
\end{itemize}

\begin{acknowledgements}
This work has been supported by the projects PRIN-INAF-MAIN-STREAM 2017 “Protoplanetary disks seen through the eyes of new-generation instruments”, PRIN-INAF 2019 "Spectroscopically Tracing the Disk Dispersal Evolution" and PRIN-INAF 2019 "Planetary systems at young ages (PLATEA)".
Funded by the European Union under the European Union’s Horizon Europe Research \& Innovation Programme 101039452 (WANDA). Views and opinions expressed are however those of the author(s) only and do not necessarily reflect those of the European Union or the European Research Council. Neither the European Union nor the granting authority can be held responsible for them.
\end{acknowledgements}

%

\begin{thebibliography}{}
\bibitem[Agra-Amboage et al.(2011)]{Agra-Amboage2011} Agra-Amboage, V., Dougados, C., Cabrit, S., et al.\ 2011, \aap, 532, A59. doi:10.1051/0004-6361/201015886
\bibitem[Allard et al.(2012)]{Allard2012} Allard, F., Homeier, D., \& Freytag, B.\ 2012, Philosophical Transactions of the Royal Society of London Series A, 370, 2765. doi:10.1098/rsta.2011.0269
\bibitem[Alcal{\'a} et al.(2014)]{Alcala2014} Alcal{\'a}, J.~M., Natta, A., Manara, C.~F., et al.\ 2014, \aap, 561, A2. doi:10.1051/0004-6361/201322254
\bibitem[Alcal{\'a} et al.(2017)]{Alcala2017} Alcal{\'a}, J.~M., Manara, C.~F., Natta, A., et al.\ 2017, \aap, 600, A20. doi:10.1051/0004-6361/201629929
\bibitem[Alcal{\'a} et al.(2019)]{Alcala2019} Alcal{\'a}, J.~M., Manara, C.~F., France, K., et al.\ 2019, \aap, 629, A108. doi:10.1051/0004-6361/201935657
\bibitem[Alcal{\'a} et al.(2021)]{Alcala2021} Alcal{\'a}, J.~M., Gangi, M., Biazzo, K., et al.\ 2021, \aap, 652, A72. doi:10.1051/0004-6361/202140918
\bibitem[Alexander et al.(2006)]{Alexander2006} Alexander, R.~D., Clarke, C.~J., \& Pringle, J.~E.\ 2006, \mnras, 369, 229. doi:10.1111/j.1365-2966.2006.10294.x
\bibitem[Alexander et al.(2014)]{Alexander2014} Alexander, R., Pascucci, I., Andrews, S., et al.\ 2014, Protostars and Planets VI, 475. doi:10.2458/azu\_uapress\_9780816531240-ch021
\bibitem[Andrews et al.(2013)]{Andrews2013} Andrews, S.~M., Rosenfeld, K.~A., Kraus, A.~L., et al.\ 2013, \apj, 771, 129. doi:10.1088/0004-637X/771/2/129
\bibitem[Antoniucci et al.(2014)]{Antoniucci2014} Antoniucci, S., Garc{\'\i}a L{\'o}pez, R., Nisini, B., et al.\ 2014, \aap, 572, A62. doi:10.1051/0004-6361/201423929
\bibitem[Akeson et al.(2005)]{Akeson2005} Akeson, R.~L., Boden, A.~F., Monnier, J.~D., et al.\ 2005, \apj, 635, 1173. doi:10.1086/497436
\bibitem[Akeson et al.(2019)]{Akeson2019} Akeson, R.~L., Jensen, E.~L.~N., Carpenter, J., et al.\ 2019, \apj, 872, 158
\bibitem[Andrews et al.(2018)]{Andrews2018} Andrews, S.~M., Huang, J., P{\'e}rez, L.~M., et al.\ 2018, \apjl, 869, L41. doi:10.3847/2041-8213/aaf741
\bibitem[Antonellini et al.(2020)]{Antonellini2020} Antonellini, S., Banzatti, A., Kamp, I., et al.\ 2020, \aap, 637, A29. doi:10.1051/0004-6361/201834077
\bibitem[Antoniucci et al.(2017)]{Antoniucci2017} Antoniucci, S., Nisini, B., Biazzo, K., et al.\ 2017, \aap, 606, A48
\bibitem[Bacciotti et al.(2018)]{Bacciotti2018} Bacciotti, F., Girart, J.~M., Padovani, M., et al.\ 2018, \apjl, 865, L12. doi:10.3847/2041-8213/aadf87
\bibitem[Baffa et al.(2001)]{Baffa2001} Baffa, C., Comoretto, G., Gennari, S., et al.\ 2001, \aap, 378, 722
\bibitem[Bai(2016)]{Bai2016} Bai, X.-N.\ 2016, \apj, 821, 80. doi:10.3847/0004-637X/821/2/80
\bibitem[Ballering \& Eisner(2019)]{Ballering2019} Ballering, N.~P. \& Eisner, J.~A.\ 2019, \aj, 157, 144. doi:10.3847/1538-3881/ab0a56
\bibitem[Banzatti et al.(2019)]{Banzatti2019} Banzatti, A., Pascucci, I., Edwards, S., et al.\ 2019, \apj, 870, 76. doi:10.3847/1538-4357/aaf1aa
\bibitem[Banzatti et al.(2022)]{Banzatti2022} Banzatti, A., Abernathy, K.~M., Brittain, S., et al.\ 2022, \aj, 163, 174. doi:10.3847/1538-3881/ac52f0
\bibitem[Baraffe et al.(2015)]{Baraffe2015} Baraffe, I., Homeier, D., Allard, F., et al.\ 2015, \aap, 577, A42. doi:10.1051/0004-6361/201425481
\bibitem[Biazzo et al.(2009)]{Biazzoetal2009} Biazzo, K., Frasca, A., Marilli, E., et al. 2009, \aap, 499, 579
\bibitem[Biazzo et al.(2012)]{Biazzo2012} Biazzo, K., Alcal{\'a}, J.~M., Covino, E., et al.\ 2012, \aap, 547, A104
\bibitem[Bouvier et al.(2007)]{Bouvier2007} Bouvier, J., Alencar, S.~H.~P., Harries, T.~J., et al.\ 2007, Protostars and Planets V, 479
\bibitem[Brown et al.(1991)]{Brown1991} Brown, S.~F., Donati, J.-F., Rees, D.~E., et al.\ 1991, \aap, 250, 463
\bibitem[Calvet \& Gullbring(1998)]{Calvet1998} Calvet, N. \& Gullbring, E.\ 1998, \apj, 509, 802. doi:10.1086/306527
\bibitem[Calvet et al.(2004)]{Calvet2004} Calvet, N., Muzerolle, J., Brice{\~n}o, C., et al.\ 2004, \aj, 128, 1294. doi:10.1086/422733
\bibitem[Carleo et al.(2018)]{Carleo2018} Carleo, I., Benatti, S., Lanza, A. F., et al.\ 2018, \aap, 613, A50
\bibitem[Cieza et al.(2009)]{Cieza2009} Cieza, L.~A., Padgett, D.~L., Allen, L.~E., et al.\ 2009, \apjl, 696, L84. doi:10.1088/0004-637X/696/1/L84
\bibitem[Cieza et al.(2012)]{Cieza2012} Cieza, L.~A., Schreiber, M.~R., Romero, G.~A., et al.\ 2012, \apj, 750, 157. doi:10.1088/0004-637X/750/2/157
\bibitem[Costigan et al.(2012)]{Costigan2012} Costigan, G., Scholz, A., Stelzer, B., et al.\ 2012, \mnras, 427, 1344. doi:10.1111/j.1365-2966.2012.22008.x
\bibitem[Costigan et al.(2014)]{Costigan2014} Costigan, G., Vink, J.~S., Scholz, A., et al.\ 2014, \mnras, 440, 3444. doi:10.1093/mnras/stu529
\bibitem[Cottaar et al.(2014)]{Cottaar2014} Cottaar, M., Covey, K.~R., Meyer, M.~R., et al.\ 2014, \apj, 794, 125. doi:10.1088/0004-637X/794/2/125
\bibitem[Currie \& Sicilia-Aguilar(2011)]{Currie2011} Currie, T. \& Sicilia-Aguilar, A.\ 2011, \apj, 732, 24. doi:10.1088/0004-637X/732/1/24
\bibitem[Cosentino et al.(2012)]{Cosentino2012} Cosentino, R., Lovis, C., Pepe, F., et al.\ 2012, \procspie, 84461V
\bibitem[Daemgen et al.(2015)]{Daemgen2015} Daemgen, S., Bonavita, M., Jayawardhana, R., et al.\ 2015, \apj, 799, 155. doi:10.1088/0004-637X/799/2/155
\bibitem[D'Alessio et al.(2000)]{Dalessio2000} D'Alessio, F., Di Cianno, A., Di Paola, A., et al.\ 2000, \procspie, 748
\bibitem[Donehew \& Brittain(2011)]{Donehew2011} Donehew, B. \& Brittain, S.\ 2011, \aj, 141, 46. doi:10.1088/0004-6256/141/2/46
\bibitem[Ercolano \& Owen(2016)]{Ercolano2016} Ercolano, B. \& Owen, J.~E.\ 2016, \mnras, 460, 3472. doi:10.1093/mnras/stw1179
\bibitem[Ercolano \& Pascucci(2017)]{Ercolano2017} Ercolano, B. \& Pascucci, I.\ 2017, Royal Society Open Science, 4, 170114. doi:10.1098/rsos.170114
\bibitem[Espaillat et al.(2014)]{Espaillat2014} Espaillat, C., Muzerolle, J., Najita, J., et al.\ 2014, Protostars and Planets VI, 497. doi:10.2458/azu\_uapress\_9780816531240-ch022
\bibitem[Espaillat et al.(2012)]{Espaillat2012} Espaillat, C., Ingleby, L., Hern{\'a}ndez, J., et al.\ 2012, \apj, 747, 103. doi:10.1088/0004-637X/747/2/103
\bibitem[Esplin et al.(2014)]{Esplin2014} Esplin, T.~L., Luhman, K.~L., \& Mamajek, E.~E.\ 2014, \apj, 784, 126
\bibitem[Esplin \& Luhman(2019)]{Esplin2019} Esplin, T.~L. \& Luhman, K.~L.\ 2019, \aj, 158, 54. doi:10.3847/1538-3881/ab2594
\bibitem[Facchini et al.(2016)]{Facchini2016} Facchini, S., Manara, C.~F., Schneider, P.~C., et al.\ 2016, \aap, 596, A38. doi:10.1051/0004-6361/201629607
\bibitem[Fang et al.(2018)]{Fang2018} Fang, M., Pascucci, I., Edwards, S., et al.\ 2018, \apj, 868, 28. doi:10.3847/1538-4357/aae780
\bibitem[Feiden(2016)]{Feiden2016} Feiden, G.~A.\ 2016, \aap, 593, A99. doi:10.1051/0004-6361/201527613
\bibitem[Fischer et al.(2011)]{Fischer2011} Fischer, W., Edwards, S., Hillenbrand, L., et al.\ 2011, \apj, 730, 73. doi:10.1088/0004-637X/730/2/73
\bibitem[Flores et al.(2021)]{Flores2021} Flores, C., Connelley, M.~S., Reipurth, B., et al.\ 2021, The 20.5th Cambridge Workshop on Cool Stars, Stellar Systems, and the Sun (CS20.5), 217. doi:10.5281/zenodo.4567056
\bibitem[France et al.(2012)]{France2012} France, K., Schindhelm, R., Herczeg, G.~J., et al.\ 2012, \apj, 756, 171. doi:10.1088/0004-637X/756/2/171
\bibitem[Francis \& van der Marel(2020)]{Francis2020} Francis, L. \& van der Marel, N.\ 2020, \apj, 892, 111. doi:10.3847/1538-4357/ab7b63
\bibitem[Frasca et al.(2003)]{Frasca2003} Frasca, A., Alcal{\'a}, J.~M., Covino, E., et al.\ 2003, \aap, 405, 149. doi:10.1051/0004-6361:20030644
\bibitem[Frasca et al.(2005)]{Frasca2005} Frasca, A., Biazzo, K., Catalano, S., et al.\ 2005, \aap, 432, 647. doi:10.1051/0004-6361:20041373
\bibitem[Frasca et al.(2015)]{Frasca2015} Frasca, A., Biazzo, K., Lanzafame, A.~C., et al.\ 2015, \aap, 575, A4. doi:10.1051/0004-6361/201424409
\bibitem[Frasca et al.(2017)]{Frasca2017} Frasca, A., Biazzo, K., Alcal{\'a}, J.~M., et al.\ 2017, \aap, 602, A33. doi:10.1051/0004-6361/201630108
\bibitem[Frasca \& Catalano(1994)]{FrascaCatalano1994} Frasca, A., \& Catalano, S. 1994, \aap, 284, 883
\bibitem[Frasca et al.(2018)]{Frasca2018} Frasca, A., Montes, D., Alcal{\`a}, J.~M., et al.\ 2018, \actaa, 68, 403. doi:10.32023/0001-5237/68.4.5
\bibitem[Gaia Collaboration et al.(2020)]{Gaia2020} Gaia Collaboration, Brown, A.~G.~A., Vallenari, A., et al.\ 2020, arXiv:2012.01533
\bibitem[Galli et al.(2018)]{Galli2018} Galli, P.~A.~B., Loinard, L., Ortiz-L{\'e}on, G.~N., et al.\ 2018, \apj, 859, 33. doi:10.3847/1538-4357/aabf91
\bibitem[Gangi et al.(2020)]{Gangi2020} Gangi, M., Nisini, B., Antoniucci, S., et al.\ 2020, \aap, 643, A32. doi:10.1051/0004-6361/202038534
\bibitem[Garufi et al.(2019)]{Garufi2019} Garufi, A., Podio, L., Bacciotti, F., et al.\ 2019, \aap, 628, A68. doi:10.1051/0004-6361/201935546
\bibitem[Giannini et al.(2019)]{Giannini2019} Giannini, T., Nisini, B., Antoniucci, S., et al.\ 2019, \aap, 631, A44
\bibitem[Grankin et al.(2008)]{Grankin2008} Grankin, K.~N., Bouvier, J., Herbst, W., et al.\ 2008, \aap, 479, 827. doi:10.1051/0004-6361:20078476
\bibitem[Gullbring et al.(1998)]{Gullbring1998} Gullbring, E., Hartmann, L., Brice{\~n}o, C., et al.\ 1998, \apj, 492, 323. doi:10.1086/305032
\bibitem[Gullbring et al.(2000)]{Gullbring2000} Gullbring, E., Calvet, N., Muzerolle, J., et al.\ 2000, \apj, 544, 927. doi:10.1086/317253
\bibitem[Gully-Santiago et al.(2017)]{Gully2017} Gully-Santiago, M.~A., Herczeg, G.~J., Czekala, I., et al.\ 2017, \apj, 836, 200. doi:10.3847/1538-4357/836/2/200
\bibitem[Guo et al.(2018)]{Guo2018} Guo, Z., Herczeg, G.~J., Jose, J., et al.\ 2018, \apj, 852, 56. doi:10.3847/1538-4357/aa9e52
\bibitem[Hartigan et al.(1995)]{Hartigan1995} Hartigan, P., Edwards, S., \& Ghandour, L.\ 1995, \apj, 452, 736. doi:10.1086/176344
\bibitem[Hartmann et al.(1994)]{Hartmann1994} Hartmann, L., Hewett, R., \& Calvet, N.\ 1994, \apj, 426, 669. doi:10.1086/174104
\bibitem[Hartmann(1998)]{Hartmann1998} Hartmann, L.\ 1998, Accretion processes in star formation / Lee Hartmann. Cambridge, UK ; New York : Cambridge University Press, 1998. (Cambridge astrophysics series ; 32)  ISBN 0521435072.
\bibitem[Hartmann et al.(1998)]{Hartmann1998bis} Hartmann, L., Calvet, N., Gullbring, E., et al.\ 1998, \apj, 495, 385. doi:10.1086/305277
\bibitem[Hartmann et al.(2016)]{Hartmann2016} Hartmann, L., Herczeg, G., \& Calvet, N.\ 2016, \araa, 54, 135. doi:10.1146/annurev-astro-081915-023347
\bibitem[Herczeg \& Hillenbrand(2008)]{Herczeg2008} Herczeg, G.~J. \& Hillenbrand, L.~A.\ 2008, \apj, 681, 594. doi:10.1086/586728
\bibitem[Herczeg \& Hillenbrand(2014)]{Herczeg2014} Herczeg, G.~J. \& Hillenbrand, L.~A.\ 2014, \apj, 786, 97. doi:10.1088/0004-637X/786/2/97
\bibitem[Hessman \& Guenther(1997)]{Hessman1997} Hessman, F.~V. \& Guenther, E.~W.\ 1997, \aap, 321, 497
\bibitem[Hildebrand(1983)]{Hildebrand1983} Hildebrand, R.~H.\ 1983, \qjras, 24, 267
\bibitem[Ingleby et al.(2013)]{Ingleby2013} Ingleby, L., Calvet, N., Herczeg, G., et al.\ 2013, \apj, 767, 112. doi:10.1088/0004-637X/767/2/112
\bibitem[Isella et al.(2009)]{Isella2009} Isella, A., Carpenter, J.~M., \& Sargent, A.~I.\ 2009, \apj, 701, 260. doi:10.1088/0004-637X/701/1/260
\bibitem[J{\"o}nsson et al.(2020)]{Jonsson2020} J{\"o}nsson, H., Holtzman, J.~A., Allende Prieto, C., et al.\ 2020, \aj, 160, 120. doi:10.3847/1538-3881/aba592
\bibitem[Jones et al.(2012)]{Jones2012} Jones, M.~G., Pringle, J.~E., \& Alexander, R.~D.\ 2012, \mnras, 419, 925. doi:10.1111/j.1365-2966.2011.19730.x
\bibitem[Kelly(2007)]{Kelly2007} Kelly, B.~C.\ 2007, \apj, 665, 1489. doi:10.1086/519947
\bibitem[K{\'o}sp{\'a}l et al.(2018)]{Kospal2018} K{\'o}sp{\'a}l, {\'A}., {\'A}brah{\'a}m, P., Zsidi, G., et al.\ 2018, \apj, 862, 44. doi:10.3847/1538-4357/aacafa
\bibitem[Kraus et al.(2011)]{Kraus2011} Kraus, A.~L., Ireland, M.~J., Martinache, F., et al.\ 2011, \apj, 731, 8. doi:10.1088/0004-637X/731/1/8
\bibitem[Kraus \& Hillenbrand(2012)]{Kraus2012} Kraus, A.~L. \& Hillenbrand, L.~A.\ 2012, \apj, 757, 141. doi:10.1088/0004-637X/757/2/141
\bibitem[Lanza et al.(2016)]{Lanza2016} Lanza, A.~F., Flaccomio, E., Messina, S., et al.\ 2016, \aap, 592, A140. doi:10.1051/0004-6361/201628382
\bibitem[Lodato et al.(2017)]{Lodato2017} Lodato, G., Scardoni, C.~E., Manara, C.~F., et al.\ 2017, \mnras, 472, 4700. doi:10.1093/mnras/stx2273
\bibitem[Long et al.(2018)]{Long2018} Long, F., Pinilla, P., Herczeg, G.~J., et al.\ 2018, \apj, 869, 17
\bibitem[Long et al.(2019)]{Long2019} Long, F., Herczeg, G.~J., Harsono, D., et al.\ 2019, \apj, 882, 49
\bibitem[L{\'o}pez-Valdivia et al.(2021)]{Lopez2021} L{\'o}pez-Valdivia, R., Sokal, K.~R., Mace, G.~N., et al.\ 2021, \apj, 921, 53. doi:10.3847/1538-4357/ac1a7b
\bibitem[Lynden-Bell \& Pringle(1974)]{Lynden1974} Lynden-Bell, D. \& Pringle, J.~E.\ 1974, \mnras, 168, 603. doi:10.1093/mnras/168.3.603
\bibitem[Manara et al.(2013)]{Manara2013} Manara, C.~F., Beccari, G., Da Rio, N., et al.\ 2013, \aap, 558, A114. doi:10.1051/0004-6361/201321866
\bibitem[Manara et al.(2016)]{Manara2016} Manara, C.~F., Rosotti, G., Testi, L., et al.\ 2016, \aap, 591, L3. doi:10.1051/0004-6361/201628549
\bibitem[Manara et al.(2017)]{Manara2017} Manara, C.~F., Frasca, A., Alcal{\'a}, J.~M., et al.\ 2017, \aap, 605, A86. doi:10.1051/0004-6361/201730807
\bibitem[Manara et al.(2020)]{Manara2020} Manara, C.~F., Natta, A., Rosotti, G.~P., et al.\ 2020, \aap, 639, A58. doi:10.1051/0004-6361/202037949
\bibitem[Manara et al.(2022)]{Manara2022} Manara, C.~F., Ansdell, M., Rosotti, G.~P., et al.\ 2022, arXiv:2203.09930
\bibitem[Mart{\'\i}n et al.(2005)]{Martin2005} Mart{\'\i}n, E.~L., Magazz{\`u}, A., Delfosse, X., et al.\ 2005, \aap, 429, 939. doi:10.1051/0004-6361:20041724
\bibitem[Mendigut{\'\i}a et al.(2011)]{Mendigutia2011} Mendigut{\'\i}a, I., Calvet, N., Montesinos, B., et al.\ 2011, \aap, 535, A99. doi:10.1051/0004-6361/201117444
\bibitem[Mendigut{\'\i}a et al.(2015)]{Mendigutia2015} Mendigut{\'\i}a, I., Oudmaijer, R.~D., Rigliaco, E., et al.\ 2015, \mnras, 452, 2837. doi:10.1093/mnras/stv1540
\bibitem[Molinari et al.(2014)]{Molinari2014} Molinari, E., Covino, S., Crimi, G., et al.\ 2014, \procspie, 9147, 91476X. doi:10.1117/12.2056390
\bibitem[Montes et al.(1995)]{Montesetal1995} Montes, D., et al. 1995, \aaps, 114, 287
\bibitem[Moultaka et al.(2004)]{Moultaka2004} Moultaka, J., Ilovaisky, S.~A., Prugniel, P., et al.\ 2004, \pasp, 116, 693. doi:10.1086/422177
\bibitem[Mulders et al.(2017)]{Mulders2017} Mulders, G.~D., Pascucci, I., Manara, C.~F., et al.\ 2017, \apj, 847, 31. doi:10.3847/1538-4357/aa8906
\bibitem[Munari et al.(2012)]{Munari2012} Munari, U., Bacci, S., Baldinelli, L., et al.\ 2012, Baltic Astronomy, 21, 13
\bibitem[Najita et al.(2007)]{Najita2007} Najita, J.~R., Carr, J.~S., Glassgold, A.~E., et al.\ 2007, Protostars and Planets V, 507
\bibitem[Nisini et al.(2018)]{Nisini2018} Nisini, B., Antoniucci, S., Alcal{\'a}, J.~M., et al.\ 2018, \aap, 609, A87. doi:10.1051/0004-6361/201730834
\bibitem[Nofi et al.(2021)]{Nofi2021} Nofi, L.~A., Johns-Krull, C.~M., L{\'o}pez-Valdivia, R., et al.\ 2021, \apj, 911, 138. doi:10.3847/1538-4357/abeab3
\bibitem[Oliva et al.(2012)]{Oliva2012} Oliva, E., Origlia, L., Maiolino, R., et al.\ 2012, \procspie, 84463T
\bibitem[Origlia et al.(2014)]{Origlia2014} Origlia, L., Oliva, E., Baffa, C., et al.\ 2014, \procspie, 91471E
\bibitem[Palla \& Stahler(2000)]{Palla2000} Palla, F. \& Stahler, S.~W.\ 2000, \apj, 540, 255. doi:10.1086/309312
\bibitem[Pascucci et al.(2022)]{Pascucci2022} Pascucci, I., Cabrit, S., Edwards, S., et al.\ 2022, arXiv:2203.10068
\bibitem[Pepe et al.(2002)]{Pepe2002} Pepe, F., Mayor, M., Galland, F., et al.\ 2002, \aap, 388, 632. doi:10.1051/0004-6361:20020433
\bibitem[Prato et al.(2008)]{Prato2008} Prato, L., Huerta, M., Johns-Krull, C.~M., et al.\ 2008, \apjl, 687, L103. doi:10.1086/593201
\bibitem[Ribas et al.(2020)]{Ribas2020} Ribas, {\'A}., Espaillat, C.~C., Mac{\'\i}as, E., et al.\ 2020, \aap, 642, A171. doi:10.1051/0004-6361/202038352
\bibitem[Rigliaco et al.(2012)]{Rigliaco2012} Rigliaco, E., Natta, A., Testi, L., et al.\ 2012, \aap, 548, A56. doi:10.1051/0004-6361/201219832
\bibitem[Rigliaco et al.(2013)]{Rigliaco2013} Rigliaco, E., Pascucci, I., Gorti, U., et al.\ 2013, \apj, 772, 60. doi:10.1088/0004-637X/772/1/60
\bibitem[Rigliaco et al.(2015)]{Rigliaco2015} Rigliaco, E., Pascucci, I., Duchene, G., et al.\ 2015, \apj, 801, 31. doi:10.1088/0004-637X/801/1/31
\bibitem[Rosotti et al.(2017)]{Rosotti2017} Rosotti, G.~P., Clarke, C.~J., Manara, C.~F., et al.\ 2017, \mnras, 468, 1631. doi:10.1093/mnras/stx595
\bibitem[Rugel et al.(2018)]{Rugel2018} Rugel, M., Fedele, D., \& Herczeg, G.\ 2018, \aap, 609, A70. doi:10.1051/0004-6361/201630111
\bibitem[Schaefer et al.(2012)]{Schaefer2012} Schaefer, G.~H., Prato, L., Simon, M., et al.\ 2012, \apj, 756, 120. doi:10.1088/0004-637X/756/2/120
\bibitem[Semel(1989)]{Semel1989} Semel, M.\ 1989, \aap, 225, 456
\bibitem[Sicilia-Aguilar et al.(2010)]{Sicilia2010} Sicilia-Aguilar, A., Henning, T., \& Hartmann, L.~W.\ 2010, \apj, 710, 597. doi:10.1088/0004-637X/710/1/597
\bibitem[Siess et al.(2000)]{Siess2000} Siess, L., Dufour, E., \& Forestini, M.\ 2000, \aap, 358, 593
\bibitem[Simon et al.(2016)]{Simon2016} Simon, M.~N., Pascucci, I., Edwards, S., et al.\ 2016, \apj, 831, 169.doi:10.3847/0004-637X/831/2/169
\bibitem[Simon et al.(2019)]{Simon2019} Simon, M., Guilloteau, S., Beck, T.~L., et al.\ 2019, \apj, 884, 42. doi:10.3847/1538-4357/ab3e3b
\bibitem[Skinner et al.(2018)]{Skinner2018} Skinner, S.~L., Schneider, P.~C., Audard, M., et al.\ 2018, \apj, 855, 143. doi:10.3847/1538-4357/aaab58
\bibitem[Smette et al.(2015)]{Smette2015} Smette, A., Sana, H., Noll, S., et al.\ 2015, \aap, 576, A77
\bibitem[Somigliana et al.(2020)]{Somigliana2020} Somigliana, A., Toci, C., Lodato, G., et al.\ 2020, \mnras, 492, 1120. doi:10.1093/mnras/stz3481
\bibitem[Tabone et al.(2022)]{Tabone2022} Tabone, B., Rosotti, G.~P., Lodato, G., et al.\ 2022, \mnras, 512, L74. doi:10.1093/mnrasl/slab124
\bibitem[Tang et al.(2014)]{Tang2014} Tang, Y.-W., Dutrey, A., Guilloteau, S., et al.\ 2014, \apj, 793, 10. doi:10.1088/0004-637X/793/1/10
\bibitem[Tazzari et al.(2021)]{Tazzari2021} Tazzari, M., Clarke, C.~J., Testi, L., et al.\ 2021, \mnras, 506, 2804. doi:10.1093/mnras/stab1808
\bibitem[Testi et al.(2022)]{Testi2022} Testi, L., Natta, A., Manara, C.~F., et al.\ 2022, \aap, 663, A98. doi:10.1051/0004-6361/202141380
\bibitem[Tripathi et al.(2017)]{Tripathi2017} Tripathi, A., Andrews, S.~M., Birnstiel, T., et al.\ 2017, \apj, 845, 44. doi:10.3847/1538-4357/aa7c62
\bibitem[Ubeira Gabellini et al.(2019)]{Ubeira2019} Ubeira Gabellini, M.~G., Miotello, A., Facchini, S., et al.\ 2019, \mnras, 486, 4638. doi:10.1093/mnras/stz1138
\bibitem[Ueda et al.(2022)]{Ueda2022} Ueda, T., Kataoka, A., \& Tsukagoshi, T.\ 2022, \apj, 930, 56. doi:10.3847/1538-4357/ac634d
\bibitem[Vitali et al.(2003)]{Vitali2003} Vitali, F., Zerbi, F.~M., Chincarini, G., et al.\ 2003, \procspie, Vol. 4841, pag. 627
\bibitem[Venuti et al.(2014)]{Venuti2014} Venuti, L., Bouvier, J., Flaccomio, E., et al.\ 2014, \aap, 570, A82. doi:10.1051/0004-6361/201423776
\bibitem[Venuti et al.(2019)]{Venuti2019} Venuti, L., Stelzer, B., Alcal{\'a}, J.~M., et al.\ 2019, \aap, 632, A46. doi:10.1051/0004-6361/201935745
\bibitem[Venuti et al.(2021)]{Venuti2021} Venuti, L., Cody, A.~M., Rebull, L.~M., et al.\ 2021, \aj, 162, 101. doi:10.3847/1538-3881/ac0536
\bibitem[Vernet et al.(2011)]{Vernet2011} Vernet, J., Dekker, H., D'Odorico, S., et al.\ 2011, \aap, 536, A105. doi:10.1051/0004-6361/201117752
\bibitem[Vorobyov \& Basu(2009)]{Vorobyov2009} Vorobyov, E.~I. \& Basu, S.\ 2009, \apj, 703, 922. doi:10.1088/0004-637X/703/1/922
\bibitem[Weber et al.(2020)]{Weber2020} Weber, M.~L., Ercolano, B., Picogna, G., et al.\ 2020, \mnras, 496, 223. doi:10.1093/mnras/staa1549
\bibitem[Weingartner \& Draine(2001)]{Weingartner2001} Weingartner, J.~C. \& Draine, B.~T.\ 2001, \apj, 548, 296. doi:10.1086/318651
\bibitem[White \& Basri(2003)]{WhiteBasri2003} White, R. J., \& Basri, G. 2003, \apj, 582, 1109
\bibitem[White \& Hillenbrand(2004)]{WhiteHillenbrand2004} White, R.~J. \& Hillenbrand, L.~A.\ 2004, \apj, 616, 998. doi:10.1086/425115
\bibitem[Zagaria et al.(2022)]{Zagaria2022} Zagaria, F., Clarke, C.~J., Rosotti, G.~P., et al.\ 2022, \mnras, 512, 3538. doi:10.1093/mnras/stac621
\bibitem[Zapata et al.(2020)]{Zapata2020} Zapata, L.~A., Rodr{\'\i}guez, L.~F., Fern{\'a}ndez-L{\'o}pez, M., et al.\ 2020, \apj, 896, 132. doi:10.3847/1538-4357/ab8fac
\bibitem[Zhu et al.(2019)]{Zhu2019} Zhu, Z., Zhang, S., Jiang, Y.-F., et al.\ 2019, \apjl, 877, L18. doi:10.3847/2041-8213/ab1f8c
\end{thebibliography}
%

\begin{appendix}
\section{Separating class II and class III stars from their EW and flux of the H$\alpha$ and Pa$\beta$ lines}\label{sec:ew_and_sub}
\begin{figure*}[!h]
\begin{center}
\includegraphics[trim=100 200 0 150, width=0.9\columnwidth,angle=90]{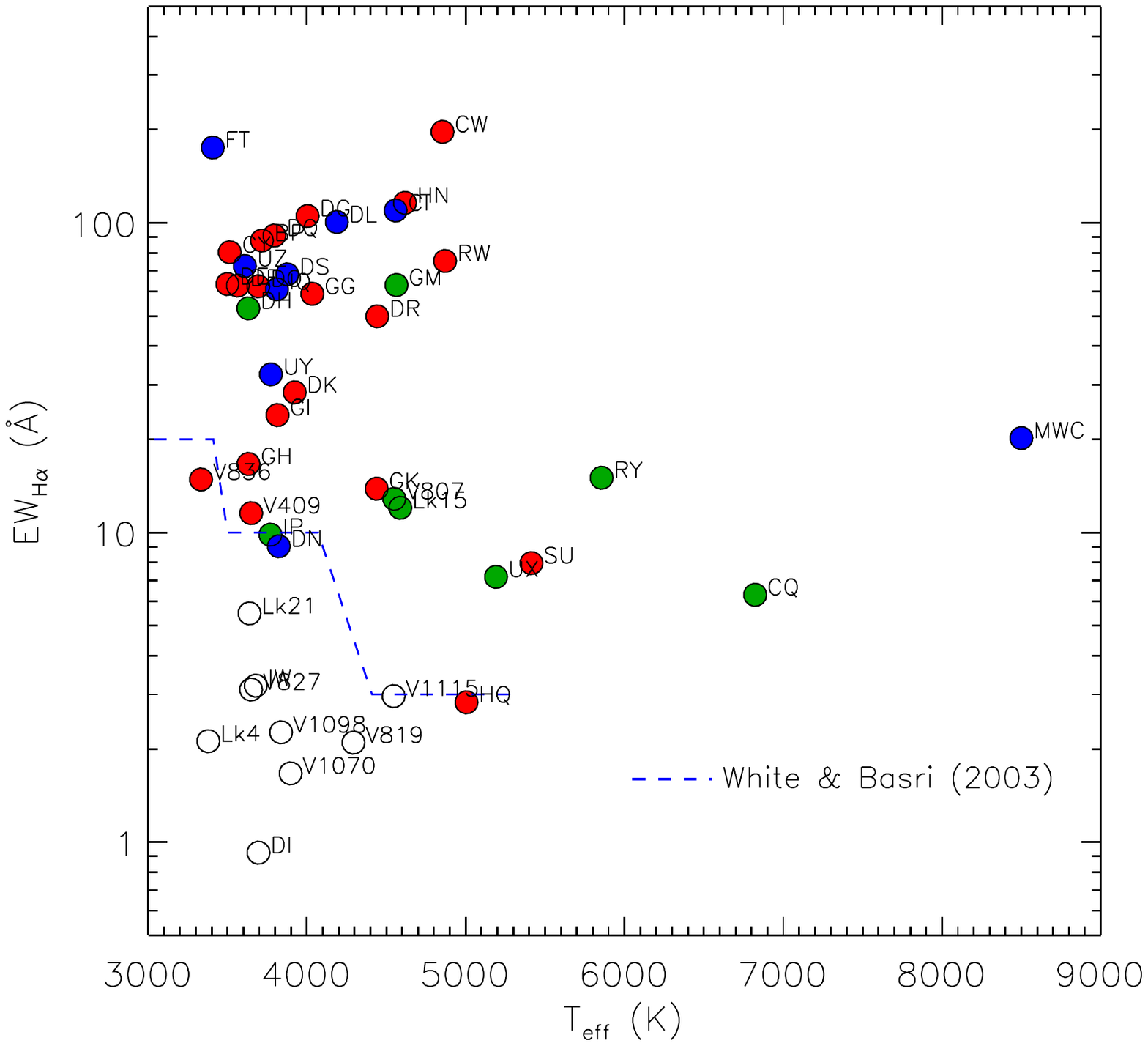}
\includegraphics[trim=100 200 0 150, width=0.9\columnwidth,angle=90]{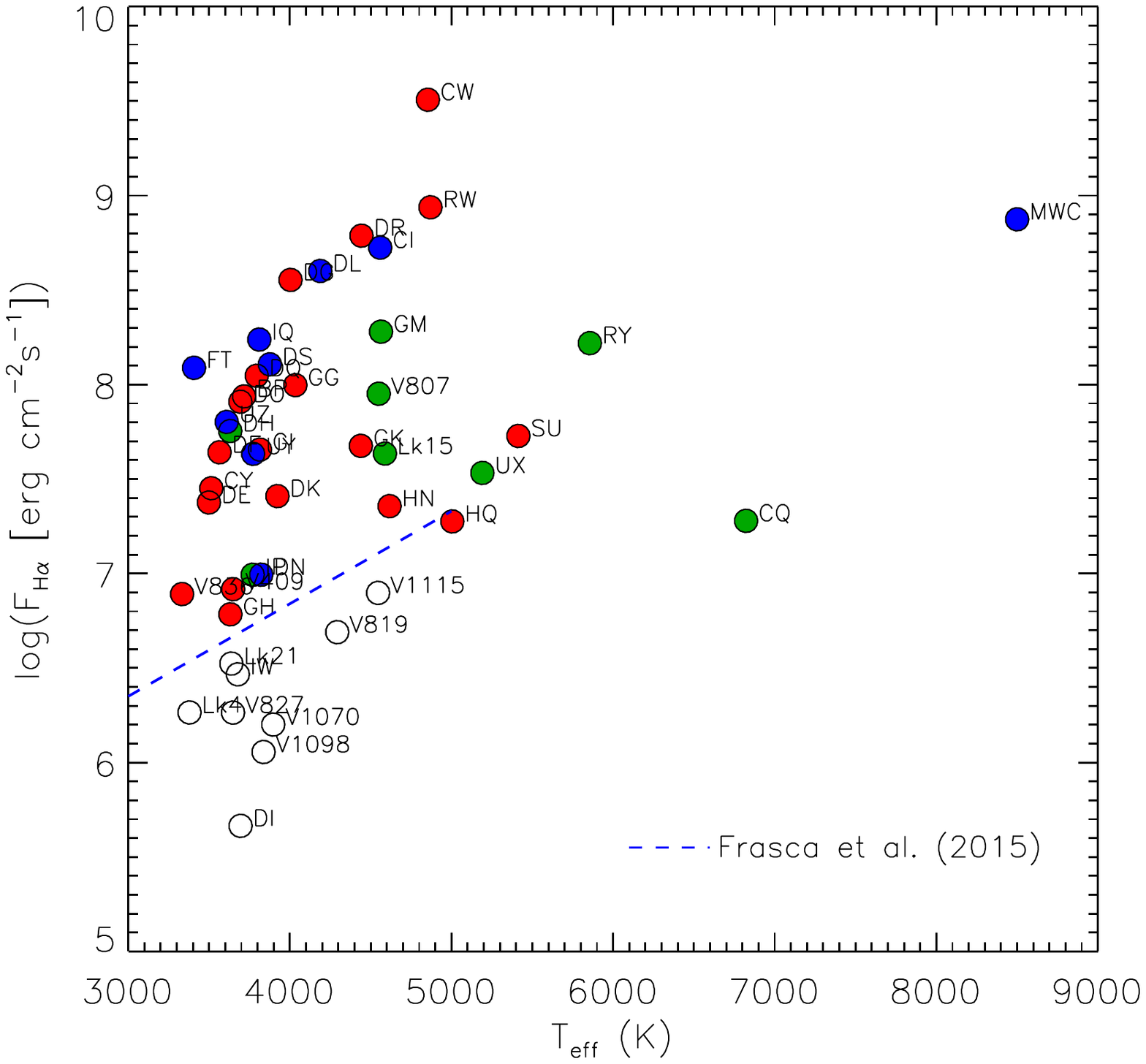}
\end{center}
\caption{Left: H$\alpha$ equivalent width vs $T_{\rm eff}$ for class\,II (filled dots) and class\,III (open dots) objects (red, green, and blue mark the position of full, transitional, and substructured disks, respectively). The dashed line defines the threshold above which most probable accretors are positioned at given spectral types, according to \cite{WhiteBasri2003}. Right: H$\alpha$ flux vs $T_{\rm eff}$ for the same objects. The dashed line represents the dividing line that was empirically defined by \cite{Frasca2015} as the upper boundary of the chromospheric flux.}
\label{fig:Halpha_EWFluxes}
\end{figure*}

\begin{figure*}[!h]
\begin{center}
\includegraphics[trim=100 200 0 150, width=0.9\columnwidth,angle=90]{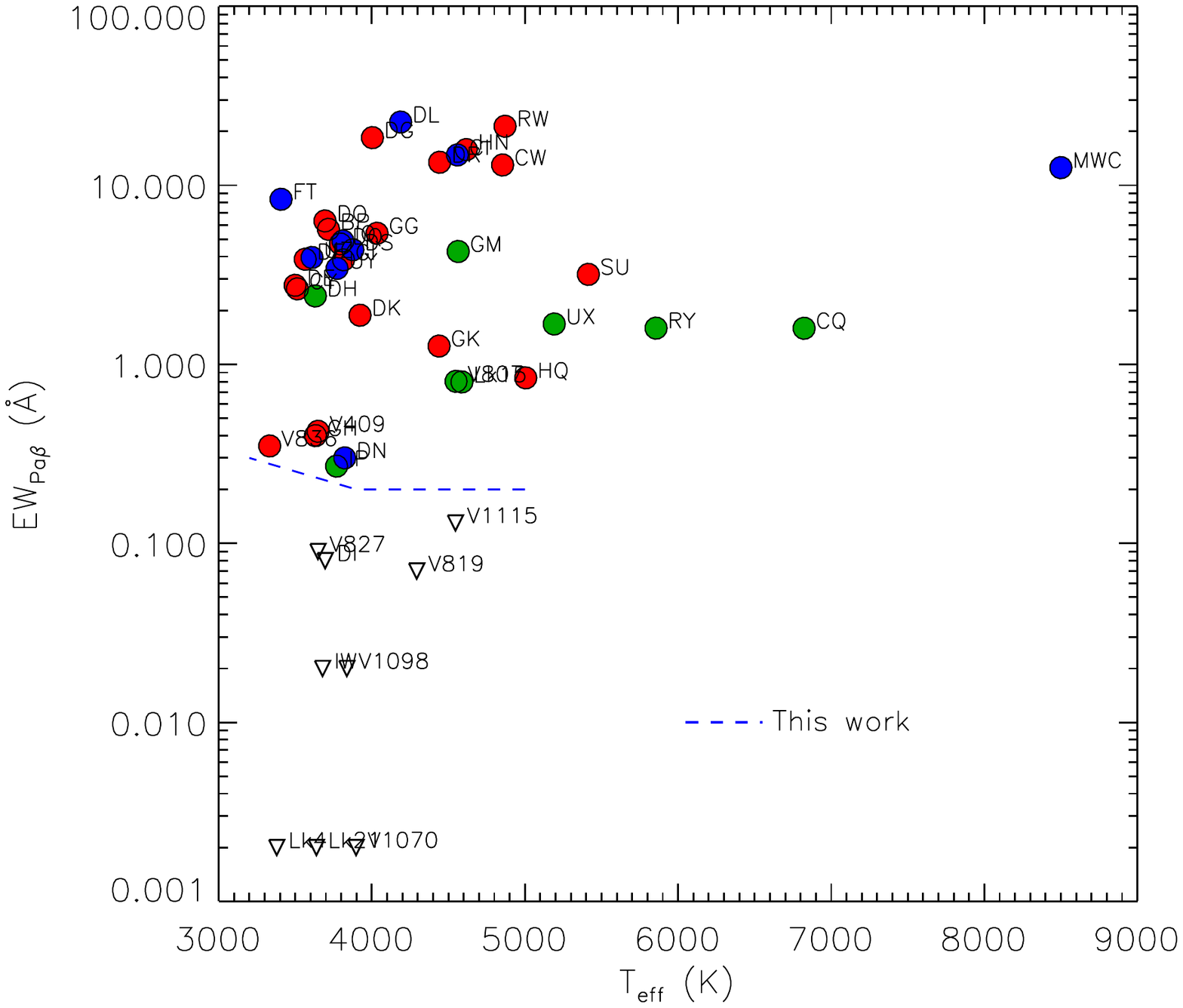}
\includegraphics[trim=100 200 0 150, width=0.9\columnwidth,angle=90]{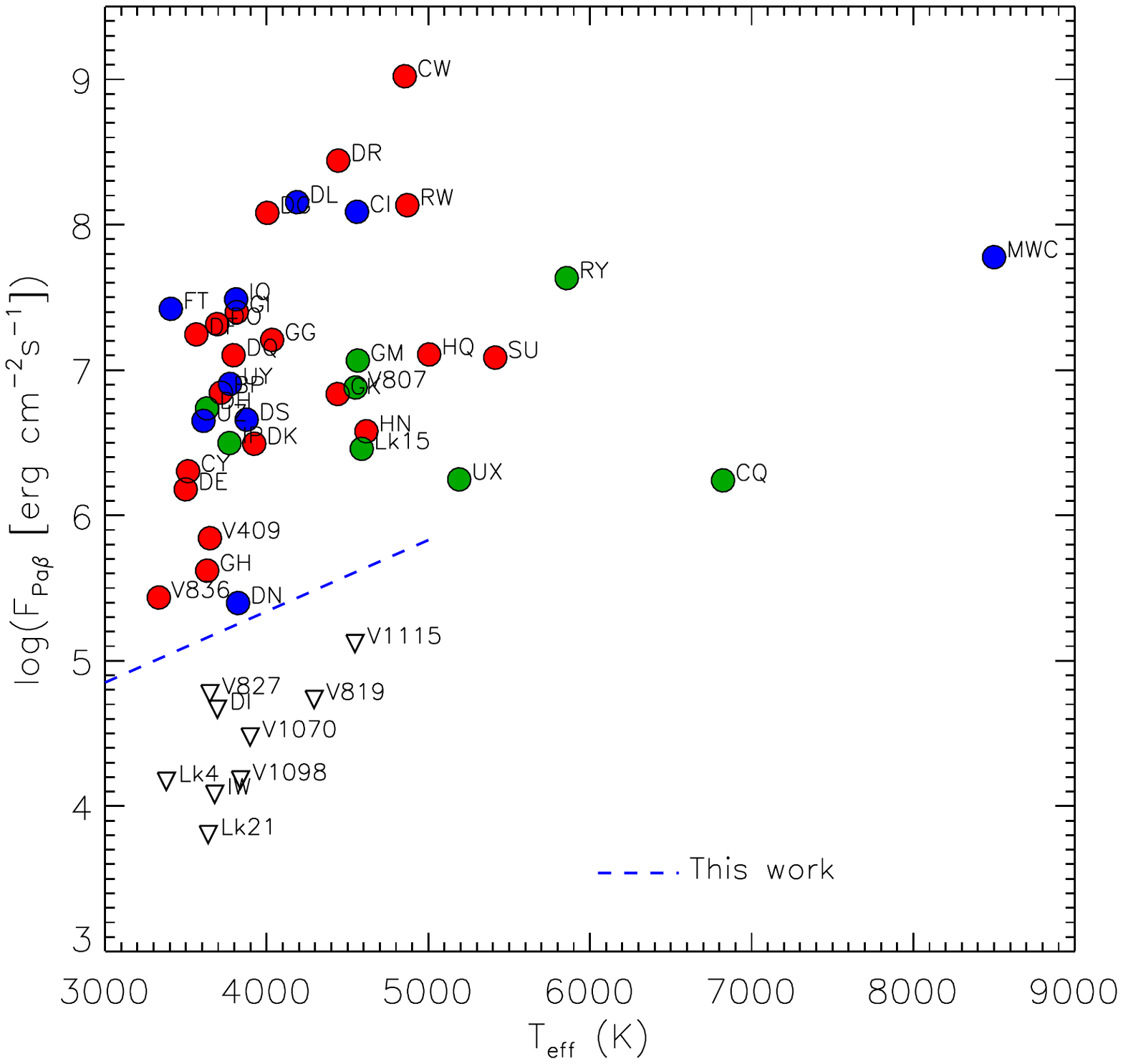}
\end{center}
\caption{Left: Pa$\beta$ equivalent width versus vs $T_{\rm eff}$ for class\,II (filled dots) and class\,III (open symbols) objects (colors as in Fig.\,\ref{fig:Halpha_EWFluxes}). Class\,III targets are shown as downward triangles because their measurements of EWs (and fluxes) in Pa$\beta$ are upper limits. The dashed line separates class\,II from class\,III objects, as empirically defined in this work. Right: Pa$\beta$ flux vs $T_{\rm eff}$ for the same objects (symbols as in the left panel). The dashed straight line, defined in this work, is shown to follow the upper envelope of the sources without accretion; this line was drawn as having the same slope of the \cite{Frasca2015} empirical relation found for the H$\alpha$ line (see the right panel of Fig.\,\ref{fig:Halpha_EWFluxes}).} 
\label{fig:Pabeta_EWFluxes}
\end{figure*}

In Figs.\,\ref{fig:Halpha_EWFluxes} and \ref{fig:Pabeta_EWFluxes}, we show the EWs and stellar fluxes of the H$\alpha$ and Pa$\beta$ lines as a function of $T_{\rm eff}$. For class\,II objects, the line fluxes were determined by integrating the emission line profile with the IRAF task \textsc{splot}, with the exception of targets earlier than K0, for which the flux measurement is strongly influenced by photospheric absorptions (see Sect.\,\ref{sec:accretionpropeties}). In these cases and for class\,III objects, for which the contribution of the photospheric flux could become important, we considered the spectral subtraction method to remove the photospheric flux and emphasize the emission of the line core (see, e.g., \citealt{FrascaCatalano1994, Montesetal1995, Biazzoetal2009, Frasca2015, Frasca2017}). In particular, we considered for each source a low-activity star observed with the GIARPS spectrograph and artificially broadened at the same $v \sin i$ of the target. Whenever a veiling $r600 > 0$ was found, it was introduced in the low-activity template before the subtraction, in order to reproduce the photospheric lines of the target best. This procedure allowed us to isolate the pure emission that often only fills in the line cores and allowed us to derive the net EWs. This is particularly important for the Pa$\beta$ line, for which no emission was present in all spectra of class\,III sources, while the same objects show a H$\alpha$ line always in emission over the continuum. For the Pa$\beta$ line, we were able to derive only upper limits because for all class\,III targets, the line is always in absorption, with a very small filling-in of the core that was only detected after the low-activity templates were subtracted.

The net $EW_{\rm H\alpha}$ shows that all class\,III targets are below the threshold established by \cite{WhiteBasri2003} for separating class\,II and class\,III objects based on H$\alpha$ EWs and spectral types. V1115\,Tau, which is also the only class\,III target with the most pronounced filling-in in the Pa$\beta$ absorption line, is close to this boundary. When translated into stellar fluxes, all class\,III targets are below the limit defined by \cite{Frasca2015} for separating classical from weak T\,Tauri stars. Performing a similar approach for the Pa$\beta$ diagnostics, we suggest in Fig.\,\ref{fig:Pabeta_EWFluxes} new empirical boundary lines to distinguish class\,III from class\,II stars within the plots $EW_{\rm Pa\beta}$ versus $T_{\rm eff}$ and $F_{\rm Pa\beta}$ versus $T_{\rm eff}$. Even though we caution that similar empirical relations do no necessarily enable us to select all classical and weak T\,Tauri stars within a sample, they can be useful for a preliminary selection of class II / III to be observed in the near-infrared. We therefore propose that a T\,Tauri star is a class\,III star if $EW_{\rm Pa\beta}<0.2$\,\AA\,for $T_{\rm eff} < 5000$\,K, with possible greater values for $T_{\rm eff} < 4000$\,K, or if the stellar flux is lower than that defined by the following equation:
\begin{equation}
\log F_{Pa\beta} {\rm [erg\,cm^{-2}\,s^{-1}]} = 4.85 +0.00049(T_{\rm eff} - 3000)
,\end{equation}
\noindent{where the slope of this relation is the same as was found by \cite{Frasca2015} for the H$\alpha$ line. We remark that we suggest these thresholds for effective temperatures in the range $\sim\,3000-5000$\,K, because we do not have a statistically significant number of targets at higher temperatures in our sample.}

\clearpage

\section{Additional tables and images}

\begin{table*}
\small
\center
\caption{\label{tab:Logbook} Logbook of GIARPS and ancillary observations.}
\begin{tabular}{lccc|cc|cccc}
\hline
\hline
          & \multicolumn{3}{c}{GIARPS/GIANO-A}             & \multicolumn{5}{c}{ANCILLARY OBSERVATIONS} \\    
          &                & 			   &			   & \multicolumn{2}{c}{LOW RES. SPECTROSCOPY} & \multicolumn{4}{c}{PHOTOMETRY}   \\
          & Obs Date   & $\rm t_{exp}$ [sec] & $\rm t_{exp}$ [sec] & Asiago   & Amici                          & \multicolumn{2}{c}{VIS} & \multicolumn{2}{c}{NIR}  \\
Source    & [YYYY-MM-DD]       & HARPS-N       & GIANO         & Obs Date & Obs Date                       & Obs Date & Instr.$\rm^{(a)}$ & Obs Date & Instr.$\rm^{(a)}$    \\
\hline   
 BP Tau          & 2020-01-26 & 3000 & 2400	& 2020-01-29 & -          & - 			& -   & 2020-01-27 & NT \\
 CI Tau          & 2018-12-09 & 3000 & 2400 & 2018-12-06 & -          & 2018-12-05 	& ANS & 2018-12-22 & RR	\\  
 CQ Tau          & 2017-11-13 & 2200 & 1800	& 2017-11-30 & -          & 2017-10-30  & ANS & 2017-11-11 & RR \\
 CW Tau          & 2018-12-08 & 3000 & 2400	& 2018-12-06 & -          & 2018-12-06	& ANS & 2018-22-12 & RR	\\
 CY Tau          & 2020-12-13 & 3060 & 2400	& 2021-01-12 & 2020-12-15 & - 			& -   & 2020-12-15 & NT \\
 DE Tau          & 2020-10-24 & 3060 & 2400 & 2020-10-29 & 2020-10-25 & 2020-10-23  & R2R & 2020-10-25 & NT \\
 DF Tau          & 2018-12-08 & 2200 & 1800	& 2018-12-06 & -          & 2018-12-10  & ANS & 2018-12-22 & RR	\\
 DG Tau          & 2017-10-29 & 2200 & 1800	& 2017-10-27 & -          & 2017-10-31 	& ANS & 2017-11-11 & RR	\\
 DH Tau          & 2019-11-02 & 3000 & 2400	& 2019-11-13 & -          & - 			& -   & 2020-01-27 & NT \\
 DI Tau	         & 2020-12-13 & 3060 & 2400	& 2021-01-12 & 2020-12-15 & 2020-12-14  & R2R & 2020-12-15 & NT \\	
 DK Tau          & 2018-12-08 & 2200 & 1800	& 2018-12-06 & -          & 2018-12-08 	& ANS & 2018-12-22 & RR	\\
 DL Tau          & 2017-10-29 & 3000 & 2400	& 2017-11-30 & -          & 2017-10-30 	& ANS & 2017-11-11 & RR	\\
 DN Tau          & 2019-11-01 & 3000 & 2400	& 2019-11-05 & -          & - 			& -   & 2020-01-27 & NT \\
 DO Tau          & 2017-11-13 & 3000 & 2400	& 2017-11-30 & -          & 2017-11-14  & ANS & 2017-11-11 & RR	\\
                 & 2020-01-26 & 3000 & 2400	& 2020-01-29 & -          & - 			& -   & 2020-01-27 & NT \\
 DQ Tau          & 2019-11-02 & 3000 & 2400	& 2019-11-13 & -          & - 			& -   & 2020-01-27 & NT \\
 DR Tau          & 2020-01-25 & 2300 & 1800	& 2020-01-29 & -          & - 			& -   & 2020-01-27 & NT \\
 DS Tau	         & 2019-11-01 & 3000 & 2400	& 2019-11-13 & -          & - 			& -   & 2020-01-27 & NT \\
 FT Tau          & 2020-01-25 & 4600 & 3600	& 2020-01-29 & -          & - 			& -   & 2020-01-27 & NT \\
 GG Tau          & 2018-12-09 & 2200 & 1800	& 2018-12-06 & -          & 2018-12-04 	& ANS & 2018-12-22 & RR	\\
 GH Tau          & 2020-01-25 & 3000 & 2400	& 2020-01-29 & -          & - 			& -   & 2020-01-27 & NT \\
 GI Tau          & 2020-01-26 & 3000 & 2400	& 2020-01-29 & -          & - 			& -   & 2020-01-27 & NT \\
 GK Tau          & 2019-11-02 & 3000 & 2400	& 2019-11-05 & -          & - 			& -   & 2020-01-27 & NT \\
 GM Aur          & 2018-12-09 & 3000 & 2400	& 2018-12-06 & -          & 2018-12-11 	& ANS & 2018-12-22 & RR	\\
 HN Tau          & 2017-10-29 & 4500 & 3600	& 2017-11-30 & -          & 2017-10-30  & ANS & 2017-11-11 & RR	\\
 HQ Tau          & 2019-11-01 & 2300 & 1800	& 2019-11-13 & -          & -	 		& -   & - 		   & -	\\
 IP Tau          & 2018-12-09 & 3000 & 2400	& 2018-12-06 & -          & 2018-12-11 	& ANS & 2018-12-22 & RR	\\
 IQ Tau          & 2019-11-02 & 3000 & 3600	& 2019-11-13 & -          & - 			& -   & 2020-01-27 & NT \\
 IW Tau			 & 2020-12-13 & 3060 & 2400 & 2020-12-16 & 2020-12-15 & 2020-12-14 	& R2R & 2020-12-15 & NT \\
 LkCa 4			 & 2020-12-13 & 3060 & 2400 & 2020-12-13 & 2020-12-15 & 2020-12-14 	& R2R & 2020-12-15 & NT \\
 LkCa 15		 & 2020-12-14 & 3060 & 2400 & 2020-12-13 & 2020-12-15 & 2020-12-13 	& R2R & 2020-12-15 & NT \\
 LkCa 21		 & 2020-12-13 & 3060 & 2400 & 2021-01-12 & 2020-12-15 & 2020-12-14	& R2R & 2020-12-15 & NT \\
 MWC480          & 2019-11-01 & 960	 & 1200	& 2019-11-05 & -          & - 			& -   & 2020-01-27 & NT \\
 RW Aur A        & 2017-11-13 & 2200 & 1800	& 2017-11-30 & -          & 2017-10-30 	& ANS & 2017-11-11 & RR	\\
 RY Tau          & 2017-11-13 & 1500 & 1200	& 2017-10-27 & -          & 2017-10-30 	& ANS & 2017-11-11 & RR	\\
 SU Aur          & 2020-10-24 & 1260 & 1200 & 2020-10-29 & 2020-10-25 & 2020-10-23	& R2R & 2020-10-25 & NT \\
 UX Tau A        & 2020-01-26 & 2280 & 1800	& 2020-01-29 & -          & - 			& -   & 2020-01-27 & NT \\
 UY Aur          & 2018-12-08 & 2200 & 1800	& 2018-12-06 & -          & 2018-12-06 	& ANS & 2018-12-22 & RR	\\
 UZ Tau E        & 2018-12-09 & 3755 & 3000	& 2018-12-06 & -          & 2018-12-10 	& ANS & 2018-12-22 & RR \\
 V409 Tau        & 2020-01-26 & 4600 & 3600	& 2020-01-29 & -          & - 			& -   & 2020-01-27 & NT \\
 V807 Tau        & 2020-10-24 & 2280 & 1800	& 2020-10-29 & 2020-10-24 & 2020-10-23	& R2R & 2020-10-25 & NT \\
 V819 Tau    	 & 2020-12-14 & 3060 & 2400	& 2021-01-12 & 2020-12-15 & 2020-12-14	& R2R & 2020-12-15 & NT \\
 V827 Tau        & 2020-12-13 & 3060 & 2400	& 2020-12-13 & 2020-12-15 & 2020-12-14	& R2R & 2020-12-15 & NT \\
 V836 Tau        & 2019-11-02 & 3000 & 2400	& 2019-11-13 & -          & - 			& -   & - 		   & -	\\
                 & 2020-01-26 & 3000 & 2400	& 2020-01-29 & -          & - 			& -   & - 		   & -	\\
 V1070 Tau       & 2020-12-13 & 3060 & 2400	& 2020-12-16 & 2020-12-15 & 2020-12-13  & R2R & 2020-12-15 & NT \\
 V1098 Tau       & 2020-10-24 & 2280 & 1800	& 2020-10-29 & 2020-10-25 & - 			& -   & 2020-10-25 & NT \\
 V1115 Tau       & 2020-12-14 & 3060 & 2400	& 2020-12-16 & 2020-12-15 & 2020-12-14  & R2R & 2020-12-15 & NT \\
\hline\end{tabular}	 
\begin{quotation}
\textbf{Notes.} (a) \textbf{ANS}: ANS collaboration telescopes; \textbf{R2R}: ROS2@REM; \textbf{NT}: NICS@TNG; \textbf{RR}: REMIR@REM. 
\end{quotation}
\end{table*}  
\begin{table*}
\small
\center
\caption{\label{tab:phot_last_run} Optical and NIR photometry taken in RUN IV.}
\begin{tabular}{lcccccccc}
\hline
\hline
           &  \multicolumn{5}{c}{VIS} &  \multicolumn{3}{c}{NIR}      \\
Source     & Obs Date     &  $g$   & $r$   & $i$   & $z$   & Obs Date   & $\rm J$   & $\rm K$   \\
\hline
           & [YYYY-MM-DD]  &  [mag] & [mag] & [mag] & [mag] &  [YYYY-MM-DD]  & [mag] & [mag] \\
\hline
\hline   
 CY Tau	   & -  		 & -      & -     & -     & -     & 2020-12-15 & 9.66  & 8.39 \\ 
 DE Tau    & 2020-10-23  & 13.50  & 12.31 & 11.34 & 10.79 & 2020-10-25 & 9.07  & 7.70 \\
 DI Tau    & 2020-12-14  & 13.48  & 12.19 & 11.31 & 10.95 & 2020-12-15 & 9.26  & 8.29 \\
 HD30171   & 2020-10-23  & 9.52   & 9.06  & 8.86  & 8.85  & 2020-10-25 & 7.83  & 7.32 \\
 IW Tau	   & 2020-12-14  & 13.13  & 11.92 & 11.12 & 10.74 & 2020-12-15 & 9.13  & 8.21 \\ 
 LkCa 4    & 2020-12-14  & 13.19  & 11.96 & 11.20 & 10.85 & -          & -     & -    \\ 
 LkCa 15   & 2020-12-13  & 13.11  & 12.01 & 11.42 & 11.20 & 2020-12-15 & 9.36  & 8.16 \\ 
 LkCa 21   & 2020-12-14  & 14.08  & 12.82 & 11.68 & 11.06 & 2020-12-15 & 9.36  & 8.36 \\
 SU Aur    & 2020-10-23  & 9.57   & 8.99  & 8.64  & 8.44  & 2020-10-25 & 7.17  & 5.82 \\
 V807 Tau  & 2020-10-23  & 11.96  & 10.91 & 10.14 & 9.85  & 2020-10-25 & 8.22  & 7.03 \\
 V819 Tau  & 2020-12-14  & 13.83  & 12.55 & 11.70 & 11.32 & 2020-12-15 & 9.33  & 8.31 \\
 V826 Tau  & 2020-10-23  & 12.32  & 11.44 & 10.71 & 10.39 & 2020-10-25 & 9.05  & 8.16 \\
 V827 Tau  & 2020-12-14  & 13.14  & 11.95 & 11.15 & 10.77 & 2020-12-15 & 9.07  & 8.16 \\
 V1070 Tau & 2020-12-13  & 12.76  & 11.71 & 10.98 & 10.49 & 2020-12-15 & 9.02  & 8.16 \\
 V1098 Tau & -           & -      & -     & -     & -     & 2020-10-25 & 8.30  & -    \\ 
 V1115 Tau & 2020-12-14  & 12.15  & 11.22 & 10.76 & 10.60 & 2020-12-15 & 9.27  & 8.48 \\
\hline\end{tabular}
\begin{quotation}
\textbf{Notes.} Typical errors in photometric magnitudes are 0.1 mag. Photometric data for the past observing RUN are reported in \citet{Gangi2020}.
\end{quotation}
\end{table*} 
  
\begin{table*}
\small
\center
\caption{\label{tab:sources_param_2Teff} Results from the 2-$\rm T_{eff}$ modeling for sources showing a strong presence of spots. $\rm T_{ROTFIT}$ is the temperature as derived by ROTFIT on the basis of the high-resolution optical HARPS-N spectra. In the modeling, $\rm T_{hot}$ is fixed to the closest value found by ROTFIT while $\rm T_{cool}$ and the weighting factors are leaved as free parameters; their best-fit values are reported in the table. Final effective temperatures, as computed as from Eq. 1, are reported in the last column.}
\begin{tabular}{lcc|cccc|c}
\hline
\hline
Source    & Obs Date & $\rm T_{ROTFIT}$ $\rm (\pm err)$ & $\rm T_{hot}$ & $\rm w_{hot}$ & $\rm T_{cool}$ & $\rm w_{cool}$ & $\rm T_{eff}$  \\
          &  [YYYY-MM-DD]  & [K] & [K]           & [\%]          & [K]            & [\%]           & [K]           \\
\hline
\multicolumn{8}{c}{Accreting sources} \\
\hline
BP Tau   & 	2020-01-26 & 4176 (84)  & 4200       &   45      &  3100       &  55        &  3717       \\       
DK Tau   & 	2018-12-08 & 4283 (155) & 4300       &   53      &  3300       &  47        &  3923       \\       
DN Tau   & 	2019-11-01 & 4060 (121) & 4000       &   72      &  3200       &  28        &  3823       \\       
DQ Tau   & 	2019-11-02 & 4044 (103) & 4000       &   68      &  3200       &  32        &  3795       \\       
DS Tau   & 	2019-11-01 & 4183 (87)  & 4200       &   61      &  3100       &  39        &  3876       \\       
GG Tau A & 	2018-12-09 & 4186 (121) & 4200       &   76      &  3300       &  24        &  4034       \\       
GI Tau   & 	2020-01-26 & 3957 (140) & 4000       &   73      &  3100       &  27        &  3815       \\       
IP Tau   & 	2018-12-09 & 4038 (119) & 4000       &   67      &  3100       &  33        &  3770       \\       
UY Aur   & 	2018-12-08 & 4086 (138) & 4100       &   58      &  3100       &  42        &  3773       \\       
V836 Tau & 	2019-11-02 & 4444 (86)  & 4400       &   11      &  3100       &  89        &  3333       \\
		 & 	2020-01-26 & 4398 (105) & 4400       &   10      &  3100       &  90        &  3313       \\       
\hline
\multicolumn{8}{c}{Non-accreting sources} \\
\hline
DI Tau	      & 2020-12-13 & 3965 (75)  &  4000      &   54    &  3200       & 46         &  3695     \\
IW Tau		  & 2020-12-13 & 4125 (92)  &  4100      &   44    &  3200       & 56         &  3678     \\
LkCa 4		  & 2020-12-13 & 4085 (152) &  4100      &   20    &  3100       & 80         &  3379     \\
V819 Tau	  & 2020-12-14 & 4486 (62)  &  4500      &   73    &  3500       & 27         &  4293     \\
V827 Tau      & 2020-12-13 & 4180 (106) &  4200      &   35    &  3200       & 65         &  3648     \\
V1070 Tau     & 2020-12-13 & 4208 (77)  &  4200      &   61    &  3200       & 39         &  3897     \\
\hline\end{tabular}	
\end{table*}

\begin{figure*}[!h]
\begin{center}
\includegraphics[trim=45 0 0 0,width=0.98\columnwidth]{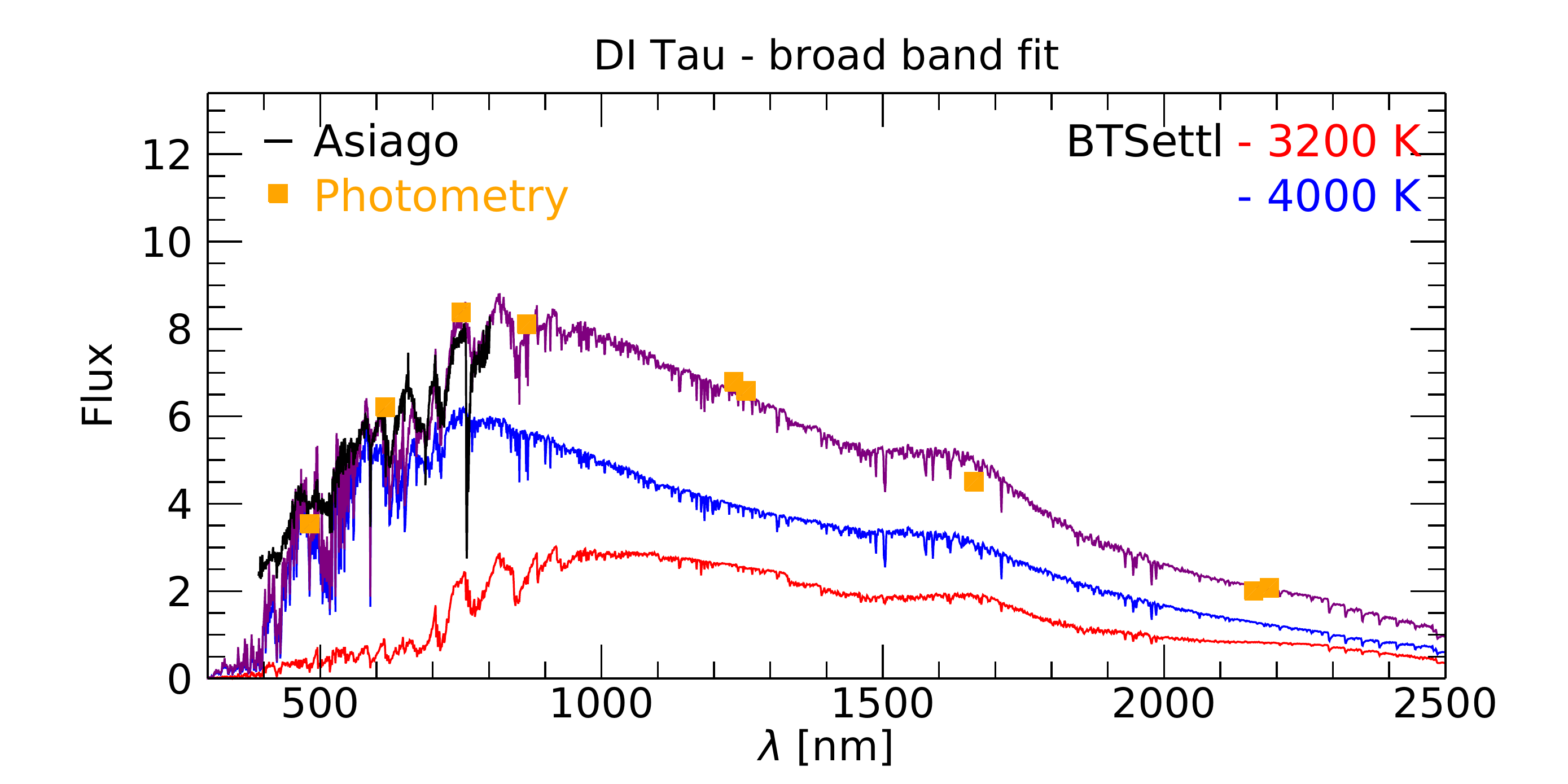}
\includegraphics[trim=45 0 0 0,width=0.98\columnwidth]{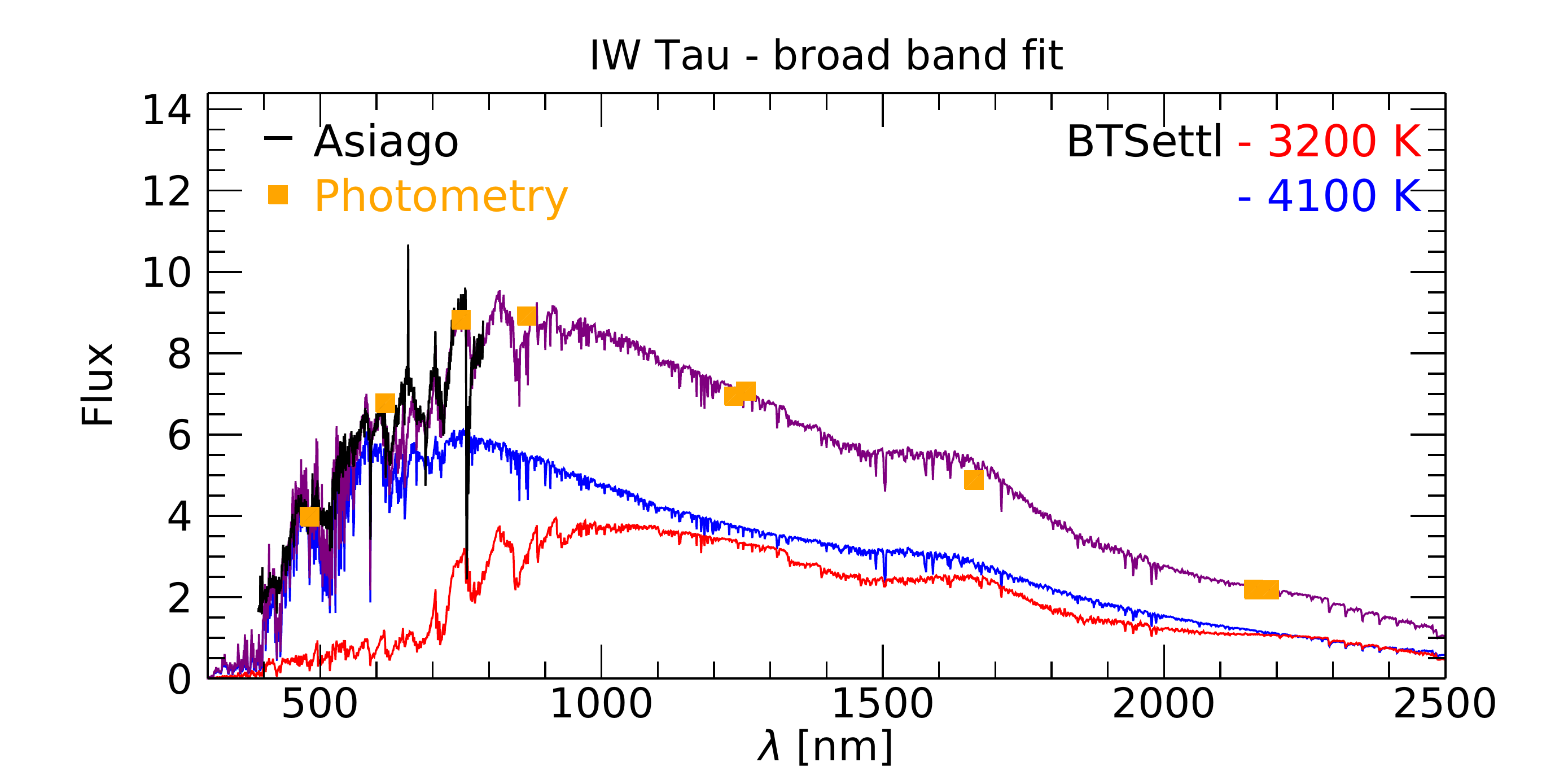}
\includegraphics[trim=45 0 0 0,width=0.98\columnwidth]{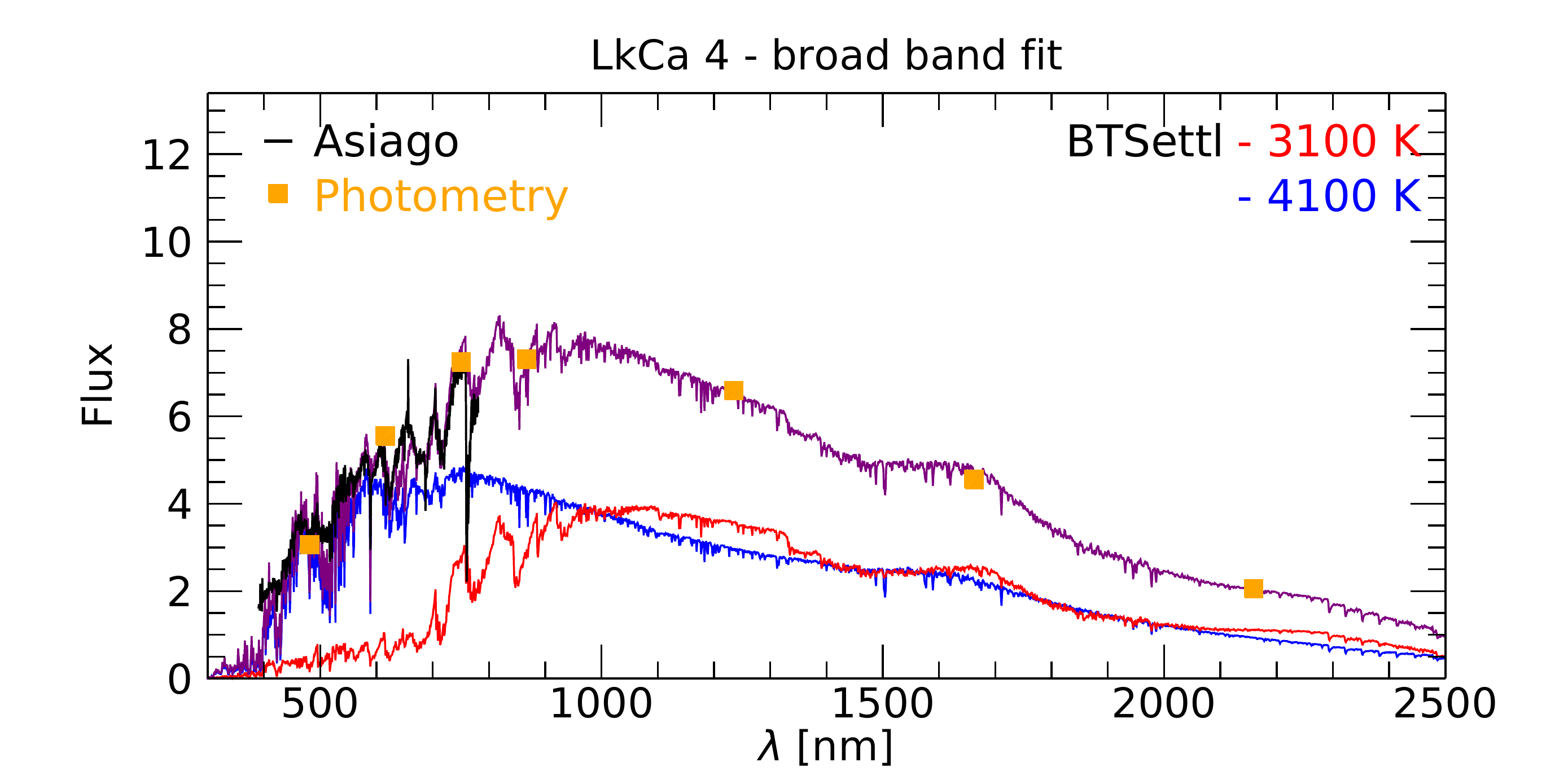}
\includegraphics[trim=45 0 0 0,width=0.98\columnwidth]{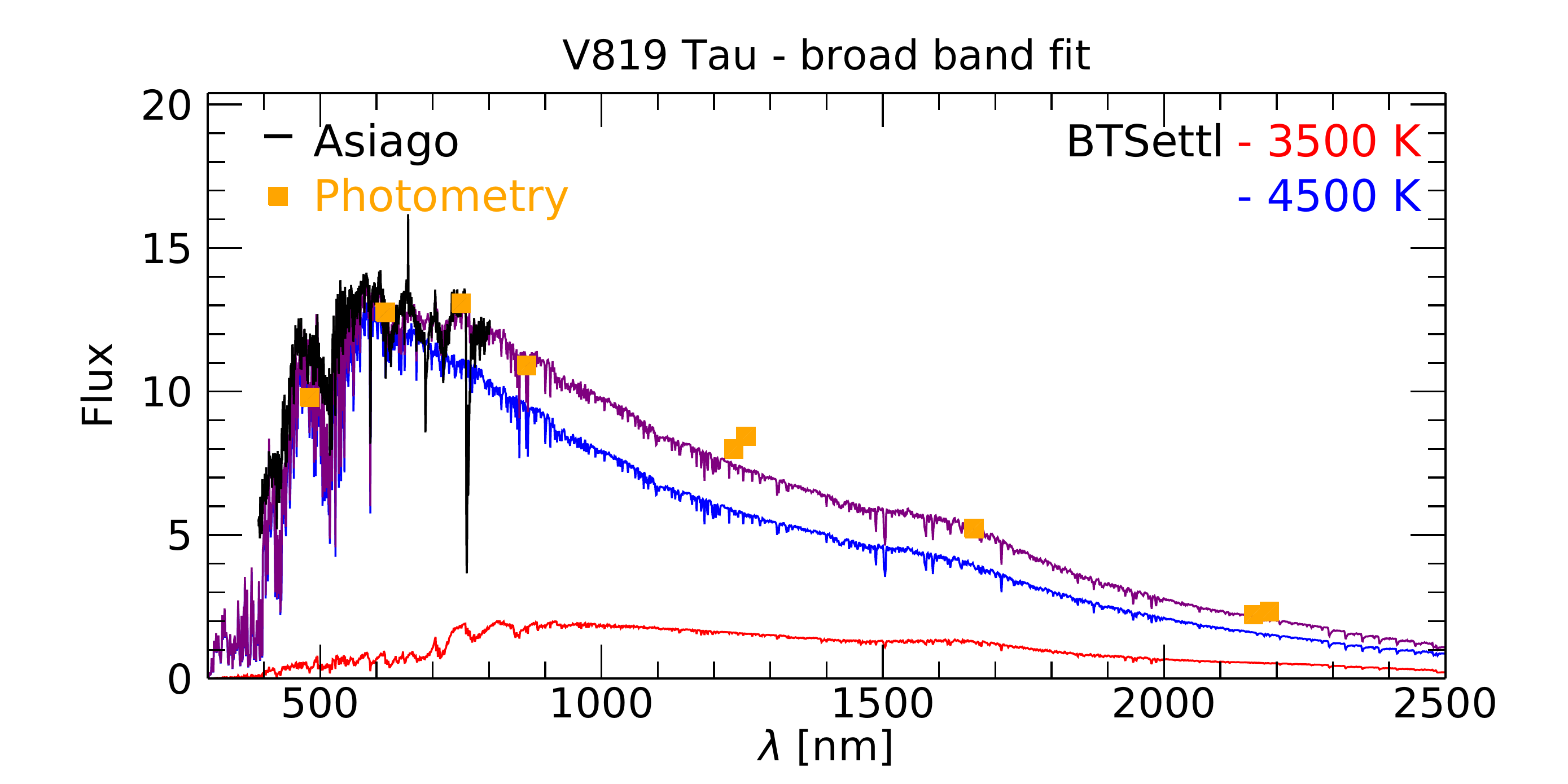}
\includegraphics[trim=45 0 0 0,width=0.98\columnwidth]{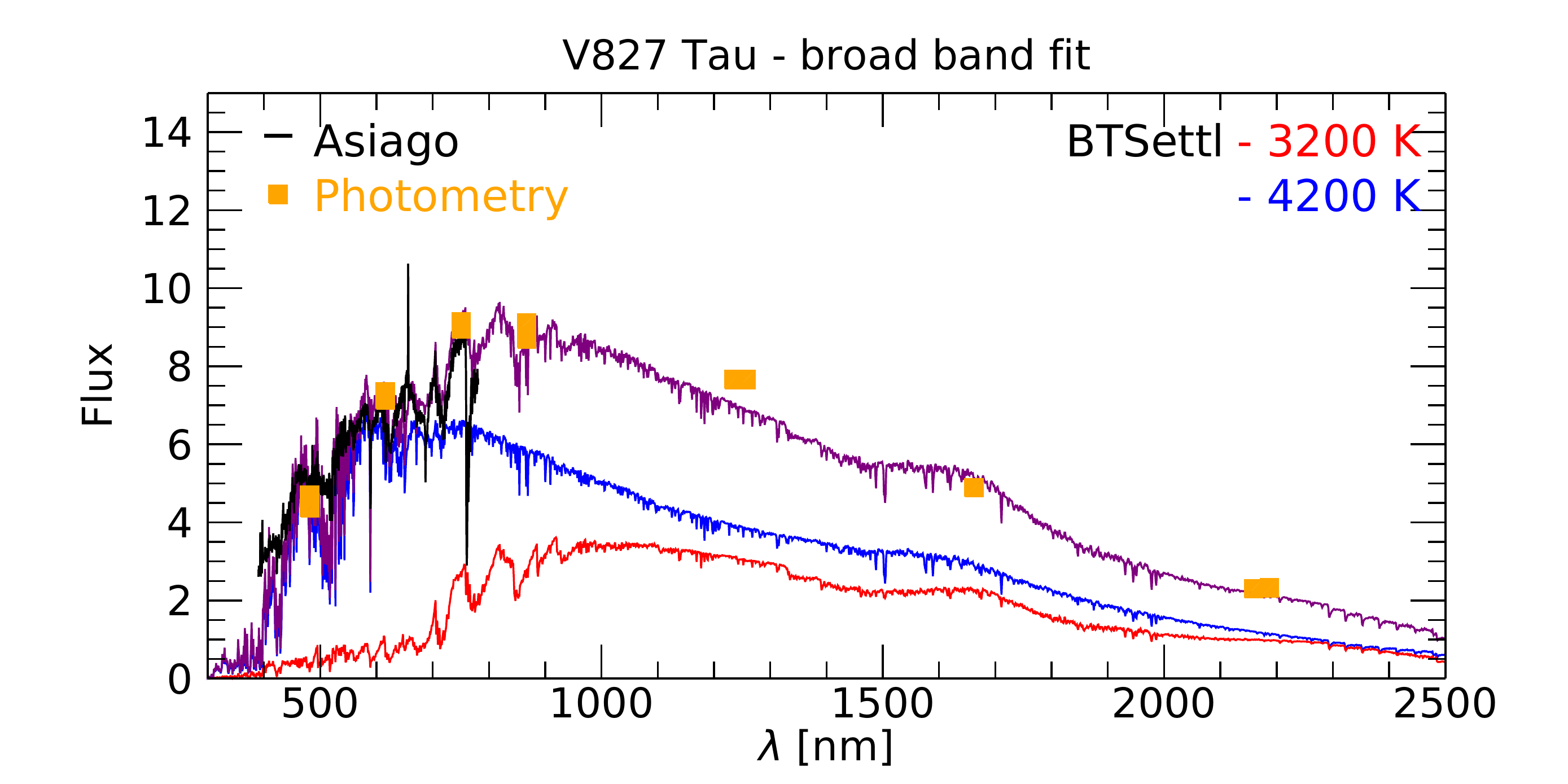}
\includegraphics[trim=45 0 0 0,width=0.98\columnwidth]{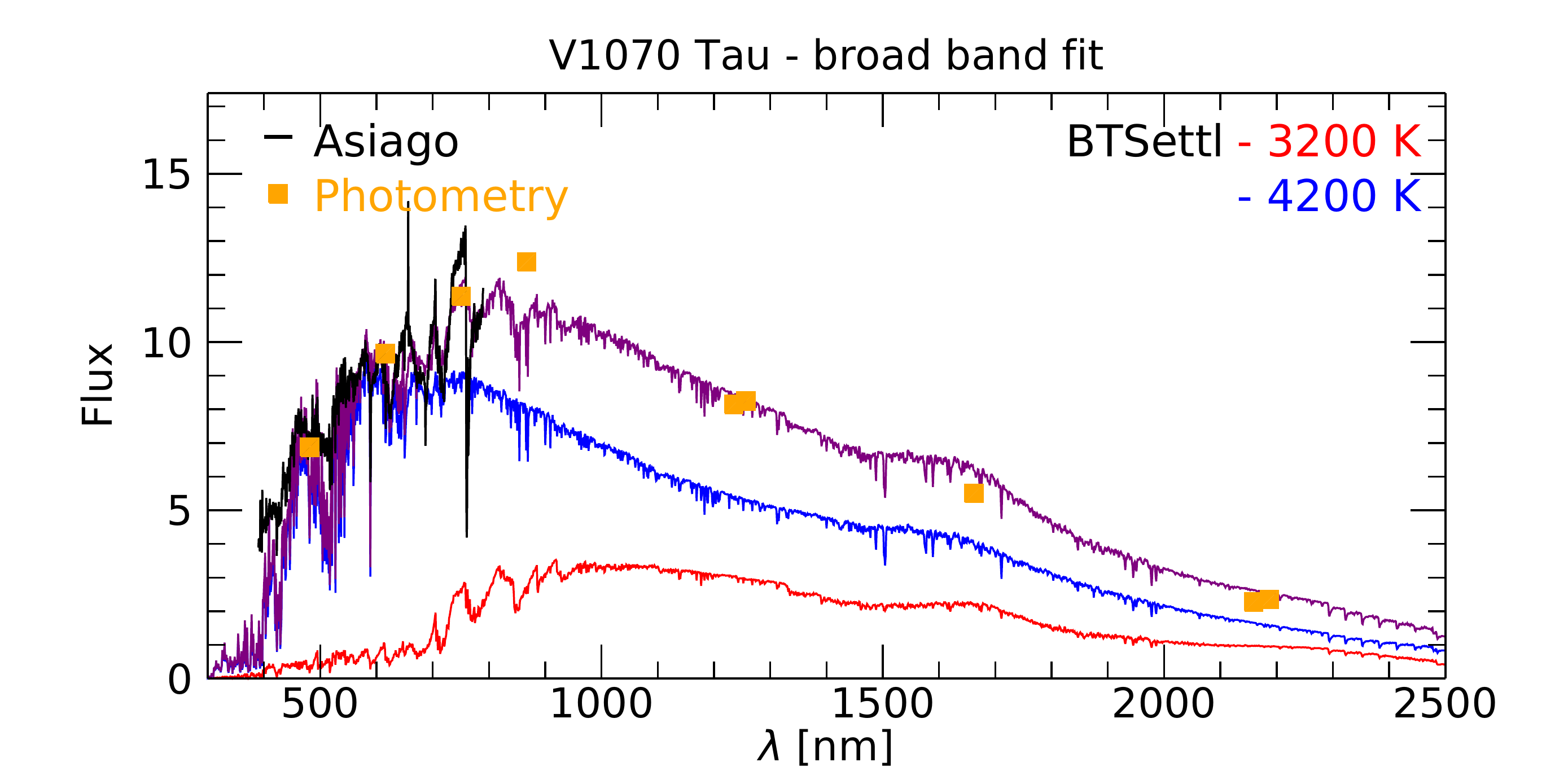}
\end{center}
\caption{\label{fig:Broad_band_fitting} 2 $\rm T_{eff}$ broadband fitting for the subsample of heavily spotted class III sources. The extinction-corrected Asiago spectrum is shown in black, and contemporaneous photometry is shown as filled orange squares. The best-fit spectrum is shown in purple, and the corresponding $\rm T_{hot}$ and $\rm T_{cool}$ BTSettl models are reported in blue and red, respectively. Flux units are $\rm 10^{-13}\ ergs^{-1}\ cm^{-2}\ nm^{-1}$.}
\end{figure*}

\begin{figure*}
\begin{center}
\includegraphics[trim=45 0 0 0,width=0.98\columnwidth]{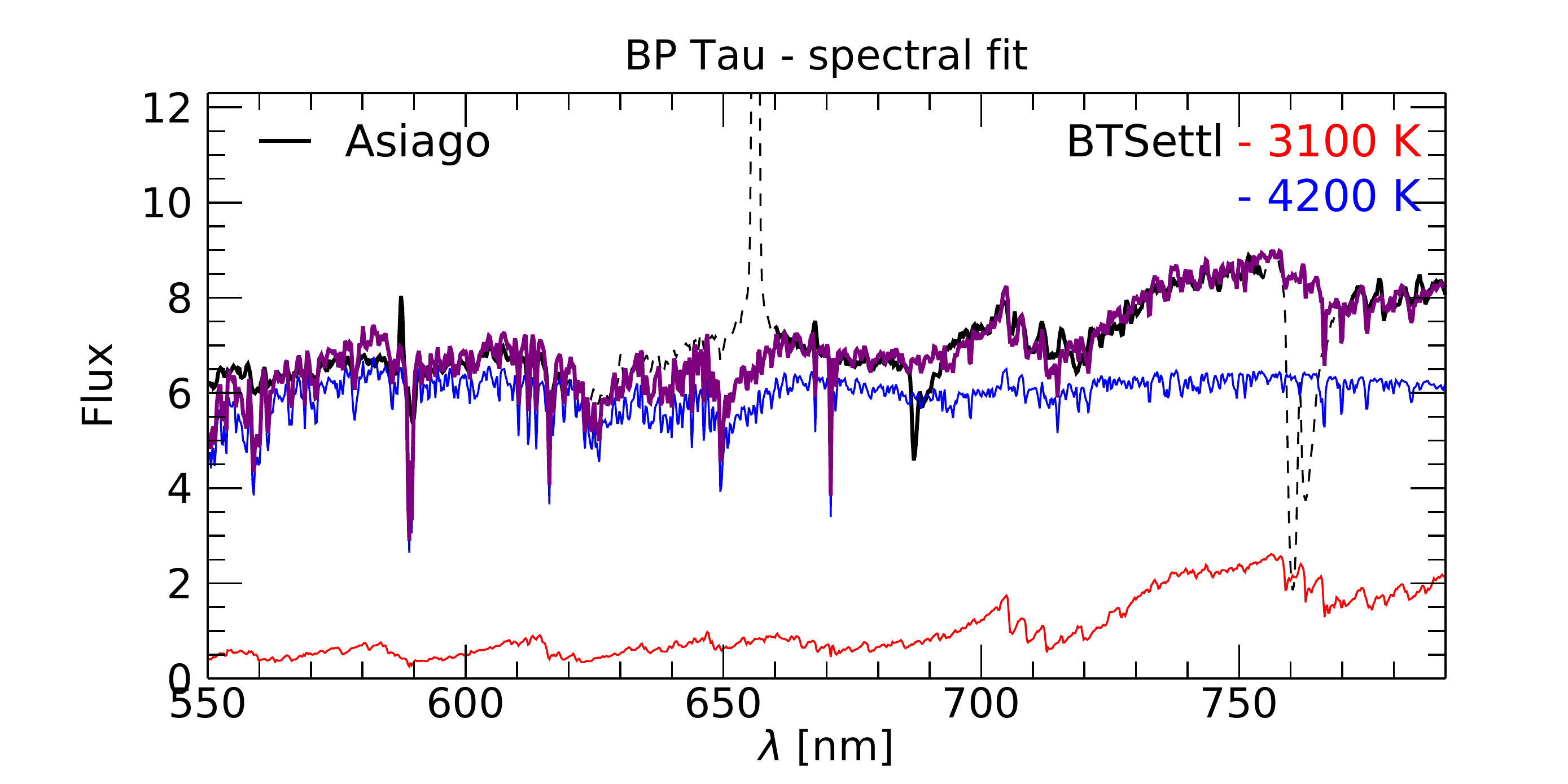}
\includegraphics[trim=45 0 0 0,width=0.98\columnwidth]{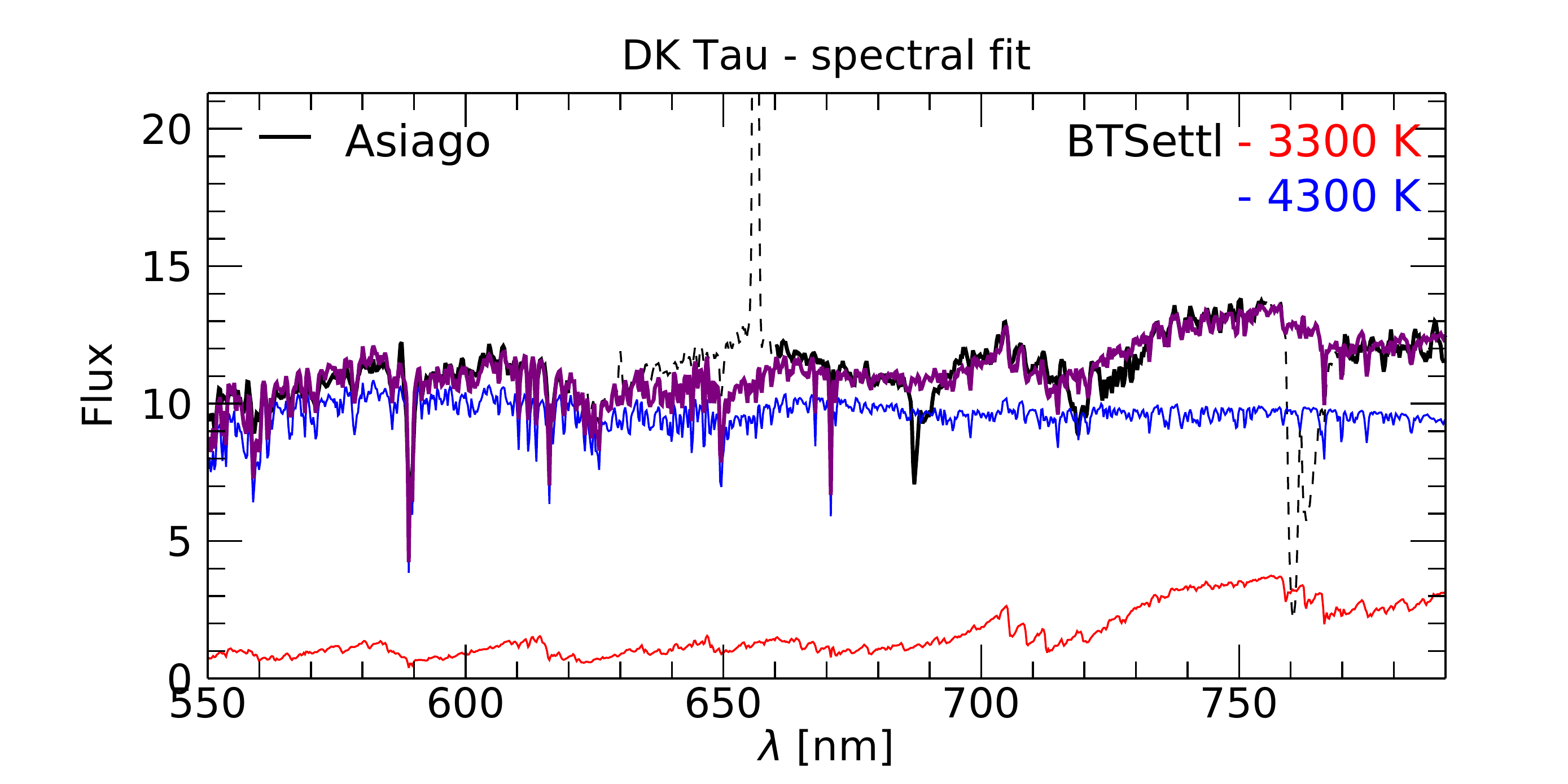}
\end{center}
\caption{\label{fig:Spectral_fitting} 2 $\rm T_{eff}$ spectral fitting for the subsample of heavily spotted class II sources. The extinction-corrected Asiago spectrum is shown in black. The best-fit spectrum is shown in purple, and the corresponding $\rm T_{hot}$ and $\rm T_{cool}$ BTSettl models are reported in blue and red, respectively. Flux units are $\rm 10^{-13}\ ergs^{-1}\ cm^{-2}\ nm^{-1}$.}
\end{figure*}

\begin{figure*}
\begin{center}
\includegraphics[trim=45 0 0 0,width=0.98\columnwidth]{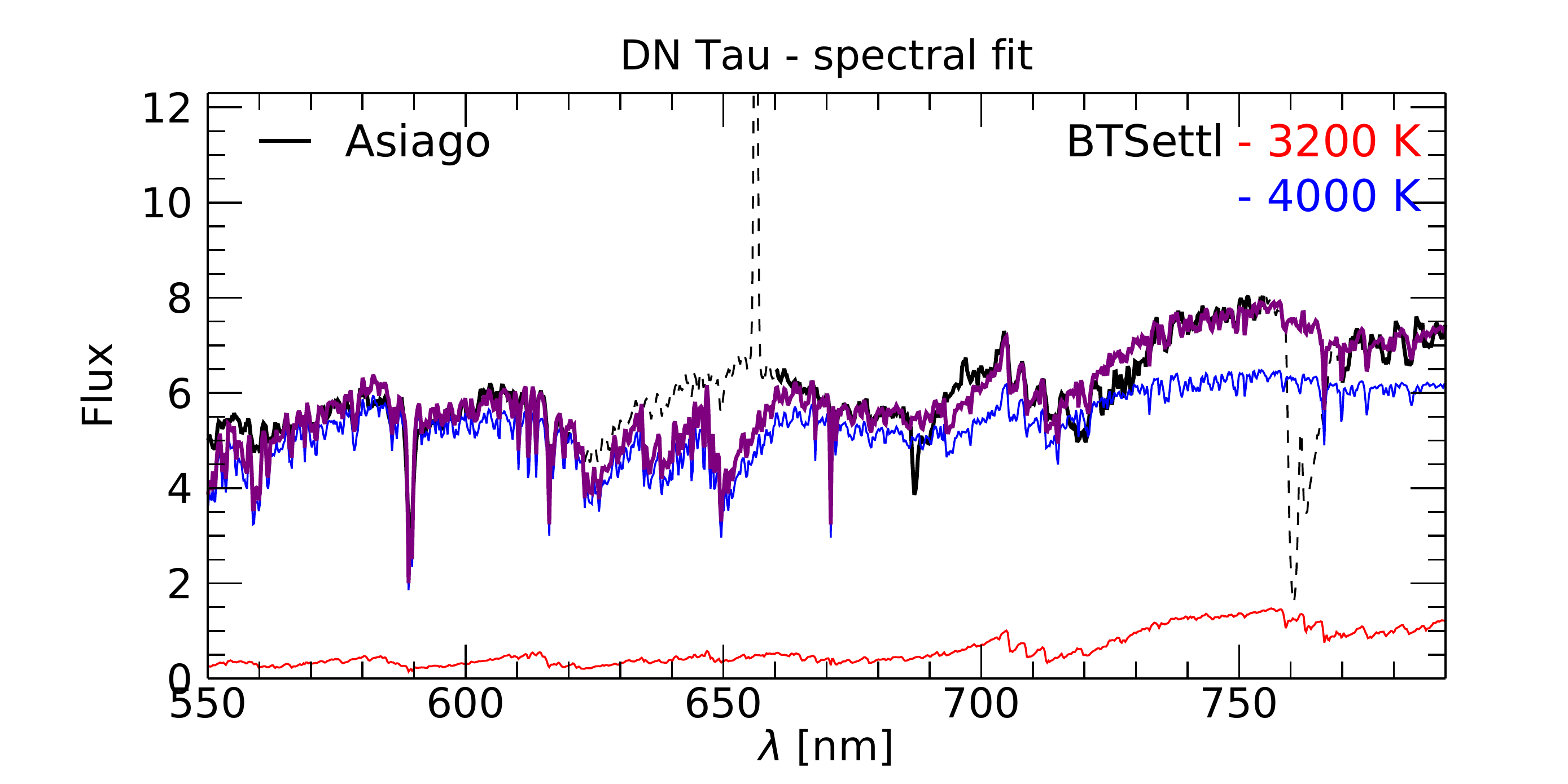}
\includegraphics[trim=45 0 0 0,width=0.98\columnwidth]{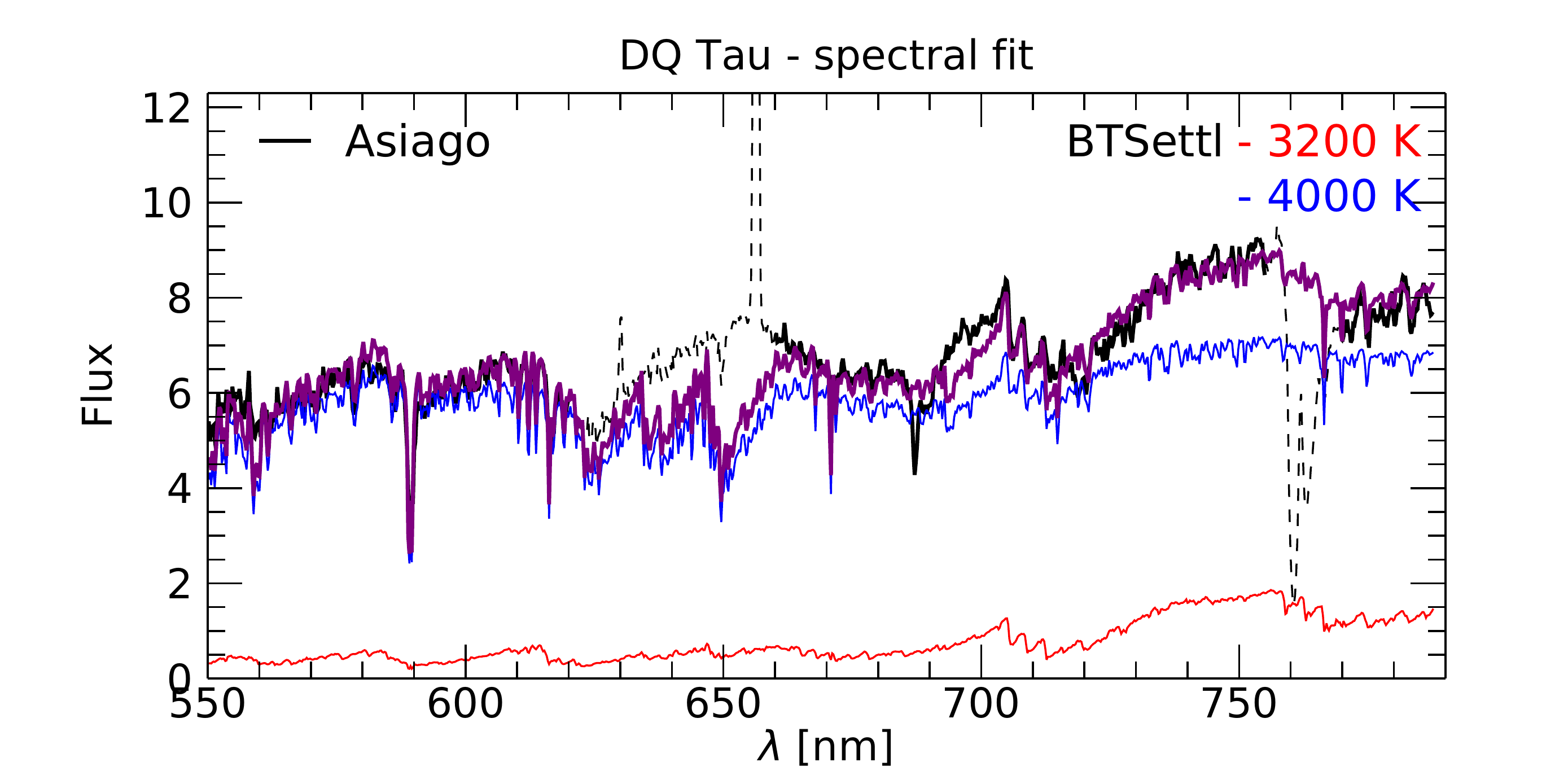}
\includegraphics[trim=45 0 0 0,width=0.98\columnwidth]{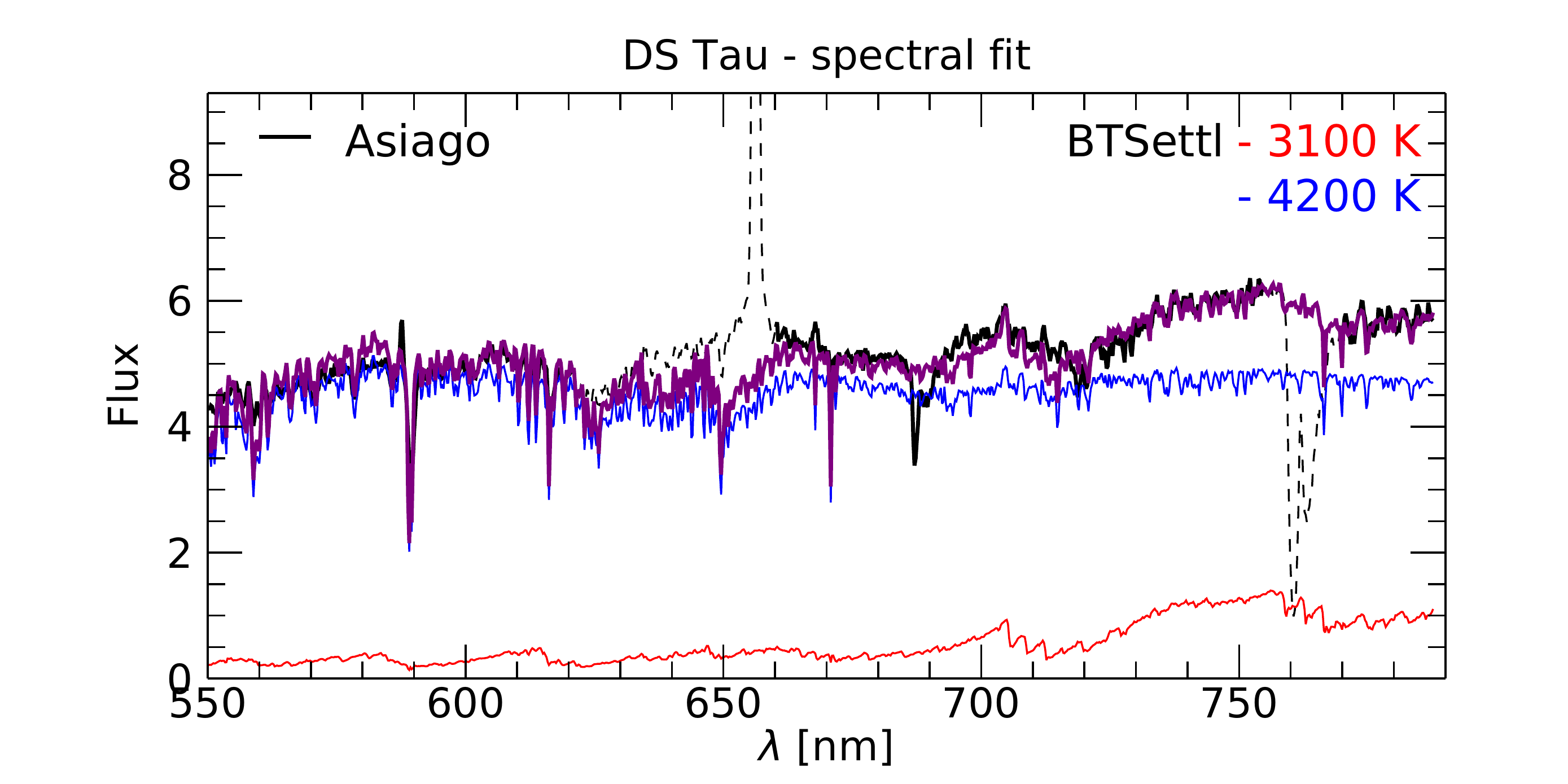}
\includegraphics[trim=45 0 0 0,width=0.98\columnwidth]{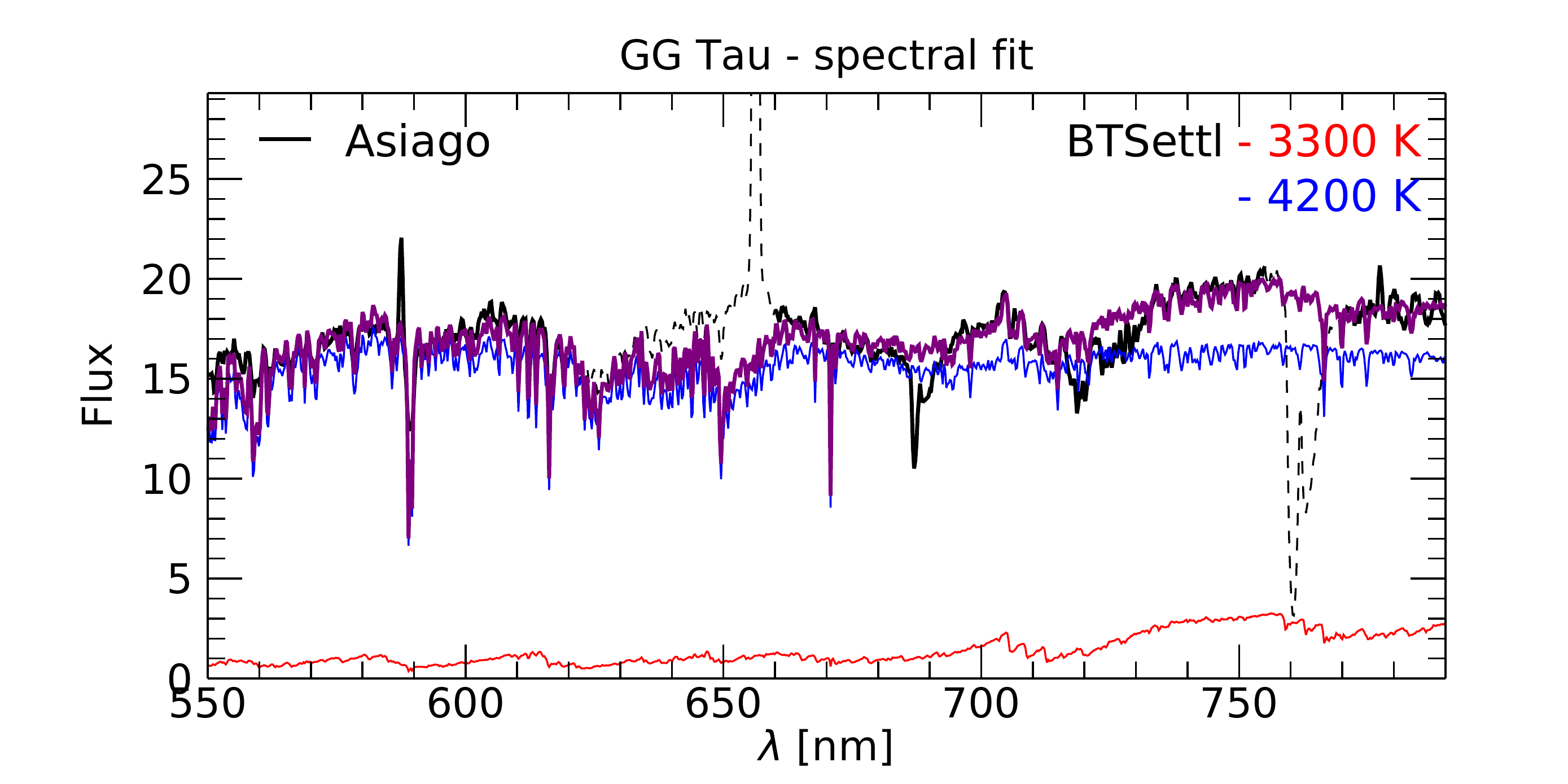}
\includegraphics[trim=45 0 0 0,width=0.98\columnwidth]{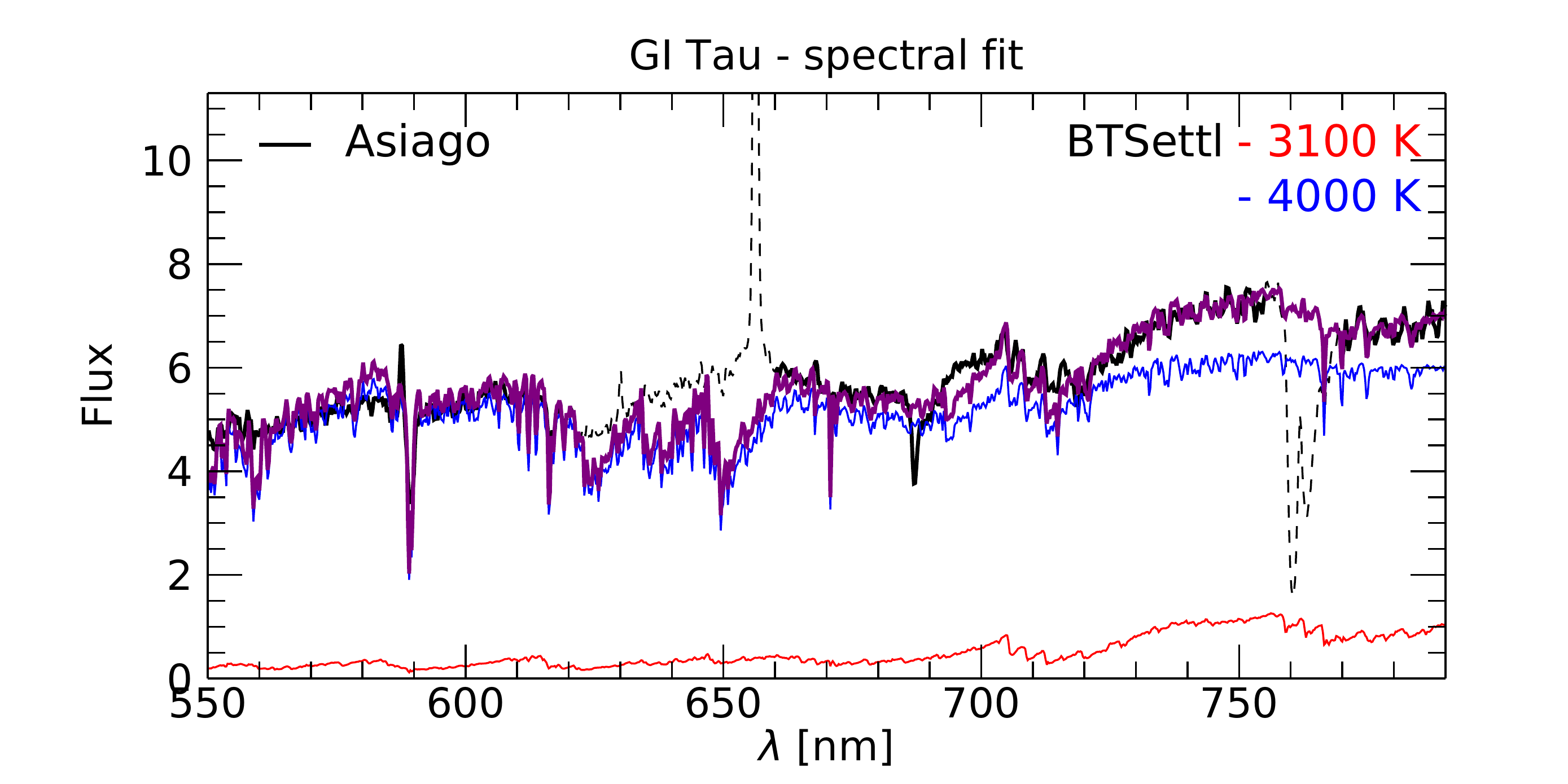}
\includegraphics[trim=45 0 0 0,width=0.98\columnwidth]{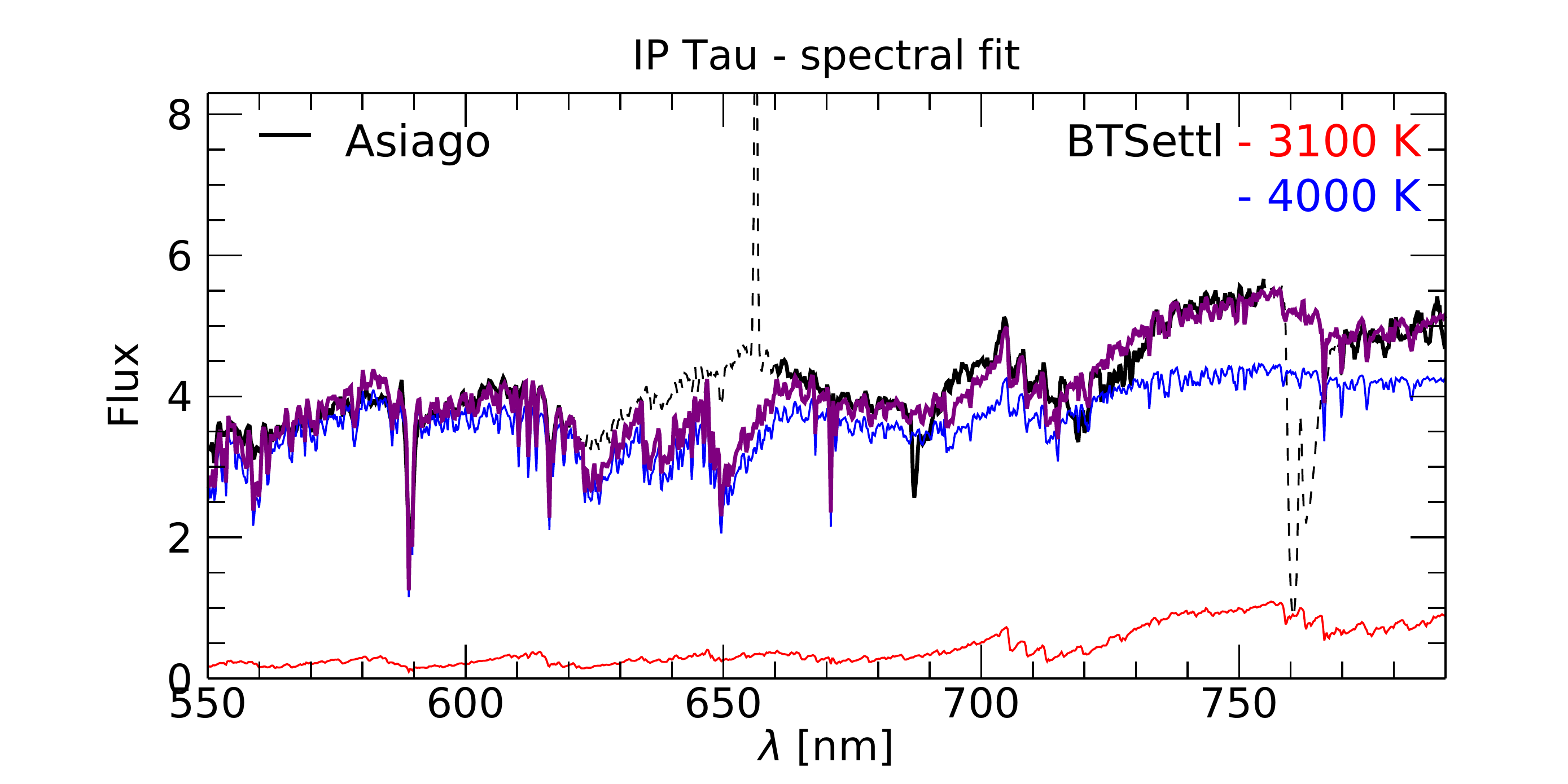}
\includegraphics[trim=45 0 0 0,width=0.98\columnwidth]{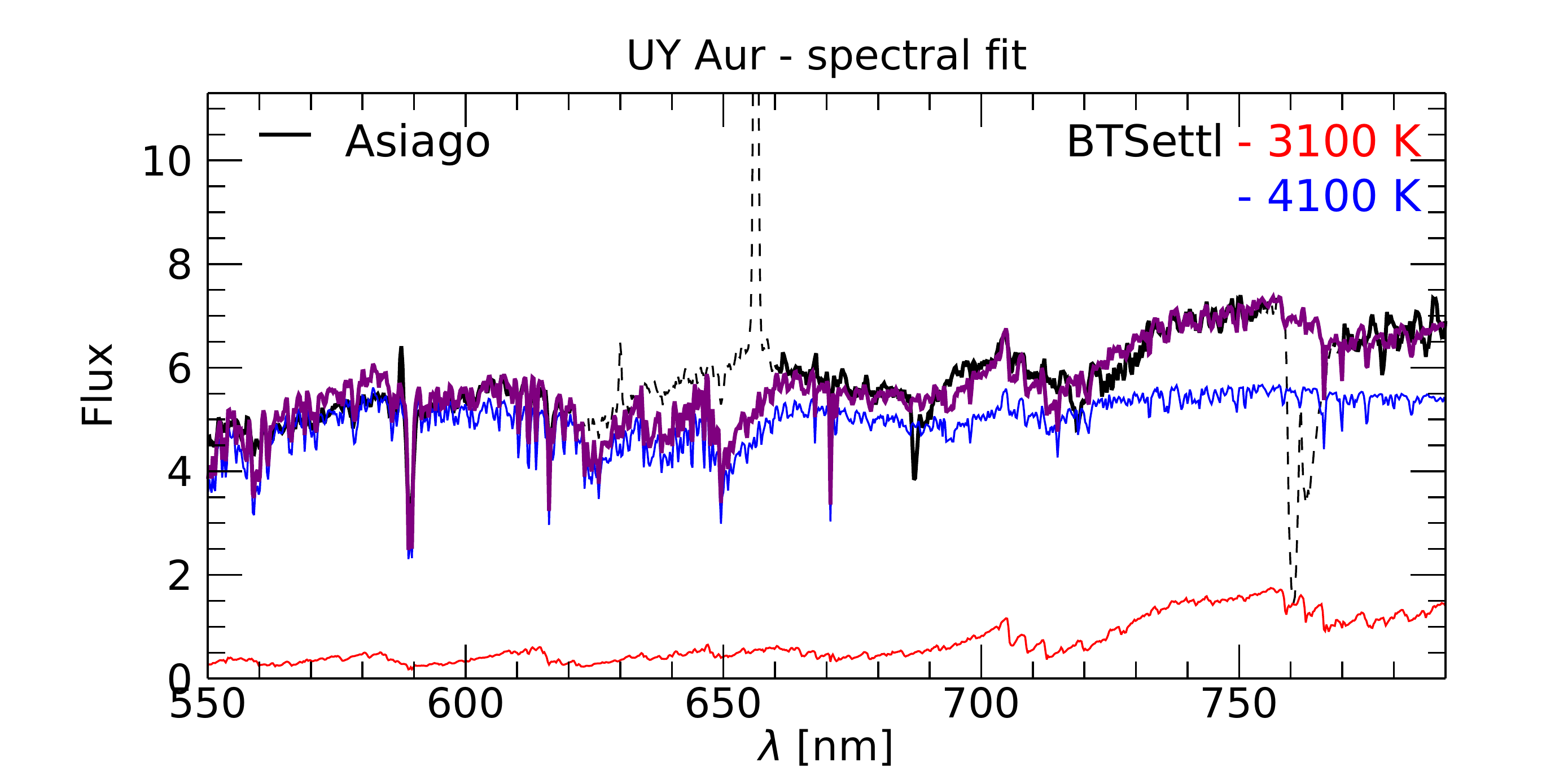}
\includegraphics[trim=45 0 0 0,width=0.98\columnwidth]{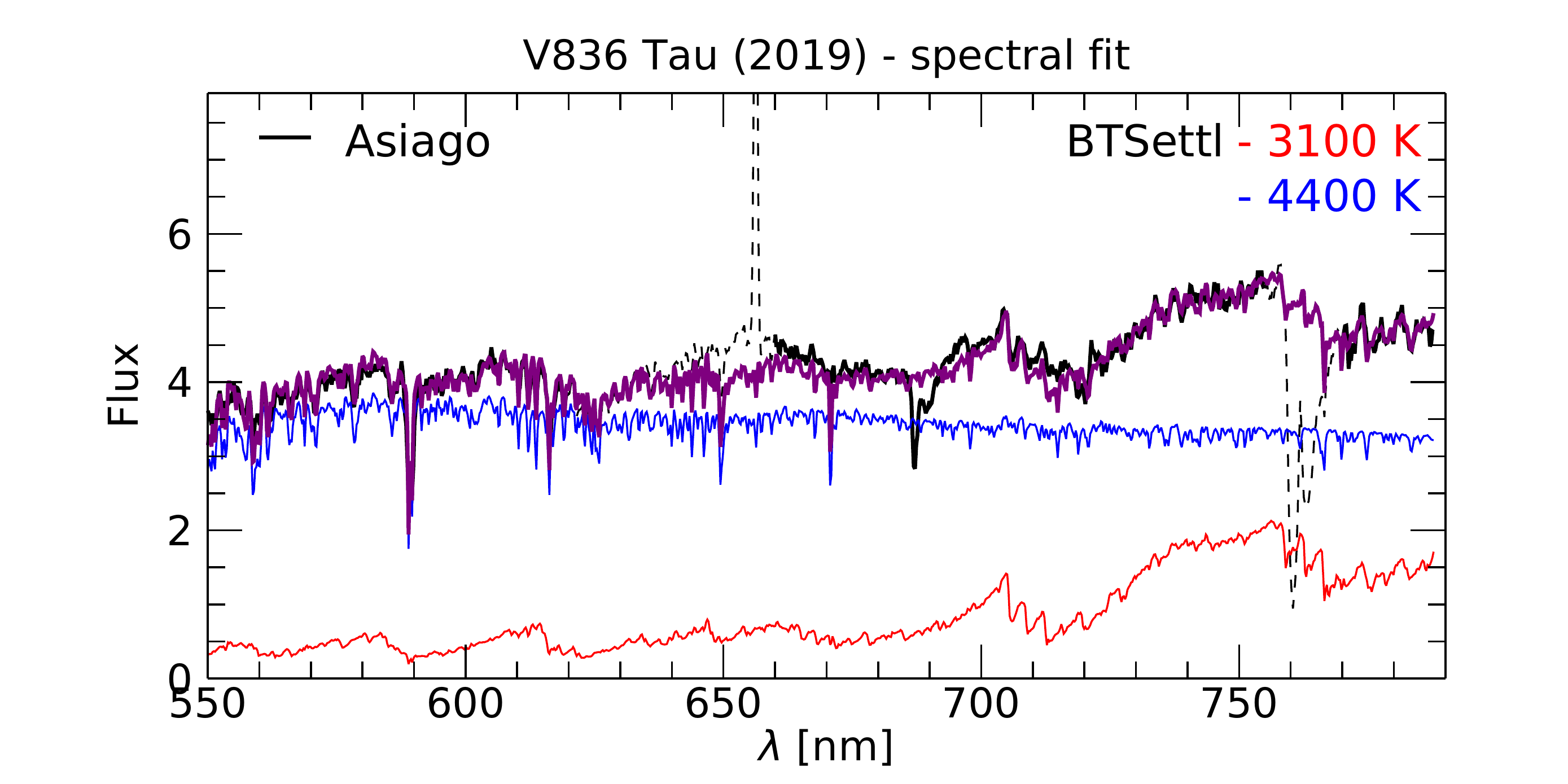}
\includegraphics[trim=45 0 0 0,width=0.98\columnwidth]{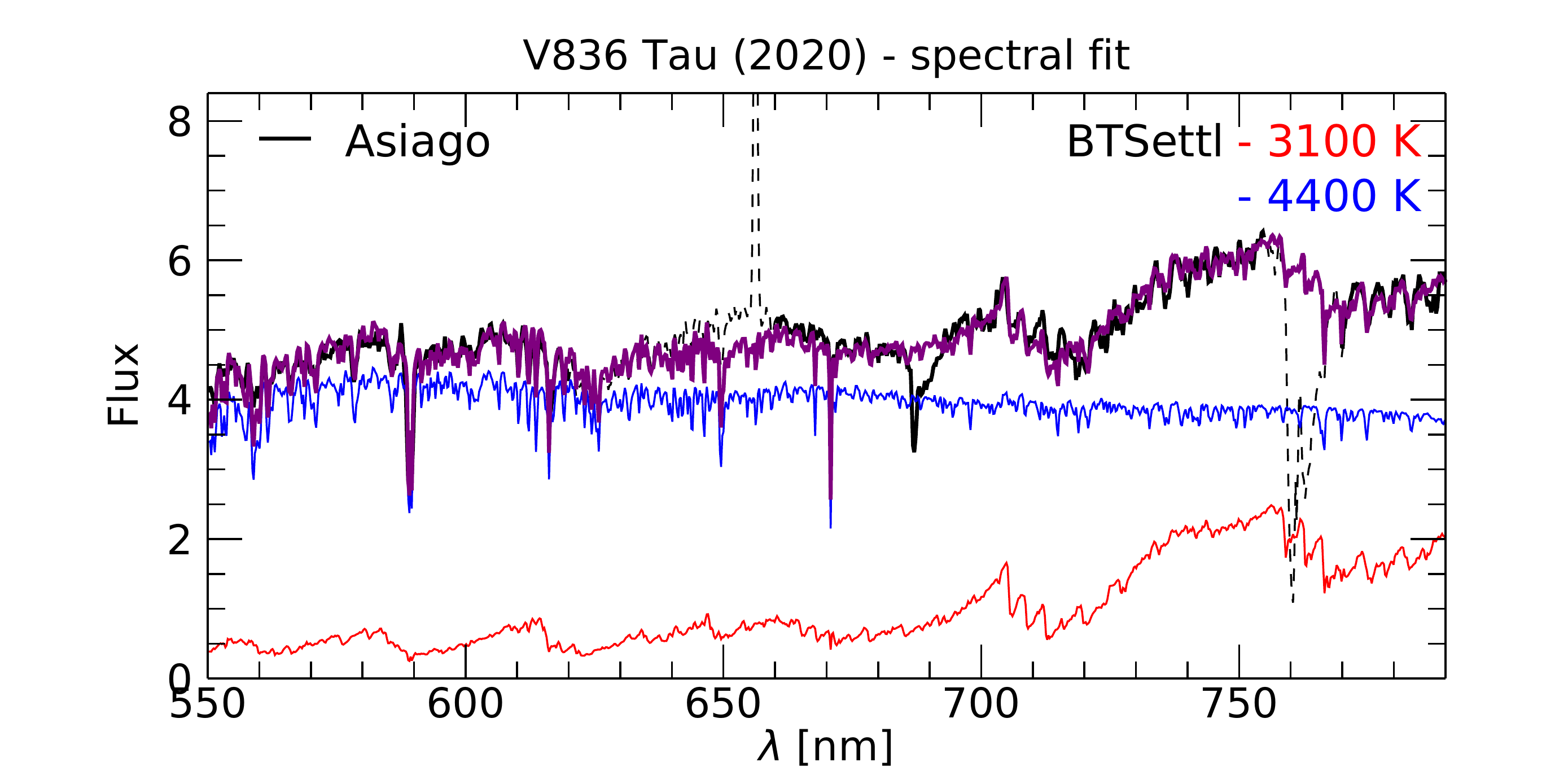}
\end{center}
\begin{center} \textbf{Fig. B.2.} continued.\end{center}
\end{figure*}

\begin{figure*}
\begin{center}
\includegraphics[trim=0 80 80 0,width=0.9\columnwidth, angle=0]{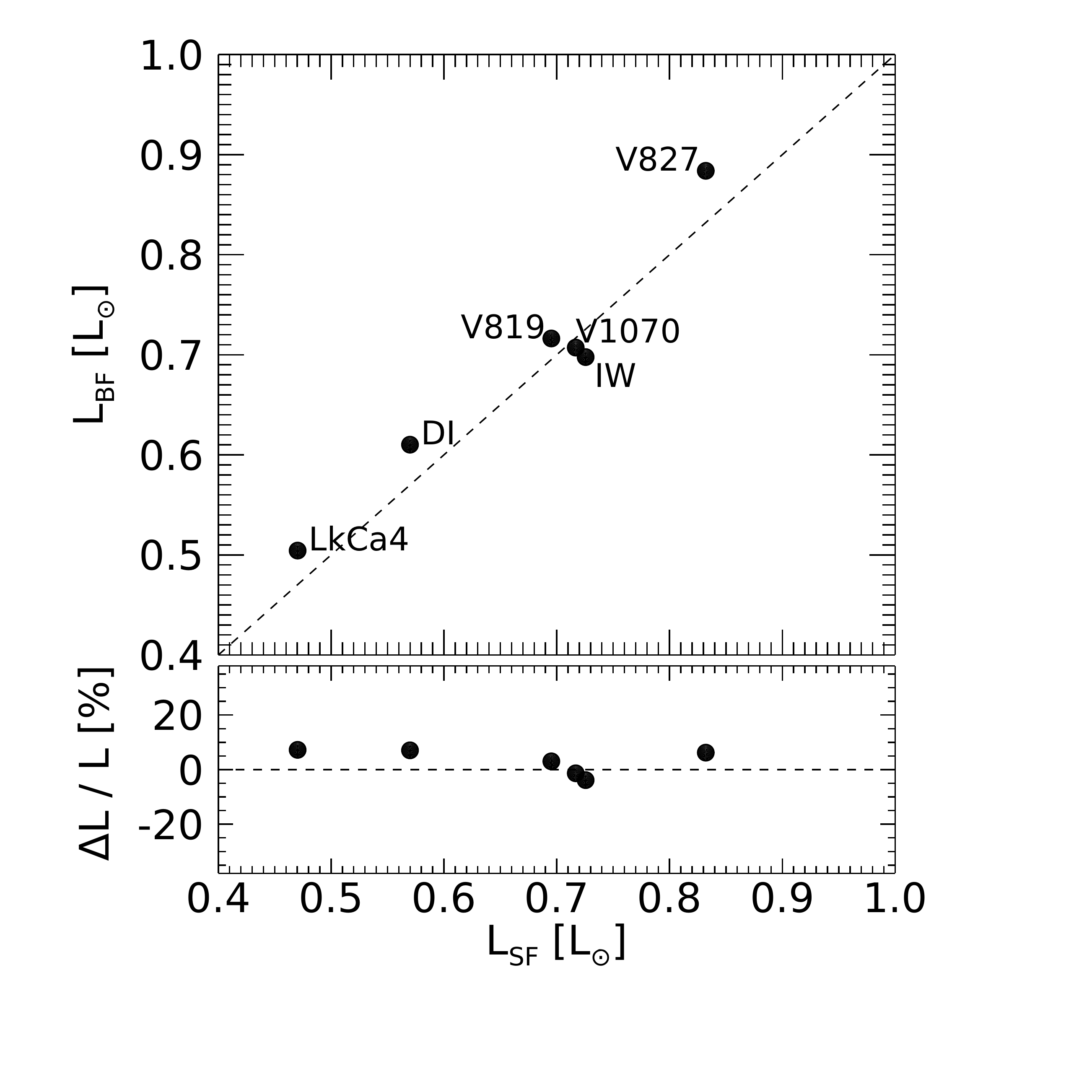}
\includegraphics[trim=0 80 80 0,width=0.9\columnwidth, angle=0]{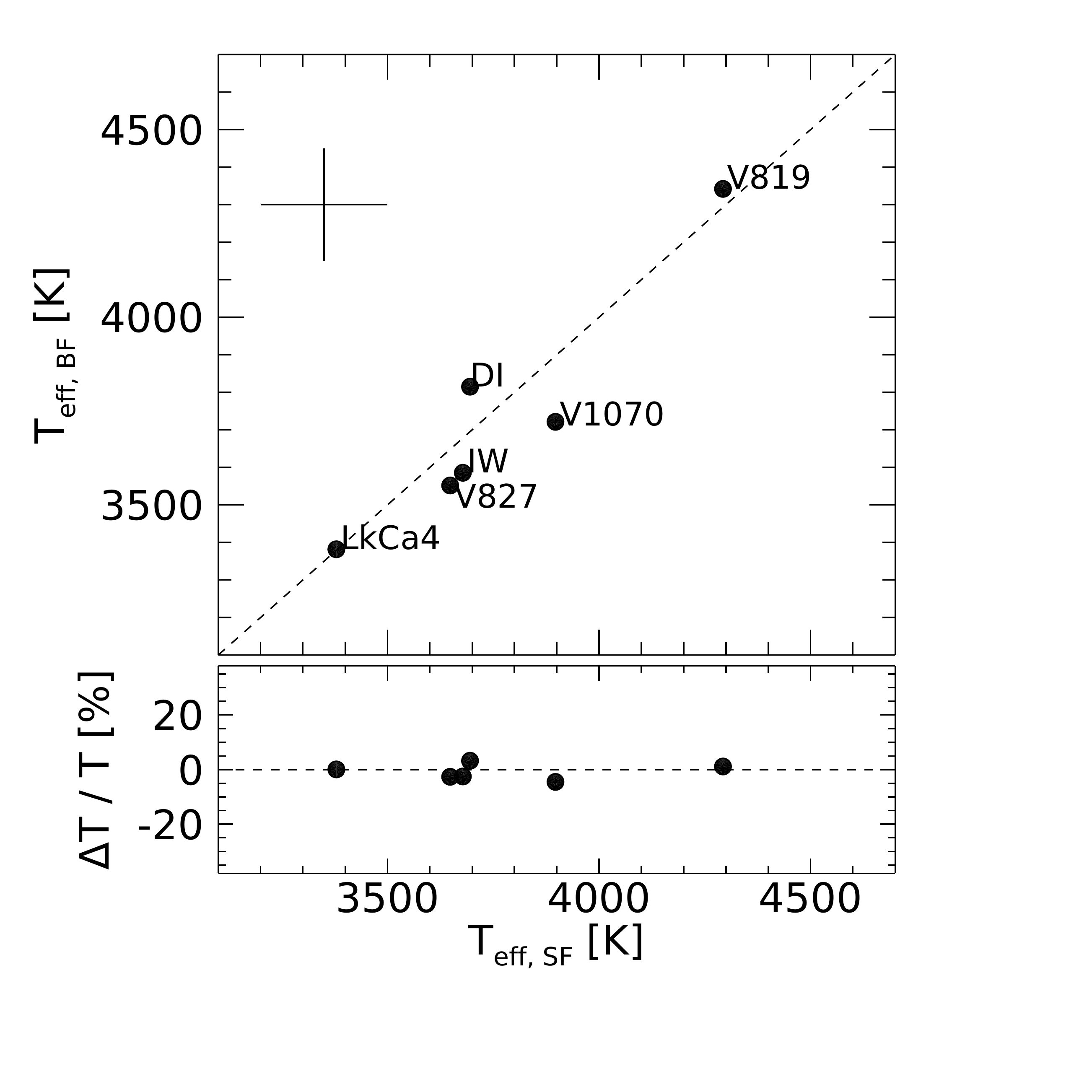}
\end{center}
\caption{\label{fig:Compare_L_Teff} Left: Stellar luminosity computed with the broadband fitting procedure ($\rm L_{BF}$) as a function of the luminosity computed with the spectral fitting ($\rm L_{SF}$) for class III sources. Right: Same comparison, but for effective temperatures. Differences are smaller than 10\% for luminosity and temperature.}
\end{figure*}

\begin{figure*}
\begin{center}
\includegraphics[trim=0 0 0 0,width=0.50\columnwidth, angle=90]{Figures/Lacc_vs_lines/BP_Tau_Lacc_2020.pdf}
\includegraphics[trim=0 0 0 0,width=0.50\columnwidth, angle=90]{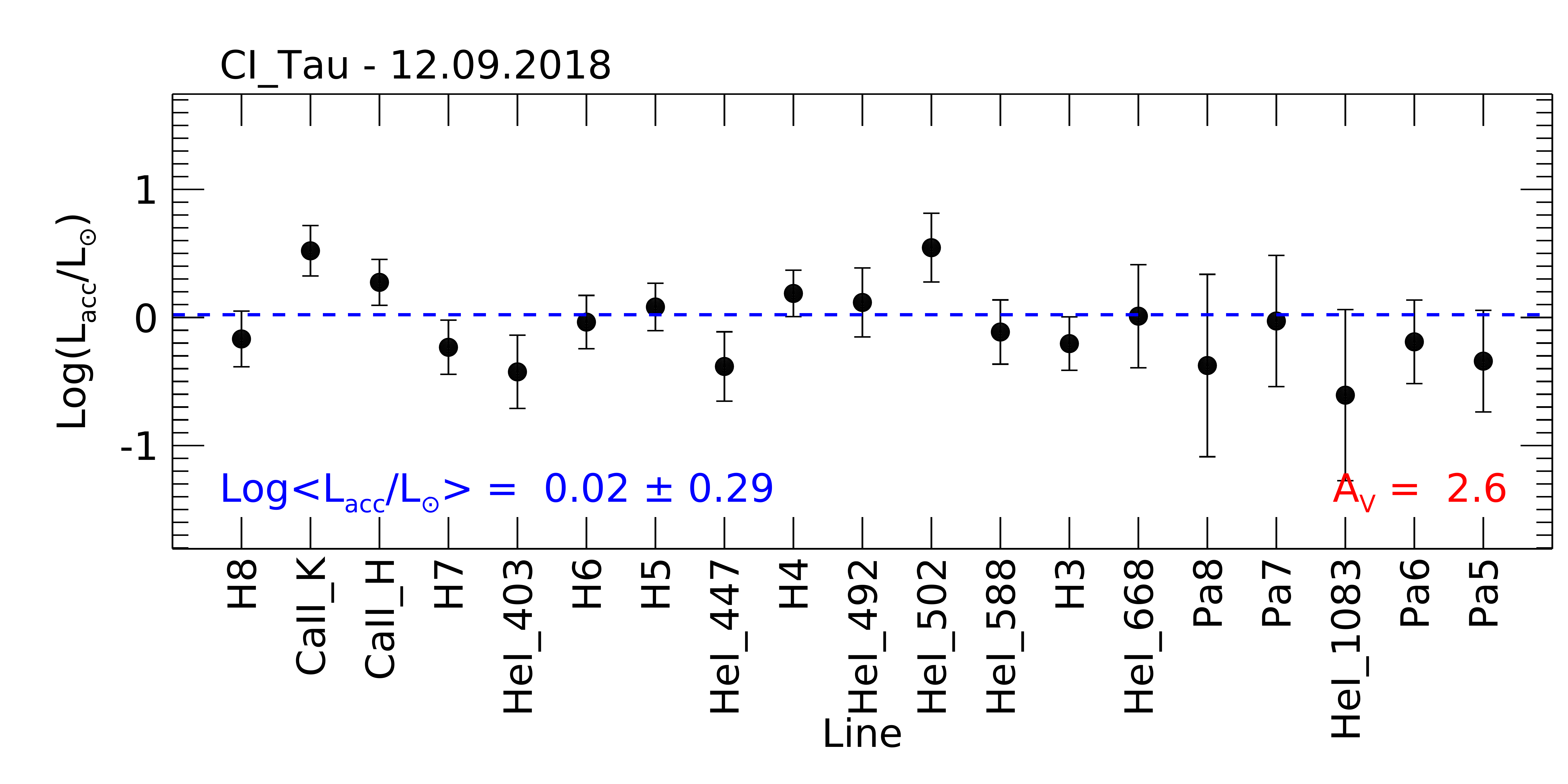}
\includegraphics[trim=0 0 0 0,width=0.50\columnwidth, angle=90]{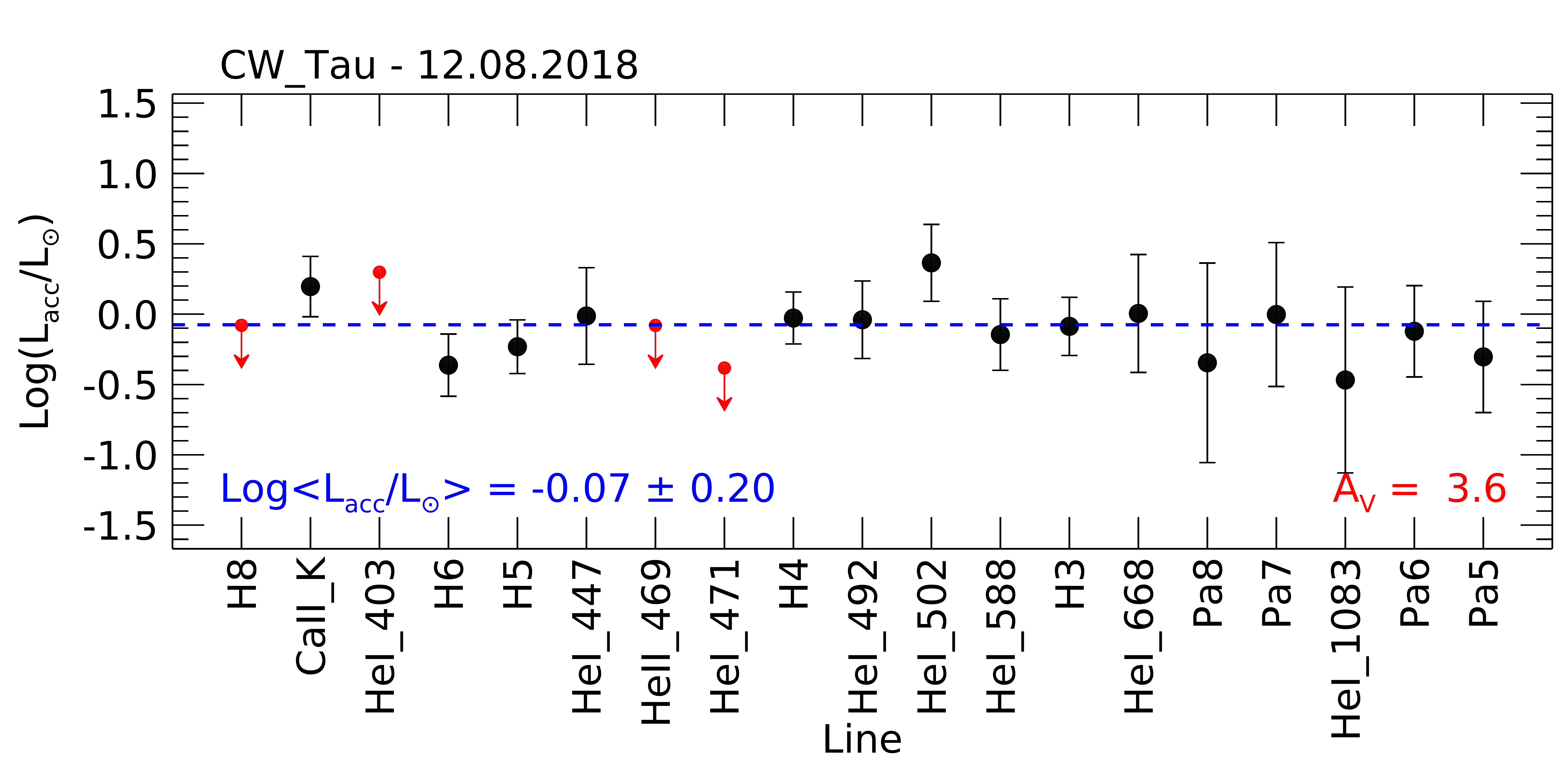}
\includegraphics[trim=0 0 0 0,width=0.50\columnwidth, angle=90]{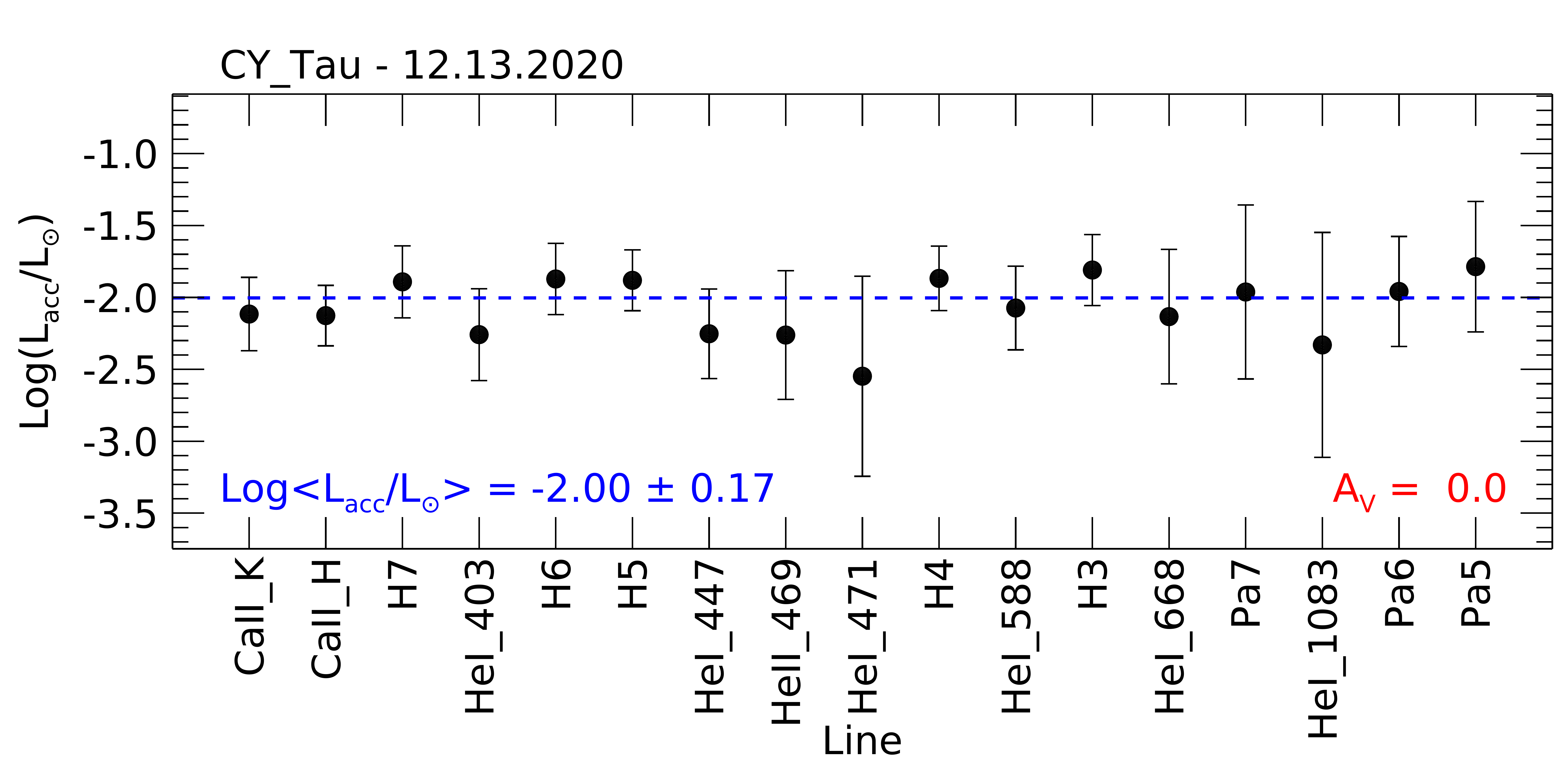}
\end{center}
\caption{\label{fig:Lacc_vs_lines} Plots of $\rm L_{acc}$ as a function of the different accretion diagnostics. The dashed horizontal blue line represents the median \lacc. Upper limits for nondetected lines are indicated with red arrows. For each panel, the target name, the observation date (MM.DD.YYYY), and the computed \lacc\ and $\rm A_v$ are also labeled.}
\end{figure*}

\begin{figure*}
\includegraphics[trim=0 0 0 0,width=1.00\columnwidth, angle=0]{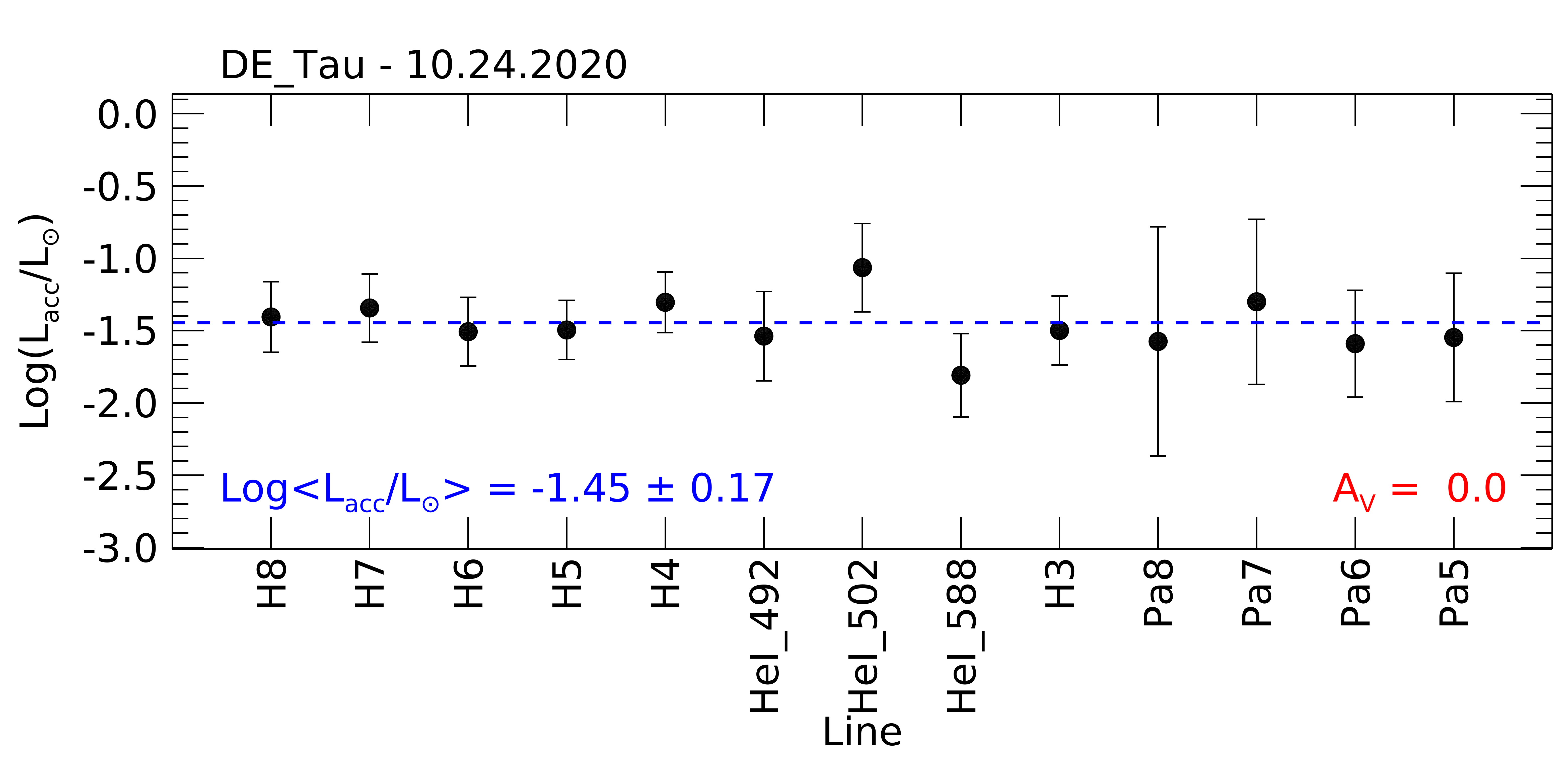}
\includegraphics[trim=0 0 0 0,width=0.50\columnwidth, angle=90]{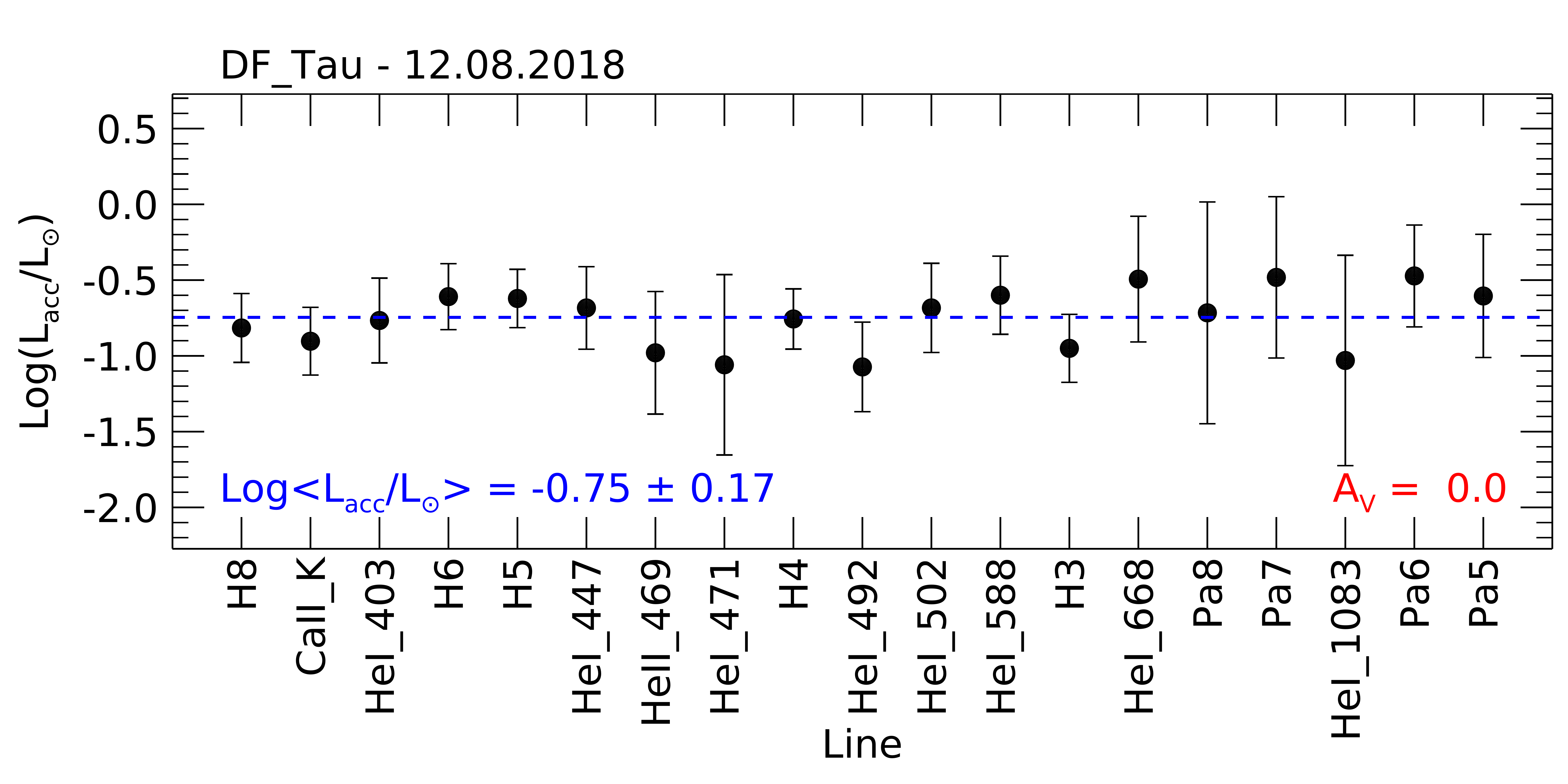}
\includegraphics[trim=0 0 0 0,width=0.50\columnwidth, angle=90]{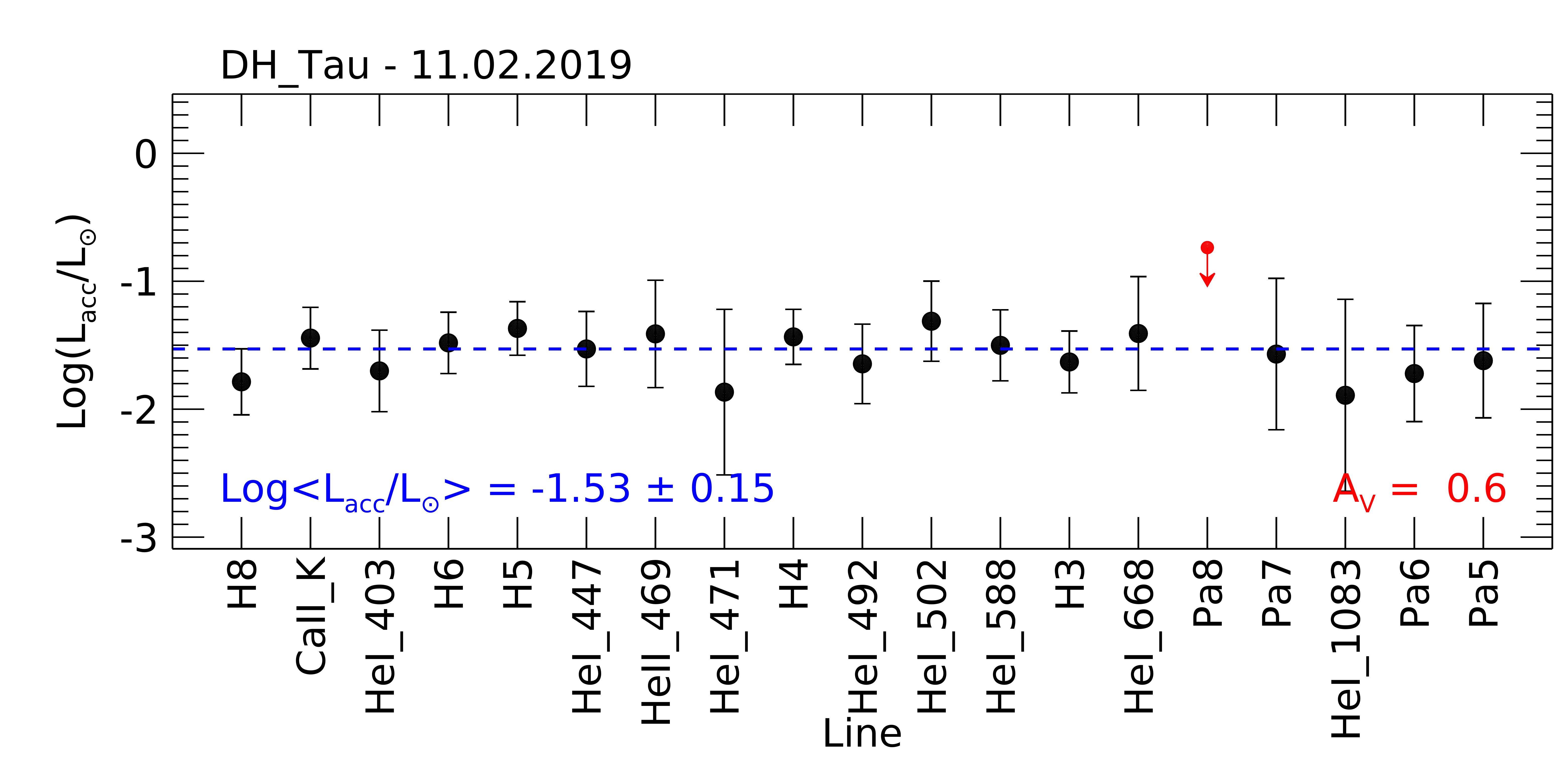}
\includegraphics[trim=0 0 0 0,width=0.50\columnwidth, angle=90]{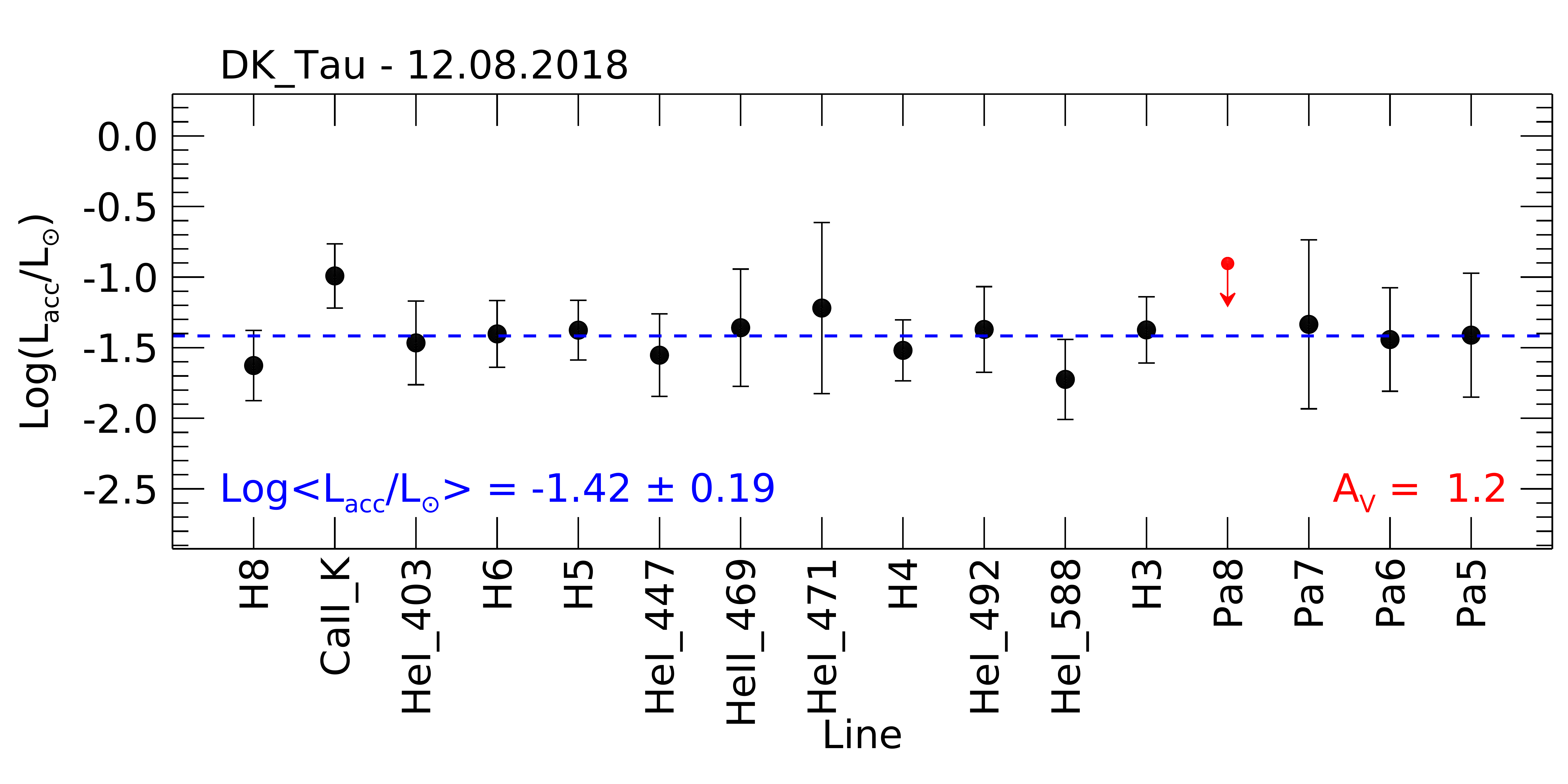}
\includegraphics[trim=0 0 0 0,width=1.00\columnwidth, angle=0]{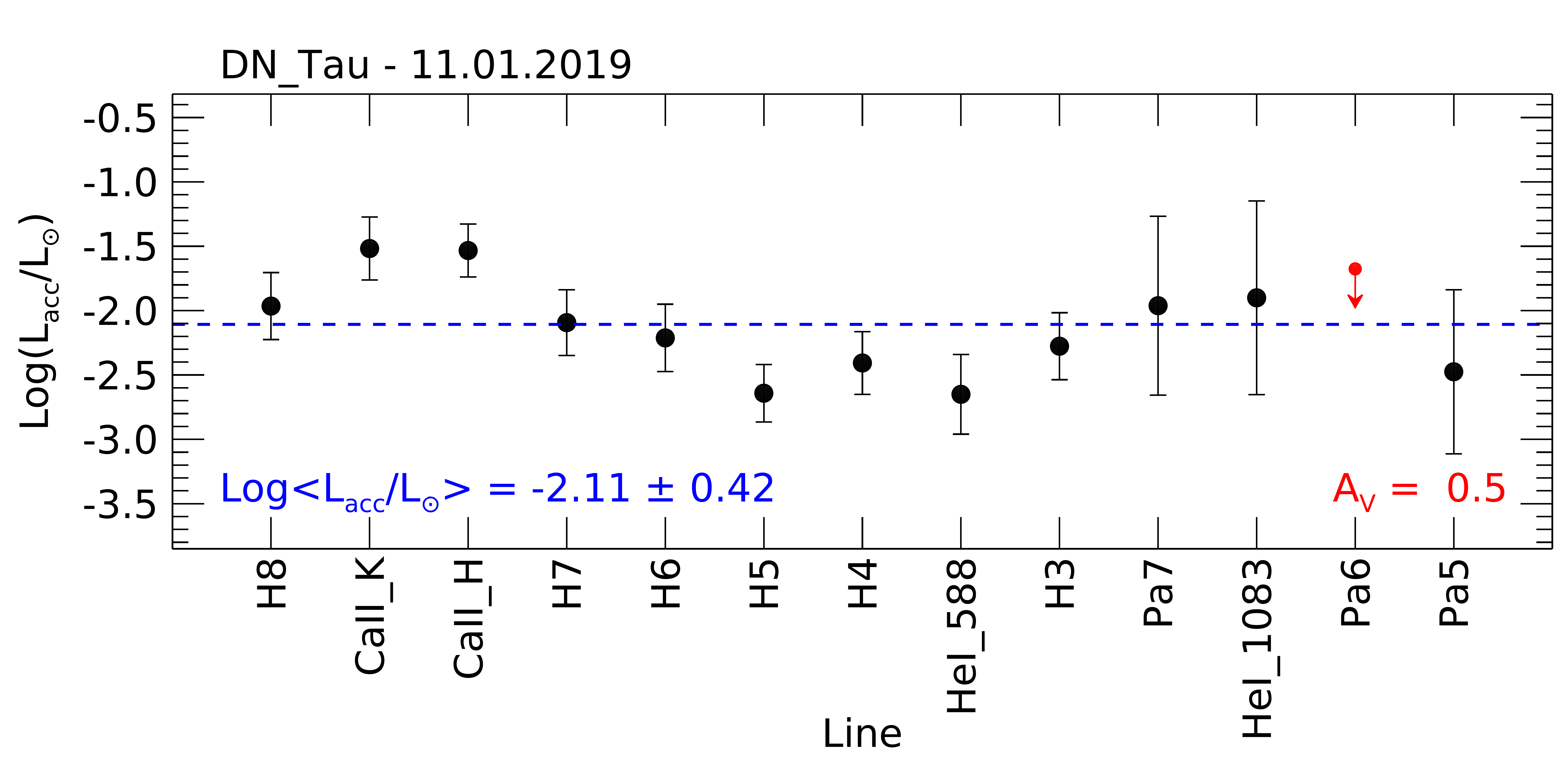}
\includegraphics[trim=0 0 0 0,width=0.50\columnwidth, angle=90]{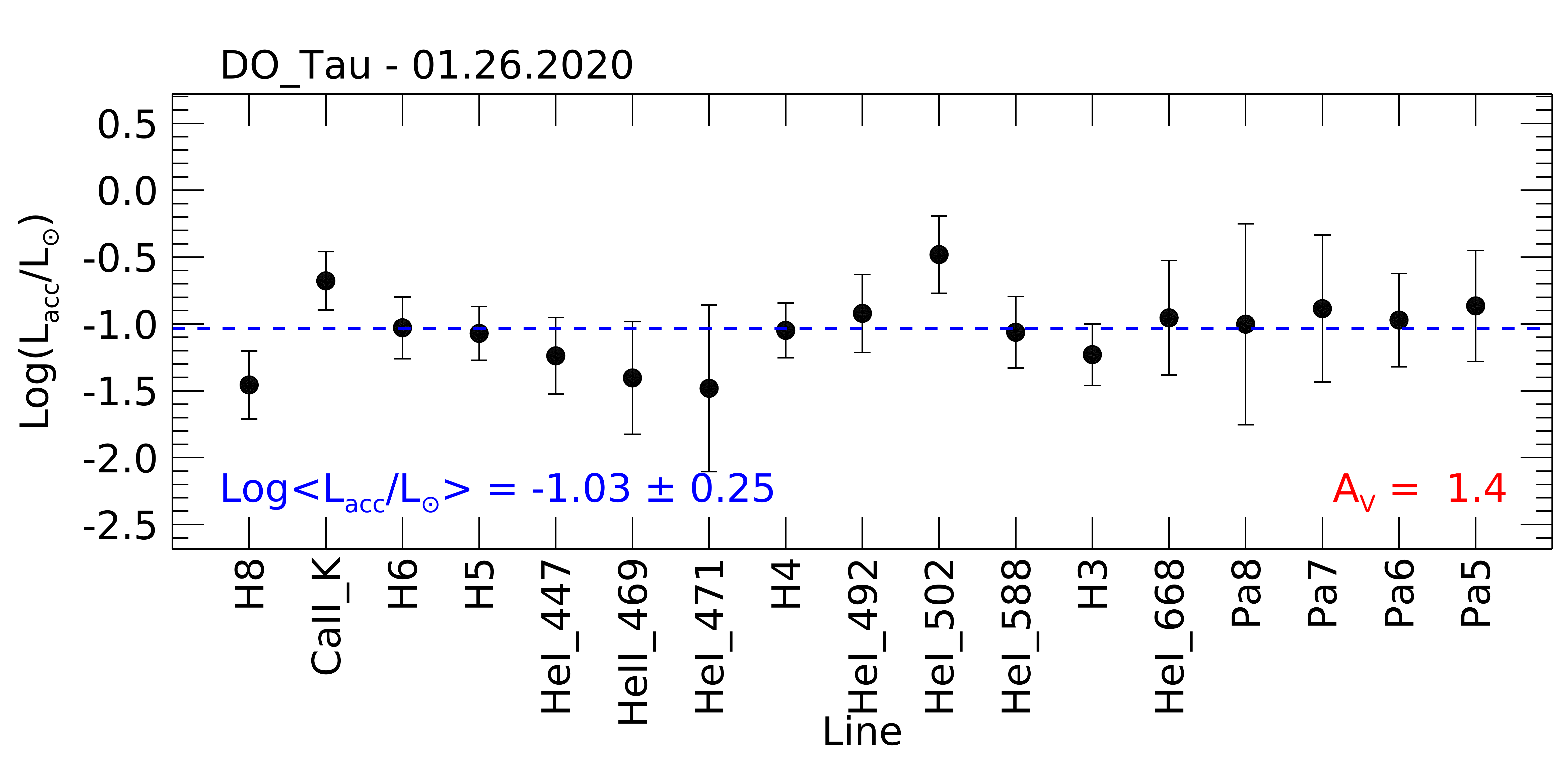}
\includegraphics[trim=0 0 0 0,width=1.00\columnwidth, angle=0]{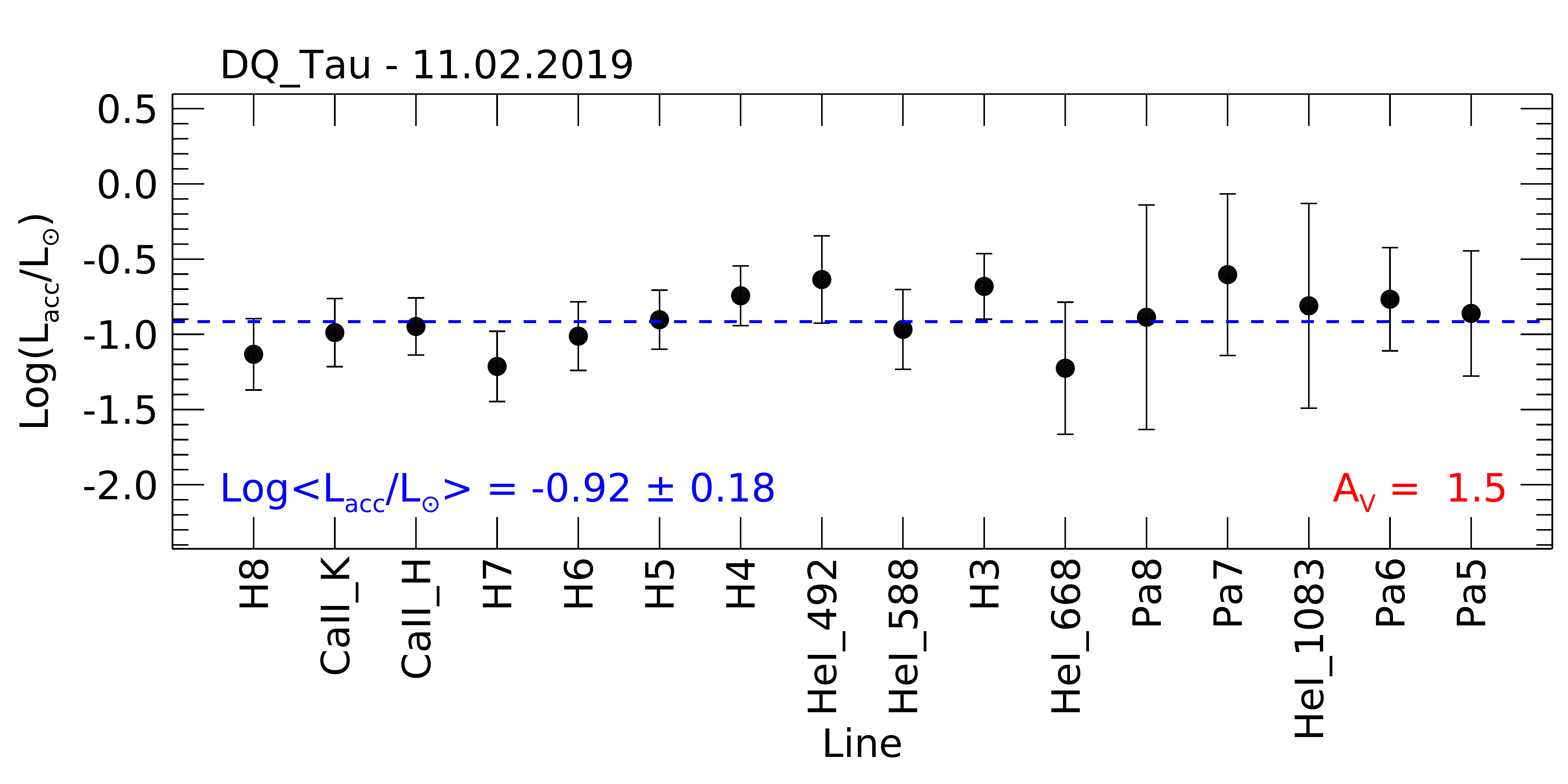}
\includegraphics[trim=0 0 0 0,width=0.50\columnwidth, angle=90]{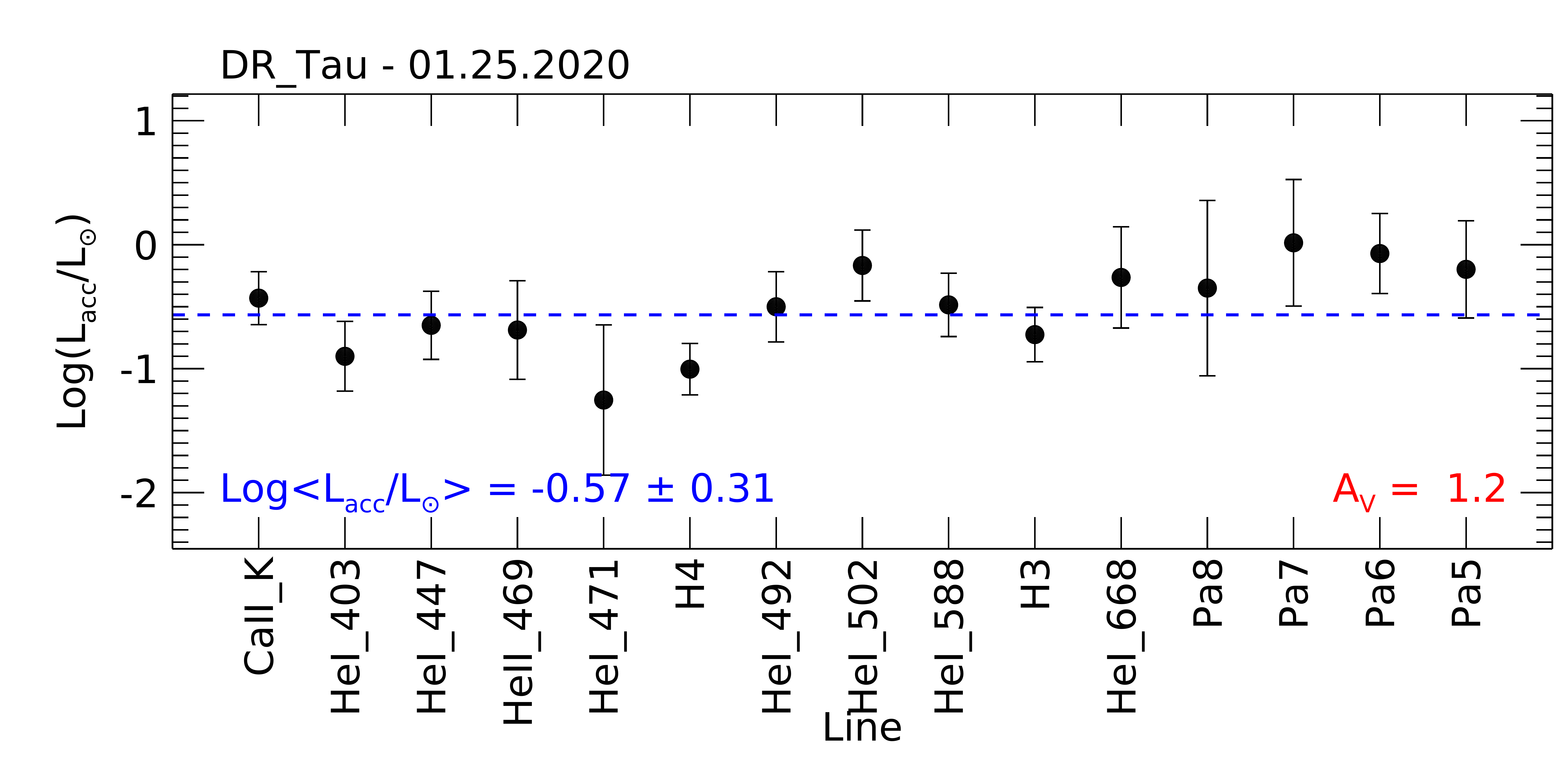}
\includegraphics[trim=0 0 0 0,width=0.50\columnwidth, angle=90]{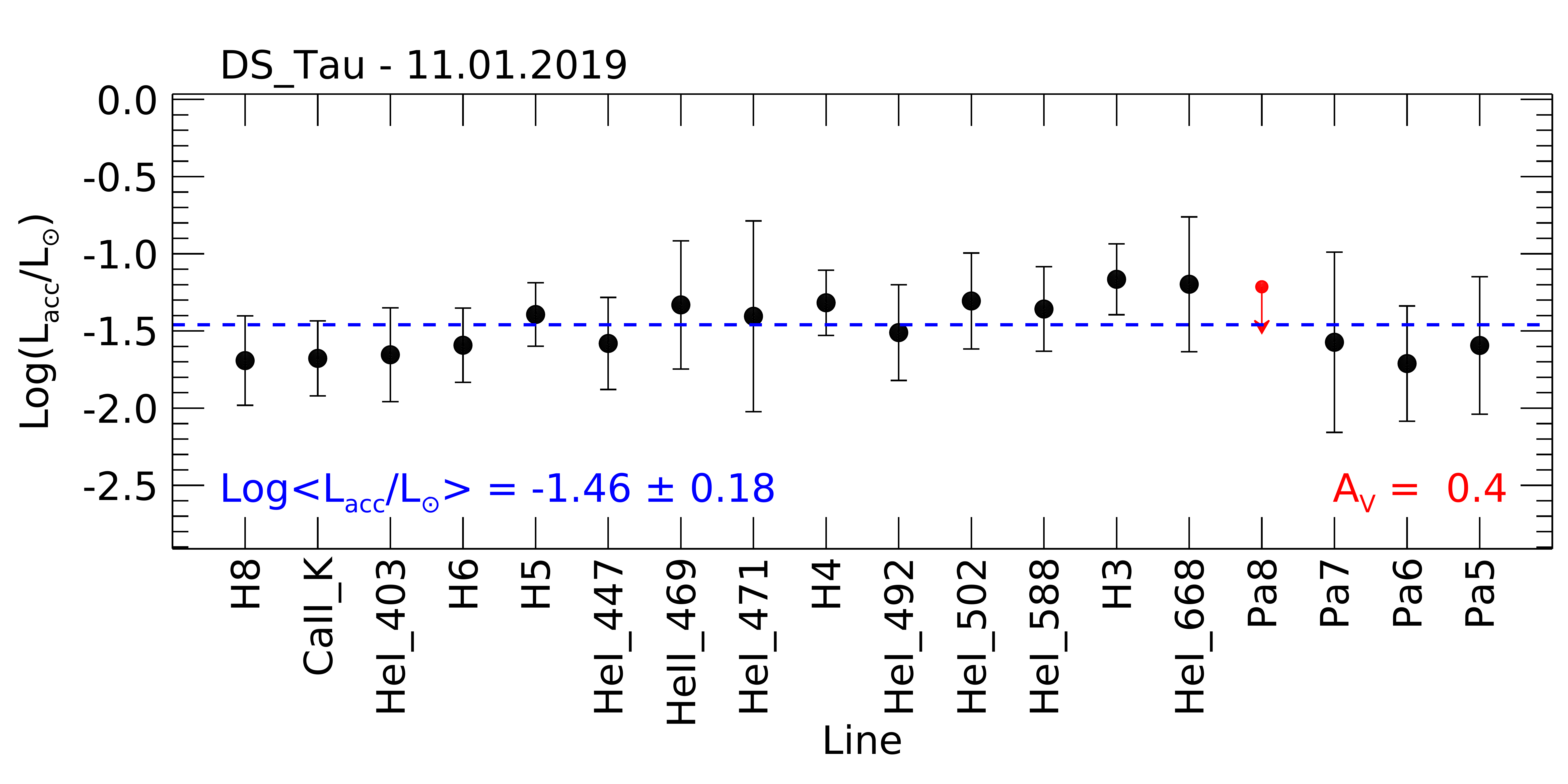}
\includegraphics[trim=0 0 0 0,width=0.50\columnwidth, angle=90]{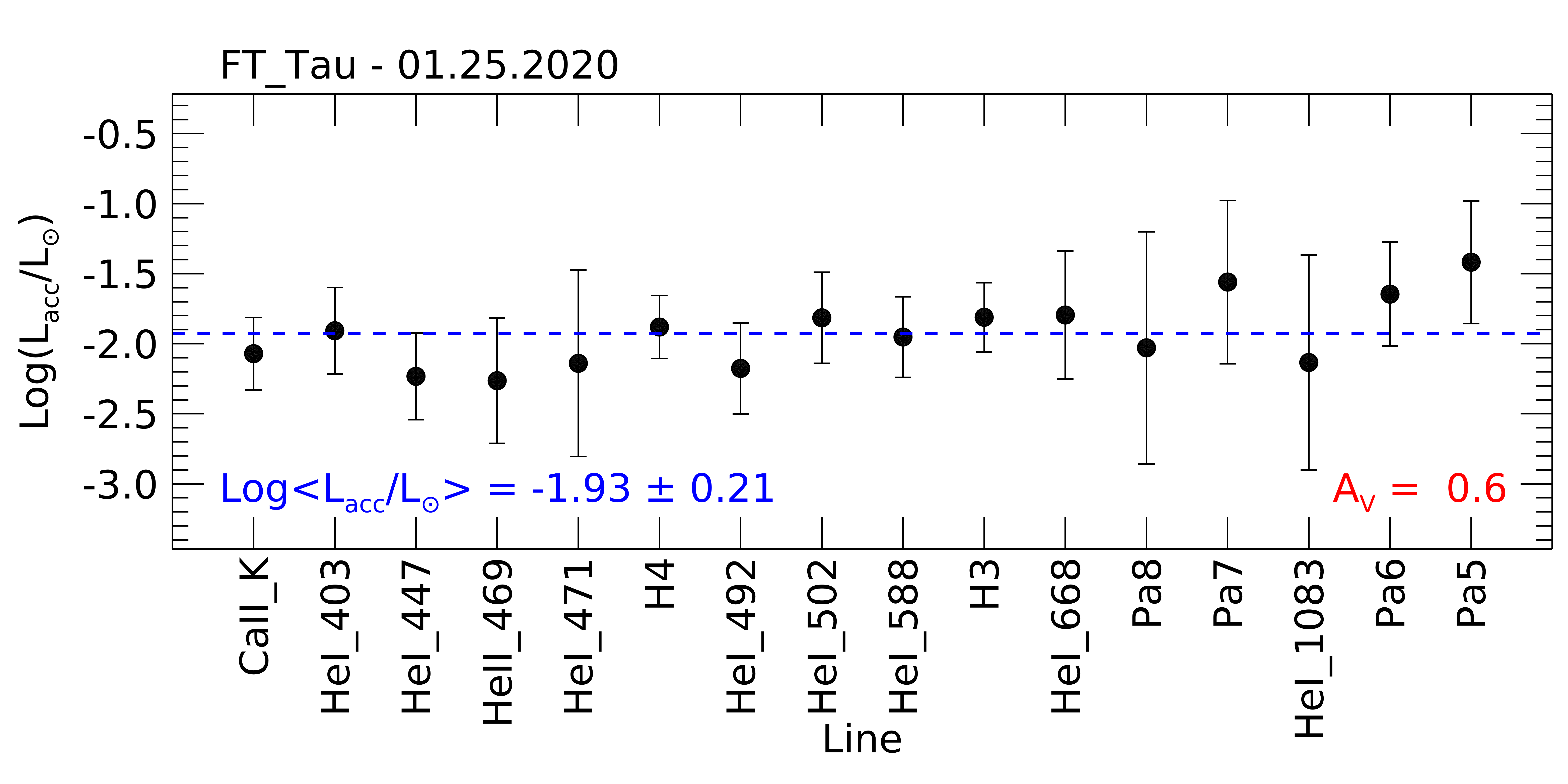}
\begin{center} \textbf{Fig. B.4.} continued.\end{center}
\end{figure*}

\begin{figure*}
\includegraphics[trim=0 0 0 0,width=1.00\columnwidth, angle=0]{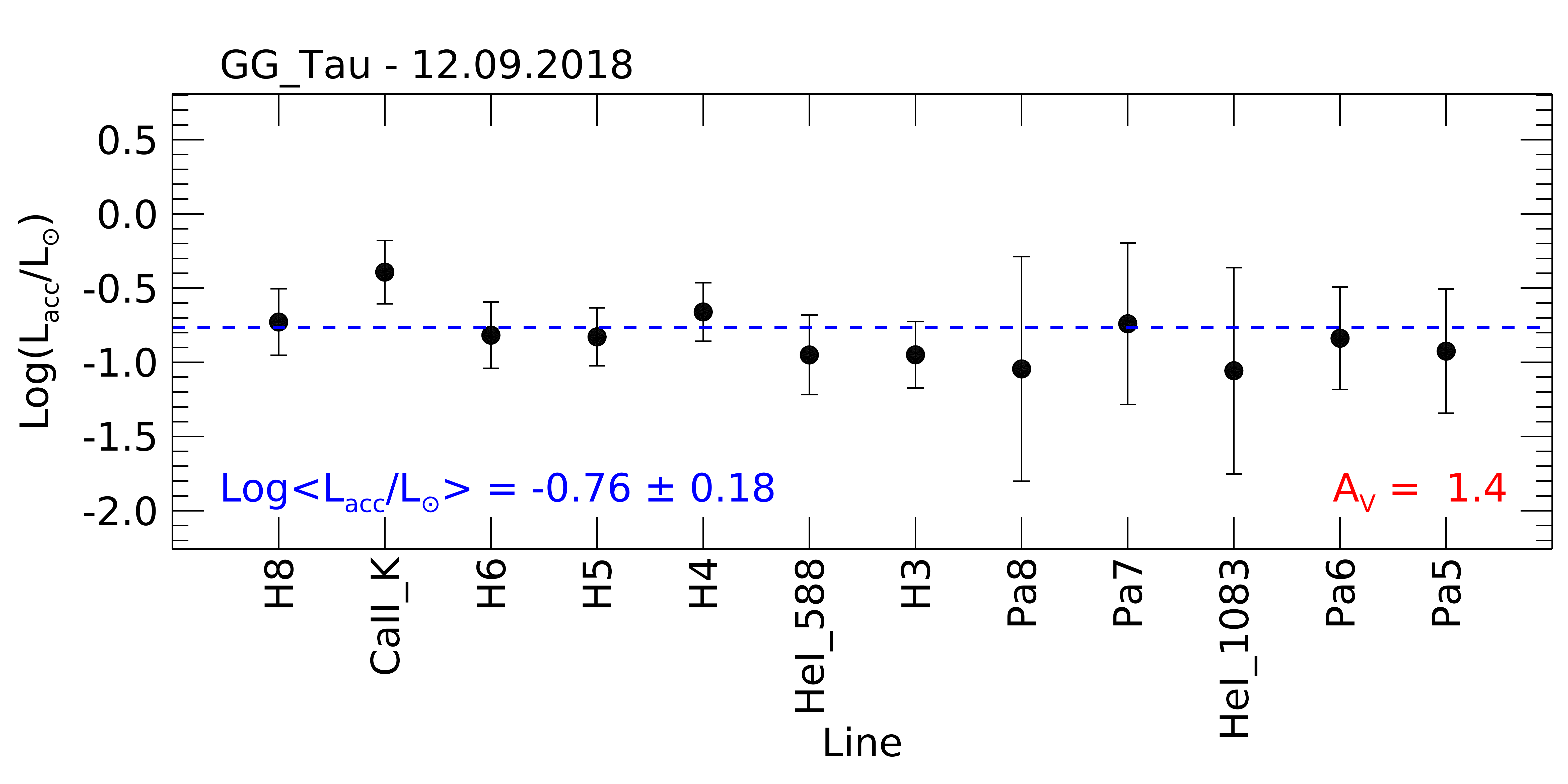}
\includegraphics[trim=0 0 0 0,width=1.00\columnwidth, angle=0]{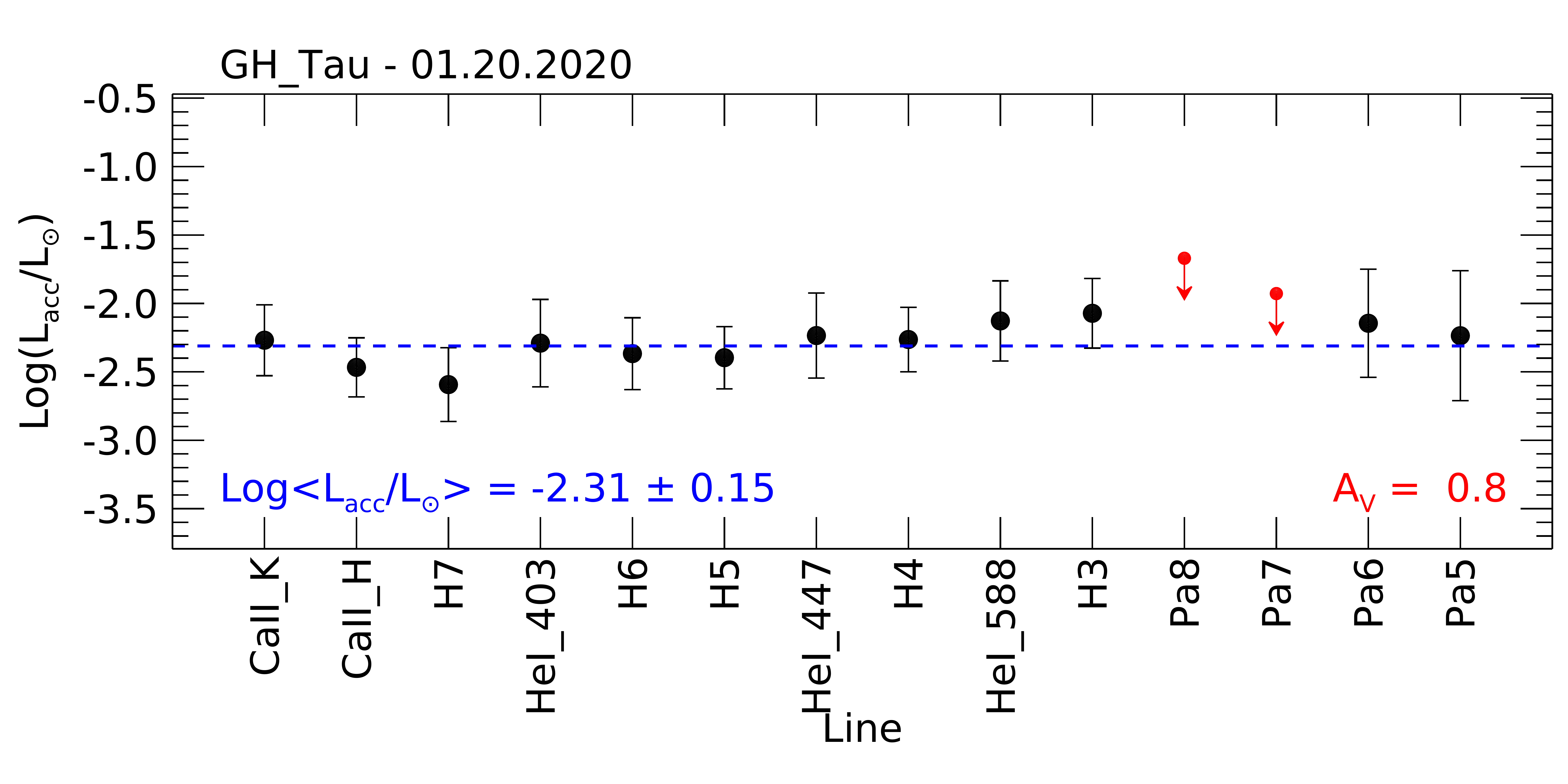}
\includegraphics[trim=0 0 0 0,width=0.50\columnwidth, angle=90]{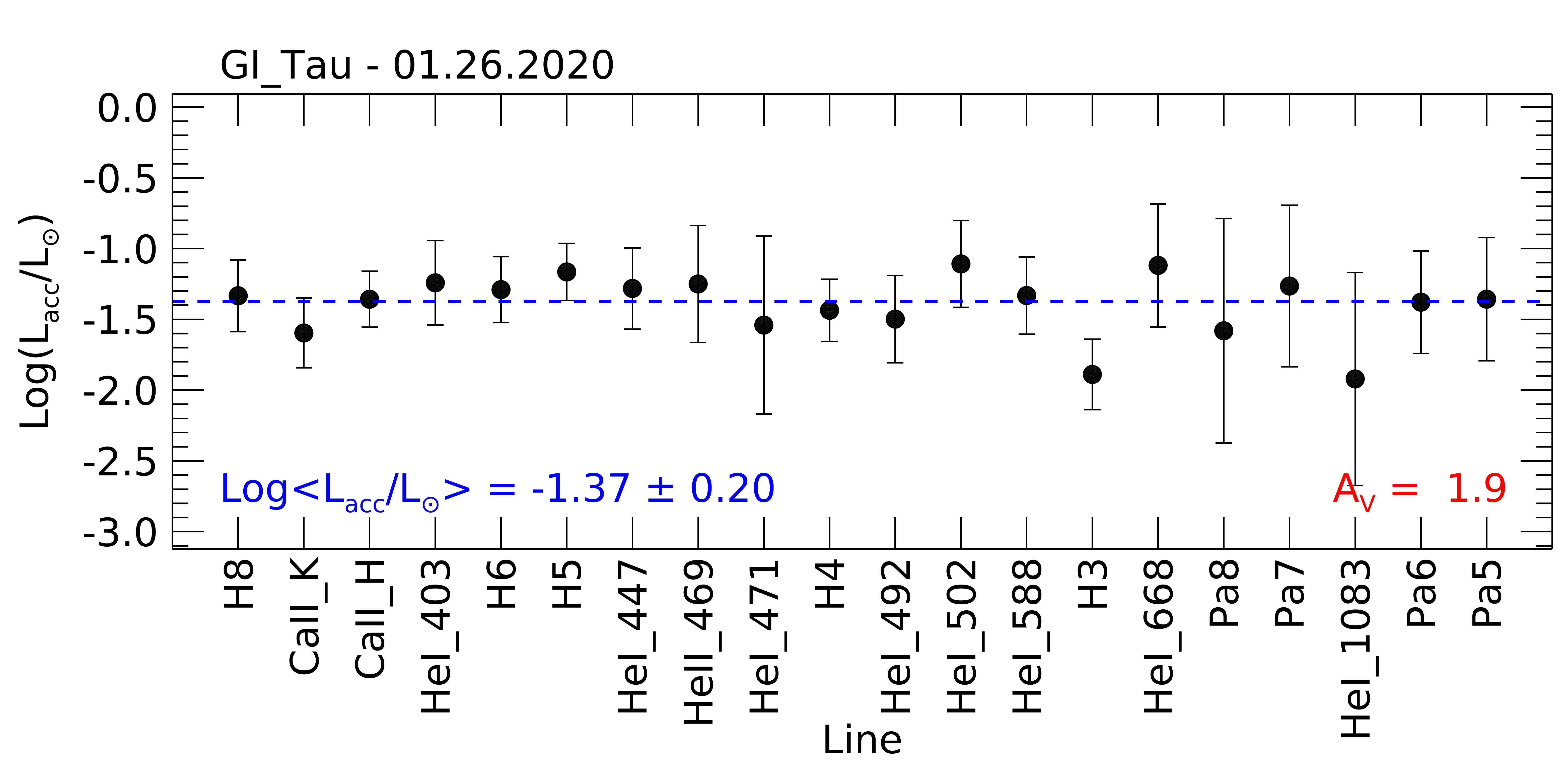}
\includegraphics[trim=0 0 0 0,width=1.00\columnwidth, angle=0]{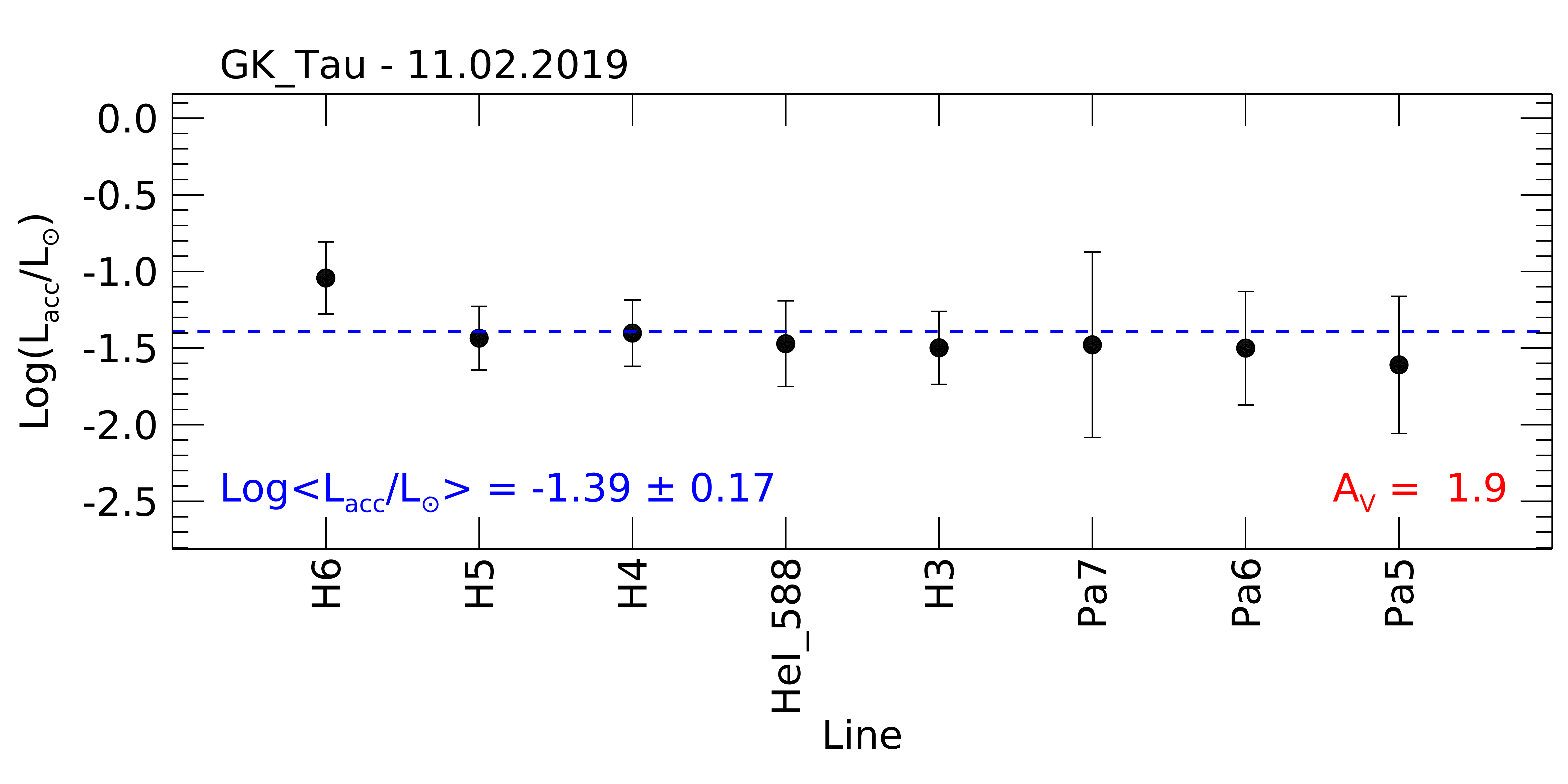}
\includegraphics[trim=0 0 0 0,width=1.00\columnwidth, angle=0]{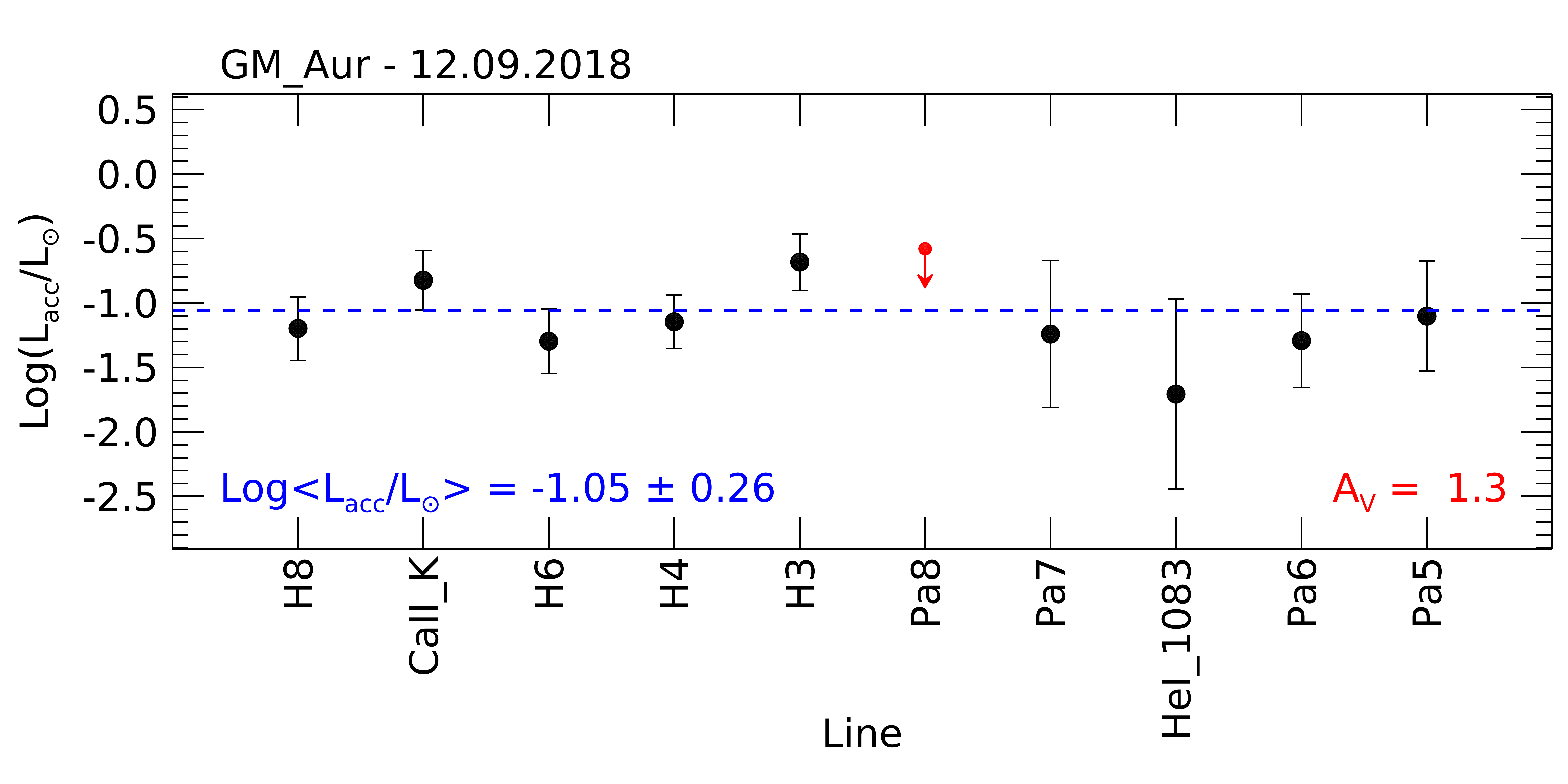}
\includegraphics[trim=0 0 0 0,width=1.00\columnwidth, angle=0]{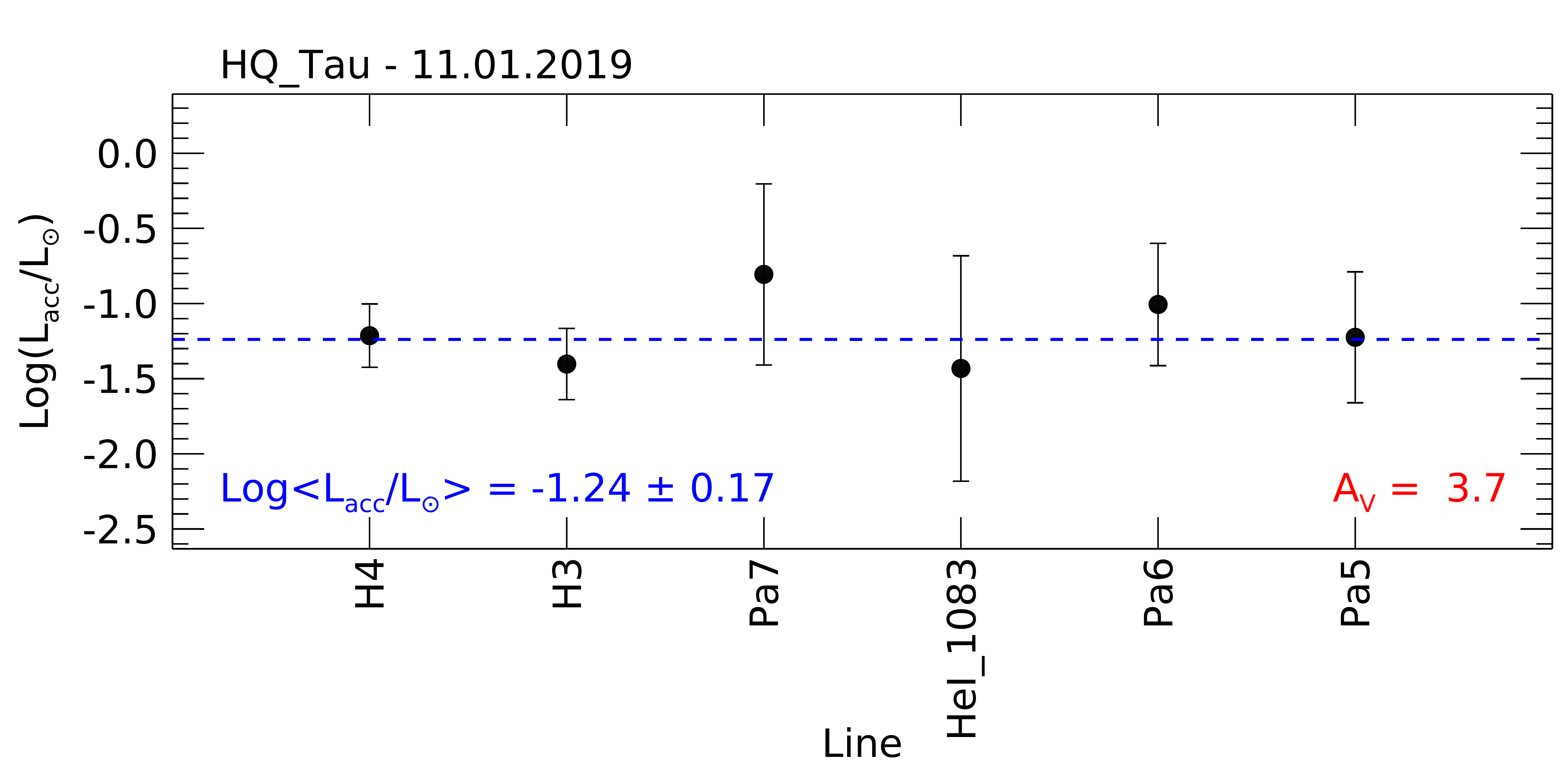}
\includegraphics[trim=0 0 0 0,width=1.00\columnwidth, angle=0]{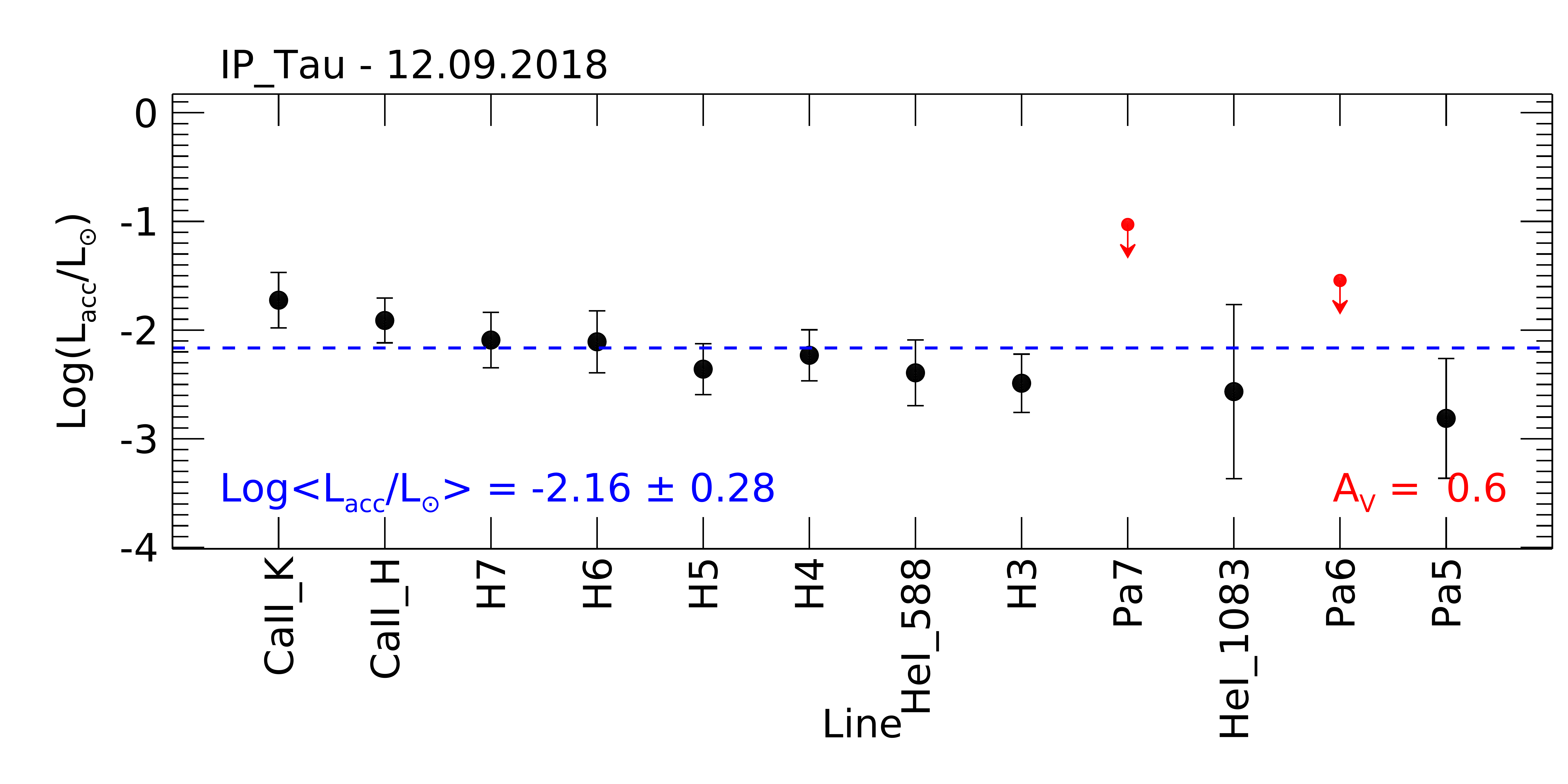}
\includegraphics[trim=0 0 0 0,width=0.50\columnwidth, angle=90]{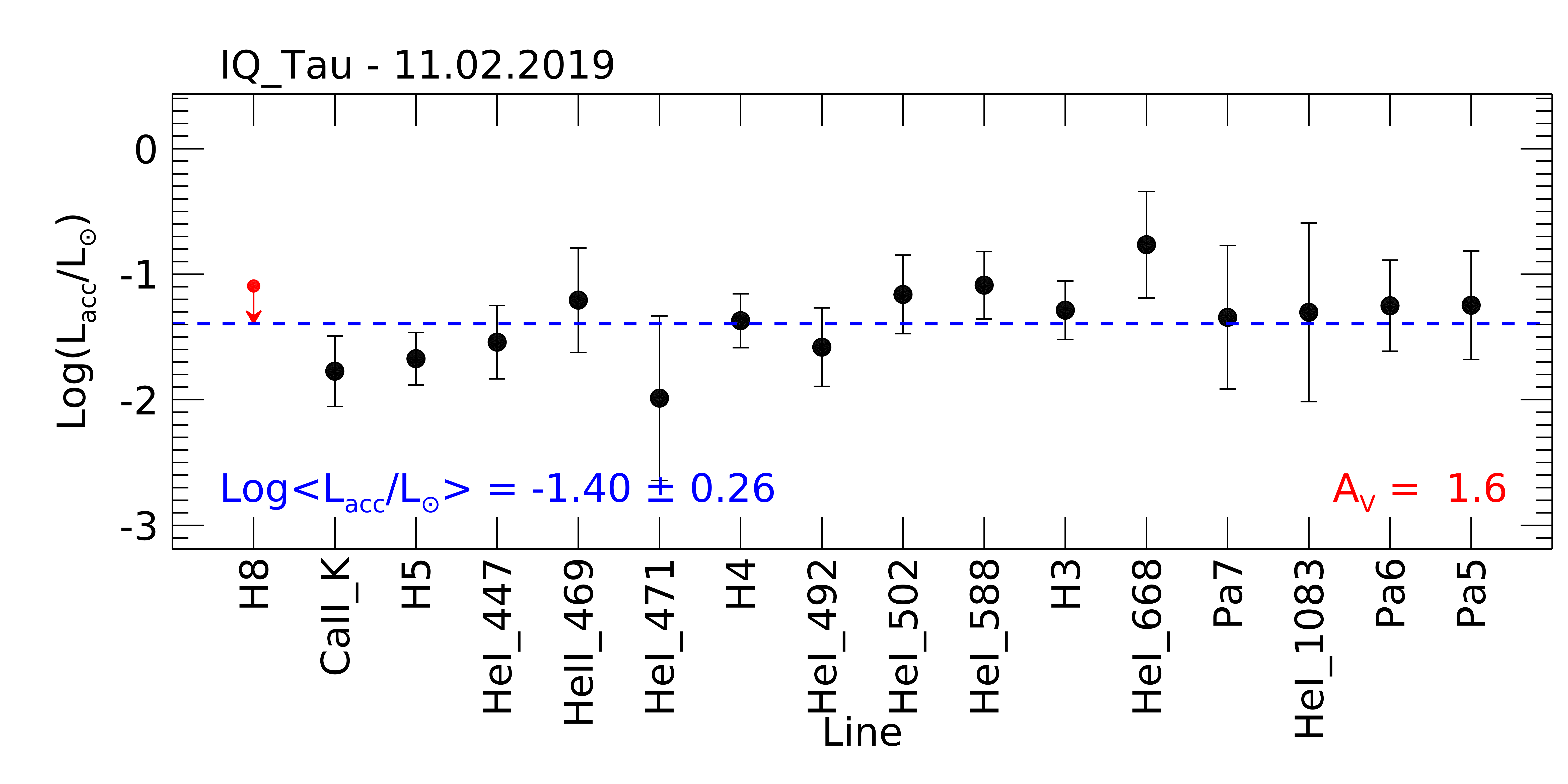}
\includegraphics[trim=0 0 0 0,width=1.00\columnwidth, angle=0]{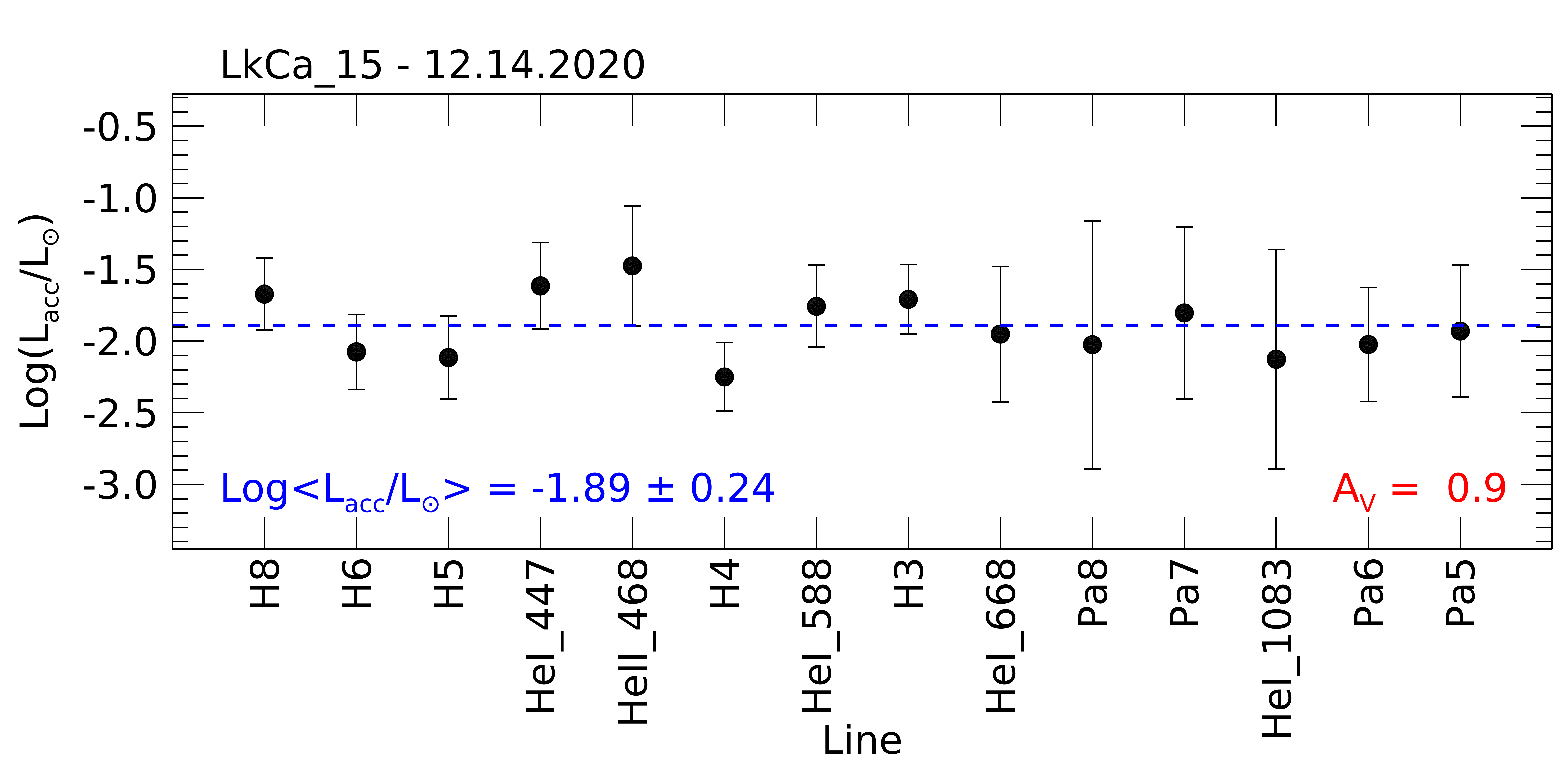}
\includegraphics[trim=0 0 0 0,width=1.00\columnwidth, angle=0]{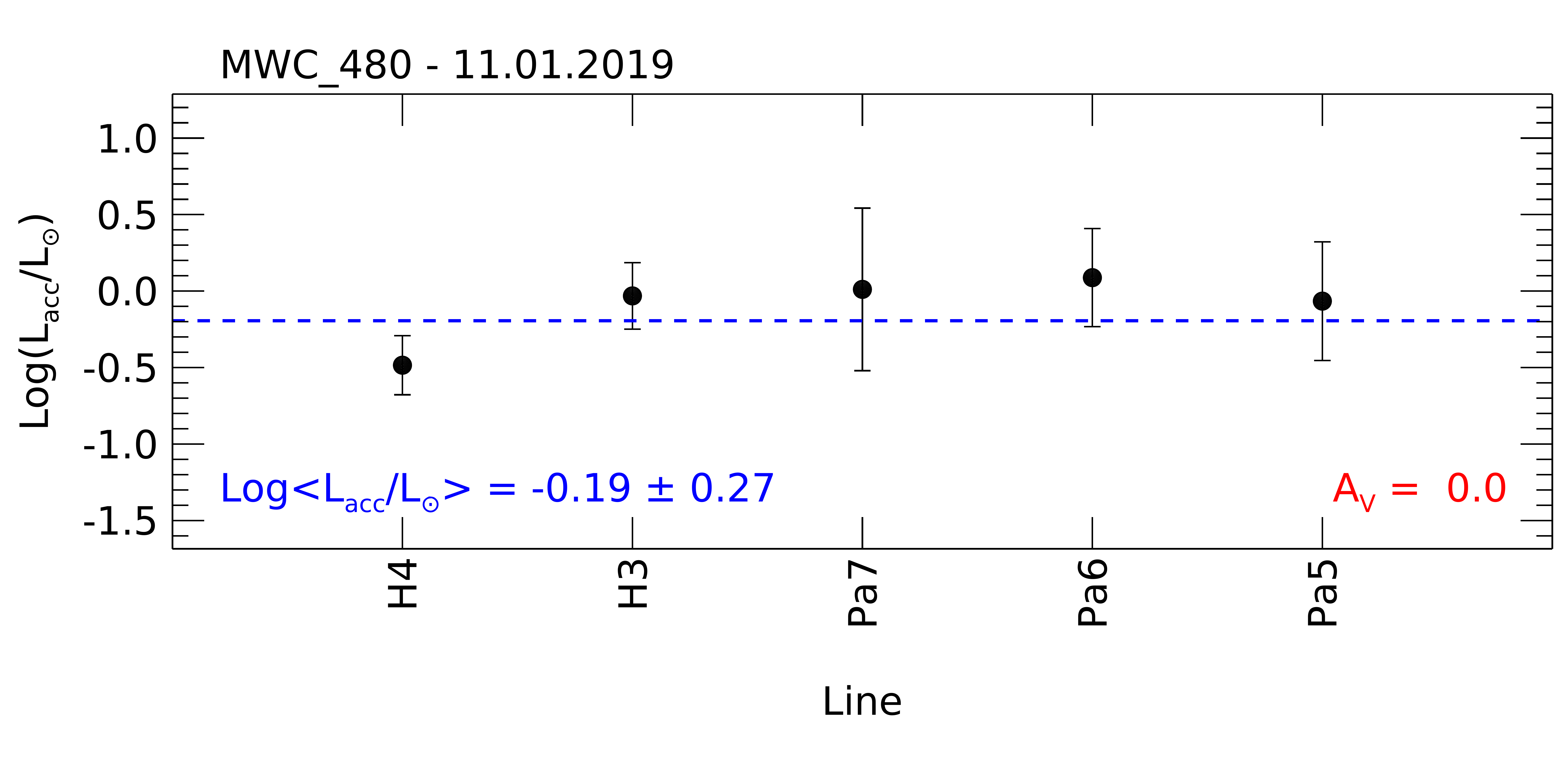}
\begin{center} \textbf{Fig. B.4.} continued.\end{center}
\end{figure*}

\begin{figure*}
\includegraphics[trim=0 0 0 0,width=1.00\columnwidth, angle=0]{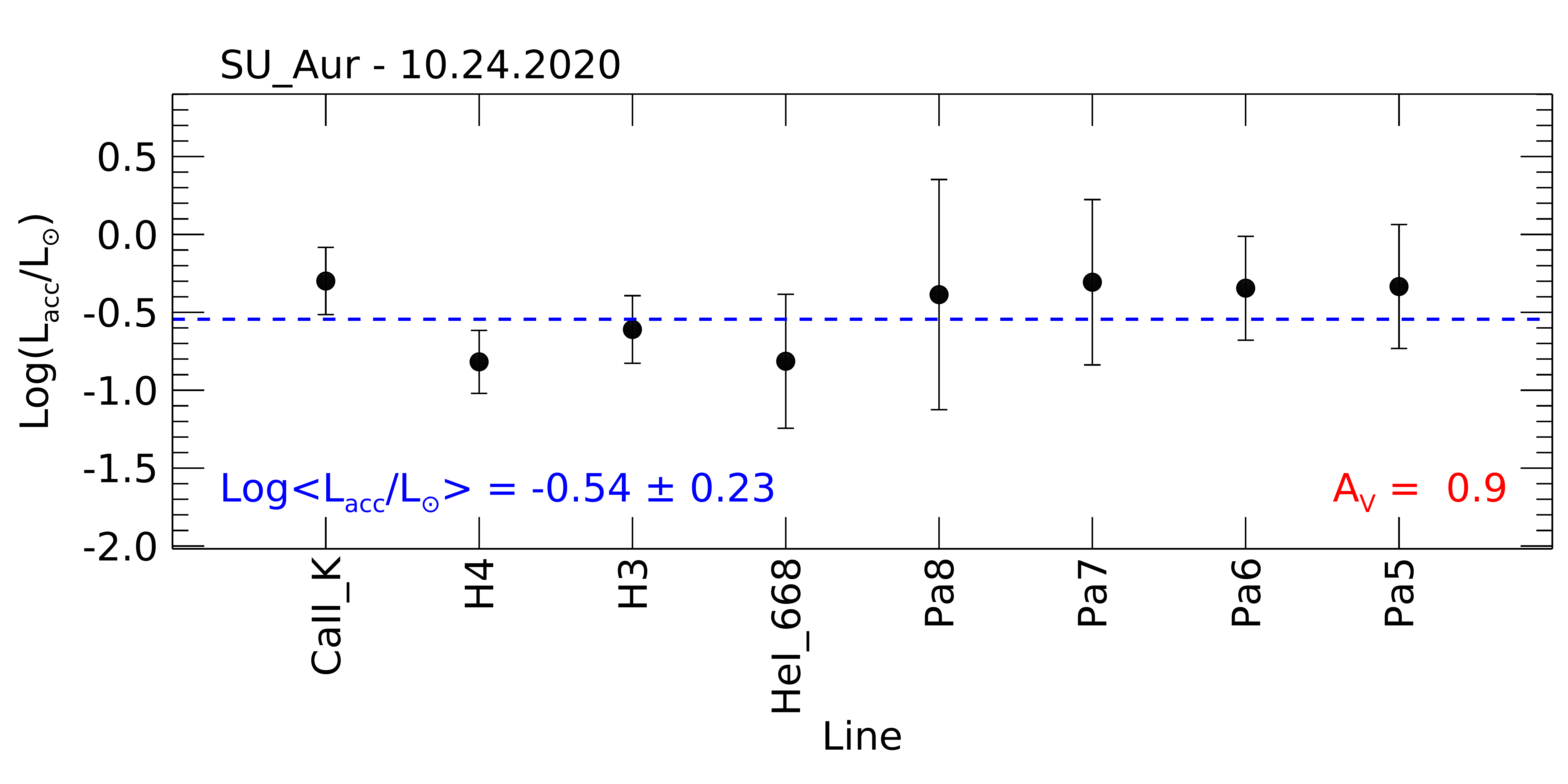}
\includegraphics[trim=0 0 0 0,width=1.00\columnwidth, angle=0]{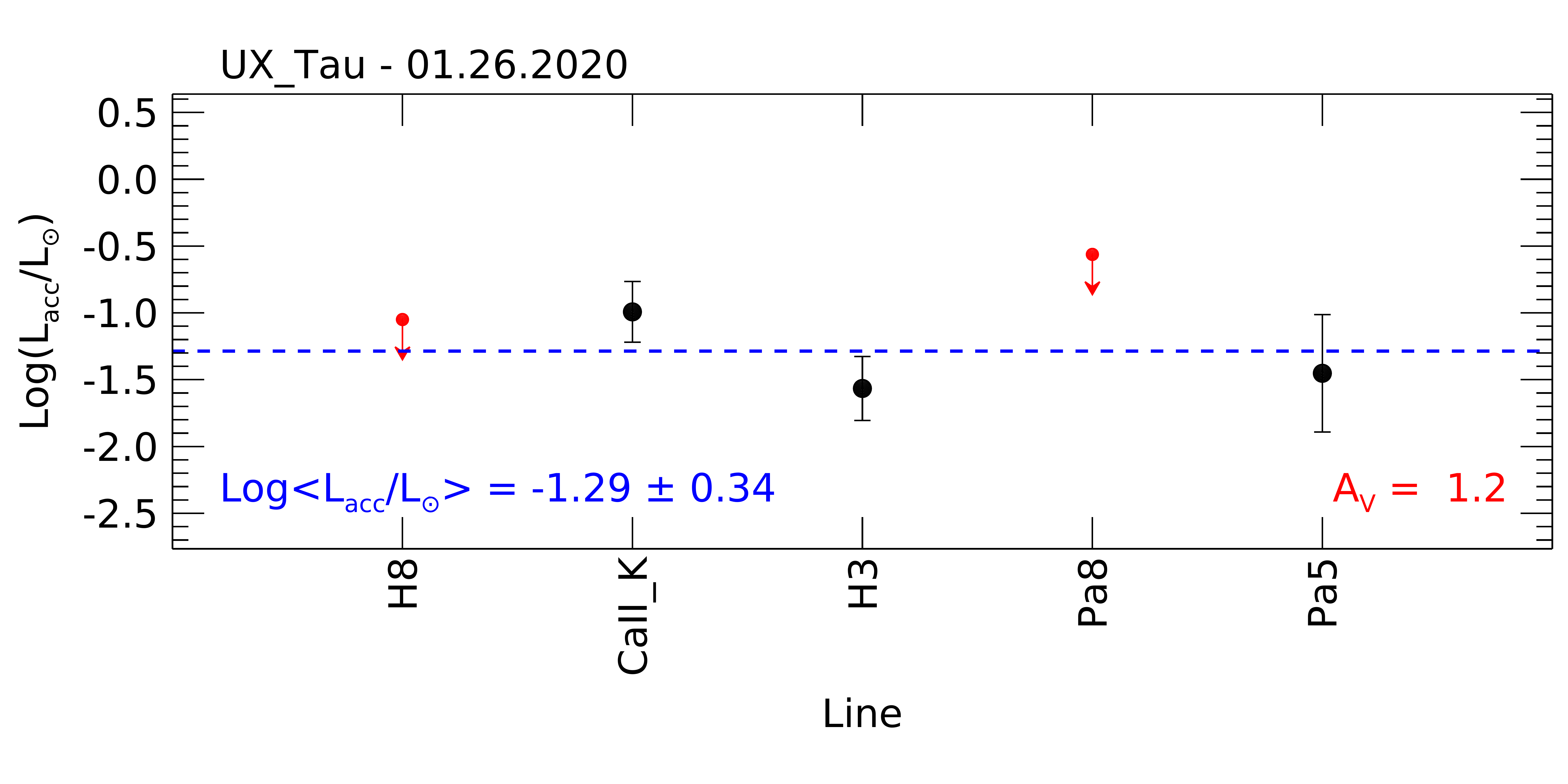}
\includegraphics[trim=0 0 0 0,width=0.50\columnwidth, angle=90]{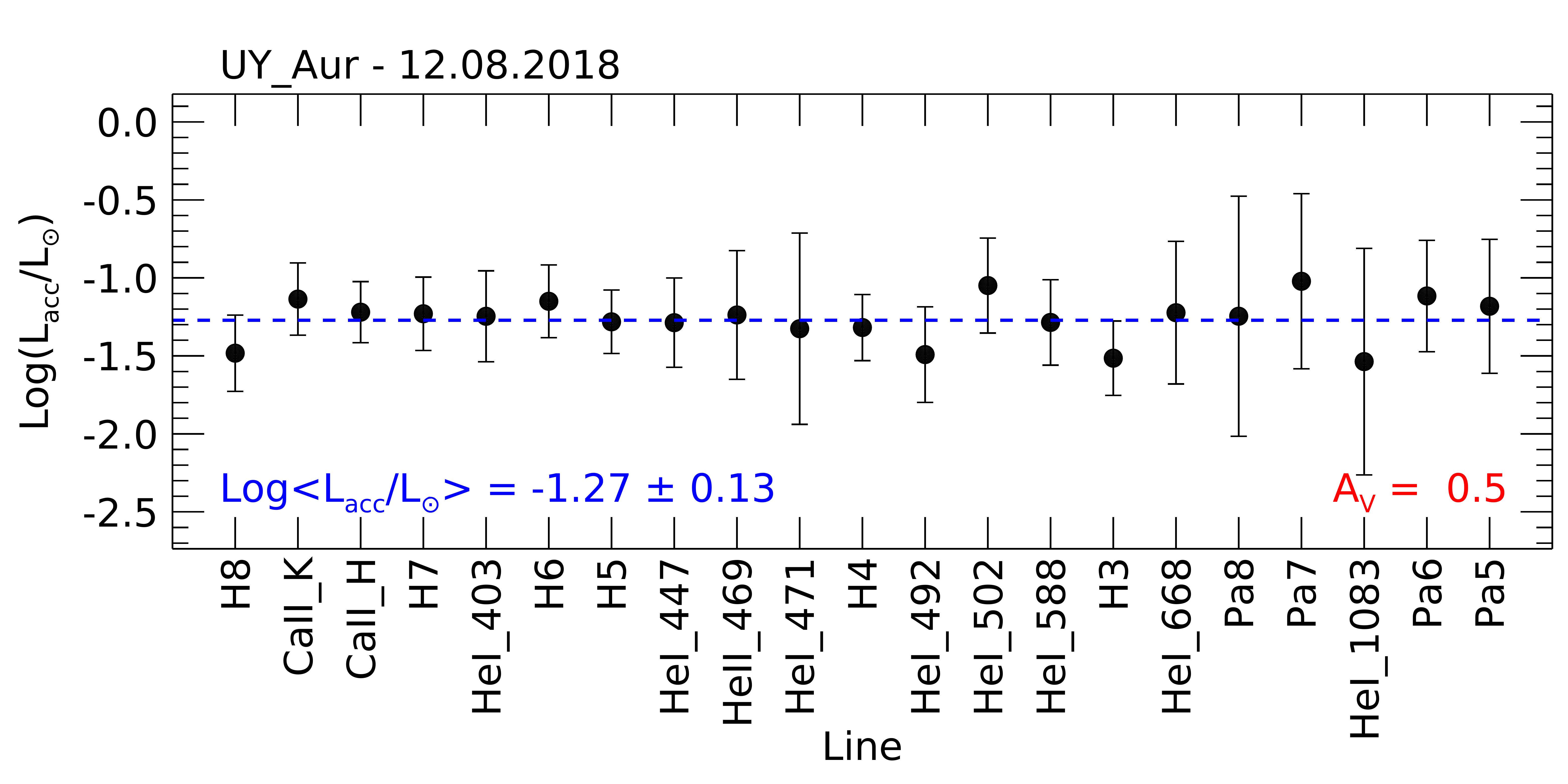}
\includegraphics[trim=0 0 0 0,width=0.50\columnwidth, angle=90]{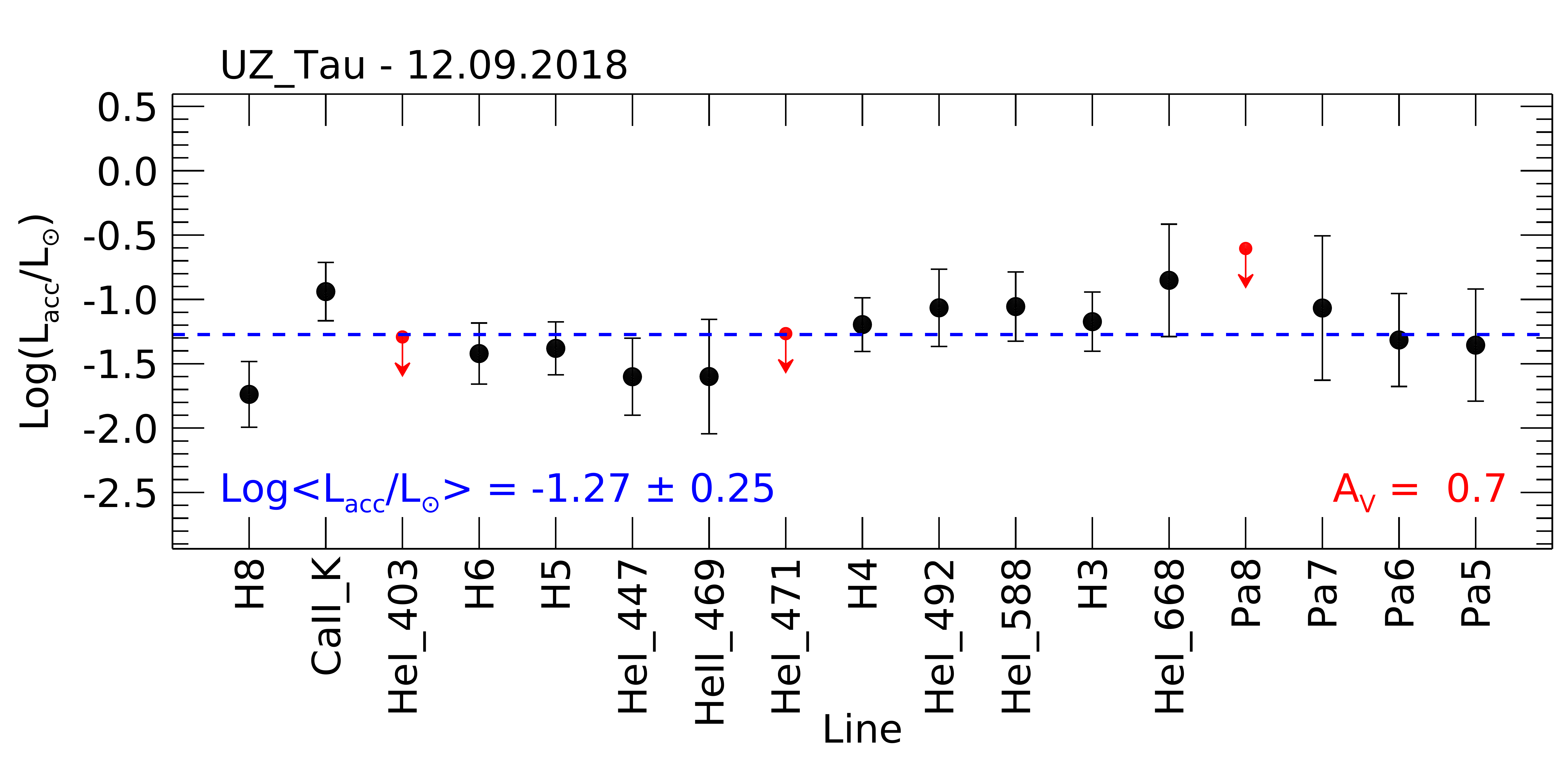}
\includegraphics[trim=0 0 0 0,width=1.00\columnwidth, angle=0]{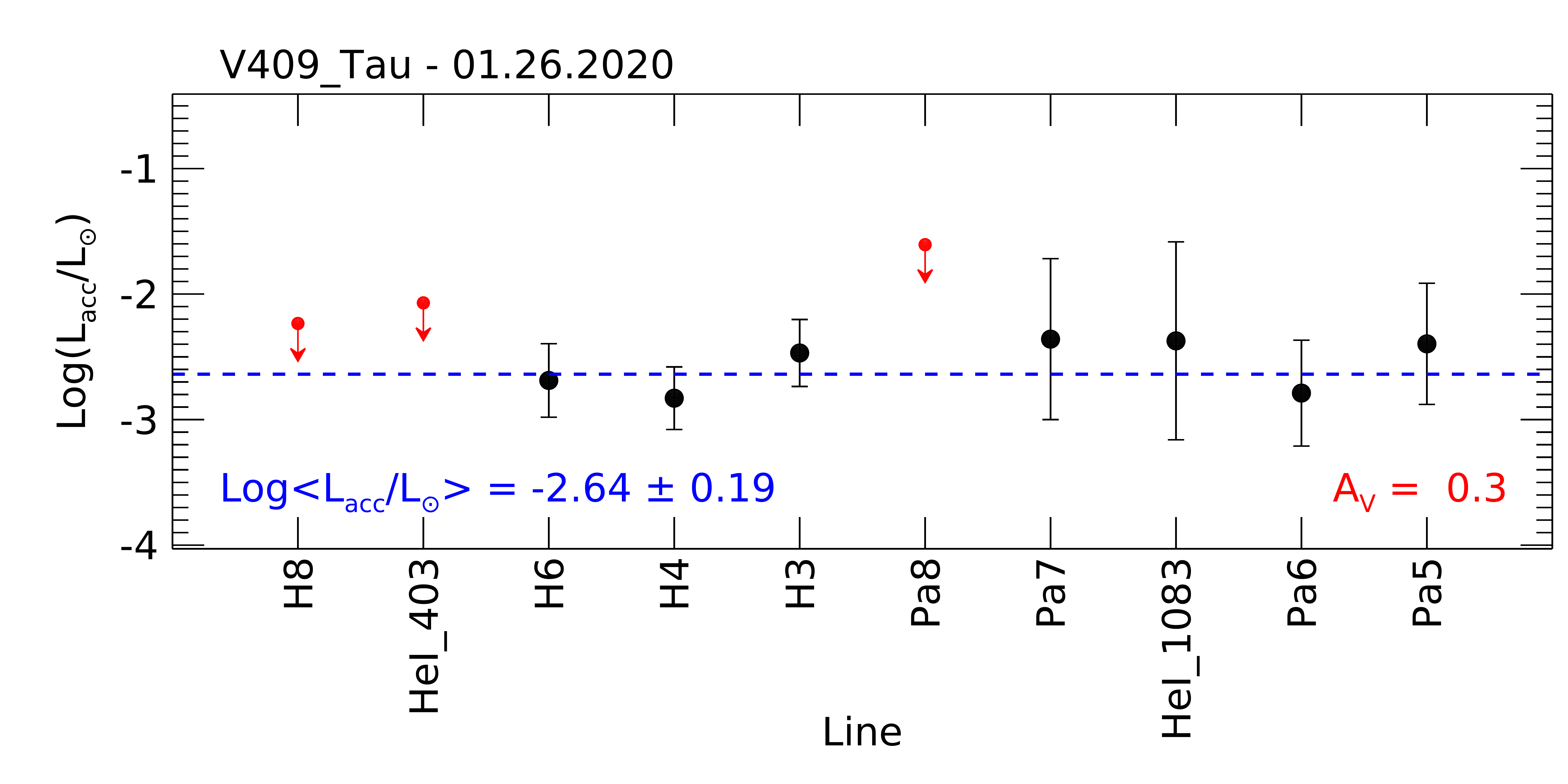}
\includegraphics[trim=0 0 0 0,width=1.00\columnwidth, angle=0]{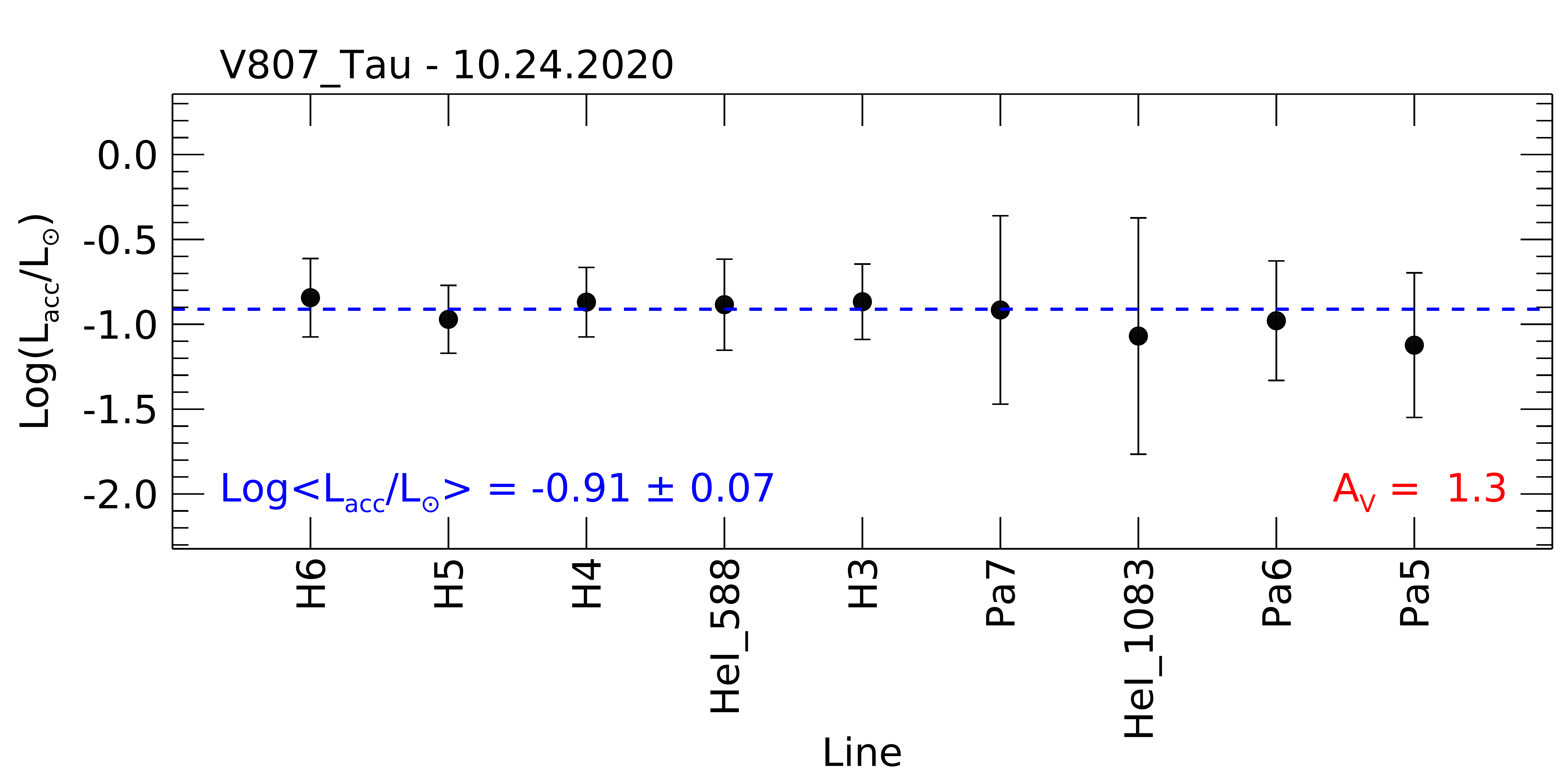}
\includegraphics[trim=0 0 0 0,width=1.00\columnwidth, angle=0]{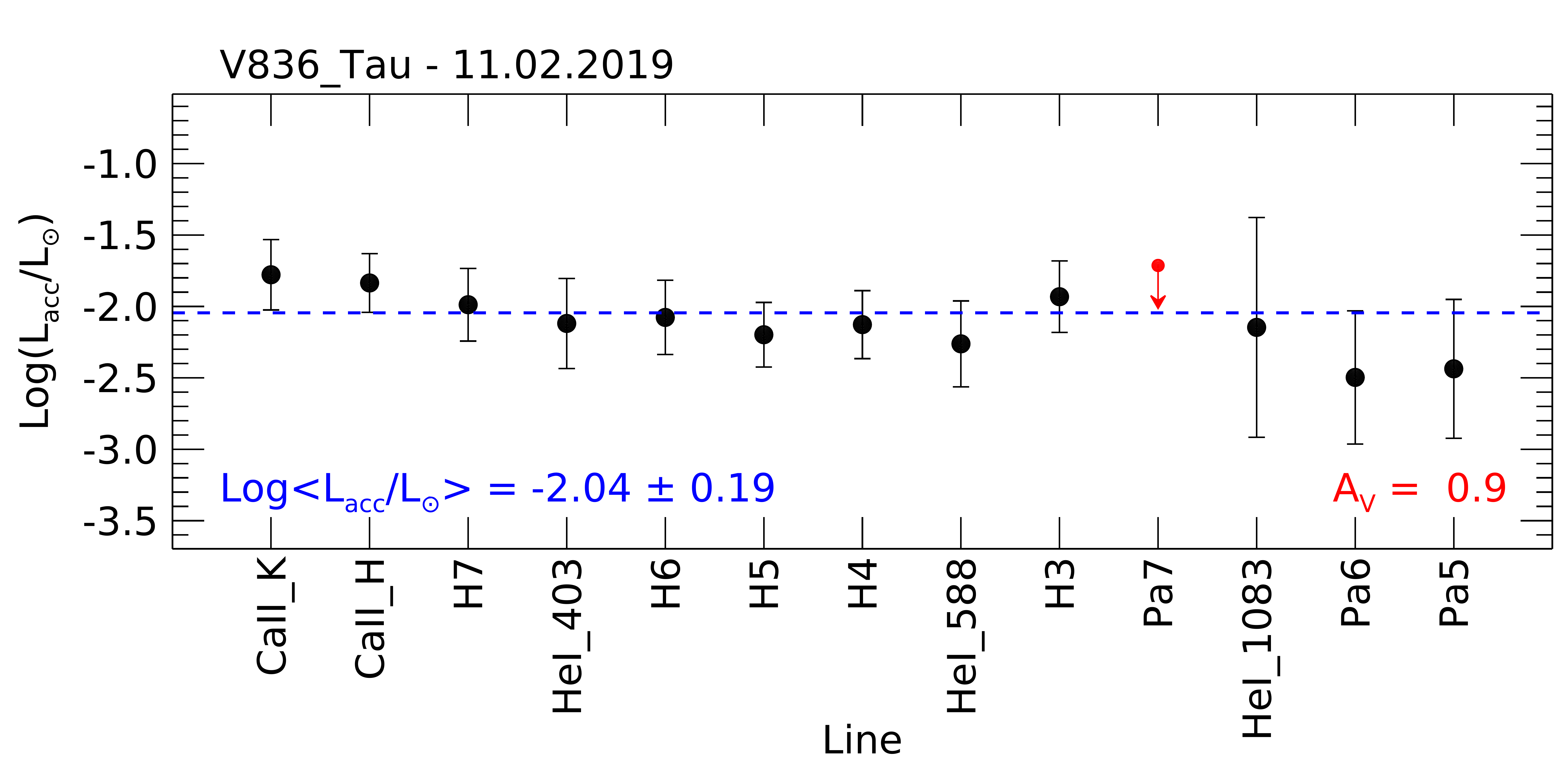}
\includegraphics[trim=0 0 0 0,width=1.00\columnwidth, angle=0]{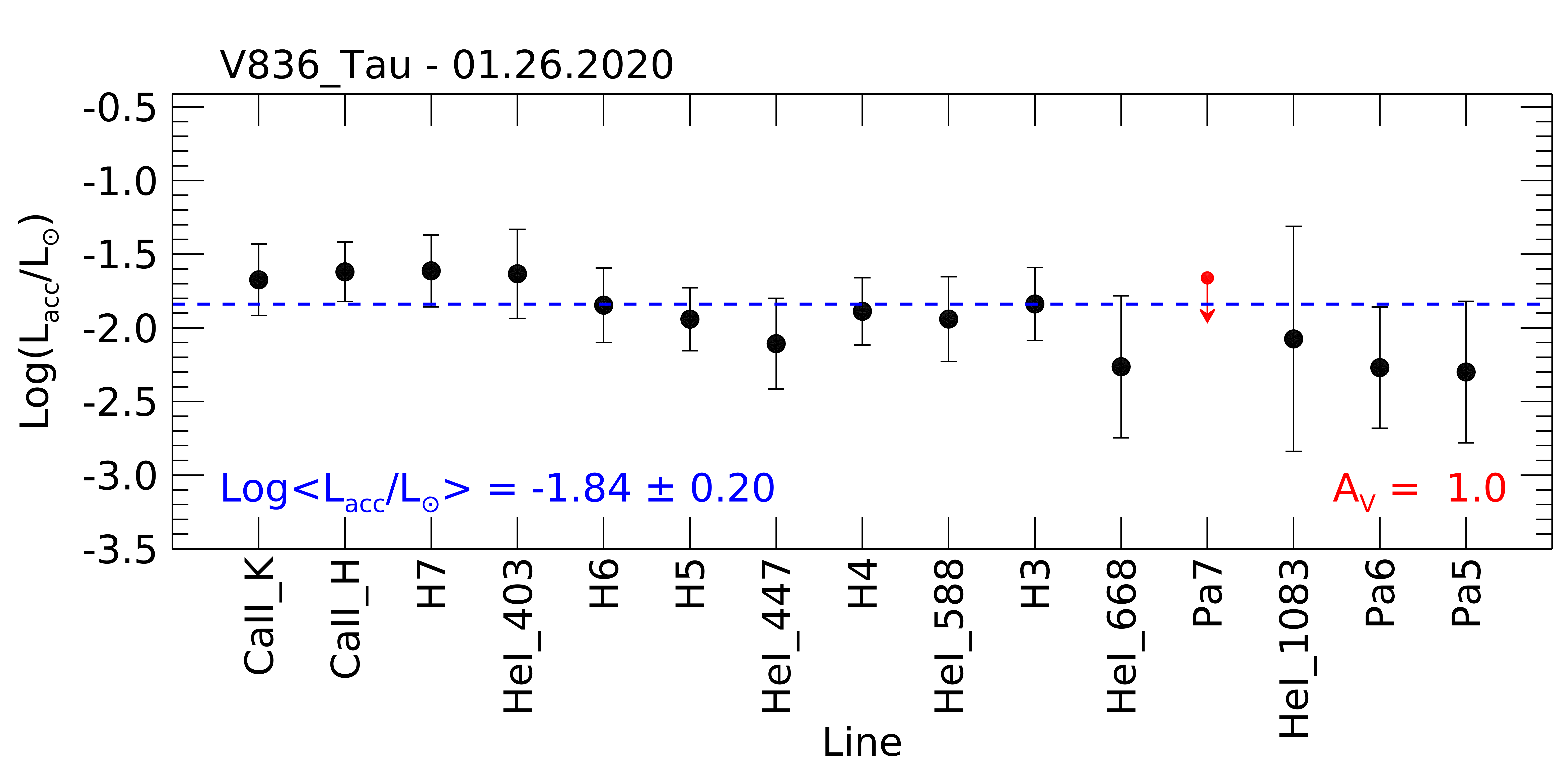}
\begin{center} \textbf{Fig. B.4.} continued.\end{center}
\end{figure*}

\clearpage

\end{appendix}

\end{document}